# On Multiple Robustness of Proximal Dynamic Treatment Regimes

Yuanshan Gao
Center for Data Science, Zhejiang University
Yang Bai
Department of Statistics and Data Science, National University of Singapore
Yifan Cui*
Center for Data Science, Zhejiang University

**Abstract**

Dynamic treatment regimes are sequential decision rules that adapt treatment according to individual time-varying characteristics and outcomes to achieve optimal effects, with applications in precision medicine, personalized recommendations, and dynamic marketing. Estimating optimal dynamic treatment regimes via sequential randomized trials might face costly and ethical hurdles, often necessitating the use of historical observational data. In this work, we utilize proximal causal inference framework for learning optimal dynamic treatment regimes when the unconfoundedness assumption fails. Our contributions are four-fold: (i) we propose three nonparametric identification methods for optimal dynamic treatment regimes; (ii) we establish the semiparametric efficiency bound for the value function of a given regime; (iii) we propose a $(K+1)$-robust method for learning optimal dynamic treatment regimes, where $K$ is the number of stages; (iv) as a by-product for marginal structural models, we establish identification and estimation of counterfactual means under a static regime. Numerical experiments validate the efficiency and multiple robustness of our proposed methods.

*The authors were supported in part by the National Key R&D Program of China (2024YFA1015600) and the National Natural Science Foundation of China (Grant No. 12471266 and U23A2064).

# 1 Introduction

In many sequential decision-making problems, actions must be chosen at a series of decision points, each informed by the evolving history of the process. Such decisions are often shaped by variability stemming from incomplete knowledge, prior experience, or the manner in which information is presented, motivating the advancement of corresponding statistical decision strategies. Dynamic Treatment Regimes (DTRs) provide a formal framework for such strategies by describing a sequence of decision rules that assign time-varying actions based on the accumulated individual history at each stage (Robins 1986, Murphy 2003, Tsiatis et al. 2019). Optimal DTRs are defined as the sequence of regimes that, if implemented for a given population, will maximize a prespecified expectation of an outcome of interest (Zhao et al. 2015, Athey & Wager 2021). Estimating such optimal regimes from historical, typically observational, data is a central statistical task, with applications ranging from precision medicine (Zhao et al. 2009, Kosorok & Laber 2019, Zhou et al. 2024) to dynamic marketing (Kopalle et al. 2023) and personalized recommendation systems (Hui et al. 2022).

The estimation for optimal DTRs has been extensively studied, with various methods being proposed. Notably, Q-learning (Murphy 2005, Goldberg & Kosorok 2012, Liu & He 2024) and A-learning (Murphy 2003, Schulte et al. 2015, Shi et al. 2018) methods, owing to their convenient modeling approaches, have found widespread application in the estimation of optimal DTRs. Nevertheless, these approaches are often prone to model misspecification, leading to the potential risk of amplifying estimation errors (Murphy 2003, Zhao et al. 2015, Tsiatis et al. 2019). Another line of work estimates optimal DTRs in a more direct way. One classical example is the dynamic regime marginal structural mean model (Robins et al. 2000, Orellana et al. 2010). However, the restrictive parametric structure of the marginal structural mean model makes its applicability highly dependent on correct model assumptions, limiting its use in broader scenarios. By contrast, value maximization/search via outcome weighted learning (Zhao et al. 2015, Zhou et al. 2017, Laha et al. 2024) directly optimizes the value function for optimal DTRs estimation, without imposing a parametric form.

To ensure the validity of the above estimation methods, several fundamental assumptions are frequently invoked for longitudinal observational data (Robins et al. 1994, Murphy 2003, Zhao et al. 2015). Among them, the sequential randomization assumption occupies a particularly pivotal position, though its plausibility is often subject to debate. This assumption states that the collected covariates are sufficiently comprehensive such that, conditional on observed history, the treatment assignment can be considered as if randomized. In practice, empirically verifying the validity of this assumption is inherently complex, as it cannot be directly confirmed on the basis of real-world data without resorting to further assumptions, which themselves may not be empirically verifiable. As a result, the possibility that time-varying unmeasured confounders remain unacknowledged must always be borne in mind.

Recently, a novel framework, known as proximal causal inference, has been proposed to address this persistent challenge (Tchetgen Tchetgen et al. 2024). Realizing the inherent difficulty in fully explaining the actual confounding mechanisms merely from observed covariates, this framework instead relies on the expectation that such measured variables can serve as at best proxies for unmeasured confounders. By leveraging two distinct types of proxy variables and confounding bridge functions (Miao et al. 2018, Cui et al. 2024), proximal causal inference achieves identification of causal effect under more general assumptions by explicitly admitting the existence of unmeasured confounders. It lends itself naturally to a variety of contexts, including mediation analysis (Dukes et al. 2023, Bai et al. 2025), complex longitudinal causal inference (Ying et al. 2023, Tchetgen Tchetgen et al. 2024), survival analysis (Ying et al. 2022, Ying 2024), and reinforcement learning (Shi et al. 2022, Bennett & Kallus 2024, Zhang & Tchetgen Tchetgen 2024).

Within the context of DTRs, Qi et al. (2024) and Shen & Cui (2023) have resolved the single-stage optimal treatment regime estimation problem under unmeasured confounding by leveraging proxy variables. For the multi-stage treatment regimes, a series of methodological advances have been developed under the framework of proximal causal inference within the setting of confounded Partially Observed Markov Decision Processes (POMDP). These include contributions by Shi et al. (2022), Bennett & Kallus (2024), and Li et al. (2025), which have extended the approach to accommodate finite- and infinite-horizon cases as well as discrete and continuous action spaces. The doubly robust and efficient estimators have also been established. However, these studies largely remain confined to the POMDP framework, which still essentially relies on the Markov assumption. In more intricate scenarios characterized by temporal dependencies among variables, where the Markov assumption may be violated, the problem of estimating optimal DTRs remains to be fully explored.

To mitigate this theoretical gap, Zhang & Tchetgen Tchetgen (2024) employ the longitudinal proximal causal inference method (Tchetgen Tchetgen et al. 2024, Ying et al. 2023) to identify and estimate optimal DTRs in scenarios accompanied by time-varying unmeasured confounders, utilizing proxy variables in conjunction with bridge functions. Specifically, they propose two distinct strategies for identifying optimal DTRs using the Q-learning approach and establish several corresponding estimation results within the categorical setting. However, implementing their methodology requires solving multiple nested Fredholm integral equations of the first kind, a class of problems that are widely recognized for their inherent ill-posedness and instability (Tikhonov et al. 1995). In particular, Q-learning, as an indirect approach we have discussed above, can exacerbate the accumulation of errors. These factors render the method considerably fragile in practice. Notably, they also do not provide a theoretical justification for the asymptotic properties of the proposed estimators. Therefore, the development of more robust and efficient methods, as well as the related theoretical assurance, for the estimation of optimal DTRs within the longitudinal proximal causal inference framework is of particular interest.

In this article, we develop several novel identification methods for optimal DTRs in scenarios where the existence of unmeasured confounders cannot be ruled out via the proximal causal inference framework. Furthermore, we establish the semiparametric efficiency bound for the value function associated with a sequence of arbitrary given regimes. Building on these results, we derive $(K+1)$-robust and efficient methods for learning optimal DTRs, together with the corresponding asymptotic analysis theory. As a by-product of our results for marginal structural models, the identification and estimation of counterfactual means under a static regime are also established. The theoretical contributions of the proposed methods are described in detail as follows.

First, we propose three nonparametric identification methods for optimal DTRs, two of which are generalizations of existing methods (Zhang & Tchetgen Tchetgen 2024), and a new one aims at reducing error amplification. As highlighted by Zhang & Tchetgen Tchetgen (2024), the nested structure of the integral equation systems employed in the definition of bridge functions will amplify statistical noise, which may give rise to an estimation procedure that is both cumbersome and highly unreliable. To alleviate this critical challenge, we propose a new proximal hybrid augmentation method that enables identification and estimation by only two subsets of bridge functions. Crucially, this approach apportions the complexity associated with solving nested integral equations between the two types of bridge functions, resulting in a marked improvement in both the robustness and flexibility of the identification and estimation processes.

Second, we provide the semiparametric efficiency bound and efficient influence function of the value function for a sequence of arbitrary given DTRs under a semiparametric model for the observed data distribution. Illuminated by the efficient influence function, we develop the multiply robust estimator for the value function with given DTRs. For a general $K$-stage setting, these estimators are consistent as long as one of the prespecified $(K+1)$ subsets of the bridge functions with size $K$ has correctly specified models. Parallels can be drawn between the proposed multiply robust estimator and that of the counterfactual mean introduced by Molina et al. (2017) as well as the doubly robust estimator of optimal DTRs proposed by Zhang et al. (2013) under the sequential randomization assumption.

Third, we prove the asymptotic properties of the optimal DTRs estimated by multiply robust value maximization. For occasions focusing on interpretability, we establish the asymptotic theory for the convergence of linear optimal DTRs estimated through the proposed efficient and multiply robust value estimator, in conjunction with a cross-fitting procedure (Chernozhukov et al. 2018, Wager 2024). Under some mild conditions, we prove that the regime parameters converge at a $n^{1/3}$-rate. However, the corresponding estimated optimal value and regret are proved to converge at a $\sqrt{n}$-rate and a rate faster than $\sqrt{n}$, respectively. The asymptotic normal distribution of the estimated optimal value is also provided.

The remainder of this article is organized as follows. In Section 2, we introduce the notations, outline the fundamental framework of longitudinal proximal causal inference, and depict the structure of optimal DTRs identification. In Section 3, we first present two identification methods that generalize Theorems 1 and 2 in Zhang & Tchetgen Tchetgen (2024), respectively. Subsequently, we establish a novel hybrid identification method, which unifies the previously introduced approaches as special cases. In Section 4, we start by developing the semiparametric inference theory for the value function. The $(K+1)$-fold multiply robust estimator for the value function is subsequently induced by the established efficient influence function. In Section 5, we present asymptotic theoretical results concerning the estimation of optimal DTRs through value maximization. In Section 6, we report the result of numerical experiments aiming at verifying the robustness and efficiency of the proposed estimators for the optimal DTRs using value maximization. The replication codes are available on GitHub. Lastly, in Section 7, we conclude the paper with a discussion of the future applications for the proposed methodologies. For the parallel results for optimal DTRs estimated by Q-learning, proofs of the main theorems, and additional numerical results, we refer to the Supplementary Material.

# 2 Preliminaries

## 2.1 Notations

We consider the problem of optimal dynamic treatment regimes estimation in the presence of time-varying unmeasured confounders under the proximal causal inference framework.

Suppose that we have $n$ independent and identically distributed (i.i.d.) samples drawn from the joint distribution of the variable sequence $\left(Y_0, (U_{k-1}, L_k, A_k, Y_k)_{k=1}^K\right)$ over stages $k = 1, 2, \ldots, K$. Specifically, $Y_0$ represents the baseline measurement of the outcome, observed before any treatment and measured confounders, and may potentially confound the effect of subsequent treatments by influencing both treatment assignments and future outcomes. At the $k^{th}$ stage, the variables proceed as follows: unmeasured confounders $U_{k-1}$ relevant to the $k^{th}$ stage are present, then measured confounders $L_k = (Z_k, W_k)$ are observed, a binary treatment $A_k \in \mathcal{A}_k = \{0, 1\}$ is assigned, and finally, the outcome $Y_k$ is measured. Without loss of generality, we suppose that $L_k$ can all serve as valid proxies. Here, $Z_k$ and $W_k$ are the treatment- and outcome-inducing proxies, respectively, whose causal structure will be introduced later in Section 2.2.

For a variable $X$ (where $X$ can be $A, Z, W, U$ or $Y$), we use $\bar{X}_k$ to denote the variable sequence from the initial stage up to the $k^{th}$ stage. Specifically, for $X = A, Z, W$, $\bar{X}_k = (X_1, X_2, \ldots, X_k)$ with $k = 1, 2, \ldots, K$; for $X = U$, $\bar{U}_{k-1} = (U_0, U_1, \ldots, U_{k-1})$ with $k = 1, 2 \ldots, K$; and for $X = Y$, $\bar{Y}_k = (Y_0, \ldots, Y_k)$ with $k = 0, 1, 2, \ldots, K$. To ensure clarity, we define $A_0, Z_0, W_0, W_{K+1}, U_{-1}, Y_{-1} = \emptyset$. We follow the convention that random variables and their realizations are correspondingly denoted with uppercase and lowercase letters, respectively. $Y_k(\bar{a}_k)$ is defined as the potential outcome that would be observed at stage $k$ if, possibly contrary to the fact, the treatment sequence were assigned as $\bar{a}_k$. For notational simplicity, we use the summation symbol "$\sum$" to represent both summation over discrete variables and integration over continuous variables. For probability density or mass functions, we use $f(x)$ instead of $f(X = x)$ to denote the value of the probability density or mass function of a random variable $X$ at realization $x$.

The regime function at the $k^{th}$ stage is denoted by $d_k(\bar{y}_{k-1}, \bar{a}_{k-1})$, which maps the accumulated history of past outcomes $\bar{y}_{k-1}$ and treatments $\bar{a}_{k-1}$ to a treatment decision in $\mathcal{A}_k$. We also denote $\mathcal{D}_k$ as the $k^{th}$ corresponding function class of candidate regimes for $d_k$, and $\mathcal{D} = (\mathcal{D}_1, \ldots, \mathcal{D}_K)$ as the set of all candidate regime sequences. Let $\mathbb{E}$ and $\mathbb{P}_n$ respectively denote the expectation and empirical expectation operators. Using $\bar{d}_K$ to denote the regime sequence $(d_1, d_2, \ldots, d_K)$, our goal is to estimate the optimal DTRs $\bar{d}_K^* = (d_1^*, d_2^*, \ldots, d_K^*)$ that maximize $\mathbb{E}[Y_K(\bar{d}_K)]$ over $\mathcal{D}$, where $Y_K(\bar{d}_K)$ is the potential outcome under the regime assignment $(A_1 = d_1, \ldots, A_K = d_K)$. Take a 2-stage setting as an example, for the individual $i$ who follows the treatment assigned by

$(d_1, d_2)$, $Y_{1i}(d_1)$ and $Y_{2i}(\bar{d}_2)$ are defined as:

$$Y_{1i}(d_1) = \sum_{a_1} Y_{1i}(a_1) I\Big(d_1(Y_{0i}) = a_1\Big),$$

$$Y_{2i}(\bar{d}_2) = \sum_{\bar{a}_2} Y_{2i}(\bar{a}_2) I\Big(d_1(Y_{0i}) = a_1\Big) I\Big(d_2\Big(Y_{0i}, Y_{1i}(a_1), a_1\Big) = a_2\Big).$$

We will focus on the case where each outcome $Y_k, k = 0, 1, \ldots, K$ is univariate; however, the extension to multivariate outcomes has no essentially theoretical obstacles, other than a slight modification to the assumptions introduced later.

## 2.2 Longitudinal Proximal Causal Inference

In the complex longitudinal study with time-varying treatments and confounders, the following sequential randomization assumption is common in the literature (Robins et al. 1994, Robins 1997, Hernán & Robins 2025): for $k = 1, 2, \ldots, K$ and any treatment sequence $\bar{a}_K$,

$$\{Y_k(\bar{a}_k), Y_{k+1}(\bar{a}_{k+1}), \ldots, Y_K(\bar{a}_K)\} \perp\!\!\!\perp A_k | (\bar{L}_k, \bar{A}_{k-1}, \bar{Y}_{k-1}).$$

However, in observational studies, the sequential randomization assumption is often violated due to the potential existence of unmeasured, time-varying confounders, $\bar{U}_k$. To mitigate this problem, Tchetgen Tchetgen et al. (2024) and Ying et al. (2023) develop the longitudinal proximal causal inference framework, which relieves the bias caused by $\bar{U}_k$ by efficiently exploiting the proxy variables $\bar{Z}_k$ and $\bar{W}_k$.

We suppose that it is possible to conceptualize joint interventions on $(\bar{A}_k, \bar{Z}_k)$, such that the potential outcomes $(Y_k(\bar{a}_k, \bar{z}_k), W_k(\bar{a}_k, \bar{z}_k))$ under a hypothetical intervention that sets $\bar{A}_k$ and $\bar{Z}_k$ to $\bar{a}_k$, $\bar{z}_k$ are well defined, for $k = 1, 2, \ldots, K$. The following assumption formalizes the causal structures of the longitudinal proximal causal inference framework:

*Assumption* 1 (Longitudinal Proximal Causal Structure). For $k = 1, 2, \ldots, K$ and any prespecified sequences of treatments and treatment-inducing proxies $\bar{a}_K$ and $\bar{z}_K$, we have:

1. $W_k(\bar{a}_k, \bar{z}_k) = W_k(\bar{a}_{k-1})$, almost surely;
2. $Y_k(\bar{a}_k, \bar{z}_k) = Y_k(\bar{a}_k)$, almost surely;

3. $\{Y_k(\bar{a}_k), Y_{k+1}(\bar{a}_{k+1}), \ldots, Y_K(\bar{a}_K), \bar{W}_k\} \perp\!\!\!\perp \{A_k, \bar{Z}_k\} | (\bar{U}_{k-1}, \bar{Y}_{k-1}, \bar{A}_{k-1} = \bar{a}_{k-1})$.

Assumptions 1.1 and 1.2 characterize valid longitudinal proxies $\bar{W}_K$ and $\bar{Z}_K$. For $k = 1, 2, \ldots, K$, Assumption 1.1 requires that $A_k$ and $\bar{Z}_k$ have no mediation path to $W_k$ other than through $\bar{A}_{K-1}$, and Assumption 1.2 implies that $\bar{Z}_k$ has no mediation path to $Y_k$ other than through $\bar{A}_k$. Assumption 1.3 necessiates that under an arbitrary prespecified treatment assignment sequence $\bar{a}_K$, for the $k^{th}$ stage, the outcome related variables $\big(Y_k(\bar{a}_k), Y_{k+1}(\bar{a}_{k+1}), \ldots, Y_K(\bar{a}_K)\big)$ and $\bar{W}_k$ are independent with treatment related variables $A_k$ and $\bar{Z}_k$, given the accumulated history $(\bar{U}_{k-1}, \bar{Y}_{k-1}, \bar{A}_{k-1} = \bar{a}_{k-1})$, which explicitly allows for the existence of the unmeasured confounders.

To convey the described causal structure more intuitively, we depict a causal Directed Acyclic Graph (DAG) that satisfies Assumption 1 when $K = 2$ in Figure 1, which closely resembles those considered by Ying et al. (2023) and Zhang & Tchetgen Tchetgen (2024).

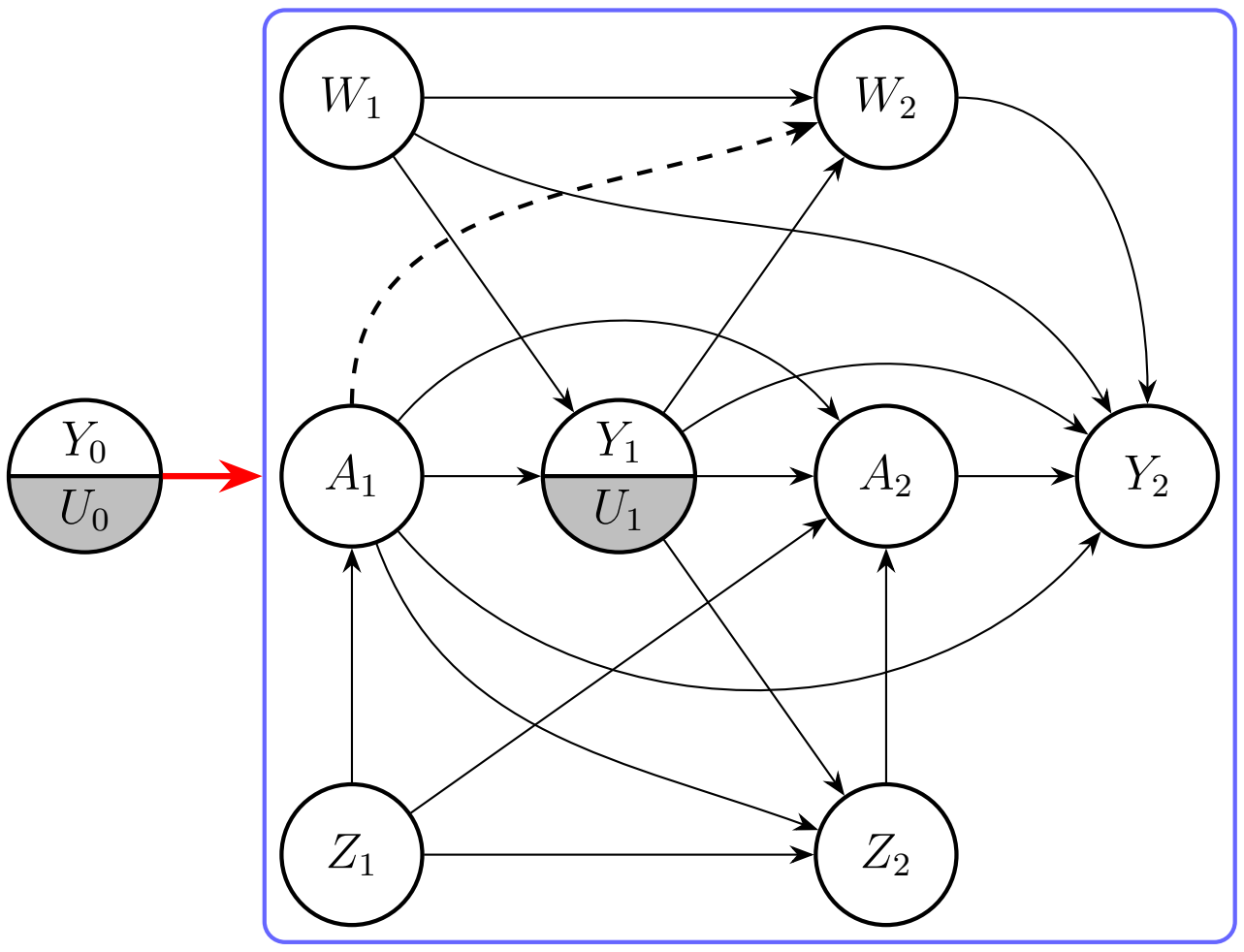

Figure 1: A DAG for the 2-stage longitudinal proximal causal inference framework that satisfies Assumptions 1. The variables $(Y_0, U_0)$ and $(Y_1, U_1)$ are each consolidated into a single node, indicating that they share the same set of child nodes. A bold red arrow and a blue rectangle are used for visual clarity: the arrow denotes that the $(Y_0, U_0)$ node has direct edges to all other nodes enclosed within the rectangle.

By Assumptions 1.1 and 1.2, $Z_1$ can only affect $Y_1$ through $A_1$, while $W_1$ can not be directly influenced by $A_1$. Moreover, $\bar{Z}_2$ can only affect $Y_2$ through $\bar{A}_2$, while $\bar{W}_2$ can not be directly influenced by $\bar{A}_2$. Here, the direct effect of $A_1$ on $W_2$ is allowed.

Throughout, we make the standard consistency assumption that for $k = 1, 2, \ldots, K$, $Y_k = Y_k(\bar{A}_k, \bar{Z}_k), W_k = W_k(\bar{A}_k, \bar{Z}_k)$ almost surely. In addition, the following longitudinal positivity assumption is also required to ensure that, within every possible subgroup, there exist individuals receiving each possible sequence of treatment assignments:

*Assumption* 2 (Sequential Positivity). For $k = 1, 2, \ldots, K$ and $\forall a \in \mathcal{A}$,

$$0 < f(A_k = a|\bar{A}_{k-1}, \bar{U}_{k-1}, \bar{Y}_{k-1}, \bar{L}_k) < 1, \text{ almost surely.}$$

The canonical occasions that Assumptions 1 and 2 above might be expected to hold are abundant in the observational studies. A good running example is the study about the joint causal effects of the disease-modifying anti-rheumatic therapy methotrexate over time among patients with rheumatoid arthritis, which was previously reanalyzed by Ying et al. (2023). The average number of tender joints at the end of follow-up is used as the outcome variable of interest, which is a well-established measure of disease progression. In their analysis, the erythrocyte sedimentation rate is used as the treatment-inducing proxy, as it is often an observational indicator obtained by physicians through diagnostic procedures before determining a treatment plan. In addition, the patient's global assessment is employed as the outcome-inducing proxy, as it is often a vague subjective assessment of personal conditions before receiving treatment, which is seldom used as a direct reference for treatment decisions. Tchetgen Tchetgen et al. (2024) also suggest the randomization in SMART as the treatment-inducing proxy variable when nonadherence appears. For more examples about the proxy variables, we refer interested readers to Lipsitch et al. (2010) and Shi et al. (2020).

To identify the causal estimands of interest under the longitudinal proximal causal inference framework, Tchetgen Tchetgen et al. (2024), Ying et al. (2023), and Zhang & Tchetgen Tchetgen (2024) also consider the following $Z$-completeness and $W$-completeness assumptions to ensure the information sufficiency of proxy variables to the unmeasured confounders, which we will now state stagewise for $l, t = 1, 2, \ldots, K$:

*Assumption* 3′ (Completeness).

1. ($Z$-Completeness at the $l^{th}$ Stage). At the $l^{th}$ stage, for any square integrable function $g_l(\bar{U}_{l-1})$,

$$\mathbb{E}[g_l(\bar{U}_{l-1})|\bar{Z}_l, \bar{y}_{l-1}, \bar{a}_l] = 0 \text{ for any } \bar{y}_{l-1}, \bar{a}_l \text{ almost surely} \iff g_l(\bar{U}_{l-1}) = 0 \text{ almost surely.}$$

2. ($W$-Completeness at the $t^{th}$ Stage). At the $t^{th}$ stage, for any square integrable function $g_t(\bar{U}_{t-1})$,

$$\mathbb{E}[g_t(\bar{U}_{t-1})|\bar{W}_t, \bar{y}_{t-1}, \bar{a}_t] = 0 \text{ for any } \bar{y}_{t-1}, \bar{a}_t \text{ almost surely} \iff g_t(\bar{U}_{t-1}) = 0 \text{ almost surely.}$$

The completeness assumption plays a fundamental role in statistical sufficiency and identification theory. Originally introduced by Lehmann & ScheffÉ (2012*a*,*b*) to characterize estimators with minimal risk within classes of unbiased estimators, this concept has since been adapted for identification purposes in econometrics and statistics. We refer interested readers to Section F of the Supplementary Material for a more comprehensive introduction to this assumption. In particular, we make the following assumption that a certain subset, indexed by $k$, of Assumption 3′ with $l, t = 1, 2, \ldots, K$ holds for convenience. The running index $k \in \{0, 1, \ldots, K\}$. When $k = 0$, we only assume that Assumption 3′.1 holds at stages 1 to $K$, and vice versa when $k = K$.

*Assumption* 3 ($k$-Completeness of Proxy Variables). Assumption 3′.1 holds at the $l^{th}$ stage, for $l = K, \ldots, k+1$; Assumption 3′.2 holds at the $t^{th}$ stage, for $t = 1, \ldots, k$.

The proximal identification relies on the so-called outcome and treatment confounding bridge functions $h_{Kl}(\bar{y}_K, \bar{w}_l, \bar{a}_K)$ and $q_{tt}(\bar{y}_{t-1}, \bar{z}_t, \bar{a}_t)$ that solve the corresponding integral equations presented below (Miao et al. 2018, Cui et al. 2024, Tchetgen Tchetgen et al. 2024, Ying et al. 2023, Zhang & Tchetgen Tchetgen 2024). For $h_{Kl}(\bar{y}_K, \bar{w}_l, \bar{a}_K)$, the first and the second subscripts $K$ and $l$ respectively indicate the maximum indices of the variables $Y$ and $W$ in its arguments; For $q_{tt}(\bar{y}_{t-1}, \bar{z}_t, \bar{a}_t)$, the first and the second subscripts $t$ respectively correspond to the maximum indices of the $A$ and $Z$ variables in its arguments. Corresponding to Ying et al. (2023), hereafter we will denote $h_{Kl}(\bar{y}_K, \bar{w}_l, \bar{a}_K)$ by $\mathcal{H}_{Kl}(\bar{a}_K)$ and $q_{tt}(\bar{y}_{t-1}, \bar{z}_t, \bar{a}_t)$ by $\mathcal{Q}_{tt}(\bar{a}_t)$ for $l, t = 1, 2, \ldots, K$ for simplicity.

For identification of optimal DTRs, we will adopt totally $K$ out of $2K$ confounding bridge functions, that is, $K-k$ outcome and $k$ treatment confounding bridge functions for $k = 0, 1, \ldots, K$. Therefore, we will assume a hybrid existence of confounding bridge functions according to the choice of $k$, which we refer to as the following $k$-existence assumption. For an arbitrary given $k \in \{0, 1, 2, \ldots, K\}$, we assume that:

*Assumption* 4 ($k$-Existence of Confounding Bridge Functions). There exist the following two subsets of confounding bridge functions:

1. There exist outcome confounding bridge functions $\mathcal{H}_{Kl}(\bar{a}_K)$, $l = K, K-1, \ldots, k+1$ that satisfy:

   For $l = K$,

$$f(Y_K = y_K | \bar{y}_{K-1}, \bar{z}_K, \bar{a}_K) = \sum_{\bar{w}_K} \mathcal{H}_{KK}(\bar{a}_K) f(\bar{w}_K | \bar{y}_{K-1}, \bar{z}_K, \bar{a}_K), \tag{1}$$

   and for $l = K-1, \ldots, k+1$,

$$\sum_{\bar{w}_{l+1}} \mathcal{H}_{K,l+1}(\bar{a}_K) f(\bar{w}_{l+1}, y_l | \bar{y}_{l-1}, \bar{z}_l, \bar{a}_l) = \sum_{\bar{w}_l} \mathcal{H}_{Kl}(\bar{a}_K) f(\bar{w}_l | \bar{y}_{l-1}, \bar{z}_l, \bar{a}_l). \tag{2}$$

2. There exist treatment confounding bridge functions $\mathcal{Q}_{tt}(\bar{a}_t)$, $t = 1, 2, \ldots, k$ that satisfy:

   For $t = 1$,

$$\frac{1}{f(a_1 | w_1, y_0)} = \sum_{z_1} \mathcal{Q}_{11}(\bar{a}_t) f(z_1 | a_1, w_1, y_0), \tag{3}$$

   and for $t = 2, \ldots, k$,

$$\frac{\sum_{\bar{z}_{t-1}} \mathcal{Q}_{t-1,t-1}(\bar{a}_{t-1}) f(\bar{z}_{t-1} | \bar{a}_{t-1}, \bar{w}_t, \bar{y}_{t-1})}{f(a_t | \bar{a}_{t-1}, \bar{w}_t, \bar{y}_{t-1})} = \sum_{\bar{z}_t} \mathcal{Q}_{tt}(\bar{a}_t) f(\bar{z}_t | \bar{a}_t, \bar{w}_t, \bar{y}_{t-1}). \tag{4}$$

Specifically, when $k = 0$, we only suppose the existence of outcome confounding bridge functions $\mathcal{H}_{Kl}(\bar{a}_K)$ for $l = 1, 2, \ldots, K$, but do not assume the existence of any treatment confounding bridge functions; vice versa when $k = K$. The regularity conditions that ensure the existence of the solutions to Equations (1)-(4) are presented in Section E of the Supplementary Material, which can be empirically verified.

## 2.3 The Structure of Optimal DTRs Identification

We now define the problem of estimating the optimal DTRs rigorously. Without loss of generality, we suppose that a greater terminal outcome $Y_K$ is always preferred. It is equivalent to treat $\sum_{i=1}^{K} Y_i$ as the optimization goal, since the outcomes observed at earlier stages, $Y_k$ for $k = 1, 2, \dots, K-1$, can also be viewed as components of the terminal outcome $Y_K$. Further elaboration is available in Chapter 6.2.1 of Tsiatis et al. (2019).

For a combination of treatment regimes $\bar{d}_K = (d_1, d_2, \dots, d_K)$, the potential outcomes $Y_k(\bar{d}_k)$, for $k = 1, 2, \dots, K$ are formally defined as:

$$Y_k(\bar{d}_k) = \sum_{\bar{a}_k} \left[ Y_k(\bar{a}_k) \prod_{j=1}^{k} I\left( d_j(\bar{Y}_{j-1}(\bar{a}_{j-1}), \bar{a}_{j-1}) = a_j \right) \right]. \tag{5}$$

By Equation (5), we define the value function of regimes $\bar{d}_K$ as:

$$V(\bar{d}_K) = \mathbb{E}[Y_K(\bar{d}_K)]. \tag{6}$$

Our ultimate goal is to find the optimal regimes $\bar{d}_K^*$ that maximize the value function:

$$\bar{d}_K^* = \underset{\bar{d}_K \in \mathcal{D}}{\operatorname{argmax}} V(\bar{d}_K).$$

This approach is generally known as the value maximization (Zhao & Laber 2014). Once the value function has been identified, we can further identify the optimal DTRs. According to Equations (5) and (6), we note the following equivalent expression of the value function:

$$V(\bar{d}_K) = \sum_{\bar{y}_K} \sum_{\bar{a}_K} \left\{ y_K \prod_{j=1}^{K} \left[ I(d_j(\bar{y}_{j-1}, \bar{a}_{j-1}) = a_j) \right] f(\bar{Y}_K(\bar{a}_K) = \bar{y}_K | Y_0 = y_0) f(Y_0 = y_0) \right\}. \tag{7}$$

As evident in Equation (7), the core part that remains to be identified is the conditional joint density of the potential outcomes:

$$f(\bar{Y}_K(\bar{a}_K) = \bar{y}_K | Y_0 = y_0).$$

This identification problem is distinct and more precise than that of functionals of the marginal distribution of $Y_K(\bar{a}_K)$ (Ying et al. 2023, Tchetgen Tchetgen et al. 2024) or $Y_1(a_1)$ (Miao et al. 2018, Cui et al. 2024, Tchetgen Tchetgen et al. 2024), since the identification results of the joint density of sequential potential outcomes imply their marginal functionals.

# 3 Proximal Identification of Optimal DTRs

## 3.1 Proximal Identification of the Joint Density

We will first present two novel nonparametric proximal identification methods for the conditional joint density $f(\bar{Y}_K(\bar{a}_K) = \bar{y}_K | Y_0 = y_0)$:

**Theorem 3.1.**

1. *Under Assumptions 1, 2,* 0*-completeness Assumption 3, and* 0*-existence Assumption 4, it follows that:*

$$f(\bar{Y}_K(\bar{a}_K) = \bar{y}_K | Y_0 = y_0) = \sum_{w_1} \mathcal{H}_{K1}(\bar{a}_K) f(w_1 | y_0). \tag{8}$$

2. *Under Assumptions 1, 2, $K$-completeness Assumption 3, and $K$-existence Assumption 4, it follows that:*

$$f(\bar{Y}_K(\bar{a}_K) = \bar{y}_K | Y_0 = y_0) = \sum_{\bar{z}_K} \mathcal{Q}_{KK}(\bar{a}_K) f(\bar{y}_K, \bar{a}_K, \bar{z}_K | y_0). \tag{9}$$

**Remark 1.** Both Theorems 3.1.1 and 3.1.2 do not rely on the additional assumption proposed by Zhang & Tchetgen Tchetgen (2024), such that they correspondingly cover Theorems 1 and 2 of Zhang & Tchetgen Tchetgen (2024). Moreover, as Theorem 2 of Zhang & Tchetgen Tchetgen (2024) only considers a two-stage, discrete-outcome setup, we further generalize it to the setting with an arbitrary finite number of stages and continuous outcomes in Theorem 3.1.2, and additionally provide closed-form expressions for the treatment-confounding bridge functions in the categorical setting. More details can be found in Sections D.2, I, and K.1 of the Supplementary Material.

We refer to the identification approach implied by Equation (8) as identification via *Proximal Outcome Regression* (POR) and the approach implied by Equation (9) as identification via *Proximal Inverse Probability Weighting* (PIPW), aligning with that defined in Tchetgen Tchetgen et al. (2024) and Ying et al. (2023). The proof of this theorem, as well as the proof of the subsequent results, will be elaborated in Section D of the Supplementary Material.

On the one hand, Identification via POR is exactly the longitudinal generalization of the identification result in Miao et al. (2018). With the help of newly designed outcome confounding bridge functions $\mathcal{H}_{K,K-1}(\bar{a}_K), \ldots, \mathcal{H}_{K1}(\bar{a}_K)$ and proxy variables, this method captures the conditional joint distribution of potential outcomes directly. Further discussions about the outcome confounding bridge functions can be found in Section J of the Supplementary Material.

On the other hand, identification via PIPW generalizes the marginal identification results established in Ying et al. (2023) and Cui et al. (2024), with essentially the same treatment bridge functions. Some distinct advantages arise for PIPW. Notably, estimating the treatment confounding bridge functions is much easier compared to estimating the outcome confounding bridge functions. Moreover, suppose there are future data for the $(K+1)^{th}$ stage, the treatment confounding bridge functions for the $k^{th}$ stage, $k = 1, 2, \ldots K$ in that case can be reused for the $(K+1)$-stage identification, which implies the potential for incremental learning (Van de Ven et al. 2022) and continual lifelong learning (Parisi et al. 2019).

A common feature shared by POR and PIPW identification methods is the requirement for solving the nested first-kind Fredholm integral equations. Because of their inherent ill-posedness (Tikhonov et al. 1995), addressing this class of equations is a key challenge to applying proximal causal inference methods in general practical settings (Miao et al. 2018, Cui et al. 2024, Ying et al. 2022, 2023, Tchetgen Tchetgen et al. 2024). While numerous methods have been developed to obtain nonparametric estimators for the initial bridge functions $\mathcal{H}_{KK}(\bar{a}_K)$ and $\mathcal{Q}_{11}(a_1)$ based on Equations (1) and (3), such as Kallus et al. (2022) and Ghassami et al. (2022), directly extending these methods to subsequent stages via Equations (2) and (4) may still substantially amplify the estimation errors. As a result, the reliability of the conclusions drawn from the solved final bridge functions $\mathcal{H}_{K1}$ or $\mathcal{Q}_{KK}$ may be called into question.

To mitigate this problem, we propose a new approach that significantly alleviates instability by combining both the outcome and treatment confounding bridge functions for identification. Our method reduces the number of nested integral equations that must be solved for each type of confounding bridge, thereby improving computational stability and flexibility. To establish this result, the following lagged proxy orthogonality assumption is needed.

*Assumption* 5 (Lagged Proxy Orthogonality at the $k^{th}$ stage). The outcome-inducing proxy $W_{k+1}$

at the $(k+1)^{th}$ stage is independent with the treatment-inducing proxy sequence $\bar{Z}_k$ until the $k^{th}$ stage, given their possible common causes $(\bar{U}_{k-1}, \bar{Y}_{k-1}, \bar{A}_{k-1})$ and mediator $A_k$ accumulated until the $k^{th}$ stage:

$$W_{k+1} \perp\!\!\!\perp \bar{Z}_k | \bar{U}_{k-1}, \bar{Y}_{k-1}, \bar{A}_k.$$

For $k = 1, 2, \ldots, K-1$, Assumption 5 at the $k^{th}$ stage is a strict subset of the additional assumption presented in Zhang & Tchetgen Tchetgen (2024). When $k = 0$ or $K$, Assumption 5 is a trivial statement which imposes no restriction on $W_1$ or $\bar{Z}_K$. We note that the correctness of Assumption 5 can be proved under Assumptions 1 and 2, except for some extreme cases. More details can be found in Section G of the Supplementary Material.

**Theorem 3.2.** *We extendingly define* $\mathcal{Q}_{00}(a_0) = \mathcal{H}_{K,K+1}(\bar{a}_K) = 1$. *For an arbitrary given* $k \in \{0, 1, \ldots, K\}$, *under Assumptions 1, 2, k-completeness Assumption 3, k-existence Assumption 4, and Assumption 5 at the* $k^{th}$ *stage, we have:*

$$f(\bar{Y}_K(\bar{a}_K) = \bar{y}_K | Y_0 = y_0) = \sum_{\bar{w}_{k+1}} \sum_{\bar{z}_k} \mathcal{H}_{K,k+1}(\bar{a}_K) \mathcal{Q}_{kk}(\bar{a}_k) f(\bar{w}_{k+1}, \bar{z}_k, \bar{y}_k, \bar{a}_k | y_0). \tag{10}$$

*Notably, Equation (10) will degenerate to Equation (8) when* $k = 0$ *or Equation (9) when* $k = K$, *respectively.*

Theorem 3.2 essentially unifies all $(K+1)$ different identification methods for a $K$-stage setting under corresponding assumptions. Both Theorems 3.1.1 and 3.1.2 are its special cases when $k = 0$ or $K$. When $\mathcal{H}_{K,k+1}$ and $\mathcal{Q}_{kk}$ simultaneously appear in the identification equation, we refer to the corresponding method as the $k^{th}$ identification via *Proximal Hybrid Augmentation* (PHA), with $k$ ranging from 1 to $K$-1. Thus, the PHA approach is a collective term that possesses a continuous structure. By PHA, we distribute the burden of solving a system of $K$ nested Fredholm integral equations between two types of bridge functions: one type is responsible for solving a nested system of $k$ equations, and the other for a system of $(K$-$k)$ equations, for $k = 1, 2, \ldots, K-1$. This property significantly enhances the flexibility of identification by allowing the choice of $k$, which can also substantially reduce solution instability.

## 3.2 Identification of the Optimal DTRs

As $(K+1)$ different identification approaches for the conditional joint density $f(\bar{Y}_K(\bar{a}_K)|Y_0 = y_0)$ have been established by Theorem 3.2, $(K+1)$ different identification results of optimal DTRs by value maximization (Corollary 3.1) can be subsequently established in the following:

**Corollary 3.1.** *For any given $k \in \{0, 1, \ldots, K\}$, we suppose that Assumptions 1, 2, $k$-completeness Assumption 3, $k$-existence Assumption 4, and Assumption 5 at the $k^{th}$ stage hold. Then, the value function of an arbitrary given DTRs $\bar{d}_K$ can be identified as:*

$$V(\bar{d}_K) = \sum_{\bar{y}_K} \sum_{\bar{a}_K} \sum_{\bar{w}_{k+1}} \sum_{\bar{z}_k} \Bigg\{ y_K \prod_{j=1}^{K} \Big[ I(d_j(\bar{y}_{j-1}, \bar{a}_{j-1}) = a_j) \Big] \times \mathcal{H}_{K,k+1}(\bar{a}_K) \mathcal{Q}_{kk}(\bar{a}_k) f(\bar{w}_{k+1}, \bar{z}_k, \bar{y}_k, \bar{a}_k) \Bigg\},$$

*and optimal DTRs $\bar{d}_K^*$ can be identified by value maximization:*

$$\bar{d}_K^* = \underset{\bar{d}_K}{\operatorname{argmax}}\, V(\bar{d}_K).$$

In particular, the cases where $k = 0$ and $k = K$ in Corollary 3.1 are also direct consequences of Theorem 3.1.

# 4 Semiparametric Efficiency and Multiply Robust Estimation

## 4.1 Semiparametric Efficiency Bound of the Value Function

For an arbitrary given sequence of regimes $\bar{d}_K(\bar{y}_{K-1}, \bar{a}_{K-1})$, we now consider the inference for its value function $V(\bar{d}_K)$ under the semiparametric model $\mathcal{M}_{sp}$, which does not restrict the observed data distribution other than the existence (but not necessary uniqueness) of treatment bridge functions $\mathcal{Q}_{tt}(\bar{a}_t)$, $t = 1, 2, \ldots, K$ that solve Equations (3) and (4).

**Theorem 4.1.** *Assume that $K$-existence Assumption 4 holds at all data-generating laws belonging to the semiparametric model $\mathcal{M}_{sp}$. In addition, suppose that at the true data-generating* *law, 0-existence Assumption 4 also holds. For notational simplicity, we denote* $\mathcal{Q}_{kk}(\bar{A}_k) = q_{kk}(\bar{Y}_{k-1}, \bar{Z}_k, \bar{A}_k)$ *and*

$$J_k(\bar{Y}_{k-1}, \bar{W}_k, \bar{a}_K)_{\bar{d}_K} = \sum_{y_K,\cdots,y_k} y_K h_{Kk}(y_K, \cdots, y_k, \bar{Y}_{k-1}, \bar{W}_k, \bar{a}_K) \\ \times \prod_{j=k+1}^{K} I(d_j(y_{j-1}, \cdots, y_k, \bar{Y}_{k-1}, \bar{a}_{j-1}) = a_j),$$

*for* $k = 1, 2, \ldots, K$*. Then, under Assumptions 1, 2, and* $K$*-completeness Assumption 3,*

$$\begin{aligned} IF_{V(\bar{d}_K)} = \sum_{\bar{a}_K} \Bigg\{ & I(\bar{d}_K(\bar{Y}_{K-1}, \bar{A}_{K-1}) = \bar{A}_K = \bar{a}_K)\mathcal{Q}_{KK}(\bar{A}_K) \\ & \times \left[ Y_K - \sum_{y_K} y_K h_{KK}(y_K, \bar{Y}_{K-1}, \bar{W}_K, \bar{a}_K) \right] \\ & + \sum_{k=1}^{K-1} I(\bar{d}_k(\bar{Y}_{k-1}, \bar{A}_{k-1}) = \bar{A}_k = \bar{a}_k)\mathcal{Q}_{kk}(\bar{A}_k) \\ & \times \left[ J_{k+1}(\bar{Y}_k, \bar{W}_{k+1}, \bar{a}_K)_{\bar{d}_K} I(d_{k+1}(\bar{Y}_k, \bar{a}_k) = a_{k+1}) - J_k(\bar{Y}_{k-1}, \bar{W}_k, \bar{a}_K)_{\bar{d}_K} \right] \\ & + J_1(Y_0, W_1, \bar{a}_K)_{\bar{d}_K} I(d_1(Y_0) = a_1) \Bigg\} - V(\bar{d}_K) \end{aligned} \tag{11}$$

*is a valid influence function for* $V(\bar{d}_K)$ *under* $\mathcal{M}_{sp}$*. Furthermore, it is the efficient influence function at the submodel where Assumption H.1 depicted in Section H of the Supplementary Material also holds, and all confounding bridge functions are unique. The corresponding semiparametric efficiency bound is given by* $\mathbb{E}[IF^2_{V(\bar{d}_K)}]$.

Theorem 4.1 presents the closed-form expression of a valid influence function and the corresponding submodels where it is an efficient influence function. The uniqueness of outcome and treatment confounding bridge functions is ensured by completeness assumptions between $Z$ and $W$, which will be discussed in Section E of the Supplementary Material.

## 4.2 Multiply Robust and Locally Efficient Estimation of Value Function

Motivated by the efficient influence function we have obtained in Section 4.1, we now present a locally efficient and $(K+1)$-fold multiply robust estimator of the value function $V(\bar{d}_K)$. The multiply robust property will guarantee the consistency of the estimator if the models of a specific subset of bridge functions are correctly specified. Specifically, we consider the following $(K+1)$ semiparametric submodels which impose parametric restrictions on different components of the observed data distribution while allowing the rest of the distribution to remain unrestricted.

For $k = 0, K$, $\mathcal{M}_0$ and $\mathcal{M}_K$ are respectively defined as:

$\mathcal{M}_0$: Assumptions 1, 2, 0-completeness Assumption 3, and 0-existence Assumption 4 hold; the models of $\mathcal{H}_{Kl}(\bar{a}_K)$, $l = 1, 2, \ldots, K$ are correctly specified;

$\mathcal{M}_K$: Assumptions 1, 2, $K$-completeness Assumption 3, and $K$-existence Assumption 4 hold; the models of $\mathcal{Q}_{tt}(\bar{a}_t)$, $t = 1, 2, \ldots, K$ are correctly specified.

For $k = 1, 2, \ldots, K-1$, $\mathcal{M}_k$ is defined as:

$\mathcal{M}_k$: Assumptions 1, 2, $k$-completeness Assumption 3, $k$-existence Assumption 4, and Assumption 5 at the $k^{th}$ stage hold; the models of $\mathcal{H}_{Kl}(\bar{a}_K)$, $l = K, K-1, \ldots, k+1$ and $\mathcal{Q}_{tt}(\bar{a}_t)$, $t = 1, 2, \ldots, k$ are correctly specified.

We will first consider the estimation within each submodel $\mathcal{M}_k$, $k = 0, 1, \ldots, K$. For $l, t = 1, 2, \ldots, K$, let $h_{Kl}(\bar{y}_K, \bar{w}_l, \bar{a}_K; \beta_l)$ and $q_{tt}(\bar{y}_{t-1}, \bar{z}_t, \bar{a}_t; \gamma_t)$ respectively denote the parametric models of $h_{Kl}(\bar{y}_K, \bar{w}_l, \bar{a}_K)$ and $q_{tt}(\bar{y}_{t-1}, \bar{z}_t, \bar{a}_t)$, which are respectively indexed by finite dimensional parameters $\beta_l$ and $\gamma_t$, for $l, t = 1, 2, \ldots, K$. According to observable Equations (1)-(4), corresponding parametric estimation equations for $\beta_l$ and $\gamma_t$, $l, t = 1, 2, \ldots, K$ can be constructed. The detailed procedures can be found in Sections B and C of the Supplementary Material. After obtaining estimations for nuisance parameters $\beta_l$ and $\gamma_t$, $l, t = 1, 2, \ldots, K$, the POR, PHA, and PIPW estimators of $V(\bar{d}_K)$ can be constructed respectively:

$$\hat{V}_{POR}(\bar{d}_K) = \sum_{\bar{a}_K} \mathbb{P}_n \Big\{ J_1(Y_0, W_1, \bar{a}_K; \hat{\beta}_1)_{\bar{d}_K} I(d_1(Y_0) = a_1) \Big\},$$

$$\hat{V}_{PHA,k}(\bar{d}_K) = \sum_{\bar{a}_K} \mathbb{P}_n \Big\{ I(\bar{d}_k(\bar{Y}_{k-1}, \bar{A}_{k-1}) = \bar{A}_k = \bar{a}_k) \mathcal{Q}_{kk}(\bar{A}_k; \hat{\gamma}_k)$$
$$\times J_{k+1}(\bar{Y}_k, \bar{W}_{k+1}, \bar{a}_K; \hat{\beta}_{k+1})_{\bar{d}_K} I(d_{k+1}(\bar{Y}_k, \bar{a}_k) = a_{k+1}) \Big\}, k = 1, \ldots, K-1,$$

$$\hat{V}_{PIPW}(\bar{d}_K) = \sum_{\bar{a}_K} \mathbb{P}_n \Big\{ Y_K \mathcal{Q}_{KK}(\bar{A}_K; \hat{\gamma}_K) I(\bar{d}_K(\bar{Y}_{K-1}, \bar{A}_{K-1}) = \bar{A}_K = \bar{a}_K) \Big\}.$$

$\hat{V}_{POR}(\bar{d}_K)$ is a Consistent and Asymptotically Normal (CAN) estimator under model $\mathcal{M}_0$, $\hat{V}_{PHA,k}(\bar{d}_K)$ is CAN under model $\mathcal{M}_k$, $k = 1, 2, \ldots, K$, and $\hat{V}_{PIPW}(\bar{d}_K)$ is CAN under model $\mathcal{M}_K$. However, correctly specifying models of outcome and treatment bridge functions can be very demanding. Even though we have offered $(K+1)$ different estimators through POR, PIPW, and PHA, it is still necessary to know the exact subsets of bridge functions with correctly specified models to construct the CAN estimators. Thus, the development of a Proximal Multiply Robust (PMR) estimator, which enables the relaxation of parametric modelling assumptions, is of interest.

**Theorem 4.2.** *Under standard regularity conditions given in Appendix B of Robins et al. (1994),*

$$\begin{aligned}
&\hat{V}_{PMR}(\bar{d}_K) \\
&= \sum_{\bar{a}_K} \mathbb{P}_n \Bigg\{ I(\bar{d}_K(\bar{Y}_{K-1}, \bar{A}_{K-1}) = \bar{A}_K = \bar{a}_K) \mathcal{Q}_{KK}(\bar{A}_K; \hat{\gamma}_K) \\
&\quad \times \Bigg[ Y_K - \sum_{y_K} y_K h_{KK}(y_K, \bar{Y}_{K-1}, \bar{W}_K, \bar{a}_K; \hat{\beta}_K) \Bigg] \\
&\quad + \sum_{k=1}^{K-1} I(\bar{d}_k(\bar{Y}_{k-1}, \bar{A}_{k-1}) = \bar{A}_k = \bar{a}_k) \mathcal{Q}_{kk}(\bar{A}_k; \hat{\gamma}_k) \\
&\quad \times \Bigg[ J_{k+1}(\bar{Y}_k, \bar{W}_{k+1}, \bar{a}_K; \hat{\beta}_{k+1})_{\bar{d}_K} I(d_{k+1}(\bar{Y}_k, \bar{a}_k) = a_{k+1}) - J_k(\bar{Y}_{k-1}, \bar{W}_k, \bar{a}_K; \hat{\beta}_k)_{\bar{d}_K} \Bigg] \\
&\quad + J_1(Y_0, W_1, \bar{a}_K; \hat{\beta}_1)_{\bar{d}_K} I(d_1(Y_0) = a_1) \Bigg\}
\end{aligned} \tag{12}$$

*is a CAN estimator of $V(\bar{d}_K)$ under the semiparametric union model $\mathcal{M}_{union} = \bigcup_{k=0}^{K} \mathcal{M}_k$. Furthermore, $\hat{V}_{PMR}(\bar{d}_K)$ is semiparametric locally efficient in $\mathcal{M}_{sp}$ at the intersection submodel $\mathcal{M}_{int} = \mathcal{M}_0 \cap \mathcal{M}_K$, when Assumption H.1 also holds.*

Though the proposed multiply robust estimator depends on models for the confounding bridge functions, only one of the $(K+1)$ models $\mathcal{M}_k$, $k = 0, 1, \ldots, K$ is correct will yield the CAN estimator $\hat{V}_{PMR}(\bar{d}_K)$ for the value function $V(\bar{d}_K)$, without necessarily knowing which one is right. It is also worth noting that the core components of the PMR estimator $\hat{V}_{PMR}(\bar{d}_K)$ closely resemble those found in the similarly $(K+1)$-fold multiply robust estimators developed in Robins et al. (1994), Tchetgen (2009) and Molina et al. (2017) for general $K$-stage complex longitudinal causal inference under the sequential randomization assumption. $\mathcal{Q}_{tt}(\bar{a}_t)$ respectively corresponds to the product of the reciprocals of propensity scores, and $J_k(\bar{Y}_{k-1}, \bar{W}_k, \bar{a}_K)_{\bar{d}_K}$ respectively corresponds to the nested structural mean models. This structural similarity suggests the profound significance of the PMR estimator $\hat{V}_{PMR}(\bar{d}_K)$ within the longitudinal proximal causal inference framework. More discussion, including another equivalent closed-form of the PMR estimator, will be deferred to Section H of the Supplementary Material.

From a qualitative perspective, the structure of the PMR estimator $\hat{V}(\bar{d}_K)_{PMR}$ can be interpreted as a stepwise procedure that systematically corrects biases in the estimation of $V(\bar{d}_K)$. The procedure commences by recursively estimating $\mathcal{H}_{Kl}(\bar{a}_K; \beta_l)$ for $l = K, K-1, \ldots, 1$ in order to construct $\hat{V}(\bar{d}_K)_{POR}$. Subsequently, it leverages $\hat{V}(\bar{d}_K)_{PHA,k}$ and exploits the recursive relationship between $J_k$ and $J_{k-1}$ to iteratively correct the bias in $\mathcal{H}_{Kk}(\bar{a}_K; \hat{\beta}_k)$ for $k = 1, 2, \ldots, K-1$. Finally, the procedure addresses the bias in $\mathcal{H}_{KK}(\bar{a}_K; \hat{\beta}_K)$ by applying $\hat{V}(\bar{d}_K)_{PIPW}$ to supplement the regression residual $\left(Y_K - \sum_{y_K} y_K h_{KK}(y_K, \bar{Y}_{K-1}, \bar{W}_K, \bar{a}_K; \hat{\beta}_K)\right)$. Thus, the PMR estimator $\hat{V}(\bar{d}_K)_{PMR}$ inherits the robustness properties of $\hat{V}(\bar{d}_K)_{POR}$, $\hat{V}(\bar{d}_K)_{PHA}$, and $\hat{V}(\bar{d}_K)_{PIPW}$ simultaneously. Moreover, it further alleviates the risk of model misspecification by leveraging $\hat{V}(\bar{d}_K)_{PHA}$ and $\hat{V}(\bar{d}_K)_{PIPW}$ to compensate for potential biases in $\hat{V}(\bar{d}_K)_{POR}$, and vice versa.

**Remark 2.** According to Theorem 4.1, the semiparametric efficiency of $\hat{V}_{PMR}(\bar{d}_K)$ does not rely on Assumption 5. Even if under some extreme cases where Assumption 5 may not hold, $\hat{V}_{PMR}(\bar{d}_K)$ is still doubly robust over the union model $\mathcal{M}'_{union} = \mathcal{M}_0 \cup \mathcal{M}_K$, and semeprametric efficient under the intersection model $\mathcal{M}_{int} = \mathcal{M}_0 \cap \mathcal{M}_K$ when Assumption *H*.1 also holds. In particular, the doubly robust estimation of the value function for static regimes, namely the Marginal Structural Mean Model (MSMM) (Ying et al. 2023), arises as a special case of $\hat{V}_{PMR}(\bar{d}_K)$. This illustrates that our proposed framework, as a unified methodology, accommodates a wide spectrum of settings ranging from static regimes to the most general dynamic treatment regimes.

# 5 Theoretical Results for Consistency of Regime Estimation

By the multiply robust estimator $\hat{V}_{PMR}(\bar{d}_K)$, $k = 1, 2, \cdots, K$, we can immediately define the corresponding multiply robust estimators for optimal DTRs $\bar{d}_K^*(\bar{y}_{K-1}, \bar{a}_{K-1})$ by value maximization as:

$$\hat{\bar{d}}_K = \underset{\bar{d}_K \in \mathcal{D}}{\operatorname{argmax}} \hat{V}_{PMR}(\bar{d}_K), \tag{13}$$

Applying machine learning methods to estimate regime functions from a large class that finite-dimensional parameters cannot index has been recommended by some researchers (Zhao et al. 2015, Ye et al. 2024). Nevertheless, considering the estimation for certain classes of regime functions with better interpretability and lower complexity is still of great interest (Kitagawa & Tetenov 2018, Athey & Wager 2021). Specifically, we will focus on estimating regimes within the linear classes $\mathcal{D}_k = \{I((\bar{Y}_{k-1}, \bar{A}_{k-1})\theta_k) > 0 : \theta_k \in \Theta_k, \|\theta_k\|_{L_2} = 1\}$, for $k = 1, 2, \ldots, K$ by value maximization, where $\theta_k$ indexes different regimes and $\Theta_k$ is a compact subset of $\mathbb{R}^p$. Let $\theta$ and $\Theta$ respectively denote $(\theta_1^T, \theta_2^T, \ldots, \theta_K^T)^T$ and $(\Theta_1^T, \Theta_2^T, \ldots, \Theta_K^T)^T$, we note that estimating the optimal regime functions $\bar{d}_K(\bar{Y}_{K-1}, \bar{A}_{K-1}; \theta_K^*)$ indexed by $\theta^*$ defined in Equation (13) is equivalent to the estimation of $\hat{\theta}$ defined below:

$$\begin{aligned} \hat{\theta} &= \underset{\theta \in \Theta}{\operatorname{argmax}} \hat{V}_{PMR}(\bar{d}_K(\bar{Y}_{K-1}, \bar{A}_{K-1}; \theta_K)) \\ &=: \underset{\theta \in \Theta}{\operatorname{argmax}} \hat{V}_{PMR}(\theta). \end{aligned} \tag{14}$$

We use $V(\theta)$ and $\hat{V}_{PMR}(\theta)$ to respectively denote the value function and its corresponding multiply robust estimator of $\bar{d}_K(\bar{y}_{K-1}, \bar{a}_{K-1}; \theta)$. To reduce the risk of overfitting, we use the cross-fitting algorithm (Athey & Wager 2021, Wager 2024) when calculating $\hat{V}_{PMR}(\theta)$, which will be narrated in Section K.2 of the Supplementary Material. We will still denote the cross-fitted estimator by $\hat{V}_{PMR}(\theta)$ to avoid overcomplicating notations. Notably, we only impose two sets of mild conditions for the estimators of bridge functions, such as the estimators of nuisance parameters converging at rates faster than $n^{1/4}$, but do not restrict the class of estimation methods. The content of conditions and their corresponding explanations will be depicted in Section A.3.1 of the Supplementary Material.

Under Conditions 1 and 2 presented in Section A.3.1 of the Supplementary Material, our main asymptotic results include that $\hat{\theta}$ converges to $\theta^*$ at $n^{1/3}$ rate, the regret converges faster than the rate of $\sqrt{n}$, and $\hat{V}_{PMR}(\hat{\theta})$ is $\sqrt{n}$-consistent to $V(\theta^*)$ as well as asymptotically normal.

**Theorem 5.1.** *Suppose Assumptions 1, 2, 0-completeness Assumption 3, and $k$-existence Assumption 4 with $k = 0$ and $K$ hold simultaneously. Let "$\xrightarrow{d}$" denote convergence in distribution. Under Conditions 1 and 2, as $n \to \infty$, we have:*

*(a).* $\|\hat{\theta} - \theta^*\|_{L_2} = O_p(n^{-1/3})$;

*(b).* $V(\theta^*) - V(\hat{\theta}) = o_p(n^{-1/2})$;

*(c).* $\sqrt{n}(\hat{V}_{PMR}(\hat{\theta}) - V(\theta^*)) \xrightarrow{d} \mathcal{N}(0, \sigma^2)$, *where* $\sigma^2 = E[IF^2_{V(\theta^*)}]$.

According to Theorem 5.1 (a) (b), while the parameters of estimated regimes $\hat{\theta}$ converge to the true parameter $\theta^*$ at a rate of $n^{1/3}$, we can still achieve a convergence rate faster than $n^{1/2}$ for the regret $V(\theta^*) - V(\hat{\theta})$. Moreover, Theorem 5.1 (c) presents an asymptotic normal distribution for the estimator $\hat{V}_{PMR}(\hat{\theta})$, which can be applied for inference. It additionally implies that, besides the required conditions of Theorem 5.1, when $K$-completeness Assumption 3 and regularity Assumption *H*.1 also hold, $\hat{V}_{PMR}(\hat{\theta})$ is an asymptotic semiparametric efficient estimator of $V(\theta^*)$.

# 6 Numerical Experiments

To evaluate the finite sample performance and compare the efficiency and robustness of the proposed proximal estimators, we conduct a series of numerical simulations. We employ the value maximization methods to estimate optimal linear DTRs. It is carried out based on POR, PHA, PIPW, PMR, a naive approach that relies on the sequential randomization assumption (denoted as SRA), and the oracle procedure that has access to the true unmeasured confounders (denoted as Oracle). More details about the implementation of these methods can be found in Section K of the Supplementary Material. The replication codes are available on GitHub.

## 6.1 Data Generating Mechanisms

Similar to Zhang & Tchetgen Tchetgen (2024), we consider a two-stage Data Generating Process (DGP), where all variables are binary. The sequential DGP for 11 variables ($Y_0, U_0, Z_1, W_1, A_1, Y_1, U_1, Z_2, W_2, A_2, Y_2$) adopts a logistic model with maximum third-order interactions, following the DAG presented in Figure 1. Assumptions 1, 2, Assumption 3′.1 for $l = 1, 2$, and Assumption 3′.2 for $t = 1, 2$ are also satisfied during the DGP. Let Bern($p$) denote the Bernoulli distribution with parameter $p$ and expit($x$) denote the sigmoid function $\frac{1}{1+e^{-x}}$. For the first stage, $U_0$ is first generated from Bern $\left(\text{expit}(0.5)\right)$. Then, $Y_0, Z_1, A_1, W_1, Y_1$ are sequentially generated according to models: $Y_0|U_0 \sim \text{Bern}\left(\text{expit}(-1 - 0.2U_0)\right)$, $Z_1|U_0, Y_0 \sim \text{Bern}\left(\text{expit}(-2 + 5U_0 + 0.1Y_0)\right)$, $A_1|Z_1, U_0, Y_0 \sim \text{Bern}\left(\text{expit}(-1+0.2Z_1+2U_0-0.25Y_0)\right)$, $W_1|U_0, Y_0 \sim \text{Bern}\left(\text{expit}(-2.2+5.2U_0+0.1Y_0)\right)$, and $Y_1|A_1, W_1, U_0, Y_0 \sim \text{Bern}\left(\text{expit}(0.1 - 0.55A_1 + 0.25W_1 + U_0 - 3Y_0 + 5A_1Y_0)\right)$.

Next, for the second stage, we first generate $U_1, W_2, Z_2, A_2$ sequentially from following models:

$$\begin{aligned}
U_1|A_1, U_0, Y_0 &\sim \text{Bern}\left(\text{expit}(0.1 + 0.15A_1 + U_0 - 0.1Y_0)\right) \\
W_2|\bar{Y}_1, \bar{U}_1, W_1 &\sim \text{Bern}\left(\text{expit}(-2 + 0.2Y_1 + 5U_1 + 0.2W_1 - 0.2U_0 - 0.2Y_0)\right) \\
Z_2|\bar{Y}_1, \bar{U}_1, A_1, Z_1 &\sim \text{Bern}\left(\text{expit}(-2 + 0.2Y_1 + 5U_1 + 0.002A_1 + 0.2Z_1 - 0.2U_0 - 0.2Y_0)\right) \\
A_2|\bar{Y}_1, \bar{U}_1, \bar{Z}_2, A_1 &\sim \text{Bern}\left(\text{expit}(-0.6 + 0.2Y_1 + 1.5U_1 - 0.5Z_2 - 0.6A_1 - 0.1Z_1 \right. \\
&\qquad \left. + 0.5U_0 + 0.2Y_0)\right)
\end{aligned}$$

Then, we generate $Y_2$ according to the model that $Y_2|Y_1, A_2, U_1, W_2, A_1, W_1, U_0, Y_0 \sim \text{Bern}\left(\text{expit}(b(\bar{Y}_1, \bar{A}_2, \bar{U}_1, \bar{W}_2))\right)$, where the function $b(\bar{Y}_1, \bar{A}_2, \bar{U}_1, \bar{W}_2)$ is defined as:

$$\begin{aligned}
&b(\bar{Y}_1, \bar{A}_2, \bar{U}_1, \bar{W}_2) \\
&= -0.25Y_1 + A_2 + 3U_1 - 0.7W_2 - 0.25A_1 - 0.7W_1 - 3U_0 - 0.25Y_0 - 4Y_1A_2 + 2A_2A_1 \\
&\quad - 2A_2Y_0 - Y_1A_2A_1 + 8Y_1A_2Y_0 + 7A_1A_2Y_0.
\end{aligned}$$

Following this DGP, we generate a random training sample of size $n = 35000$ and repeat the experiments 500 times.

As emphasized in Section 2, the core of estimating optimal DTRs hinges on estimating the conditional joint density $f(Y_2(\bar{a}_2), Y_1(a_1)|Y_0)$ from the generated data. To achieve this goal, estimating the outcome and treatment confounding bridge functions is necessary. We note that the outcome confounding bridge functions $\mathcal{H}_{2l}(\bar{a}_2)$, $l = 1, 2$ admit closed-form expressions in the categorical setting, as highlighted in Zhang & Tchetgen Tchetgen (2024), Appendix D. We also complement the closed-form expressions for the treatment confounding bridge functions $\mathcal{Q}_{tt}(\bar{a}_t)$, $t = 1, 2$, which is presented in Section K.1 of the Supplementary Material. Exploiting these expressions, the estimation focus shifts to the joint distribution of the observed data, which we obtain by maximum likelihood.

We consider scenarios where the models $\mathcal{M}_0$, $\mathcal{M}_1$, and $\mathcal{M}_2$ are misspecified to varying degrees, following Wang & Tchetgen Tchetgen (2018), to further evaluate the robustness of our proposed methods. Specifically, we substitute the correctly estimated bridge functions with randomly generated pseudo bridge functions $\mathcal{H}^{\dagger}_{22}(\bar{a}_2)$, $\mathcal{H}^{\dagger}_{21}(\bar{a}_2)$, $\mathcal{Q}^{\dagger}_{11}(a_1)$, and $\mathcal{Q}^{\dagger}_{22}(\bar{a}_2)$ to mimic misspecification.

1. All correct: all bridge functions are estimated from the true training sample.
2. $\mathcal{M}_0$ correct: $\mathcal{Q}_{11}(a_1)$ and $\mathcal{Q}_{22}(\bar{a}_2)$ are estimated from the true training sample, while pseudo bridge functions $\mathcal{H}^{\dagger}_{22}(\bar{a}_2), \mathcal{H}^{\dagger}_{21}(\bar{a}_2)$ are used as the estimators of $\mathcal{H}_{22}(\bar{a}_2), \mathcal{H}_{21}(\bar{a}_2)$.
3. $\mathcal{M}_1$ correct: $\mathcal{Q}_{11}(a_1)$ and $\mathcal{H}_{22}(\bar{a}_2)$ are estimated from the true training sample, while pseudo bridge functions $\mathcal{H}^{\dagger}_{21}(\bar{a}_2)$, $\mathcal{Q}^{\dagger}_{22}(\bar{a}_2)$ are used as the estimators of $\mathcal{H}_{21}(\bar{a}_2)$, $\mathcal{Q}_{22}(\bar{a}_2)$.
4. $\mathcal{M}_2$ correct: $\mathcal{H}_{22}(\bar{a}_2)$ and $\mathcal{H}_{21}(\bar{a}_2)$ are estimated from the true training sample, while pseudo bridge functions $\mathcal{Q}^{\dagger}_{11}(a_1)$, $\mathcal{Q}^{\dagger}_{22}(\bar{a}_2)$ are used as the estimators of $\mathcal{Q}_{11}(a_1)$, $\mathcal{Q}_{22}(\bar{a}_2)$.
5. All wrong: pseudo bridge functions $\mathcal{H}^{\dagger}_{22}(\bar{a}_2), \mathcal{H}^{\dagger}_{21}(\bar{a}_2)$, $\mathcal{Q}^{\dagger}_{11}(a_1)$, and $\mathcal{Q}^{\dagger}_{22}(\bar{a}_2)$ are used as the estimators of $\mathcal{H}_{22}(\bar{a}_2), \mathcal{H}_{21}(\bar{a}_2)$, $\mathcal{Q}_{11}(a_1)$, and $\mathcal{Q}_{22}(\bar{a}_2)$.

After obtaining the estimated bridge functions, we estimate optimal linear DTRs based on POR, PHA, PIPW, PMR, SRA, and Oracle via value maximization. A common assessment metric used to evaluate the performance of estimated regimes is to calculate their corresponding regrets $V(\bar{d}_2^*) - V(\hat{\bar{d}}_2)$. However, due to the finite number of combinations of categorical variables, there may be equivalence classes among different sequences of regimes. This fact may lead to the coincidence that a sequence of estimated regimes has low regret even if it is estimated by maximizing an incorrectly estimated value function. To maintain accuracy of evaluation in such cases, we compute an additional evaluation metric, the overall error $|V(\bar{d}_2^*) - \hat{V}(\hat{\bar{d}}_2)|$. This is done to evaluate whether the estimated DTRs are obtained based on an accurate estimation of the decision values. We will use both assessment metrics to evaluate the performance and verify the robustness of estimators conducted by POR, PHA, PIPW, PMR, SRA, and Oracle in the following sections.

## 6.2 Numerical Results of Value Maximization

The optimal linear DTRs are estimated through value maximization using a linear programming approach, which will be elaborated in Section K.3 of the Supplementary Material. The optimization is performed with the solver provided by Gurobi. Any element $(d_1, d_2)$ of the linear regime function class $\mathcal{D}_L$ for two-stage DTRs with parameters $(\theta_1, \theta_2)$ can be denoted as:

$$d_1(y_0; \theta_1) = I(\theta_{10} + \theta_{11} y_0 > 0),$$

$$d_2(y_0, y_1, a_1; \theta_2) = I(\theta_{20} + \theta_{21} y_0 + \theta_{22} y_1 + \theta_{23} a_1 > 0),$$

where $\theta_{10}$ and $\theta_{20}$ are additionally introduced intercept terms, and $\theta_1 = (\theta_{10}, \theta_{11})^T$ along with $\theta_2 = (\theta_{20}, \theta_{21}, \theta_{22}, \theta_{23})^T$ respectively have $L_2$ norms of 1. From the true data-generating distribution, we can calculate the true value function $V(\bar{d}_2), \bar{d}_2 \in \mathcal{D}_L$, and obtain that the true optimal value $V(\bar{d}_2^*)$ for linear DTRs is 0.4414.

Let $\epsilon$ denote $10^{-10}$, and each number that smaller than $10^{-10}$ will be denoted as $< \epsilon$. Table 1 presents the calculated regrets for linear DTRs estimated by POR, PHA, PIPW, and POR via value maximization under each scenario. We also report the standard error of each result in the bracket behind, along with the corresponding Root Mean Squared Error (RMSE) in the lower part of the table. The minimum regret, overall error, and RMSE within each scenario are highlighted in bold. Here, the SRA and Oracle estimators yield identical results across all five scenarios, as they do not rely on bridge functions. For the SRA estimator, its regret is 0.0029, the standard error is $< \epsilon$, and the RMSE is 0.0029; For the Oracle estimator, its regret, standard error, and RMSE are all $< \epsilon$. We next present the overall errors of POR, PHA, PIPW, and PMR estimators based on value maximization for linear DTRs in the Table 2. For the SRA estimator, its overall error is 0.2190, the standard error is 0.0001, and the RMSE is 0.2190; For the Oracle estimator, its overall error is 0.0049, the standard error is 0.0002, and the RMSE is 0.0065.

Table 1: Regret of linear DTRs estimated by value maximization

| Model | | | Regret (n = 35000, repetition = 500) | | | |
|---|---|---|---|---|---|---|
| $\mathcal{M}_0$ | $\mathcal{M}_1$ | $\mathcal{M}_2$ | *POR* | *PHA* | *PIPW* | *PMR* |
| Regret | | | | | | |
| ✓ | ✓ | ✓ | $\boldsymbol{< \epsilon\ (< \epsilon)}$ | $\boldsymbol{< \epsilon\ (< \epsilon)}$ | 0.0001 (0.00007) | 0.0001 (0.00005) |
| ✓ | × | × | $\boldsymbol{< \epsilon\ (< \epsilon)}$ | $\boldsymbol{< \epsilon\ (< \epsilon)}$ | 0.1185 ($< \epsilon$) | 0.0005 (0.0001) |
| × | ✓ | × | 0.2865 ($< \epsilon$) | $\boldsymbol{< \epsilon\ (< \epsilon)}$ | 0.1185 ($< \epsilon$) | 0.0004 (0.0001) |
| × | × | ✓ | 0.2865 ($< \epsilon$) | 0.2172 ($< \epsilon$) | **0.0001 (0.00007)** | 0.0005 (0.0001) |
| × | × | × | 0.2865 ($< \epsilon$) | 0.1575 (0.0003) | **0.1185 ($\boldsymbol{< \epsilon}$)** | 0.2510 (0.002) |
| RMSE | | | | | | |
| ✓ | ✓ | ✓ | $\boldsymbol{< \epsilon}$ | $\boldsymbol{< \epsilon}$ | 0.0016 | 0.0012 |
| ✓ | × | × | $\boldsymbol{< \epsilon}$ | $\boldsymbol{< \epsilon}$ | 0.1185 | 0.0030 |
| × | ✓ | × | 0.2865 | $\boldsymbol{< \epsilon}$ | 0.1185 | 0.0029 |
| × | × | ✓ | 0.2865 | 0.2172 | **0.0016** | 0.0031 |
| × | × | × | 0.2865 | 0.1577 | **0.1185** | 0.2541 |

From Tables 1 and 2, we can see that: POR and PIPW estimators only have small regrets and overall errors when $\mathcal{M}_0$ and $\mathcal{M}_2$ are correct, respectively. The PHA estimator yields small regret when either $\mathcal{M}_0$ or $\mathcal{M}_1$ is correct, but small overall error only when $\mathcal{M}_1$ is correct. We suggest that this coincidence is due to the equivalence across different classes of regimes. As predicted by Theorem 4.2, the PMR estimator yields small regret and overall errors when one, but not necessarily more than one, of the models $\mathcal{M}_0, \mathcal{M}_1, \mathcal{M}_2$ is correct. We also note that the overall errors of the PMR estimator are the smallest across the first three scenarios, which proves its efficiency and robustness. All of the POR, PHA, PIPW, and PMR estimators perform significantly better than the SRA estimator when their corresponding models are correct, illustrating the effectiveness of our methods in the presence of unmeasured confounding.

Table 2: Overall error of linear DTRs estimated by value maximization

| Model | | | Overall Error (n = 35000, repetition = 500) | | | |
|---|---|---|---|---|---|---|
| $\mathcal{M}_0$ | $\mathcal{M}_1$ | $\mathcal{M}_2$ | *POR* | *PHA* | *PIPW* | *PMR* |
| Overall Error | | | | | | |
| ✓ | ✓ | ✓ | 0.0010 (0.0003) | 0.0007 (0.0003) | 0.0024 (0.0003) | **0.0002 (0.0014)** |
| ✓ | × | × | 0.0010 (0.0003) | 0.7353 (0.0011) | 1.0309 (0.0025) | **0.0005 (0.0004)** |
| × | ✓ | × | 0.2942 (0.00003) | 0.0007 (0.0003) | 1.0309 (0.0025) | **0.0003 (0.0004)** |
| × | × | ✓ | 0.2942 (0.00003) | 1.5204 (0.0026) | **0.0024 (0.0003)** | 0.0033 (0.0005) |
| × | × | × | **0.2942 (0.00003)** | 1.4659 (0.0041) | 1.0309 (0.0025) | 0.3337 (0.0337) |
| RMSE | | | | | | |
| ✓ | ✓ | ✓ | **0.0057** | 0.0058 | 0.0075 | 0.0310 |
| ✓ | × | × | **0.0057** | 0.7357 | 1.0325 | 0.0086 |
| × | ✓ | × | 0.2942 | **0.0058** | 1.0325 | 0.0087 |
| × | × | ✓ | 0.2942 | 1.5215 | **0.0075** | 0.0111 |
| × | × | × | **0.2942** | 1.4688 | 1.0325 | 0.8249 |

# 7 Discussion

Based on the longitudinal proximal causal inference framework, Zhang & Tchetgen Tchetgen (2024) propose two identification approaches for optimal $K$-stage DTRs when unmeasured confounders are present. However, since these methods necessitate solving a system consisting of $K$ nested first-kind Fredholm integral equations, either recursively or iteratively, the resulting identification tends not to be robust or efficient.

In this article, we first establish two identification methods that correspondingly generalize the two approaches in Zhang & Tchetgen Tchetgen (2024) by relaxing the additional assumption, respectively. Then, we propose a novel Proximal Hybrid Augmentation (PHA) identification method for optimal DTRs. By distributing the burden of solving nested integral equations between two types of bridge functions, this strategy significantly enhances both the flexibility and the robustness of identification methods for optimal DTRs. Furthermore, we derive a semiparametric efficiency bound for the value function $V(\bar{d}_K)$ of an arbitrary given sequence of regimes $\bar{d}_K$ and provide the closed-form expression for its associated efficient influence function. Building on these, we obtain a class of Proximal Multiply Robust (PMR) estimators for the optimal DTRs and establish their corresponding asymptotic properties. For regimes with application scenarios focusing on interpretablity, such as linear regimes, we establish a $n^{1/3}$ convergence rate for the regime parameters, a convergence rate for the regret that is strictly faster than $n^{-1/2}$, and $\sqrt{n}$-consistent asymptotic normality for $\hat{V}_{PMR}(\hat{\bar{d}}_K)$, without requiring $\sqrt{n}$-consistent estimation of each nuisance parameter. The efficiency and robustness of our methods are both verified by numerical experiments.

Our methodology can be extended in several promising directions. First, it can be readily applied to identifying and estimating optimal DTRs that target maximizing other distributional parameters of $\bar{Y}_K(\bar{d}_K)$ when unmeasured confounders are present. Examples include maximizing the multistage outcome quantile (Wang et al. 2018) and the restricted mean survival time (Royston & Parmar 2011).

Second, our methods can be extended to settings involving both unmeasured confounders and conditionally independent censored outcomes. This includes the single-stage proximal survival analysis scenarios studied by Ying et al. (2022), as well as multi-stage survival analysis problems, such as the inventory-sales problem discussed by Bian et al. (2024), which will see broad applications in medical and management science contexts.

Third, while proximal causal inference relaxes the traditional sequential randomization assumption, it still relies on certain untestable assumptions, such as conditional independence requirements. In more general settings where these assumptions may not hold, for example, when invalid proxies are present, recent methodological advances, such as Yu et al. (2025), offer effective solutions. Adopting such methods could further enhance the robustness of our approach to complex scenarios and expand the range of its practical applications.

Finally, estimating bridge functions, especially the recursive estimation of the outcome confounding bridge functions, remains a significant challenge. Building on the semiparametric efficient influence function derived in Section 4.1, as well as the CGMM and CGEL methods discussed in Sections B and C of the Supplementary Material, future research can explore the development of stable and efficient machine learning techniques for learning bridge functions. This fruitful avenue will broaden the range of applicability and enhance the robustness of our proposed methods.

# Supplement To "On Multiple Robustness of Proximal Dynamic Treatment Regimes"

### Abstract

In this supplementary material, we provide parallel results for optimal DTRs identified and estimated by Q-learning and the multiply robust identification of joint density, which can be readily applied to a broad range of new occasions for optimal DTRs estimation. We also depict detailed proofs of the main results presented in the paper, along with additional discussions and explanations for interested readers. Moreover, we supply complementary details omitted from the main text to ensure completeness and transparency. In Section A, we depict the multiply robust identification and estimation results of optimal DTRs by Q-learning. We also establish the parallel asymptotic results for the estimated optimal DTRs by Q-learning with those of estimated optimal DTRs by value maximization in the main paper. In Section B, we supplement the details of parametric estimation for outcome and treatment bridge functions that we omit in the main paper. In Section C, we introduce an algorithm for estimating outcome confounding bridge functions using CGMM. In Section D, we present the proofs of the main results. In Section E, we discuss the conditions necessary to ensure the existence and uniqueness of bridge functions. In Section F, we supply a further introduction to the completeness assumptions we adopted frequently in the main paper. In Section G, we offer further discussion on the lagged proxy orthogonality Assumption 5. In Section H, we examine the relationships between the proposed multiply robust estimator and previous related work. In Section I, we investigate the identification of optimal dynamic treatment regimes within an extended decision space. In Section J, we provide additional details and exploration regarding outcome confounding bridge functions. In Section K, we include supplementary details about the simulation studies.

# A Optimal DTRs Identified and Estimated by Q-Learning

In this section, we will present extra results for optimal DTRs identified and estimated by Q-learning. Unlike estimating optimal DTRs via value maximization, Q-learning-related approaches are much easier to implement, which admits their capability to estimate global optimal DTRs.

## A.1 Identifying Optimal DTRs via Q-Learning

We first equivalently define the optimal DTRs $\bar{d}_K^*$ by a stepwise maximization approach through Q-Learning, according to Tsiatis et al. (2019), Chapter 7.2, and Zhang & Tchetgen Tchetgen (2024). For $k = K$:

$$\begin{aligned} Q_K(\bar{y}_{K-1}, \bar{a}_{K-1}; a_K) &= \mathbb{E}[Y_K(\bar{a}_K) | \bar{Y}_{K-1}(\bar{a}_{K-1}) = \bar{y}_{K-1}], \\ d_K^*(\bar{y}_{K-1}, \bar{a}_{K-1}) &= \underset{a_K}{\operatorname{argmax}} \; Q_K(\bar{y}_{K-1}, \bar{a}_{K-1}; a_K). \end{aligned} \tag{1}$$

For $k = 1, 2, \ldots, K-1$:

$$\begin{aligned} Q_k(\bar{y}_{k-1}, \bar{a}_{k-1}; a_k) &= \mathbb{E}\left[\max_{a_{k+1}} \; Q_{k+1}(\bar{Y}_k(\bar{a}_k), \bar{a}_k; a_{k+1}) | \bar{Y}_{k-1}(\bar{a}_{k-1}) = \bar{y}_{k-1}\right], \\ d_k^*(\bar{y}_{k-1}, \bar{a}_{k-1}) &= \underset{a_k}{\operatorname{argmax}} \; Q_k(\bar{y}_{k-1}, \bar{a}_{k-1}; a_k). \end{aligned} \tag{2}$$

Here, the $k^{th}$ Q function $Q_k(\cdot;\cdot)$ takes the historical state-treatment pair $(\bar{y}_{k-1}, \bar{a}_{k-1})$ along with treatment $a_k$ as arguments for $k = 1, 2, \ldots, K$. Though value maximization and Q-learning have long been regarded as two distinct approaches in traditional reinforcement learning and dynamic treatment regime paradigms (Zhao & Laber 2014, Zhao et al. 2015, Sutton & Barto 2018, Tsiatis et al. 2019), we note that the identification problems implied by these approaches can essentially be unified from the density perspective by restating the definition of Q-functions implied by Equations (1)-(2): For $k = K$,

$$\begin{aligned} & Q_K(\bar{y}_{K-1}, \bar{a}_{K-1}; a_K) = \sum_{y_K} y_K f(Y_K(\bar{a}_K) = y_K | \bar{Y}_{K-1}(\bar{a}_{K-1}) = \bar{y}_{K-1}), \\ & \text{for } k = 1, 2, \ldots, K-1, \\ & Q_k(\bar{y}_{k-1}, \bar{a}_{k-1}; a_k) = \sum_{y_k} \max_{a_{k+1}} \; Q_{k+1}(\bar{y}_k, \bar{a}_k; a_{k+1}) f(Y_k(\bar{a}_k) = y_k | \bar{Y}_{k-1}(\bar{a}_{k-1}) = \bar{y}_{k-1}). \end{aligned} \tag{3}$$

Thus, the identification result of the joint density presented in Section 3.1 can be readily applied to Q-learning. Below is another corollary of Theorem 3.2 about the identification of the optimal DTRs via Q-learning:

**Corollary A.1.** *For any given $k \in \{0, 1, \ldots, K\}$, we suppose that Assumptions 1, 2, k-completeness Assumption 3, k-existence Assumption 4, and Assumption 5 at the $k^{th}$ stage hold. Then, the optimal DTRs can be identified by Q-learning:*

*For $s = K$,*

$$Q_K(\bar{y}_{K-1}, \bar{a}_{K-1}; a_K) = \frac{\sum_{y_K} y_K \sum_{\bar{w}_{k+1}} \sum_{\bar{z}_k} \mathcal{H}_{K,k+1}(\bar{a}_K) \mathcal{Q}_{kk}(\bar{a}_k) f(\bar{w}_{k+1}, \bar{z}_k, \bar{y}_k, \bar{a}_k)}{\sum_{y_K} \sum_{\bar{w}_{k+1}} \sum_{\bar{z}_k} \mathcal{H}_{K,k+1}(\bar{a}_K) \mathcal{Q}_{kk}(\bar{a}_k) f(\bar{w}_{k+1}, \bar{z}_k, \bar{y}_k, \bar{a}_k)},$$

$$d^*_K(\bar{y}_{K-1}, \bar{a}_{K-1}) = \operatorname*{argmax}_{a_K} Q_K(\bar{y}_{K-1}, \bar{a}_{K-1}; a_K),$$

*and for $s = K-1, K-2, \ldots, 1$,*

$$Q_s(\bar{y}_{s-1}, \bar{a}_{s-1}; a_s) = \sum_{y_s} \Bigg\{ \max_{a_{s+1}} Q_{s+1}(\bar{y}_s, \bar{a}_s; a_{s+1}) \times \frac{\sum_{y_K, \ldots, y_{s+1}} \sum_{\bar{w}_{k+1}} \sum_{\bar{z}_k} \mathcal{H}_{K,k+1}(\bar{a}_K) \mathcal{Q}_{kk}(\bar{a}_k) f(\bar{w}_{k+1}, \bar{z}_k, \bar{y}_k, \bar{a}_k)}{\sum_{y_K, \ldots, y_s} \sum_{\bar{w}_{k+1}} \sum_{\bar{z}_k} \mathcal{H}_{K,k+1}(\bar{a}_K) \mathcal{Q}_{kk}(\bar{a}_k) f(\bar{w}_{k+1}, \bar{z}_k, \bar{y}_k, \bar{a}_k)} \Bigg\},$$

$$d^*_s(\bar{y}_{s-1}, \bar{a}_{s-1}) = \operatorname*{argmax}_{a_s} Q_s(\bar{y}_{s-1}, \bar{a}_{s-1}; a_s).$$

Corollary A.1 offers $(K+1)$ methods for identifying the optimal DTRs by Q-learning. In particular, the cases where $k = 0$ and $k = K$ in Corollary A.1 are also direct consequences of Theorem 3.1. Q-learning methods typically benefit from standard optimization tools but rely on a stepwise and indirect estimation process that can introduce additional errors and make them highly sensitive to model misspecification (Qian & Murphy 2011, Nahum-Shani et al. 2012, Laber et al. 2014, Tsiatis et al. 2019). This key observation motivates the derivation of the following multiply robust Q-learning method.

## A.2 Multiply Robust Identification of Conditional Joint Density and Q Functions

As we have obtained the multiply robust estimator $\hat{V}_{PMR}(\bar{d}_K)$ for the value function $V(\bar{d}_K)$ in the main paper, one may wonder whether there exist multiply robust identification or estimators for the Q functions as well. According to Equation (3), this hinges on the existence of multiply robust identification of the conditional joint density of potential outcomes, which we will subsequently present. By comparing the form of $\hat{V}_{PMR}(\bar{d}_K)$ (Equation (12)) with the definition of value function (Equation (7)), we first note the following multiply robust identification of the conditional joint density $f(\bar{Y}_K(\bar{a}_K) = \bar{y}_K|Y_0 = y_0)$:

**Proposition A.1.** *Under the semiparametric union model* $\mathcal{M}_{union} = \bigcup_{k=0}^{K} \mathcal{M}_k$, *the conditional joint density* $f(\bar{Y}_K(\bar{a}_K) = \bar{y}_K|Y_0 = y_0)$ *can be identified by:*

$$
\begin{aligned}
&f(\bar{Y}_K(\bar{a}_K) = \bar{y}_K|Y_0 = y_0) \\
&= \sum_{\bar{z}_K} \mathcal{Q}_{KK}(\bar{a}_K) \left( f(\bar{y}_K, \bar{z}_K, \bar{a}_K|y_0) - \sum_{\bar{w}_K} \mathcal{H}_{KK}(\bar{a}_K) f(\bar{w}_K, \bar{z}_K, \bar{y}_{K-1}, \bar{a}_K|y_0) \right) \\
&\quad + \sum_{k=1}^{K-1} \sum_{\bar{z}_k} \Bigg\{ \mathcal{Q}_{kk}(\bar{a}_k) \\
&\qquad \times \left( \sum_{\bar{w}_{k+1}} \mathcal{H}_{K,k+1}(\bar{a}_K) f(\bar{w}_{k+1}, \bar{z}_k, \bar{y}_k, \bar{a}_k|y_0) - \sum_{\bar{w}_k} \mathcal{H}_{Kk}(\bar{a}_K) f(\bar{w}_k, \bar{z}_k, \bar{y}_{k-1}, \bar{a}_k|y_0) \right) \Bigg\} \\
&\quad + \sum_{w_1} \mathcal{H}_{K1}(\bar{a}_K) f(w_1|y_0).
\end{aligned} \tag{4}
$$

This identification result is multiply robust in that, as long as any one of the submodels $\mathcal{M}_k$, $k = 0, 1, \ldots, K$ (but without the necessity to know which one) is correct, Equation (4) remains valid. It essentially implies the multiply robust identification of the conditional joint distribution of $\bar{Y}_K(\bar{d}_K)$ under the longitudinal proximal causal inference framework.

We use $\bar{\mathcal{H}}_{KK}$ and $\bar{\mathcal{Q}}_{KK}$ to denote the sequences of outcome and treatment confounding bridge functions $\mathcal{Q}_{tt}(\bar{a}_t)$ and $\mathcal{H}_{Kl}(\bar{a}_K)$ for $l, t = 1, 2, \ldots, K$. According to Proposition A.1, we have the following multiply robust identification and corresponding estimation methods for the Q functions $Q_k(\bar{y}_{k-1}, \bar{a}_{k-1}; a_k)$, $k = 1, 2, \ldots, K$:

**Corollary A.2.** *Under the semiparametric union model* $\mathcal{M}_{union} = \bigcup_{k=0}^{K} \mathcal{M}_k$*, if we denote the multiply robust identification form of the conditional joint density* $f(\bar{Y}_K(\bar{a}_K) = \bar{y}_K | Y_0 = y_0)$ *presented in Equation (4) as* $g(\bar{y}_K, \bar{a}_K; \bar{\mathcal{H}}_{KK}, \bar{\mathcal{Q}}_{KK})$*, Q functions* $Q_k(\bar{y}_{k-1}, \bar{a}_{k-1}; a_k)$*,* $k = 1, 2, \ldots, K$ *can then be identified by:*

*For* $k = K$*,*

$$Q_K(\bar{y}_{K-1}, \bar{a}_{K-1}; a_K) = \frac{\sum_{y_K} y_K \cdot g(\bar{y}_K, \bar{a}_K; \bar{\mathcal{H}}_{KK}, \bar{\mathcal{Q}}_{KK})}{\sum_{y_K} g(\bar{y}_K, \bar{a}_K; \bar{\mathcal{H}}_{KK}, \bar{\mathcal{Q}}_{KK})}, \tag{5}$$

*and for* $k = K-1, \ldots, 1$*,*

$$Q_k(\bar{y}_{k-1}, \bar{a}_{k-1}; a_k) = \sum_{y_k} \max_{a_{k+1}} Q_{k+1}(\bar{y}_k, \bar{a}_k; a_{k+1}) \frac{\sum_{y_K, \ldots, y_{k+1}} g(\bar{y}_K, \bar{a}_K; \bar{\mathcal{H}}_{KK}, \bar{\mathcal{Q}}_{KK})}{\sum_{y_K, \ldots, y_k} g(\bar{y}_K, \bar{a}_K; \bar{\mathcal{H}}_{KK}, \bar{\mathcal{Q}}_{KK})}. \tag{6}$$

**Corollary A.3.** *Under the semiparametric union model* $\mathcal{M}_{union} = \bigcup_{k=0}^{K} \mathcal{M}_k$*, suppose that we have sequentially estimated* $\mathcal{H}_{Kl}(\bar{a}_K; \hat{\beta}_l)$ *and* $\mathcal{Q}_{tt}(\bar{a}_t; \hat{\gamma}_t)$*,* $l, t = 1, 2, \ldots, K$ *by methods mentioned in Section 4.2. Besides, we also assume that we have consistently estimated joint densities* $f(\bar{y}_K, \bar{z}_K, \bar{a}_K)$ *and* $f(\bar{w}_k, \bar{z}_k, \bar{y}_{k-1}, \bar{a}_k)$ *as* $\hat{f}(\bar{y}_K, \bar{z}_K, \bar{a}_K)$ *and* $\hat{f}(\bar{w}_k, \bar{z}_k, \bar{y}_{k-1}, \bar{a}_k)$*, respectively, for* $k = 1, 2, \ldots, K$*.*

*Let* $\hat{g}(\bar{y}_K, \bar{a}_K; \hat{\bar{\gamma}}_K, \hat{\bar{\beta}}_K)$ *be the plug-in estimator for* $g(\bar{y}_K, \bar{a}_K; \bar{\mathcal{H}}_{KK}, \bar{\mathcal{Q}}_{KK})$ *with estimated joint densities and bridge functions, the following estimators for Q functions are consistent over the union model* $\mathcal{M}_{union} = \bigcup_{k=0}^{K} \mathcal{M}_k$*:*

*For* $k = K$*,*

$$\hat{Q}_{K,PMR}(\bar{y}_{K-1}, \bar{a}_{K-1}; a_K) = \frac{\sum_{y_K} y_K \cdot \hat{g}(\bar{y}_K, \bar{a}_K; \hat{\bar{\gamma}}_K, \hat{\bar{\beta}}_K)}{\sum_{y_K} \hat{g}(\bar{y}_K, \bar{a}_K; \hat{\bar{\gamma}}_K, \hat{\bar{\beta}}_K)}, \tag{7}$$

*and for* $k = K-1, \ldots, 1$*,*

$$\hat{Q}_{k,PMR}(\bar{y}_{k-1}, \bar{a}_{k-1}; a_k) = \sum_{y_k} \max_{a_{k+1}} \hat{Q}_{k+1,PMR}(\bar{y}_k, \bar{a}_k; a_{k+1}) \frac{\sum_{y_K, \ldots, y_{k+1}} \hat{g}(\bar{y}_K, \bar{a}_K; \hat{\bar{\gamma}}_K, \hat{\bar{\beta}}_K)}{\sum_{y_K, \ldots, y_k} \hat{g}(\bar{y}_K, \bar{a}_K; \hat{\bar{\gamma}}_K, \hat{\bar{\beta}}_K)}. \tag{8}$$

## A.3 Theoretical Results for Consistency of Regime Estimation

### A.3.1 Conditions

We denote by $\hat{h}_{Kl}$ and $\hat{q}_{tt}$ the estimators for $h_{Kl}$ and $q_{tt}$, respectively. Below are the conditions required to establish the asymptotic results in the main paper.

*Condition* 1.

1. The support of $\bar{Y}_K$ is bounded.
2. The outcome and treatment confounding functions $\mathcal{H}_{Kl}(\bar{a}_K)$ and $\mathcal{Q}_{tt}(\bar{a}_t)$ serving as solutions to Equations (1)-(4) are smooth and bounded, for $l, t = 1, 2, \dots, K$. Moreover, one of the following conditions holds:
   (a) $\mathcal{H}_{Kl}(\bar{a}_K) \geq 0$, for $l = K-1, \dots, 1$.
   (b) For each index $l$, $K-1 \geq l \geq 1$, there exist a positive constant $c_l$, such that for every $(\bar{y}_K, \bar{a}_K)$, $f(Y_K(\bar{a}_K) = y_K, \dots, Y_l(\bar{a}_l) = y_l | \bar{a}_{l-1}, \bar{y}_{l-1}) \geq c_l > 0$.
3. The value function $V(\theta)$ is twice continuously differentiable in a neighborhood of $\theta^*$.
4. $\theta^*$ is unique and lies at interior points of the compact set $\Theta$.

*Condition* 2. For $k = 1, 2, \dots, K$ and any prespecified tretament sequence $\bar{a}_K$,

1. $P(|(\bar{Y}_{k-1}(\bar{a}_{k-1}), \bar{d}^*_{k-1}(\bar{Y}_{k-2}(\bar{a}_{k-2}))\theta^*_k| \leq \delta) \leq c_k \delta$.
2. $\left\| \sum_{y_K, \dots, y_k} y_K \left[ \hat{h}_{Kk}(y_K, \cdots, y_k, \bar{Y}_{k-1}, \bar{W}_k, \bar{a}_K) - h_{Kk}(y_K, \cdots, y_k, \bar{Y}_{k-1}, \bar{W}_k, \bar{a}_K) \right] \right\|_{L_2} = o_p(n^{-1/4})$, and $\left\| \hat{q}_{kk}(\bar{Y}_{k-1}, \bar{Z}_k, \bar{a}_k) - q_{kk}(\bar{Y}_{k-1}, \bar{Z}_k, \bar{a}_k) \right\|_{L_2} = o_p(n^{-1/4})$. Here, we use $\| \cdot \|_{L_2}$ to denote the $L_2$ norm.

Conditions 1.1, 1.2, and 1.3 are standard regularity conditions to guarantee the uniform convergence (Van der Vaart 2000). Condition 1.4 is common in literature (Van der Vaart 2000, Delsol & Van Keilegom 2020, Zhou et al. 2024), ensuring the identifiability or uniqueness of parameters within the optimal regimes $\bar{d}^*_K$. Condition 2.1, also known as the margin condition, is frequently used in research about classification (Tsybakov 2004), statistical learning theory (Koltchinskii 2006, Steinwart & Scovel 2007), and optimal treatment assignment (Qian & Murphy 2011, Zhao & Cui 2025). The general form of this class of conditions is $P(|(\bar{Y}_{k-1}(\bar{d}^*_{k-1}), \bar{d}^*_{k-1})\theta^*_k| \leq$ $\delta) \leq (c_k\delta)^b$, where $b$ controls the rate of $P(|(\bar{Y}_{k-1}(\bar{d}^*_{k-1}), \bar{d}^*_{k-1})\theta^*_k| \leq \delta)$ approaching zero. As discussed in Zhao & Cui (2025), when $b = 0$, this condition is trivial for positing no effective restriction; When $b = \infty$, it bounds $(\bar{Y}_{k-1}(\bar{d}^*_{k-1}), \bar{d}^*_{k-1})\theta^*_k$ away from zero. We exploit this condition with $b = 1$ here to ensure the regularity of optimal regimes $\bar{d}^*_K$ as well as the fast convergence rate. Condition 2.1 generally holds as long as $(\bar{Y}_{k-1}(\bar{d}^*_{k-1}), \bar{d}^*_{k-1})\theta^*_k$ is continuously distributed with bounded density. Condition 2.2 restricts that for $l, t = 1, 2, \ldots, K$, the nuisance parameters $\mathcal{H}_{Kl}(\bar{a}_K)$ and $\mathcal{Q}_{tt}(\bar{a}_t)$ must be consistently estimated with convergence rate faster than $n^{-1/4}$. For cases with categorical outcomes, as shown in Zhang & Tchetgen Tchetgen (2024), this can be achieved by numerous existing machine learning methods (Mastouri et al. 2021, Ghassami et al. 2022, Kallus et al. 2022, Kompa et al. 2022), under certain nonparametric regularity assumptions.

### A.3.2 Asymptotic Results for Optimal DTRs Estimated by Q-Learning

We first denote the estimated optimal DTRs by Q-learning via the multiply robust Q-functions $\hat{Q}_{k,PMR}(\bar{y}_{k-1}, \bar{a}_{k-1}; a_k)$ as:

$$\begin{aligned}\hat{d}_{k,Q} &= \operatorname*{argmax}_{a_k} \hat{Q}_{k,PMR}(\bar{y}_{k-1}, \bar{a}_{k-1}; a_k) \\ &= I(\hat{Q}_{k,PMR}(\bar{y}_{k-1}, \bar{a}_{k-1}; 1) > \hat{Q}_{k,PMR}(\bar{y}_{k-1}, \bar{a}_{k-1}; 0)), \text{ for } k = 1, 2, \ldots, K.\end{aligned} \tag{9}$$

Different from value maximization, Q-learning cuts down optimization complexity by estimating optimal DTRs through backward induction. As a result, through Q-learning, the problem of estimating optimal DTRs can be transformed into the stage-wise estimation of the Q-functions, making it more likely to be used in practice for estimating globally optimal DTRs. This feature makes this approach especially appealing when the performance of estimated regimes is highlighted. Under some mild conditions, we establish the following convergence results for the optimal DTRs obtained via Q-learning as well:

**Theorem A.1.** *Suppose that Assumptions 1, 2, $0$-completeness (or $K$-completeness) Assumption 3, and $k$-existence Assumption 4 with $k = 0$ and $K$ hold simultaneously. We also assume that Condtions 1.1, 1.2 and 2.2 hold, and $f(\bar{y}_K, \bar{z}_K, \bar{a}_K)$ and $f(\bar{w}_k, \bar{z}_k, \bar{y}_{k-1}, \bar{a}_k)$, $k = 1, 2, \ldots, K$ are consistently estimated. Let "$\xrightarrow{p}$" denote convergence in probability, we have:*

1. $\mathbb{P}_n\{\max_{a_1} \hat{Q}_{1,PMR}(Y_0; a_1)\} \xrightarrow{p} V(\bar{d}^*_{K,Q})$;
2. $V(\hat{\bar{d}}_{K,Q}) \xrightarrow{p} V(\bar{d}^*_{K,Q})$.

## A.4 Numerical Results of Q-Learning

Although the optimal linear DTRs estimated using the value maximization approach have several strong theoretical properties, they are generally not globally optimal. Therefore, more general optimal DTRs estimated based on Q-learning are highly appreciated in scenarios where the performance of the estimated DTRs is of primary concern. We calculate from the true data-generating distribution that the global optimal value $V(\bar{d}^*_2)$ is 0.4535, which is slightly larger than that of optimal linear DTRs. The estimation procedure of optimal DTRs via Q-learning follows Corollary A.3 and Equation (9). We report our simulation results in Tables 1 and 2, with the same format as Tables 1 and 2. Table 1 presents the regret for DTRs estimated by POR,

Table 1: Regret of DTRs estimated by Q-learning

| Model | | | Regret (n = 35000, repetition = 500) | | | |
|---|---|---|---|---|---|---|
| $\mathcal{M}_0$ | $\mathcal{M}_1$ | $\mathcal{M}_2$ | *POR* | *PHA* | *PIPW* | *PMR* |
| Regret | | | | | | |
| ✓ | ✓ | ✓ | **0.0001 (0.00004)** | 0.0001 (0.00005) | 0.0003 (0.00009) | 0.0002 (0.00007) |
| ✓ | × | × | **0.0001 (0.00004)** | 0.0121 ($<\epsilon$) | 0.1306 ($<\epsilon$) | 0.0012 (0.0002) |
| × | ✓ | × | 0.2306 ($<\epsilon$) | **0.0001 (0.00005)** | 0.1306 ($<\epsilon$) | 0.0023 (0.0003) |
| × | × | ✓ | 0.2306 ($<\epsilon$) | 0.2293 ($<\epsilon$) | **0.0003 (0.00009)** | 0.0060 (0.0003) |
| × | × | × | 0.2306 ($<\epsilon$) | 0.1710 ($<\epsilon$) | **0.1306** ($<\epsilon$) | 0.1917 ($<\epsilon$) |
| RMSE | | | | | | |
| ✓ | ✓ | ✓ | **0.0009** | 0.0011 | 0.0019 | 0.0015 |
| ✓ | × | × | **0.0009** | 0.0121 | 0.1306 | 0.0046 |
| × | ✓ | × | 0.2306 | **0.0011** | 0.1306 | 0.0063 |
| × | × | ✓ | 0.2306 | 0.2293 | **0.0019** | 0.0096 |
| × | × | × | 0.2306 | 0.1710 | **0.1306** | 0.1917 |

PHA, PIPW, and PMR based on Q-learning. For the SRA estimator, its regret is 0.0751, the standard error is $< \epsilon$, and the RMSE is 0.0751; For the Oracle estimator, its regret, standard error, and RMSE are all $< \epsilon$. Table 2 presents the overall errors of Q-learning estimators based on POR, PHA, PIPW, and PMR. For the SRA estimator, its overall error is 0.1663, the standard error is 0.0002, and the RMSE is 0.1664; For the Oracle estimator, its overall error is 0.0006, the standard error is 0.0002, and the RMSE is 0.0044.

Table 2: Overall error of DTRs estimated by Q-learning

| Model | | | Overall Error (n = 35000, repetition = 500) | | | |
|---|---|---|---|---|---|---|
| $\mathcal{M}_0$ | $\mathcal{M}_1$ | $\mathcal{M}_2$ | *POR* | *PHA* | *PIPW* | *PMR* |
| Overall Error | | | | | | |
| ✓ | ✓ | ✓ | 0.0007 (0.0003) | **0.0006 (0.0003)** | 0.0013 (0.0003) | 0.0010 (0.0015) |
| ✓ | × | × | **0.0007 (0.0003)** | 0.7232 (0.0011) | 1.0187 (0.0025) | 0.0011 (0.0004) |
| × | ✓ | × | 0.3378 (0.00008) | **0.0006 (0.0003)** | 1.0188 (0.0025) | 0.0006 (0.0004) |
| × | × | ✓ | 0.3378 (0.00008) | 1.5094 (0.0026) | **0.0013 (0.0003)** | 0.0241 (0.0025) |
| × | × | × | **0.3378 (0.00008)** | 1.4651 (0.0029) | 1.0187 (0.0025) | 12.4728 (0.1162) |
| RMSE | | | | | | |
| ✓ | ✓ | ✓ | **0.0056** | 0.0057 | 0.0065 | 0.0330 |
| ✓ | × | × | **0.0056** | 0.7236 | 1.0202 | 0.0093 |
| × | ✓ | × | 0.3378 | **0.0057** | 1.0204 | 0.0098 |
| × | × | ✓ | 0.3378 | 1.5105 | **0.0065** | 0.0614 |
| × | × | × | **0.3378** | 1.4665 | 1.0202 | 12.7408 |

Similar conclusions as Section 6.2 can be drawn from Tables 1 and 2. When the corresponding models are correctly specified, POR, PHA, PIPW, and PMR estimators all demonstrate significantly better performance than the SRA estimator and are very close to the Oracle estimator, indicating the effectiveness of our methods when unmeasured confounding is present. Specifically, POR, PHA, and PIPW estimators achieve small regrets and overall errors only when $\mathcal{M}_0$, $\mathcal{M}_1$, and $\mathcal{M}_2$ are valid, respectively. Notably, the PMR estimator stands out by producing low regret and overall error whenever any single model among $\mathcal{M}_0$, $\mathcal{M}_1, \mathcal{M}_2$ is correct, which is consistent with Corollary A.3. Such robustness demonstrates the effectiveness of the PMR estimator in addressing various model misspecifications and reliably recovering the true optimal DTRs in intricate situations.

# B Estimation of Parameters

In this section, we will present the estimation details of parameters of bridge functions $\mathcal{H}_{Kl}$ and $\mathcal{Q}_{tt}$ for $l, t = 1, 2, \ldots, K$, respectively. For $l, t = 1, 2, \ldots, K$, let $h_{Kl}(\bar{y}_K, \bar{w}_l, \bar{a}_K; \beta_l)$ and $q_{tt}(\bar{y}_{t-1}, \bar{z}_t, \bar{a}_t; \gamma_t)$ respectively denote the parametric models of $h_{Kl}(\bar{y}_K, \bar{w}_l, \bar{a}_K)$ and $q_{tt}(\bar{y}_{t-1}, \bar{z}_t, \bar{a}_t)$, which are respectively indexed by finite dimensional parameters $\beta_l$ and $\gamma_t$, for $l, t = 1, 2, \ldots, K$.

Deriving estimators for $\beta_l$ is very challenging. Zhang & Tchetgen Tchetgen (2024) suggest the following estimation equations for the case where all of the outcome variables are of finite categories:

For $l = K$,

$$\mathbb{P}_n\Big\{ \Big[I(Y_K = y_K) - h_{KK}(y_K, \bar{Y}_{K-1}, \bar{W}_K, \bar{A}_K; \beta_K)\Big]\, m_K(\bar{Y}_{K-1}, \bar{A}_K, \bar{Z}_K)\Big\} = 0, \tag{10}$$

and for $l = K-1, \ldots, 1$ as well as each $(a_K, \ldots, a_{l+1}) \in (\mathcal{A}_K, \ldots, \mathcal{A}_{l+1})$,

$$\begin{aligned}\mathbb{P}_n\Big\{ \Big[&I(Y_l = y_l) h_{K,l+1}(y_K, \cdots, y_{l+1}, \bar{Y}_l, \bar{W}_{l+1}, a_K, \cdots, a_{l+1}, \bar{A}_l; \beta_{l+1}) \\ &-h_{K,l}(y_K, \cdots, y_l, \bar{Y}_{l-1}, \bar{W}_l, a_K, \cdots, a_{l+1}, \bar{A}_l; \beta_l)\Big]\, m_l(\bar{Y}_{l-1}, \bar{A}_l, \bar{Z}_l)\Big\} = 0.\end{aligned} \tag{11}$$

Here, we use $m_l(\bar{y}_{l-1}, \bar{a}_l, \bar{z}_l)$, which is of the same dimension as $\beta_l$, to denote the instrumental functions that respectively transform the conditional moment restrictions into unconditional moment restrictions, for $l = 1, 2, \ldots, K$. For fixed $\bar{a}_K$ and $y_K, \ldots, y_1$, Equations (10) and (11) are ordinary unconditional moment restrictions of a finite number, which can be estimated following the algorithms presented by Ying et al. (2023), Kallus et al. (2022), Ghassami et al. (2022).

For general cases with continuous outcome variables, the conditional densities, such as the $f(y_K|\bar{y}_{K-1}, \bar{a}_K, \bar{z}_K)$ and $f(y_l|\bar{w}_{l+1}, \bar{y}_{l-1}, \bar{z}_l, \bar{a}_l)$, can be respectively estimated from the sample through various techniques, like the kernel methods or maximum likelihood estimation (Chen 2017, Sheather 2004, Kim & Scott 2012, Silverman 2018). By plugging their estimators into

Equations (1) and (2), we get the following empirical conditional moment restrictions:

For $l = K$,

$$\mathbb{P}_n\Big\{h_{KK}(y_K, \bar{Y}_{K-1}, \bar{W}_K, \bar{A}_K; \beta_K) - \hat{f}(y_K|\bar{Y}_{K-1}, \bar{A}_K, \bar{Z}_K)|\bar{Y}_{K-1}, \bar{A}_K, \bar{Z}_K\Big\} = 0, \tag{12}$$

and for $l = K-1, \ldots, 1$,

$$\begin{aligned}\mathbb{P}_n\Big\{&h_{K,l+1}(y_K, \cdots, y_l, \bar{Y}_{l-1}, \bar{W}_{l+1}, a_K, \cdots, a_{l+1}, \bar{A}_l; \beta_{l+1})\hat{f}(y_l|\bar{W}_{l+1}, \bar{Y}_{l-1}, \bar{Z}_l, \bar{A}_l)\\ &- h_{Kl}(y_K, \cdots, y_l, \bar{Y}_{l-1}, \bar{W}_l, a_K, \cdots, a_{l+1}, \bar{A}_l; \beta_l)|\bar{Y}_{l-1}, \bar{Z}_l, \bar{A}_l\Big\} = 0.\end{aligned} \tag{13}$$

Equations (12) and (13) are often referred to as a continuum of conditional moment restrictions (Carrasco & Florens 2000, Carrasco et al. 2007, Chaussé 2011, Carrasco 2012, Carrasco & Kotchoni 2017, Kotchoni & Carrasco 2019), indexed by $y_K, \ldots, y_l$ after fixing $\bar{a}_K$, for $l = 1, 2, \ldots, K$. Then, Generalized Method of Moment for a Continuum (CGMM) (Carrasco & Florens 2000, Carrasco et al. 2007, Kotchoni & Carrasco 2019) or Generalized Empirical Likelihood method for a Continuum (CGEL) (Chaussé 2011) can be used to recursively obtain the estimation of $\beta_l$, for $l = 1, 2, \ldots, K$. The introduction of the corresponding algorithm is deferred to Section C of the Supplementary Material.

The procedure of constructing the estimators for $\gamma_t$ can be similarly specified according to Ying et al. (2023). To make it more concrete, for $t = 1, 2, \ldots, K$, we denote by $n_t(\bar{y}_{t-1}, \bar{w}_t, \bar{a}_t)$ the instrumental functions which are of the same dimensions as $\gamma_t$, and $n_{t,+}(\bar{y}_{t-1}, \bar{w}_t, \bar{a}_{t-1}) = n_t(\bar{y}_{t-1}, \bar{w}_t, \bar{a}_{t-1}, 1) + n_t(\bar{y}_{t-1}, \bar{w}_t, \bar{a}_{t-1}, 0)$. The following iterative estimation procedure can be applied:

For $t = 1$,

$$\mathbb{P}_n\{q_{11}(Y_0, Z_1, A_1; \gamma_1)n_1(Y_0, W_1, A_1) - n_{0,+}(Y_0, W_1)\} = 0, \tag{14}$$

and for $t = 2, 3, \ldots, K$,

$$\begin{aligned}\mathbb{P}_n\{&q_{tt}(\bar{Y}_{t-1}, \bar{Z}_t, \bar{A}_t; \gamma_t)n_t(\bar{Y}_{t-1}, \bar{W}_t, \bar{A}_t)\\ &- q_{t-1,t-1}(\bar{Y}_{t-2}, \bar{Z}_{t-1}, \bar{A}_{t-1}; \gamma_{t-1})n_{t,+}(\bar{Y}_{t-1}, \bar{W}_t, \bar{A}_{t-1})\} = 0.\end{aligned} \tag{15}$$

The efficient choices of $n_t(\bar{y}_{t-1}, \bar{w}_t, \bar{a}_t)$, $t = 1, 2, \ldots, K$ can be found in Ying et al. (2023), Appendix E.1, which we do not further discuss here.

# C Estimating Outcome Confounding Bridge Functions by CGMM

We now discuss the estimation of outcome confounding bridge functions in the setting with general continuous outcome variables. As we have stated in Section 4, after respectively estimating the conditional densities $f(y_K|\bar{y}_{K-1}, \bar{a}_K, \bar{z}_K)$ and $f(y_l|\bar{w}_{l+1}, \bar{y}_{l-1}, \bar{z}_l, \bar{a}_l)$ from the sample and plugging them into Equations (1) and (2), we get the following empirical conditional moment restrictions:

For $l = K$,

$$\mathbb{P}_n\Big\{h_{KK}(y_K, \bar{Y}_{K-1}, \bar{W}_K, \bar{A}_K; \beta_K) - \hat{f}(y_K|\bar{Y}_{K-1}, \bar{A}_K, \bar{Z}_K)|\bar{Y}_{K-1}, \bar{A}_K, \bar{Z}_K\Big\} = 0, \tag{16}$$

and for $l = K-1, \ldots, 1$,

$$\begin{aligned}\mathbb{P}_n\Big\{&h_{K,l+1}(y_K, \cdots, y_l, \bar{Y}_{l-1}, \bar{W}_{l+1}, a_K, \cdots, a_{l+1}, \bar{A}_l; \beta_{l+1})\hat{f}(y_l|\bar{W}_{l+1}, \bar{Y}_{l-1}, \bar{Z}_l, \bar{A}_l)\\ &- h_{Kl}(y_K, \cdots, y_l, \bar{Y}_{l-1}, \bar{W}_l, a_K, \cdots, a_{l+1}, \bar{A}_l; \beta_l)|\bar{Y}_{l-1}, \bar{Z}_l, \bar{A}_l\Big\} \\ = 0.&\end{aligned} \tag{17}$$

In these equations, $a_k$, $k = 1, 2, \ldots, K$ is fixed for they are all binary variables with finite categories. However, $y_0, y_1, \ldots, y_K$ are all continuous variables, which impose essentially infinite conditional moment restrictions for each $l$, $l = 1, 2, \ldots, K$. Estimation problems with moment restrictions of this type, which are often referred to as a continuum of conditional moment restrictions (Carrasco & Florens 2000, Carrasco et al. 2007, Chaussé 2011, Carrasco 2012, Carrasco & Kotchoni 2017, Kotchoni & Carrasco 2019), are known for their non-uniqueness, ill-posedness, and instability. In this section, we will discuss the possibility of applying the Generalized Method of Moment for a Continuum (CGMM), proposed by (Carrasco & Florens 2000, Carrasco et al. 2007, Kotchoni & Carrasco 2019) to recursively handle the estimation of $\beta_l$, for $l = 1, 2, \ldots, K$.

Before introducing the estimation procedure, we start by considering the estimation of the interested parameters with the following finite moment restrictions by the usual GMM:

$$\mathbb{E}[g_t(X, \theta_0)] = 0, \qquad t = \frac{1}{m}, \frac{2}{m}, \ldots, 1.$$

where $g_t$ is a real-valued function. Recall that the usual definition of the over-identified GMM estimator is obtained by a minimization:

$$\hat{\theta}_n = \underset{\theta}{\operatorname{argmin}} \left\| B_n \Big( \frac{1}{n} \sum_{i=1}^{n} g(x_i, \theta) \Big) \right\| \tag{18}$$

$$= \underset{\theta}{\operatorname{argmin}} [\bar{g}_n(\theta)]' A_n [\bar{g}_n(\theta)], \tag{19}$$

where the norm is determined in the Hilbert space and $B_n$ converges to a linear operator $B$; $A_n$ is a given random, positive definite symmetric $m \times m$ matrix related with the empirical operator $B_n$, and $\bar{g}_n(\theta)$ is a $m$-vector with $j$-th element $\bar{g}_{\frac{j}{m}}(\theta) = \frac{1}{n} \sum_{i=1}^{n} g_{\frac{j}{m}}(X_i, \theta)$. Now, assume that the full continuum of moment is available. The empirical counterpart of Equation (19) defines the following GMM estimator:

$$\hat{\theta}_n = \underset{\theta}{\operatorname{argmin}} \int_0^T \int_0^T \bar{g}_t(\theta) a_n(t,s) \bar{g}_s(\theta) dt ds, \tag{20}$$

where $\bar{g}_t(\theta) = \frac{1}{n} \sum_{i=1}^{n} g_t(X_i, \theta)$ and $a_n(t,s)$ converges to $a(t,s)$ characterized by:

$$\|B\varphi\| = \int_0^T \int_0^T \varphi(s) a(t,s) \varphi(t) ds dt.$$

Equation (20) is intuitively the limit of the usual GMM quadratic form depicted in Equation (19) as the interval between observations goes to zero, and the corresponding estimation problem is the estimation task defined by continuous moment restrictions:

$$\mathbb{E}[g_t(X, \theta_0)] = 0, \qquad \forall t \in [0, T].$$

To achieve efficiency, the operator $B$ must be defined as the inverse of the square root of the asymptotic covariance operator of $\sqrt{n}\bar{g}_n(\theta)$, $K$, which is formally defined by:

$$(K\psi)(t) = \int_0^T \mathbb{E}[g_s g_t] \psi(s) ds.$$

Optimal GMM estimation is based on the use of $K^{-1}$, which is the counterpart of the inverse of the covariance matrix in the finite-dimensional framework. However, because $K$ is a compact operator and is not invertible on the full reference space, the objective function is ill-posed because it can be written as $\langle \bar{g}_n(\theta), K^{-1}\bar{g}_n(\theta)\rangle$, where the second term of the inner product is the solution to $Kx = \bar{g}_n(\theta)$. Here, we use $\langle , \rangle$ to denote the inner product between functions defined on $[0, T]$. We have to use a regularized estimator of $K^{-1}$, denoted $(K_n^{\alpha_n})^{-1}$. This operator is constructed in the following way. Let $K_n$ be the estimator of $K$ with $\mathbb{E}[g_s g_t]$ being substituted by $\frac{1}{n}\sum_{i=1}^{n} g_s(X_i; \hat{\theta}_0) g_t(X_i; \hat{\theta}_0)$, where $\hat{\theta}_0$ is a a first stage consistent estimate. $\hat{\theta}_0$ can be obtained by setting $K$ as the identity operator, which can lead to an estimator that is consistent but may not be efficient. The $(K_n^{\alpha_n})^{-1}$ can be defined as:

$$(K_n^{\alpha_n})^{-1} = (\alpha_n I + K_n^2)^{-1} K_n,$$

by the Tikhonov regularization approach Tikhonov (1963), Tikhonov et al. (1995), but we can estimate it in another way. We first estimate the $n$ eigenvalues, $\mu_j^{(n)}$, and eigenfunctions, $\phi_j^{(n)}$, of $K_n$ by solving the functional equation $K_n \phi = \mu\phi$. The eigenvalues $\mu_j^{(n)}$ are perturbed by the smoothing parameter $\alpha_n \in \mathbb{R}^+$ and replaced by $\frac{(\mu_j^{(n)})^2 + \alpha_n}{\mu_j^{(n)}}$. Then, the operator $(K_n^{\alpha_n})^{-1}$ satisfies:

$$((K_n^{\alpha_n})^{-1}\psi)(t) = \sum_{j=1}^{n} \frac{\mu_j^{(n)}}{(\mu_j^{(n)})^2 + \alpha_n} \langle \psi, \phi_j^n \rangle \phi_j^{(n)}(t),$$

and the optimal GMM estimator satisfies the following condition:

$$\begin{aligned}
\hat{\theta}_n &= \arg\min_{\theta} \|(K_n^{\alpha_n})^{-\frac{1}{2}} \bar{g}_n(\theta)\| \\
&= \arg\min_{\theta} \sum_{j=1}^{n} \frac{\mu_j^{(n)}}{\left(\mu_j^{(n)}\right)^2 + \alpha_n} \left\langle \phi_j^{(n)}, \bar{g_n}(\theta) \right\rangle^2 .
\end{aligned}$$

Above, we have introduced the baseline framework of CGMM, which we can now apply to our estimation procedure. By the algorithms introduced by Carrasco & Florens (2000) and Carrasco et al. (2007), we can recursively estimate $\beta_K, \beta_{K-1}, \ldots, \beta_1$ from Equations (16) and (17), following the algorithms we depicted below:

**Algorithm 1** CGMM Estimation for $\bar{\beta}_K$

**Input** the Data Set $\{Y_0, (Z_k, W_k, A_k, Y_k)_{k=1}^K\}$

**Step 1.** Obtain the estimator for the density function $f(y_K|\bar{Y}_{K-1}, \bar{A}_K, \bar{Z}_K)$ (Chen 2017, Sheather 2004, Kim & Scott 2012, Silverman 2018) and denote it as $\hat{f}(y_K|\bar{Y}_{K-1}, \bar{A}_K, \bar{Z}_K)$.

**Step 2.** Transform the conditional moment restriction (16) into unconditional moment restriction by the instrumental function $\exp\{t^T(\bar{Y}_{K-1}, \bar{A}_K, \bar{Z}_K)\}$:

$$\mathbb{P}_n\Big\{\big[h_{KK}(y_K, \bar{Y}_{K-1}, \bar{W}_K, \bar{A}_K; \beta_K) - \hat{f}(y_K|\bar{Y}_{K-1}, \bar{A}_K, \bar{Z}_K)\big] \exp\{t^T(\bar{Y}_{K-1}, \bar{A}_K, \bar{Z}_K)\}\Big\} = 0, \tag{21}$$

and denote the left-hand side of Equation (21) as $\mathbb{P}_n\{g_K(y_K, t; X_K; \beta_K)\}$, where $X_K = (\bar{Y}_{K-1}, \bar{W}_K, \bar{A}_K, \bar{Z}_K)$.

**Step 3.** Obtain the initial estimator $\tilde{\beta}_K$ by:

$$\tilde{\beta}_K = \arg\min_{\beta} \|I\bar{g}_K(y_K, t; X_K; \beta)\|,$$

where $I$ is the identity operator.

**Step 4.** Denote the empirical corresponding covariate operator by $K_{n,K}$, which is defined as:

$$(K_{n,K}\ \psi)(y_K^{(1)}, t^{(1)}) = \int_{\mathcal{Y}_K \times \mathcal{T}} \mathbb{E}[g_K(y_K^{(1)}, t^{(1)}; X_K; \tilde{\beta}_K) g_K(y_K^{(2)}, t^{(2)}; X_K; \tilde{\beta}_K)]\psi(y_K^{(2)}, t^{(2)}) d(y_K^{(2)}, t^{(2)}).$$

Calculate its corresponding eigenvalues and eigenfunctions $(\mu_{K,j}^{(n)}, \phi_{K,j}^{(n)})$, $j = 1, 2, \ldots, n$.

**Step 5.** Estimate $\hat{\beta}_k$ through:

$$\hat{\beta}_K = \arg\min_{\beta} \sum_{j=1}^{n} \frac{\mu_j^{(n)}}{\left(\mu_j^{(n)}\right)^2 + \alpha_n} \left\langle \phi_j^{(n)}, \mathbb{P}_n\{g_K(y_K, t; X_K; \beta)\} \right\rangle^2.$$

**For** $l = K-1, K-2, \ldots, 1$, Repeat **Step 6** to **Step 10**:

**Step 6.** Obtain the estimator for the density function $f(y_l|\bar{W}_{l+1}, \bar{Y}_{l-1}, \bar{A}_l, \bar{Z}_l)$ (Chen 2017, Sheather 2004, Kim & Scott 2012, Silverman 2018) and denote it as $\hat{f}(y_l|\bar{W}_{l+1}, \bar{Y}_{l-1}, \bar{A}_l, \bar{Z}_l)$.

**Step 7.** Transform the conditional moment restriction (17) into unconditional moment restriction by the instrumental function $\exp\{t^T(\bar{Y}_{l-1}, \bar{A}_l, \bar{Z}_l)\}$:

$$\begin{aligned}\mathbb{P}_n\Big\{\big[&h_{K,l+1}(y_K, \cdots, y_l, \bar{Y}_{l-1}, \bar{W}_{l+1}, a_K, \cdots, a_{l+1}, \bar{A}_l; \hat{\beta}_{l+1}) \hat{f}(y_l|\bar{W}_{l+1}, \bar{Y}_{l-1}, \bar{Z}_l, \bar{A}_l) \\ &- h_{Kl}(y_K, \cdots, y_l, \bar{Y}_{l-1}, \bar{W}_l, a_K, \cdots, a_{l+1}, \bar{A}_l; \beta_l)\big] \exp\{t^T(\bar{Y}_{l-1}, \bar{A}_l, \bar{Z}_l)\}\Big\} = 0,\end{aligned} \tag{22}$$

and denote the left-hand side of Equation (22) as $\mathbb{P}_n\{g_l(y_K, \ldots, y_l, t; X_l; \beta_l)\}$, where $X_K = (\bar{Y}_{l-1}, \bar{W}_{l+1}, \bar{A}_l, \bar{Z}_l)$, and $\hat{\beta}_{l+1}$ is already estimated in the last iteration.

**Step 8.** Obtain the initial estimator $\tilde{\beta}_l$ by:

$$\tilde{\beta}_l = \arg\min_{\beta} \|I\bar{g}_l(y_K, \ldots, y_l, t; X_l; \beta_l)\|,$$

where $I$ is the identity operator.

**Step 9.** Denote the empirical corresponding covariate operator by $K_{n,l}$, which is defined as:

$$\begin{aligned}
&(K_{n,l}\ \psi)(y_K^{(1)}, \ldots, y_l^{(1)}, t^{(1)}) \\
&= \int_{\mathcal{Y}_K \times \ldots, \mathcal{Y}_l \times \mathcal{T}} \mathbb{E}\Big[g_l(y_K^{(1)}, \ldots, y_l^{(1)}, t^{(1)}; X_l; \tilde{\beta}_l) g_l(y_K^{(2)}, \ldots, y_l^{(2)}, t^{(2)}; X_l; \tilde{\beta}_l)\Big] \\
&\quad \times \psi(y_K^{(2)}, \ldots, y_l^{(2)}, t^{(2)}) d(y_K^{(2)}, \ldots, y_l^{(2)}, t^{(2)}).
\end{aligned}$$

Calculate its corresponding eigenvalues and eigenfunctions $(\mu_{l,j}^{(n)}, \phi_{l,j}^{(n)})$, $j = 1, 2, \ldots, n$.

**Step 10.** Estimate $\hat{\beta}_l$ through:

$$\hat{\beta}_l = \arg\min_{\beta} \sum_{j=1}^{n} \frac{\mu_{l,j}^{(n)}}{\left(\mu_{l,j}^{(n)}\right)^2 + \alpha_n} \left\langle \phi_{l,j}^{(n)}, \mathbb{P}_n\{g_l(y_K, \ldots, y_l, t; X_l; \beta)\} \right\rangle^2.$$

**Output** $\hat{\beta}_l$, $l = 1, 2, \ldots, K$.

**End.**

---

There is also another possible estimation strategy for $\bar{\beta}_K$, the CGEL (Chaussé 2011). We refer interested readers to (Chaussé 2011) for a further exploration. The detailed discussion is beyond the scope of this article.

# D Proofs of the Main Results

## D.1 Proof of Theorem 3.1.1

Under Assumption 5 with $k = 1, 2, \ldots, K - 1$, the following latent equations of outcome confounding bridge functions for $l = 1, 2, \ldots, K$ will share the same solution space with observable Equations (1) and (2):

For $l = K$,

$$f(y_K|\bar{y}_{K-1}, \bar{u}_{K-1}, \bar{a}_K) = \sum_{\bar{w}_K} \mathcal{H}_{KK}(\bar{a}_K) f(\bar{w}_K|\bar{y}_{K-1}, \bar{u}_{K-1}, \bar{a}_K), \tag{23}$$

and for $l = K-1, K-2, \dots, 1$,

$$\sum_{\bar{w}_{l+1}} \mathcal{H}_{K,l+1}(\bar{a}_K) f(\bar{w}_{l+1}, y_l | \bar{y}_{l-1}, \bar{u}_{l-1}, \bar{a}_l) = \sum_{\bar{w}_l} \mathcal{H}_{Kl}(\bar{a}_K) f(\bar{w}_l | \bar{y}_{l-1}, \bar{u}_{l-1}, \bar{a}_l). \tag{24}$$

The proof of this fact is elaborated in Zhang & Tchetgen Tchetgen (2024), Appendix C, which they rely on to establish their Theorem 1, thus omitted. We will prove a more detailed version of Theorem 3.1.1 in this section:

**Theorem D.1.** *Suppose that there exist the outcome bridge functions $\mathcal{H}_{Kl}(\bar{a}_K) = h_{Kl}(\bar{y}_K, \bar{w}_l, \bar{a}_K)$, $l = 1, 2, \dots, K$ that satisfy:*

*For $l = K$,*

$$f(y_K | \bar{y}_{K-1}, \bar{z}_K, \bar{a}_K) = \sum_{\bar{w}_K} \mathcal{H}_{KK}(\bar{a}_K) f(\bar{w}_K | \bar{y}_{K-1}, \bar{z}_K, \bar{a}_K), \tag{25}$$

*and for $l = K-1, K-2, \dots, 1$,*

$$\sum_{\bar{w}_{l+1}} \mathcal{H}_{K,l+1}(\bar{a}_K) f(\bar{w}_{l+1}, y_l | \bar{y}_{l-1}, \bar{z}_l, \bar{a}_l) = \sum_{\bar{w}_l} \mathcal{H}_{Kl}(\bar{a}_K) f(\bar{w}_l | \bar{y}_{l-1}, \bar{z}_l, \bar{a}_l). \tag{26}$$

*Then, under Assumptions 1, 2, and 0-completeness Assumption 3, we have:*

*For $l = K, K-1, \dots, 1$,*

$$f(Y_K(\bar{a}_K) = y_K, \dots, Y_l(\bar{a}_l) = y_l | \bar{a}_{l-1}, \bar{u}_{l-1}, \bar{y}_{l-1}) = \sum_{\bar{w}_l} \mathcal{H}_{Kl}(\bar{a}_K) f(\bar{w}_l | \bar{a}_{l-1}, \bar{u}_{l-1}, \bar{y}_{l-1}). \tag{27}$$

*Specifically, when $l = 1$, it follows that*

$$f(\bar{Y}_K(\bar{a}_K) = \bar{y}_K | y_0) = \sum_{w_1} \mathcal{H}_{K1}(\bar{a}_K) f(w_1 | y_0). \tag{28}$$

*Proof.* We prove this conclusion by backward induction. We first prove that for $l = K, K-1$, Equation (27) holds. Note that

$$\begin{aligned}
& f(y_K | \bar{a}_K, \bar{z}_K, \bar{y}_{K-1}) \\
&= \sum_{\bar{u}_{K-1}} f(y_K | \bar{a}_K, \bar{u}_{K-1}, \bar{y}_{K-1}) f(\bar{u}_{K-1} | \bar{a}_K, \bar{z}_K, \bar{y}_{K-1}) \\
&= \sum_{\bar{w}_K} \mathcal{H}_{KK}(\bar{a}_K) f(\bar{w}_K | \bar{y}_{K-1}, \bar{z}_K, \bar{a}_K) \\
&= \sum_{\bar{u}_{K-1}} \sum_{\bar{w}_K} \mathcal{H}_{KK}(\bar{a}_K) f(\bar{w}_K | \bar{y}_{K-1}, \bar{u}_{K-1}, \bar{a}_K) f(\bar{u}_{K-1} | \bar{a}_K, \bar{z}_K, \bar{y}_{K-1}),
\end{aligned}$$

By Assumption 3′.1 for $l = K$, we have

$$\begin{aligned}
&f(Y_K(\bar{a}_K) = y_K | \bar{a}_{K-1}, \bar{u}_{K-1}, \bar{y}_{K-1}) \\
&= f(y_K | \bar{a}_K, \bar{u}_{K-1}, \bar{y}_{K-1}) \\
&= \sum_{\bar{w}_K} \mathcal{H}_{KK}(\bar{a}_K) f(\bar{w}_K | \bar{y}_{K-1}, \bar{u}_{K-1}, \bar{a}_K),
\end{aligned}$$

which prove Equation (27) when $l = K$. For $l = K - 1$,

$$\begin{aligned}
&f(Y_K(\bar{a}_K) = y_K, Y_{K-1}(\bar{a}_{K-1}) = y_{K-1} | \bar{a}_{K-1}, \bar{z}_{K-1}, \bar{y}_{K-2}) \\
&= f(Y_K(\bar{a}_K) = y_K | \bar{a}_{K-1}, \bar{z}_{K-1}, \bar{y}_{K-1}) f(y_{K-1} | \bar{a}_{K-1}, \bar{z}_{K-1}, \bar{y}_{K-2}) \\
&= \sum_{\bar{u}_{K-1}} f(Y_K(\bar{a}_K) = y_K | \bar{u}_{K-1}, \bar{a}_{K-1}, \bar{z}_{K-1}, \bar{y}_{K-1}) f(\bar{u}_{K-1} | \bar{a}_{K-1}, \bar{z}_{K-1}, \bar{y}_{K-1}) f(y_{K-1} | \bar{a}_{K-1}, \bar{z}_{K-1}, \bar{y}_{K-2}) \\
&= \sum_{\bar{u}_{K-1}} f(y_K | \bar{u}_{K-1}, \bar{a}_K, \bar{y}_{K-1}) f(\bar{u}_{K-1} | \bar{a}_{K-1}, \bar{z}_{K-1}, \bar{y}_{K-1}) f(y_{K-1} | \bar{a}_{K-1}, \bar{z}_{K-1}, \bar{y}_{K-2}) \\
&= \sum_{\bar{u}_{K-1}} \sum_{\bar{w}_K} \mathcal{H}_{KK}(\bar{a}_K) f(\bar{w}_K | \bar{y}_{K-1}, \bar{u}_{K-1}, \bar{a}_K) f(\bar{u}_{K-1} | \bar{a}_{K-1}, \bar{z}_{K-1}, \bar{y}_{K-1}) f(y_{K-1} | \bar{a}_{K-1}, \bar{z}_{K-1}, \bar{y}_{K-2}) \\
&= \sum_{\bar{u}_{K-1}} \sum_{\bar{w}_K} \mathcal{H}_{KK}(\bar{a}_K) f(\bar{w}_K, \bar{u}_{K-1} | \bar{a}_{K-1}, \bar{z}_{K-1}, \bar{y}_{K-1}) f(y_{K-1} | \bar{a}_{K-1}, \bar{z}_{K-1}, \bar{y}_{K-2}) \\
&= \sum_{\bar{w}_K} \mathcal{H}_{KK}(\bar{a}_K) f(\bar{w}_K, y_{K-1} | \bar{a}_{K-1}, \bar{z}_{K-1}, \bar{y}_{K-2}) \\
&= \sum_{\bar{w}_{K-1}} \mathcal{H}_{K,K-1}(\bar{a}_K) f(\bar{w}_{K-1} | \bar{y}_{K-2}, \bar{z}_{K-1}, \bar{a}_{K-1}).
\end{aligned}$$

Also note that

$$\begin{aligned}
&f(Y_K(\bar{a}_K) = y_K, Y_{K-1}(\bar{a}_{K-1}) = y_{K-1} | \bar{a}_{K-1}, \bar{z}_{K-1}, \bar{y}_{K-2}) \\
&= \sum_{\bar{u}_{K-2}} f(Y_K(\bar{a}_K) = y_K, Y_{K-1}(\bar{a}_{K-1}) = y_{K-1} | \bar{a}_{K-1}, \bar{u}_{K-2}, \bar{z}_{K-1}, \bar{y}_{K-2}) f(\bar{u}_{K-2} | \bar{a}_{K-1}, \bar{z}_{K-1}, \bar{y}_{K-2}) \\
&= \sum_{\bar{u}_{K-2}} f(Y_K(\bar{a}_K) = y_K, Y_{K-1}(\bar{a}_{K-1}) = y_{K-1} | \bar{a}_{K-2}, \bar{u}_{K-2}, \bar{y}_{K-2}) f(\bar{u}_{K-2} | \bar{a}_{K-1}, \bar{z}_{K-1}, \bar{y}_{K-2}),
\end{aligned}$$

and

$$\begin{aligned}
&\sum_{\bar{w}_{K-1}} \mathcal{H}_{K,K-1}(\bar{a}_K) f(\bar{w}_{K-1} | \bar{y}_{K-2}, \bar{z}_{K-1}, \bar{a}_{K-1}) \\
&= \sum_{\bar{u}_{K-2}} \sum_{\bar{w}_{K-1}} \mathcal{H}_{K,K-1}(\bar{a}_K) f(\bar{w}_{K-1} | \bar{u}_{K-2}, \bar{y}_{K-2}, \bar{z}_{K-1}, \bar{a}_{K-1}) f(\bar{u}_{K-2} | \bar{y}_{K-2}, \bar{z}_{K-1}, \bar{a}_{K-1}) \\
&= \sum_{\bar{u}_{K-2}} \sum_{\bar{w}_{K-1}} \mathcal{H}_{K,K-1}(\bar{a}_K) f(\bar{w}_{K-1} | \bar{u}_{K-2}, \bar{y}_{K-2}, \bar{a}_{K-2}) f(\bar{u}_{K-2} | \bar{y}_{K-2}, \bar{z}_{K-1}, \bar{a}_{K-1}),
\end{aligned}$$

thus, by Assumption 3′.1 for $l = K-1$,

$$f(Y_K(\bar{a}_K) = y_K, Y_{K-1}(\bar{a}_{K-1}) = y_{K-1} | \bar{a}_{K-2}, \bar{u}_{K-2}, \bar{y}_{K-2})$$
$$= \sum_{\bar{w}_{K-1}} \mathcal{H}_{K,K-1}(\bar{a}_K) f(\bar{w}_{K-1} | \bar{u}_{K-2}, \bar{y}_{K-2}, \bar{a}_{K-2}).$$

Suppose that for $l = k+1$ ($1 \le k \le K-1$), Equation (27) holds, such that

$$f(Y_K(\bar{a}_K) = y_K, \dots, Y_{k+1}(\bar{a}_{k+1}) = y_{k+1} | \bar{a}_k, \bar{u}_k, \bar{y}_k) = \sum_{\bar{w}_{k+1}} \mathcal{H}_{K,k+1}(\bar{a}_K) f(\bar{w}_{k+1} | \bar{a}_k, \bar{u}_k, \bar{y}_k).$$

We now prove that it also holds for $l = k$. Note that

$$f(Y_K(\bar{a}_K) = y_K, \dots, Y_k(\bar{a}_k) = y_k | \bar{a}_k, \bar{z}_k, \bar{y}_{k-1})$$
$$= f(Y_K(\bar{a}_K) = y_K, \dots, Y_{k+1}(\bar{a}_{k+1}) = y_{k+1} | \bar{a}_k, \bar{z}_k, \bar{y}_k) f(y_k | \bar{a}_k, \bar{z}_k, \bar{y}_{k-1})$$
$$= \sum_{\bar{u}_k} f(Y_K(\bar{a}_K) = y_K, \dots, Y_{k+1}(\bar{a}_{k+1}) = y_{k+1} | \bar{a}_k, \bar{u}_k, \bar{y}_k) f(\bar{u}_k | \bar{a}_k, \bar{z}_k, \bar{y}_k) f(y_k | \bar{a}_k, \bar{z}_k, \bar{y}_{k-1})$$
$$= \sum_{\bar{u}_k} \sum_{\bar{w}_{k+1}} \mathcal{H}_{K,k+1}(\bar{a}_K) f(\bar{w}_{k+1} | \bar{a}_k, \bar{u}_k, \bar{y}_k) f(\bar{u}_k | \bar{a}_k, \bar{z}_k, \bar{y}_k) f(y_k | \bar{a}_k, \bar{z}_k, \bar{y}_{k-1})$$
$$= \sum_{\bar{w}_{k+1}} \mathcal{H}_{K,k+1}(\bar{a}_K) f(\bar{w}_{k+1}, y_k | \bar{a}_k, \bar{z}_k, \bar{y}_{k-1})$$
$$= \sum_{\bar{w}_k} \mathcal{H}_{Kk}(\bar{a}_K) f(\bar{w}_k | \bar{a}_k, \bar{z}_k, \bar{y}_{k-1}).$$

Again,

$$f(Y_K(\bar{a}_K) = y_K, \dots, Y_k(\bar{a}_k) = y_k | \bar{a}_k, \bar{z}_k, \bar{y}_{k-1})$$
$$= \sum_{\bar{u}_{k-1}} f(Y_K(\bar{a}_K) = y_K, \dots, Y_k(\bar{a}_k) = y_k | \bar{u}_{k-1}, \bar{a}_k, \bar{z}_k, \bar{y}_{k-1}) f(\bar{u}_{k-1} | \bar{a}_k, \bar{z}_k, \bar{y}_{k-1})$$
$$= \sum_{\bar{u}_{k-1}} f(Y_K(\bar{a}_K) = y_K, \dots, Y_k(\bar{a}_k) = y_k | \bar{u}_{k-1}, \bar{a}_{k-1}, \bar{y}_{k-1}) f(\bar{u}_{k-1} | \bar{a}_k, \bar{z}_k, \bar{y}_{k-1}),$$

and

$$\sum_{\bar{w}_k} \mathcal{H}_{Kk}(\bar{a}_K) f(\bar{w}_k | \bar{a}_k, \bar{z}_k, \bar{y}_{k-1})$$
$$= \sum_{\bar{u}_{k-1}} \sum_{\bar{w}_k} \mathcal{H}_{Kk}(\bar{a}_K) f(\bar{w}_k | \bar{u}_{k-1}, \bar{a}_k, \bar{z}_k, \bar{y}_{k-1}) f(\bar{u}_{k-1} | \bar{a}_k, \bar{z}_k, \bar{y}_{k-1})$$
$$= \sum_{\bar{u}_{k-1}} \sum_{\bar{w}_k} \mathcal{H}_{Kk}(\bar{a}_K) f(\bar{w}_k | \bar{u}_{k-1}, \bar{a}_{k-1}, \bar{y}_{k-1}) f(\bar{u}_{k-1} | \bar{a}_k, \bar{z}_k, \bar{y}_{k-1}).$$

Thus, we have proved that

$$f(Y_K(\bar{a}_K) = y_K, \ldots, Y_k(\bar{a}_k) = y_k|\bar{u}_{k-1}, \bar{a}_{k-1}, \bar{y}_{k-1}) = \sum_{\bar{w}_k} \mathcal{H}_{Kk}(\bar{a}_K) f(\bar{w}_k|\bar{u}_{k-1}, \bar{a}_{k-1}, \bar{y}_{k-1}) \quad (29)$$

also holds for $l = k$. By backward induction, this is sufficient to show that Equation (27) holds for $l = K, K-1, \ldots, 1$. Specifically, when $l = 1$, we have:

$$f(Y_K(\bar{a}_K) = y_K, \ldots, Y_1(a_1) = y_1|u_0, y_0) = \sum_{w_1} \mathcal{H}_{K1}(\bar{a}_K) f(w_1|u_0, y_0),$$

which implies Equation (28).

## D.2 Proof of Theorem 3.1.2

*Proof.* We first prove that under Assumptions 1, 2, and $K$-completeness Assumption 3, treatment confounding bridge functions that solve Equations (3) and (4) are also solutions to corresponding latent Equations:

For $k = 1$,

$$\frac{1}{f(a_1|u_0, y_0)} = \sum_{z_1} q_{11}(y_0, z_1, a_1) f(z_1|a_1, u_0, y_0), \quad (30)$$

and for $k = 2, 3, \cdots, K$,

$$\frac{\sum_{\bar{z}_{k-1}} q_{k-1,k-1}(\bar{y}_{k-2}, \bar{z}_{k-1}, \bar{a}_{k-1}) f(\bar{z}_{k-1}|\bar{a}_{k-1}, \bar{u}_{k-1}, \bar{y}_{k-1})}{f(a_k|\bar{a}_{k-1}, \bar{u}_{k-1}, \bar{y}_{k-1})} = \sum_{\bar{z}_k} q_{kk}(\bar{y}_{k-1}, \bar{z}_k, \bar{a}_k) f(\bar{z}_k|\bar{a}_k, \bar{u}_{k-1}, \bar{y}_{k-1}), \quad (31)$$

which is restated formally in the following lemma:

*Lemma* D.2. *Suppose that there exist functions* $q_{11}, q_{22}, \cdots, q_{KK}$ *that satisfy*

$$\frac{1}{f(a_1|w_1, y_0)} = \sum_{z_1} q_{11}(y_0, z_1, a_1) f(z_1|a_1, w_1, y_0)$$

*and for* $k = 2, 3, \cdots, K$,

$$\frac{\sum_{\bar{z}_{k-1}} q_{k-1,k-1}(\bar{y}_{k-2}, \bar{z}_{k-1}, \bar{a}_{k-1}) f(\bar{z}_{k-1}|\bar{a}_{k-1}, \bar{w}_k, \bar{y}_{k-1})}{f(a_k|\bar{a}_{k-1}, \bar{w}_k, \bar{y}_{k-1})} = \sum_{\bar{z}_k} q_{kk}(\bar{y}_{k-1}, \bar{z}_k, \bar{a}_k) f(\bar{z}_k|\bar{a}_k, \bar{w}_k, \bar{y}_{k-1})$$

*If Assumptions 1, 2, and $K$-completeness Assumption 3 hold, then* $q_{11}, \cdots, q_{KK}$ *solve the respective integral Equations (30) and (31) as well.*

*Proof.* For $q_{kk}$, $k = 2, \ldots, K$, we have:

$$
\begin{aligned}
&\sum_{\bar{u}_{k-1}} \frac{\sum_{\bar{z}_{k-1}} q_{k-1,k-1}(\bar{y}_{k-2}, \bar{z}_{k-1}, \bar{a}_{k-1}) f(\bar{z}_{k-1}|\bar{a}_{k-1}, \bar{u}_{k-1}, \bar{y}_{k-1})}{f(a_k|\bar{a}_{k-1}, \bar{u}_{k-1}, \bar{y}_{k-1})} f(\bar{u}_{k-1}|\bar{a}_k, \bar{w}_k, \bar{y}_{k-1}) \\
&= \sum_{\bar{u}_{k-1}} \frac{\sum_{\bar{z}_{k-1}} q_{k-1,k-1}(\bar{y}_{k-2}, \bar{z}_{k-1}, \bar{a}_{k-1}) f(\bar{z}_{k-1}|\bar{a}_{k-1}, \bar{u}_{k-1}, \bar{w}_k, \bar{y}_{k-1})}{f(a_k|\bar{a}_{k-1}, \bar{u}_{k-1}, \bar{w}_k, \bar{y}_{k-1})} f(\bar{u}_{k-1}|\bar{a}_k, \bar{w}_k, \bar{y}_{k-1}) \\
&= \frac{\sum_{\bar{z}_{k-1}} q_{k-1,k-1}(\bar{y}_{k-2}, \bar{z}_{k-1}, \bar{a}_{k-1}) f(\bar{z}_{k-1}|\bar{a}_{k-1}, \bar{w}_k, \bar{y}_{k-1})}{f(a_k|\bar{a}_{k-1}, \bar{w}_k, \bar{y}_{k-1})} \\
&= \sum_{\bar{z}_k} q_{kk}(\bar{y}_{k-1}, \bar{z}_k, \bar{a}_k) f(\bar{z}_k|\bar{a}_k, \bar{w}_k, \bar{y}_{k-1}) \\
&= \sum_{\bar{u}_{k-1}} \sum_{\bar{z}_k} q_{kk}(\bar{y}_{k-1}, \bar{z}_k, \bar{a}_k) f(\bar{z}_k|\bar{a}_k, \bar{u}_{k-1}, \bar{y}_{k-1}) f(\bar{u}_{k-1}|\bar{a}_k, \bar{w}_k, \bar{y}_{k-1})
\end{aligned}
$$

Then, for $q_{11}$, we have:

$$
\begin{aligned}
&\sum_{u_0} \frac{1}{f(a_1|u_0, y_0)} f(u_0|w_1, a_1, y_0) \\
&= \sum_{u_0} \frac{f(u_0, a_1|w_1, y_0) f(u_0|w_1, y_0)}{f(a_1|w_1, y_0) f(u_0, a_1|w_1, y_0)} \\
&= \sum_{u_0} \frac{f(u_0|w_1, y_0)}{f(a_1|w_1, y_0)} \\
&= \frac{1}{f(a_1|w_1, y_0)} \\
&= \sum_{z_1} q_{11}(y_0, z_1, a_1) f(z_1|a_1, w_1, y_0) \\
&= \sum_{u_0} \sum_{z_1} q_{11}(y_0, z_1, a_1) f(z_1|a_1, u_0, y_0) f(u_0|w_1, a_1, y_0)
\end{aligned}
$$

By $K$-completeness Assumption 3, we have

$$
\begin{aligned}
&\frac{\sum_{\bar{z}_{k-1}} q_{k-1,k-1}(\bar{y}_{k-2}, \bar{z}_{k-1}, \bar{a}_{k-1}) f(\bar{z}_{k-1}|\bar{a}_{k-1}, \bar{u}_{k-1}, \bar{y}_{k-1})}{f(a_k|\bar{a}_{k-1}, \bar{u}_{k-1}, \bar{y}_{k-1})} \\
&= \sum_{\bar{z}_k} q_{kk}(\bar{y}_{k-1}, \bar{z}_k, \bar{a}_k) f(\bar{z}_k|\bar{a}_k, \bar{u}_{k-1}, \bar{y}_{k-1}),
\end{aligned}
$$

and

$$
\frac{1}{f(a_1|u_0, y_0)} = \sum_{z_1} q_{11}(y_0, z_1, a_1) f(z_1|a_1, u_0, y_0).
$$

almost surely, which ensures the equivalence between the solution sets of Equations (3) and (4) and the solution sets of Equations (30) and (31).

To prove the identification result, we further need the following transition lemma:

*Lemma* D.3. *Under Assumptions 1, 2, and $K$-completeness Assumption 3, we suppose that there exist treatment confounding bridge functions $q_{11},\ldots,q_{KK}$ that solve Equations (3) and (4). Then, the following equation holds for $k=1,2,\ldots,K$:*

$$
\begin{aligned}
&f(\bar{Y}_K(\bar{a}_K)=y_K,\cdots,Y_k(\bar{a}_k)=y_k|\bar{u}_{k-1},\bar{y}_{k-1},\bar{a}_k)\sum_{\bar{z}_k} q_{kk}(\bar{a}_k,\bar{y}_{k-1},\bar{z}_k)f(\bar{z}_k|\bar{a}_k,\bar{u}_{k-1},\bar{y}_{k-1})\\
&=\sum_{u_k} f(Y_K(\bar{a}_K)=y_K,\cdots,Y_{k+1}(\bar{a}_{k+1})=y_{k+1}|\bar{u}_k,\bar{y}_k,\bar{a}_{k+1})f(u_k|\bar{u}_{k-1},\bar{y}_k,\bar{a}_k)f(a_{k+1}|\bar{a}_k,\bar{u}_k,\bar{y}_k)\\
&\quad\times f(Y_k=y_k|\bar{u}_{k-1},\bar{y}_{k-1},\bar{a}_k)\sum_{\bar{z}_{k+1}} q_{k+1,k+1}(\bar{a}_{k+1},\bar{y}_k,\bar{z}_{k+1})f(\bar{z}_{k+1}|\bar{a}_{k+1},\bar{u}_k,\bar{y}_k).
\end{aligned}
\tag{32}
$$

*Proof.*

$$f(Y_K(\bar{a}_K) = y_K, \cdots, Y_k(\bar{a}_k) = y_k|\bar{u}_{k-1}, \bar{y}_{k-1}, \bar{a}_k) \sum_{\bar{z}_k} q_{kk}(\bar{a}_k, \bar{y}_{k-1}, \bar{z}_k) f(\bar{z}_k|\bar{a}_k, \bar{u}_{k-1}, \bar{y}_{k-1})$$

$$= f(Y_K(\bar{a}_K) = y_K, \cdots, Y_k(\bar{a}_k) = y_k|\bar{u}_{k-1}, \bar{y}_{k-1}, \bar{a}_k, \bar{z}_k) \sum_{\bar{z}_k} q_{kk}(\bar{a}_k, \bar{y}_{k-1}, \bar{z}_k) f(\bar{z}_k|\bar{a}_k, \bar{u}_{k-1}, \bar{y}_{k-1})$$

$$= f(Y_K(\bar{a}_K) = y_K, \cdots, Y_k(\bar{a}_k) = y_k, \bar{z}_k|\bar{u}_{k-1}, \bar{y}_{k-1}, \bar{a}_k) \sum_{\bar{z}_k} q_{kk}(\bar{a}_k, \bar{y}_{k-1}, \bar{z}_k)$$

$$= \sum_{u_k} f(Y_K(\bar{a}_K) = y_K, \cdots, Y_{k+1}(\bar{a}_{k+1}) = y_{k+1}, \bar{z}_k|\bar{u}_k, \bar{y}_k, \bar{a}_k) f(u_k|\bar{u}_{k-1}, \bar{y}_k, \bar{a}_k)$$

$$\times f(Y_k = y_k|\bar{u}_{k-1}, \bar{y}_{k-1}, \bar{a}_k) \sum_{\bar{z}_k} q_{kk}(\bar{a}_k, \bar{y}_{k-1}, \bar{z}_k)$$

$$= \sum_{u_k} f(Y_K(\bar{a}_K) = y_K, \cdots, Y_{k+1}(\bar{a}_{k+1}) = y_{k+1}|\bar{u}_k, \bar{y}_k, \bar{a}_k) f(u_k|\bar{u}_{k-1}, \bar{y}_k, \bar{a}_k)$$

$$\times \sum_{\bar{z}_k} q_{kk}(\bar{a}_k, \bar{y}_{k-1}, \bar{z}_k) f(\bar{z}_k|\bar{u}_k, \bar{y}_k, \bar{a}_k)$$

$$= \sum_{u_k} f(Y_K(\bar{a}_K) = y_K, \cdots, Y_{k+1}(\bar{a}_{k+1}) = y_{k+1}|\bar{u}_k, \bar{y}_k, \bar{a}_k) f(u_k|\bar{u}_{k-1}, \bar{y}_k, \bar{a}_k) f(Y_k = y_k|\bar{u}_{k-1}, \bar{y}_{k-1}, \bar{a}_k)$$

$$\times f(a_{k+1}|\bar{a}_k, \bar{u}_k, \bar{y}_k) \frac{\sum_{\bar{z}_k} q_{kk}(\bar{a}_k, \bar{y}_{k-1}, \bar{z}_k) f(\bar{z}_k|\bar{u}_k, \bar{y}_k, \bar{a}_k)}{f(a_{k+1}|\bar{a}_k, \bar{u}_k, \bar{y}_k)}$$

$$= \sum_{u_k} f(Y_K(\bar{a}_K) = y_K, \cdots, Y_{k+1}(\bar{a}_{k+1}) = y_{k+1}|\bar{u}_k, \bar{y}_k, \bar{a}_k) f(u_k|\bar{u}_{k-1}, \bar{y}_k, \bar{a}_k) f(Y_k = y_k|\bar{u}_{k-1}, \bar{y}_{k-1}, \bar{a}_k)$$

$$\times f(a_{k+1}|\bar{a}_k, \bar{u}_k, \bar{y}_k) \sum_{\bar{z}_{k+1}} q_{k+1,k+1}(\bar{a}_{k+1}, \bar{y}_k, \bar{z}_{k+1}) f(\bar{z}_{k+1}|\bar{u}_k, \bar{y}_k, \bar{a}_{k+1})$$

$$= \sum_{u_k} f(Y_K(\bar{a}_K) = y_K, \cdots, Y_{k+1}(\bar{a}_{k+1}) = y_{k+1}|\bar{u}_k, \bar{y}_k, \bar{a}_{k+1}) f(u_k|\bar{u}_{k-1}, \bar{y}_k, \bar{a}_k) f(a_{k+1}|\bar{a}_k, \bar{u}_k, \bar{y}_k)$$

$$\times f(Y_k = y_k|\bar{u}_{k-1}, \bar{y}_{k-1}, \bar{a}_k) \sum_{\bar{z}_{k+1}} q_{k+1,k+1}(\bar{a}_{k+1}, \bar{y}_k, \bar{z}_{k+1}) f(\bar{z}_{k+1}|\bar{u}_k, \bar{y}_k, \bar{a}_{k+1}).$$

The first, second, third, and seventh equality follow from Assumption 1.3, and the sixth equality follows from Lemma D.2.

By Lemma D.3, we have

$$
\begin{aligned}
&f(Y_K(\bar{a}_K)=y_K,\cdots,Y_1(a_1)=y_1|y_0)\\
&=\sum_{u_0} f(u_0|y_0)f(a_1|u_0,y_0)f(Y_K(\bar{a}_K)=y_K,\cdots,Y_1(a_1)=y_1|u_0,y_0)\sum_{z_1} q_{11}(a_1,y_0,z_1)f(z_1|a_1,u_0,y_0)\\
&=\sum_{u_0} f(u_0|y_0)f(a_1|u_0,y_0)\sum_{u_1} f(u_1|u_0,\bar{y}_1,a_1)f(y_1|u_0,y_0,a_1)f(a_2|a_1,\bar{u}_1,\bar{y}_1)\\
&\qquad\times f(Y_K(\bar{a}_K)=y_K,\cdots,Y_2(a_1,a_2)=y_2|\bar{u}_1,\bar{y}_1,\bar{a}_2)\sum_{\bar{z}_2} q_{22}(\bar{a}_2,\bar{y}_1,\bar{z}_2)f(\bar{z}_2|\bar{a}_2,\bar{u}_1,\bar{y}_1)\\
&=\sum_{\bar{u}_1} f(u_0|y_0)f(a_1|u_0,y_0)f(u_1|u_0,\bar{y}_1,a_1)f(y_1|u_0,y_0,a_1)f(a_2|a_1,\bar{u}_1,\bar{y}_1)\\
&\qquad\times f(Y_K(\bar{a}_K)=y_K,\cdots,Y_2(a_1,a_2)=y_2|\bar{u}_1,\bar{y}_1,\bar{a}_2)\sum_{\bar{z}_2} q_{22}(\bar{a}_2,\bar{y}_1,\bar{z}_2)f(\bar{z}_2|\bar{a}_2,\bar{u}_1,\bar{y}_1)\\
&=\cdots\\
&=\sum_{\bar{u}_{K-1}} f(Y_K(\bar{a}_K)=y_K|\bar{u}_{K-1},\bar{y}_{K-1},\bar{a}_K)\sum_{\bar{z}_K} q_{KK}(\bar{a}_K,\bar{y}_{K-1},\bar{z}_K)f(\bar{z}_K|\bar{a}_K,\bar{u}_{K-1},\bar{y}_{K-1})\\
&\qquad\times\prod_{j=2}^{K}\{f(a_j|\bar{a}_{j-1},\bar{u}_{j-1},\bar{y}_{j-1})f(u_{j-1}|\bar{u}_{j-2},\bar{y}_{j-1},\bar{a}_{j-1})f(y_{j-1}|\bar{u}_{j-2},\bar{y}_{j-2},\bar{a}_{j-1})\}f(a_1|u_0,y_0)f(u_0|y_0)\\
&=\sum_{\bar{z}_K}\sum_{\bar{u}_{K-1}} q_{KK}(\bar{a}_K,\bar{y}_{K-1},\bar{z}_K)f(Y_K=y_K,\bar{z}_K|\bar{u}_{K-1},\bar{y}_{K-1},\bar{a}_K)f(\bar{a}_K,\bar{u}_{K-1},\bar{y}_{K-1}|y_0)\\
&=\sum_{\bar{z}_K} q_{KK}(\bar{a}_K,\bar{y}_{K-1},\bar{z}_K)f(\bar{y}_K,\bar{z}_K,\bar{a}_K|y_0).
\end{aligned}
$$

Here, the first equality follows from the proximal g-formula and Assumption 1.3, the fourth and fifth equality follow from the iterative usage of Lemma D.3.

## D.3 Proof of Theorem 3.2

*Proof.* We first prove the conclusion under the 2-stage setting. Note that under Assumptions 1, 2, 0-completeness Assumption 3, and Assumption 5 at the first stage, suppose that there exists function $h_{22}$ solving Equation (1), then by Lemma 1 in Appendix A of Zhang & Tchetgen Tchetgen (2024), we have:

$$
\begin{aligned}
f(Y_2(a_1,a_2)=y_2,Y_1(a_1)=y_1|y_0)&=\sum_{u_0}\sum_{u_1} f(Y_2=y_2|\bar{y}_1,\bar{u}_1,\bar{a}_2)f(u_1|\bar{y}_1,a_1,u_0)f(y_1|y_0,a_1,u_0)f(u_0|y_0)\\
&=\sum_{\bar{w}_2}\sum_{u_0} h_{22}(\bar{y}_2,\bar{w}_2,\bar{a}_2)f(y_1|y_0,u_0,a_1)f(\bar{w}_2|\bar{y}_1,a_1,u_0)f(u_0|y_0),
\end{aligned}
$$

which is also elaborated in the Appendices A.1 and A.2 of Zhang & Tchetgen Tchetgen (2024).

It follows from the above result that

$$
\begin{aligned}
f(\bar{Y}_2(a_1, a_2) = y_2|y_0) =& \sum_{\bar{w}_2}\sum_{u_0} h_{22}(\bar{y}_2, \bar{w}_2, \bar{a}_2) f(y_1|y_0, u_0, a_1) f(\bar{w}_2|\bar{y}_1, a_1, u_0) f(u_0|y_0) \\
=& \sum_{\bar{w}_2}\sum_{u_0} h_{22}(\bar{y}_2, \bar{w}_2, \bar{a}_2) f(\bar{w}_2, y_1|y_0, a_1, u_0) f(u_0|y_0) \\
=& \sum_{\bar{w}_2}\sum_{u_0}\sum_{z_1} q_{11}(a_1, y_0, z_1) f(z_1|a_1, u_0, y_0) f(a_1|u_0, y_0) h_{22}(\bar{y}_2, \bar{w}_2, \bar{a}_2) f(u_0|y_0) \\
& \times f(\bar{w}_2, y_1|y_0, a_1, u_0) \\
=& \sum_{\bar{w}_2}\sum_{u_0}\sum_{z_1} q_{11}(a_1, y_0, z_1) f(a_1|u_0, y_0) h_{22}(\bar{y}_2, \bar{w}_2, \bar{a}_2) f(\bar{w}_2, y_1, z_1|y_0, a_1, u_0) f(u_0|y_0) \\
=& \sum_{\bar{w}_2}\sum_{z_1} q_{11}(a_1, y_0, z_1) h_{22}(\bar{y}_2, \bar{w}_2, \bar{a}_2) f(\bar{w}_2, y_1, z_1, a_1|y_0),
\end{aligned}
$$

where the third equality is due to Equation (3), and the fourth equality follows from Assumptions 1 and 5 ($\{\bar{W}_2, Y_1\} \perp\!\!\!\perp Z_1|A_1, U_0, Y_0$).

Finally,

$$
\begin{aligned}
f(Y_1(a_1) = y_1|y_0) =& \sum_{y_2} f(Y_2(a_1, a_2) = y_2, Y_1(a_1) = y_1|y_0) \\
=& \sum_{y_2}\sum_{\bar{w}_2}\sum_{z_1} h_{22}(\bar{y}_2, \bar{w}_2, \bar{a}_2) q_{11}(a_1, y_0, z_1) f(\bar{w}_2, y_1, a_1, z_1|y_0).
\end{aligned}
$$

For the general $K$-stage setting, by Assumption 1 and $k$-completeness Assumption 3, we suppose that there exist treatment confounding bridge functions $\mathcal{Q}_{tt}(\bar{a}_t)$, $t = 1, 2, \ldots, k$ that solve

Equations (3) and (4) respectively, then:

$$
\begin{aligned}
&f(\bar{Y}_K(\bar{a}_K) = \bar{y}_K|y_0) \\
&= \sum_{u_0} f(Y_K(\bar{a}_K) = y_K, \cdots, Y_1(a_1) = y_1|u_0, y_0, a_1) f(u_0|y_0) \\
&= \sum_{u_0} \sum_{z_1} f(Y_K(\bar{a}_K) = y_K, \cdots, Y_2(\bar{a}_2) = y_2, y_1|u_0, y_0, a_1) f(a_1|u_0, y_0) \mathcal{Q}_{11}(a_1) f(z_1|a_1, u_0, y_0) f(u_0|y_0) \\
&= \sum_{u_0} \sum_{z_1} f(Y_K(\bar{a}_K) = y_K, \cdots, Y_2(\bar{a}_2) = y_2, y_1, z_1|u_0, y_0, a_1) f(a_1, u_0|y_0) \mathcal{Q}_{11}(a_1) \\
&= \sum_{\bar{u}_1} \sum_{z_1} f(Y_K(\bar{a}_K) = y_K, \cdots, Y_2(\bar{a}_2) = y_2, z_1|\bar{u}_1, \bar{y}_1, a_1) f(u_1, y_1|u_0, a_1, y_0) f(a_1, u_0|y_0) \mathcal{Q}_{11}(a_1) \\
&= \sum_{\bar{u}_1} f(Y_K(\bar{a}_K) = y_K, \cdots, Y_2(\bar{a}_2) = y_2|\bar{u}_1, \bar{y}_1, a_1) f(u_1, y_1|u_0, a_1, y_0) f(a_1, u_0|y_0) \\
&\quad \times \sum_{z_1} \mathcal{Q}_{11}(a_1) f(z_1|\bar{u}_1, \bar{y}_1, a_1) \\
&= \sum_{\bar{u}_1} f(Y_K(\bar{a}_K) = y_K, \cdots, Y_2(\bar{a}_2) = y_2|\bar{u}_1, \bar{y}_1, a_1) f(u_1, y_1|u_0, a_1, y_0) f(a_1, u_0|y_0) \\
&\quad \times f(a_2|a_1, \bar{u}_1, \bar{y}_1) \sum_{\bar{z}_2} \mathcal{Q}_{22}(\bar{a}_2) f(\bar{z}_2|\bar{a}_2, \bar{u}_1, \bar{y}_1) \\
&= \sum_{\bar{u}_1} \sum_{\bar{z}_2} f(Y_K(\bar{a}_K) = y_K, \cdots, Y_3(\bar{a}_3) = y_3, y_2, \bar{z}_2|\bar{u}_1, \bar{y}_1, \bar{a}_2) f(\bar{a}_2, \bar{u}_1, y_1|y_0) \mathcal{Q}_{22}(\bar{a}_2) \\
&= \sum_{\bar{u}_2} \sum_{\bar{z}_2} f(Y_K(\bar{a}_K) = y_K, \cdots, Y_3(\bar{a}_3) = y_3, \bar{z}_2|\bar{u}_2, \bar{y}_2, \bar{a}_2) f(y_2, u_2|\bar{u}_1, \bar{y}_1, \bar{a}_2) f(\bar{a}_2, \bar{u}_1, y_1|y_0) \mathcal{Q}_{22}(\bar{a}_2) \\
&= \sum_{\bar{u}_2} f(Y_K(\bar{a}_K) = y_K, \cdots, Y_3(\bar{a}_3) = y_3|\bar{u}_2, \bar{y}_2, \bar{a}_2) f(y_2, u_2|\bar{u}_1, \bar{y}_1, \bar{a}_2) f(\bar{a}_2, \bar{u}_1, y_1|y_0) \\
&\quad \times \sum_{\bar{z}_2} \mathcal{Q}_{22}(\bar{a}_2) f(\bar{z}_2|\bar{u}_2, \bar{y}_2, \bar{a}_2) \\
&= \sum_{\bar{u}_2} f(Y_K(\bar{a}_K) = y_K, \cdots, Y_4(\bar{a}_4) = y_4, y_3|\bar{u}_2, \bar{y}_2, \bar{a}_3) f(y_2, u_2|\bar{u}_1, \bar{y}_1, \bar{a}_2) f(\bar{a}_2, \bar{u}_1, y_1|y_0) \\
&\quad \times f(a_3|\bar{a}_2, \bar{u}_2, \bar{y}_2) \sum_{\bar{z}_3} \mathcal{Q}_{33}(\bar{a}_3) f(\bar{z}_3|\bar{u}_2, \bar{y}_2, \bar{a}_3) \\
&= \sum_{\bar{u}_2} \sum_{\bar{z}_3} f(Y_K(\bar{a}_K) = y_K, \cdots, Y_4(\bar{a}_4) = y_4, y_3, \bar{z}_3|\bar{u}_2, \bar{y}_2, \bar{a}_3) f(\bar{y}_2, \bar{u}_2, \bar{a}_3|y_0) \mathcal{Q}_{33}(\bar{a}_3) \\
&= \cdots \\
&= \sum_{\bar{u}_{k-1}} \sum_{\bar{z}_k} f(Y_K(\bar{a}_K) = y_K, \cdots, Y_{k+1}(\bar{a}_{k+1}) = y_{k+1}, y_k, \bar{z}_k|\bar{u}_{k-1}, \bar{y}_{k-1}, \bar{a}_k) f(\bar{y}_{k-1}, \bar{u}_{k-1}, \bar{a}_k|y_0) \mathcal{Q}_{kk}(\bar{a}_k).
\end{aligned}
$$

We further suppose that there exist outcome confounding bridge functions $\mathcal{H}_{Kl}(\bar{a}_K)$, $l = K, K -$

$1, \ldots, k+1$ that solve Equations (1) and (2) respectively. Note that:

$$
\begin{aligned}
&\sum_{\bar{u}_{k-1}} \sum_{\bar{z}_k} f(Y_K(\bar{a}_K) = y_K, \cdots, Y_{k+1}(\bar{a}_{k+1}) = y_{k+1}, y_k, \bar{z}_k | \bar{u}_{k-1}, \bar{y}_{k-1}, \bar{a}_k) f(\bar{y}_{k-1}, \bar{u}_{k-1}, \bar{a}_k | y_0) \mathcal{Q}_{kk}(\bar{a}_k) \\
&= \sum_{\bar{z}_k} f(Y_K(\bar{a}_K) = y_K, \cdots, Y_{k+1}(\bar{a}_{k+1}) = y_{k+1}, y_k, \bar{z}_k | \bar{y}_{k-1}, \bar{a}_k) f(\bar{y}_{k-1}, \bar{a}_k | y_0) \mathcal{Q}_{kk}(\bar{a}_k) \\
&= \sum_{\bar{u}_k} \sum_{\bar{z}_k} f(Y_K(\bar{a}_K) = y_K, \cdots, Y_{k+1}(\bar{a}_{k+1}) = y_{k+1}, \bar{z}_k | \bar{y}_k, \bar{u}_k, \bar{a}_k) \\
&\quad \times f(y_k, \bar{u}_k | \bar{y}_{k-1}, \bar{a}_k) f(\bar{y}_{k-1}, \bar{a}_k | y_0) \mathcal{Q}_{kk}(\bar{a}_k) \\
&= \sum_{\bar{u}_k} \sum_{\bar{z}_k} f(Y_K(\bar{a}_K) = y_K, \cdots, Y_{k+1}(\bar{a}_{k+1}) = y_{k+1} | \bar{y}_k, \bar{u}_k, \bar{a}_k) f(\bar{z}_k | \bar{y}_k, \bar{u}_k, \bar{a}_k) \\
&\quad \times f(y_k, \bar{u}_k | \bar{y}_{k-1}, \bar{a}_k) f(\bar{y}_{k-1}, \bar{a}_k | y_0) \mathcal{Q}_{kk}(\bar{a}_k),
\end{aligned}
$$

and

$$
\begin{aligned}
&f(Y_K(\bar{a}_K) = y_K, \cdots, Y_{k+1}(\bar{a}_{k+1}) = y_{k+1} | \bar{y}_k, \bar{u}_k, \bar{a}_k) \\
&= f(Y_K(\bar{a}_K) = y_K, \cdots, Y_{k+2}(\bar{a}_{k+2}) = y_{k+2} | \bar{y}_{k+1}, \bar{u}_k, \bar{a}_{k+1}) f(y_{k+1} | \bar{y}_k, \bar{u}_k, \bar{a}_{k+1}) \\
&= \sum_{u_{k+1}} f(Y_K(\bar{a}_K) = y_K, \cdots, Y_{k+2}(\bar{a}_{k+2}) = y_{k+2} | \bar{y}_{k+1}, \bar{u}_{k+1}, \bar{a}_{k+1}) \\
&\quad \times f(u_{k+1} | \bar{y}_{k+1}, \bar{u}_k, \bar{a}_{k+1}) f(y_{k+1} | \bar{y}_k, \bar{u}_k, \bar{a}_{k+1}) \\
&= \cdots \\
&= \sum_{u_{K-1}, \ldots, u_{k+1}} f(y_{k+1} | \bar{y}_k, \bar{u}_k, \bar{a}_{k+1}) \prod_{j=k+2}^{K} \Big[ f(y_j | \bar{y}_{j-1}, \bar{u}_{j-1}, \bar{a}_j) f(u_{j-1} | \bar{y}_{j-1}, \bar{u}_{j-2}, \bar{a}_{j-1}) \Big].
\end{aligned}
$$

We will first consider the situation where Assumption 5 holds for $k = K, K-1, \ldots, k$. By the definition of outcome confounding bridge functions and Lemma 6 in Appendix C of Zhang &

Tchetgen Tchetgen (2024), we have:

$$\sum_{\bar{u}_k}\sum_{\bar{z}_k} f(Y_K(\bar{a}_K)=y_K,\cdots,Y_{k+1}(\bar{a}_{k+1})=y_{k+1}|\bar{y}_k,\bar{u}_k,\bar{a}_k)f(\bar{z}_k|\bar{y}_k,\bar{u}_k,\bar{a}_k)$$

$$\times f(y_k,\bar{u}_k|\bar{y}_{k-1},\bar{a}_k)f(\bar{y}_{k-1},\bar{a}_k|y_0)\mathcal{Q}_{kk}(\bar{a}_k)$$

$$=\sum_{\bar{u}_{K-1}} f(y_{k+1}|\bar{y}_k,\bar{u}_k,\bar{a}_{k+1})\prod_{j=k+2}^{K}\left[f(y_j|\bar{y}_{j-1},\bar{u}_{j-1},\bar{a}_j)f(u_{j-1}|\bar{y}_{j-1},\bar{u}_{j-2},\bar{a}_{j-1})\right]$$

$$\times\sum_{\bar{z}_k} f(\bar{z}_k|\bar{y}_k,\bar{u}_k,\bar{a}_k)f(y_k,\bar{u}_k|\bar{y}_{k-1},\bar{a}_k)f(\bar{y}_{k-1},\bar{a}_k|y_0)\mathcal{Q}_{kk}(\bar{a}_k)$$

$$=\sum_{\bar{u}_{K-1}}\sum_{\bar{w}_K}\mathcal{H}_{KK}(\bar{a}_K)f(\bar{w}_K|\bar{y}_{K-1},\bar{u}_{K-1},\bar{a}_K)f(u_{K-1}|\bar{y}_{K-1},\bar{u}_{K-2},\bar{a}_{K-1})$$

$$\times f(y_{k+1}|\bar{y}_k,\bar{u}_k,\bar{a}_{k+1})\prod_{j=k+2}^{K-1}\left[f(y_j|\bar{u}_{j-1},\bar{y}_{j-1},\bar{a}_j)f(u_{j-1}|\bar{y}_{j-1},\bar{u}_{j-2},\bar{a}_{j-1})\right]$$

$$\times\sum_{\bar{z}_k} f(\bar{z}_k|\bar{y}_k,\bar{u}_k,\bar{a}_k)f(y_k,\bar{u}_k|\bar{y}_{k-1},\bar{a}_k)f(\bar{y}_{k-1},\bar{a}_k|y_0)\mathcal{Q}_{kk}(\bar{a}_k)$$

$$=\sum_{\bar{u}_{K-2}}\sum_{\bar{w}_K}\mathcal{H}_{KK}(\bar{a}_K)f(\bar{w}_K,y_{K-1}|\bar{y}_{K-2},\bar{u}_{K-2},\bar{a}_{K-1})f(u_{K-2}|\bar{y}_{K-2},\bar{u}_{K-3},\bar{a}_{K-2})$$

$$\times f(y_{k+1}|\bar{y}_k,\bar{u}_k,\bar{a}_{k+1})\prod_{j=k+2}^{K-2}\left[f(y_j|\bar{u}_{j-1},\bar{y}_{j-1},\bar{a}_j)f(u_{j-1}|\bar{y}_{j-1},\bar{u}_{j-2},\bar{a}_{j-1})\right]$$

$$\times\sum_{\bar{z}_k} f(\bar{z}_k|\bar{y}_k,\bar{u}_k,\bar{a}_k)f(y_k,\bar{u}_k|\bar{y}_{k-1},\bar{a}_k)f(\bar{y}_{k-1},\bar{a}_k|y_0)\mathcal{Q}_{kk}(\bar{a}_k)$$

$$=\sum_{\bar{u}_{K-3}}\sum_{\bar{w}_{K-1}}\mathcal{H}_{K,K-1}(\bar{a}_K)f(\bar{w}_{K-1}|\bar{y}_{K-2},\bar{u}_{K-3},\bar{a}_{K-2})$$

$$\times f(y_{k+1}|\bar{y}_k,\bar{u}_k,\bar{a}_{k+1})\prod_{j=k+2}^{K-2}\left[f(y_j|\bar{u}_{j-1},\bar{y}_{j-1},\bar{a}_j)f(u_{j-1}|\bar{y}_{j-1},\bar{u}_{j-2},\bar{a}_{j-1})\right]$$

$$\times\sum_{\bar{z}_k} f(\bar{z}_k|\bar{y}_k,\bar{u}_k,\bar{a}_k)f(y_k,\bar{u}_k|\bar{y}_{k-1},\bar{a}_k)f(\bar{y}_{k-1},\bar{a}_k|y_0)\mathcal{Q}_{kk}(\bar{a}_k)$$

$$=\cdots$$

$$=\sum_{\bar{u}_k}\sum_{\bar{w}_{k+2}}\mathcal{H}_{K,k+2}(\bar{a}_K)f(\bar{w}_{k+2},y_{k+1}|\bar{y}_k,\bar{u}_k,\bar{a}_{k+1})$$

$$\times\sum_{\bar{z}_k} f(\bar{z}_k|\bar{y}_k,\bar{u}_k,\bar{a}_k)f(y_k,\bar{u}_k|\bar{y}_{k-1},\bar{a}_k)f(\bar{y}_{k-1},\bar{a}_k|y_0)\mathcal{Q}_{kk}(\bar{a}_k)$$

$$=\sum_{\bar{u}_k}\sum_{\bar{w}_{k+1}}\mathcal{H}_{K,k+1}(\bar{a}_K)f(\bar{w}_{k+1}|\bar{y}_k,\bar{u}_k,\bar{a}_k)$$

$$\times\sum_{\bar{z}_k} f(\bar{z}_k|\bar{y}_k,\bar{u}_k,\bar{a}_k)f(y_k,\bar{u}_k|\bar{y}_{k-1},\bar{a}_k)f(\bar{y}_{k-1},\bar{a}_k|y_0)\mathcal{Q}_{kk}(\bar{a}_k)$$

$$=\sum_{\bar{u}_k}\sum_{\bar{w}_{k+1}}\sum_{\bar{z}_k}\mathcal{H}_{K,k+1}(\bar{a}_K)\mathcal{Q}_{kk}(\bar{a}_k)f(\bar{w}_{k+1},\bar{y}_k,\bar{u}_k,\bar{z}_k,\bar{a}_k|y_0)$$

$$=\sum_{\bar{w}_{k+1}}\sum_{\bar{z}_k}\mathcal{H}_{K,k+1}(\bar{a}_K)\mathcal{Q}_{kk}(\bar{a}_k)f(\bar{w}_{k+1},\bar{y}_k,\bar{z}_k,\bar{a}_k|y_0),$$

which ends the proof. The eighth equality follows from Assumption 5.

To further relax the dependence of Assumption 5 for Theorem 3.2 and simplify the proof above, we use the conclusion drawn from the proving process of Theorem D.1. By Equation (29), we note that the correctness of

$$f(Y_K(\bar{a}_K) = y_K, \ldots, Y_{k+1}(\bar{a}_{k+1}) = y_{k+1}|\bar{y}_k, \bar{u}_k, \bar{a}_k) = \sum_{\bar{w}_{k+1}} \mathcal{H}_{K,k+1}(\bar{a}_K) f(\bar{w}_{k+1}|\bar{u}_k, \bar{a}_k, \bar{y}_k)$$

does not depend on Assumption 5 with $k = K, K-1, \ldots, k+1$, which directly leads to:

$$\begin{aligned}
&\sum_{\bar{u}_k}\sum_{\bar{z}_k} f(Y_K(\bar{a}_K) = y_K, \cdots, Y_{k+1}(\bar{a}_{k+1}) = y_{k+1}|\bar{y}_k, \bar{u}_k, \bar{a}_k) f(\bar{z}_k|\bar{y}_k, \bar{u}_k, \bar{a}_k) \\
&\quad \times f(y_k, \bar{u}_k|\bar{y}_{k-1}, \bar{a}_k) f(\bar{y}_{k-1}, \bar{a}_k|y_0) \mathcal{Q}_{kk}(\bar{a}_k) \\
&= \sum_{\bar{u}_k}\sum_{\bar{z}_k}\sum_{\bar{w}_{k+1}} \mathcal{H}_{K,k+1}(\bar{a}_K) f(\bar{w}_{k+1}|\bar{u}_k, \bar{a}_k, \bar{y}_k) f(y_k, \bar{u}_k|\bar{y}_{k-1}, \bar{a}_k) f(\bar{y}_{k-1}, \bar{a}_k|y_0) f(\bar{z}_k|\bar{y}_k, \bar{u}_k, \bar{a}_k) \\
&\quad \times \mathcal{Q}_{kk}(\bar{a}_k) \\
&= \sum_{\bar{w}_{k+1}}\sum_{\bar{z}_k} \mathcal{H}_{K,k+1}(\bar{a}_K) \mathcal{Q}_{kk}(\bar{a}_k) f(\bar{w}_{k+1}, \bar{y}_k, \bar{z}_k, \bar{a}_k|y_0).
\end{aligned}$$

This fact implies that we only need Assumption 5 to hold at the $k^{th}$ stage to prove the correctness of the $k^{th}$ PHA identification method.

## D.4 Proof of Theorem 4.1

*Proof.* For a given sequence of regimes $\bar{d}_K$, to find the efficient influence function for $V(\bar{d}_K)$, we need to first find a random variable $G$ with mean 0 and

$$\frac{\partial V_x(\bar{d}_K)}{\partial x}|_{x=0} = \mathbb{E}[GS(\mathcal{O}; x)]|_{x=0}, \tag{33}$$

where $S(\mathcal{O}; x) = \partial \log f(\mathcal{O}; x)/\partial x$, and $V_x(\bar{d}_K)$ is the parameter of interest $V(\bar{d}_K)$ under a regular parametric submodel in $\mathcal{M}_{sp}$ indexed by $x$ that includes the true data generating mechanism at $x = 0$.

As we have assumed that treatment bridge functions $\mathcal{Q}_{tt}(\bar{a}_t)$, $t = 1, 2, \ldots, K$ that respectively solve Equations (3) and (4) exist in every data-generating mechanism of $\mathcal{M}_{sp}$, we have the following moment conditions:

For any $a_1, a_2, \ldots, a_K \in \{0, 1\}$, For $t = 1$,

$$\frac{\partial}{\partial x}\mathbb{E}_x[q_{11_x}(Y_0, Z_1, A_1)I(A_1 = a_1)|W_1, Y_0]|_{x=0} = 0, \tag{34}$$

and for $t = 2, 3, \cdots, K$,

$$\begin{aligned}&\frac{\partial}{\partial x}\mathbb{E}_x[q_{t-1,t-1_x}(\bar{Y}_{t-2}, \bar{Z}_{t-1}, \bar{A}_{t-1})|\bar{W}_t, \bar{Y}_{t-1}, \bar{A}_{t-1}]|_{x=0}\\&= \frac{\partial}{\partial x}\mathbb{E}_x[q_{tt_x}(\bar{Y}_{t-1}, \bar{Z}_t, \bar{A}_t)I(A_t = a_t)|\bar{W}_t, \bar{Y}_{t-1}, \bar{A}_{t-1}]|_{x=0}.\end{aligned} \tag{35}$$

Also note that moment condition (34) is equivalent to

$$\begin{aligned}&\mathbb{E}[\frac{\partial q_{11_x}(Y_0, Z_1, A_1)}{\partial x}|_{x=0}f(A_1|W_1, Y_0)|A_1, W_1, Y_0]\\&= \mathbb{E}[-q_{11}(Y_0, Z_1, A_1)f(A_1|W_1, Y_0)S(Z_1, A_1, W_1, Y_0)\\&\quad + q_{11}(Y_0, Z_1, A_1)f(A_1|W_1, Y_0)S(W_1, Y_0)|A_1, W_1, Y_0],\end{aligned} \tag{36}$$

and moment condition (35) is equivalent to

$$\begin{aligned}&\mathbb{E}[\frac{\partial q_{t-1,t-1_x}(\bar{Y}_{t-2}, \bar{Z}_{t-1}, \bar{A}_{t-1})}{\partial x}|_{x=0}|\bar{A}_{t-1}, \bar{W}_t, \bar{Y}_{t-1}]\\&- \mathbb{E}[\frac{\partial q_{tt_x}(\bar{Y}_{t-1}, \bar{Z}_t, \bar{A}_t)}{\partial x}|_{x=0}f(A_t|\bar{A}_{t-1}, \bar{W}_t, \bar{Y}_{t-1})|\bar{A}_t, \bar{W}_t, \bar{Y}_{t-1}]\\&= \mathbb{E}[q_{tt}(\bar{Y}_{t-1}, \bar{Z}_t, \bar{A}_t)f(A_t|\bar{A}_{t-1}, \bar{W}_t, \bar{Y}_{t-1})S(\bar{Z}_t, A_t|\bar{A}_{t-1}, \bar{W}_t, \bar{Y}_{t-1})|\bar{A}_t, \bar{W}_t, \bar{Y}_{t-1}]\\&- \mathbb{E}[q_{t-1,t-1}(\bar{Y}_{t-2}, \bar{Z}_{t-1}, \bar{A}_{t-1})S(\bar{Z}_t, A_t|\bar{A}_{t-1}, \bar{W}_t, \bar{Y}_{t-1})|\bar{A}_{t-1}, \bar{W}_t, \bar{Y}_{t-1}].\end{aligned} \tag{37}$$

By Theorem 3.1.2, we have that

$$\begin{aligned}\frac{\partial}{\partial x}V_x(\bar{d}_K)|_{x=0} &= \frac{\partial}{\partial x}\mathbb{E}_x[Y_K I(\bar{A}_K = \bar{d}_K(\bar{Y}_{K-1}, \bar{A}_{K-1}))q_{KK_x}(\bar{Y}_{K-1}, \bar{Z}_K, \bar{A}_K)]|_{x=0}\\&= \mathbb{E}[Y_K I(\bar{A}_K = \bar{d}_K(\bar{Y}_{K-1}, \bar{A}_{K-1}))\frac{\partial}{\partial x}q_{KK_x}(\bar{Y}_{K-1}, \bar{Z}_K, \bar{A}_K)|_{x=0}]\\&+ \mathbb{E}[Y_K I(\bar{A}_K = \bar{d}_K(\bar{Y}_{K-1}, \bar{A}_{K-1}))q_{KK}(\bar{Y}_{K-1}, \bar{Z}_K, \bar{A}_K)S(\bar{Y}_K, \bar{A}_K, \bar{Z}_K)].\end{aligned}$$

For the first term, we note that

$$\begin{aligned}&\mathbb{E}[Y_K I(\bar{A}_K = \bar{d}_K(\bar{Y}_{K-1}, \bar{A}_{K-1}))\frac{\partial}{\partial x}q_{KK_x}(\bar{Y}_{K-1}, \bar{Z}_K, \bar{A}_K)|_{x=0}]\\&= \sum_{\bar{y}_k}\sum_{\bar{a}'_k}\sum_{\bar{a}_k}\sum_{\bar{z}_k} y_K I(\bar{a}'_K = \bar{a}_K)I(\bar{d}_K(\bar{y}_{K-1}, \bar{a}_{K-1}) = \bar{a}_K)\frac{\partial}{\partial x}q_{KK_x}(\bar{y}_{K-1}, \bar{z}_K, \bar{a}_K)f(\bar{y}_K, \bar{z}_K, \bar{a}'_K).\end{aligned}$$

For given $\bar{a}_K$, exploiting the moment condition stated in Equation (35), we have:

$$
\begin{aligned}
&\sum_{\bar{y}_K}\sum_{\bar{z}_K} y_K I(\bar{d}_K(\bar{y}_{K-1},\bar{a}_{K-1})=\bar{a}_K)\frac{\partial}{\partial x}q_{KK_x}(\bar{y}_{K-1},\bar{z}_K,\bar{a}_K)f(\bar{y}_K,\bar{z}_K,\bar{a}_K)\\
&=\sum_{\bar{y}_K}\sum_{\bar{z}_K} y_K I(\bar{d}_K(\bar{y}_{K-1},\bar{a}_{K-1})=\bar{a}_K)\frac{\partial}{\partial x}q_{KK_x}(\bar{y}_{K-1},\bar{z}_K,\bar{a}_K)\\
&\quad\times\sum_{\bar{w}_K} h_{KK}(\bar{y}_K,\bar{w}_K,\bar{a}_K)f(\bar{w}_K,\bar{y}_{K-1},\bar{z}_K,\bar{a}_K)\\
&=\sum_{\bar{y}_K}\sum_{\bar{z}_K}\sum_{\bar{w}_K} y_K I(\bar{d}_K(\bar{y}_{K-1},\bar{a}_{K-1})=\bar{a}_K)\frac{\partial}{\partial x}q_{KK_x}(\bar{y}_{K-1},\bar{z}_K,\bar{a}_K)\\
&\quad\times f(\bar{z}_K,a_K|\bar{w}_K,\bar{y}_{K-1},\bar{a}_{K-1})h_{KK}(\bar{y}_K,\bar{w}_K,\bar{a}_K)f(\bar{w}_K,\bar{y}_{K-1},\bar{a}_{K-1})\\
&=\sum_{\bar{y}_K}\sum_{\bar{z}_{K-1}}\sum_{\bar{w}_K} y_K I(\bar{d}_K(\bar{y}_{K-1},\bar{a}_{K-1})=\bar{a}_K)\frac{\partial}{\partial x}q_{K-1,K-1_x}(\bar{y}_{K-2},\bar{z}_{K-1},\bar{a}_{K-1})\\
&\quad\times f(\bar{z}_{K-1}|\bar{w}_K,\bar{y}_{K-1},\bar{a}_{K-1})h_{KK}(\bar{y}_K,\bar{w}_K,\bar{a}_K)f(\bar{w}_K,\bar{y}_{K-1},\bar{a}_{K-1})\\
&+\mathbb{E}\Bigg[\sum_{y_K} y_K q_{K-1,K-1}(\bar{Y}_{K-2},\bar{Z}_{K-1},\bar{A}_{K-1})h_{KK}(y_K,\bar{Y}_{K-1},\bar{W}_K,\bar{a}_K)\\
&\quad\times I(d_K(\bar{Y}_{K-1},\bar{A}_{K-1})=a_K)I(\bar{A}_{K-1}=\bar{d}_{K-1}(\bar{Y}_{K-2},\bar{A}_{K-2})=\bar{a}_{K-1})\\
&\quad\times S(\bar{Z}_{K-1},\bar{W}_K,\bar{Y}_{K-1},\bar{A}_{K-1})\Bigg]\\
&-\mathbb{E}\Bigg[\sum_{y_K} y_K q_{KK}(\bar{Y}_{K-1},\bar{Z}_K,\bar{A}_K)h_{KK}(y_K,\bar{Y}_{K-1},\bar{W}_K,\bar{a}_K)\\
&\quad\times I(\bar{A}_K=\bar{d}_K(\bar{Y}_{K-1},\bar{A}_{K-1})=\bar{a}_K)S(\bar{Z}_K,\bar{W}_K,\bar{Y}_{K-1},\bar{A}_K)\Bigg]
\end{aligned}
$$

For the first term, which contains the derivative for treatment confounding bridge function $q_{K-1,K-1}$, we repeatedly use the moment condition implied by Equation (35), and first only focus on terms that contain the derivative for treatment bridge function $q_{tt}$, for $t = K-1, K-2, \ldots, 2$.

We will finally arrive at:

$$
\begin{aligned}
&\sum_{\bar{y}_K}\sum_{\bar{z}_2}\sum_{\bar{w}_3} y_K I(\bar{d}_K(\bar{y}_{K-1},\bar{a}_{K-1})=\bar{a}_K)\frac{\partial}{\partial x}q_{22_x}(\bar{y}_1,\bar{z}_2,\bar{a}_2)\\
&\quad\times f(\bar{w}_3,\bar{z}_2,\bar{y}_2,\bar{a}_2)h_{K3}(\bar{y}_K,\bar{w}_3,\bar{a}_K)\\
&=\sum_{\bar{y}_K}\sum_{\bar{z}_2}\sum_{\bar{w}_2} y_K I(\bar{d}_K(\bar{y}_{K-1},\bar{a}_{K-1})=\bar{a}_K)\frac{\partial}{\partial x}q_{22_x}(\bar{y}_1,\bar{z}_2,\bar{a}_2)\\
&\quad\times f(\bar{w}_2,\bar{z}_2,\bar{y}_1,\bar{a}_2)h_{K2}(\bar{y}_K,\bar{w}_2,\bar{a}_K)\\
&=\sum_{\bar{y}_K}\sum_{z_1}\sum_{\bar{w}_2} y_K I(\bar{d}_K(\bar{y}_{K-1},\bar{a}_{K-1})=\bar{a}_K)\frac{\partial}{\partial x}q_{11_x}(y_0,z_1,a_1)\\
&\quad\times f(z_1,\bar{w}_2,\bar{y}_1,a_1)h_{K2}(\bar{y}_K,\bar{w}_2,\bar{a}_K)\\
&+\mathbb{E}\Bigg[\sum_{y_K,\ldots,y_2} y_K q_{11}(Y_0,Z_1,A_1)h_{K2}(y_K,\cdots,y_2,\bar{Y}_1,\bar{W}_2,\bar{a}_K)\\
&\quad\times I(d_1(Y_0)=A_1=a_1)\prod_{j=2}^{K} I(d_j(y_{j-1},\cdots,y_2,\bar{Y}_1,\bar{a}_{j-1})=a_j)S(Z_1,\bar{W}_2,\bar{Y}_2,A_1)\Bigg]\\
&-\mathbb{E}\Bigg[\sum_{y_K,\ldots,y_2} y_K q_{22}(\bar{Y}_1,\bar{Z}_2,A_1)h_{K2}(y_K,\cdots,y_2,\bar{Y}_1,\bar{W}_2,\bar{a}_K)\\
&\quad\times I(\bar{d}_2(\bar{Y}_1,\bar{A}_1)=\bar{A}_2=\bar{a}_2)\prod_{j=3}^{K} I(d_j(y_{j-1},\cdots,y_2,\bar{Y}_1,\bar{a}_{j-1})=a_j)\\
&\quad\times S(\bar{Z}_2,\bar{W}_2,\bar{Y}_1,\bar{A}_2)\Bigg].
\end{aligned}
$$

For the first term, by the moment condition presented in Equation (34) we have:

$$
\begin{aligned}
&\sum_{\bar{y}_K}\sum_{z_1}\sum_{\bar{w}_2} y_K I(\bar{d}_K(\bar{y}_{K-1},\bar{a}_{K-1})=\bar{a}_K)\frac{\partial}{\partial x}q_{11_x}(y_0,z_1,a_1) \\
&\quad\times f(z_1,\bar{w}_2,\bar{y}_1,a_1)h_{K2}(\bar{y}_K,\bar{w}_2,\bar{a}_K) \\
&=\sum_{\bar{y}_K}\sum_{z_1}\sum_{w_1} y_K I(\bar{d}_K(\bar{y}_{K-1},\bar{a}_{K-1})=\bar{a}_K)\frac{\partial}{\partial x}q_{11_x}(y_0,z_1,a_1) \\
&\quad\times f(z_1,w_1,y_0,a_1)h_{K1}(\bar{y}_K,w_1,\bar{a}_K) \\
&=\mathbb{E}\Bigg[\sum_{y_K,\dots,y_1} y_K\frac{\partial}{\partial x}q_{11_x}(Y_0,Z_1,A_1)I(A_1=a_1)h_{K1}(y_K,\cdots,y_1,Y_0,W_1,\bar{a}_K) \\
&\quad\times I(d_1(Y_0)=a_1)\prod_{j=2}^{K}I(d_j(y_{j-1},\cdots,y_1,Y_0,\bar{a}_{j-1})=a_j)\Bigg] \\
&=-\mathbb{E}\Bigg[\sum_{y_K,\dots,y_1} y_K q_{11}(Y_0,Z_1,A_1)I(A_1=a_1)h_{K1}(y_K,\cdots,y_1,Y_0,W_1,\bar{a}_K) \\
&\quad\times S(A_1,Z_1,W_1,Y_0) \\
&\quad\times I(d_1(Y_0)=a_1)\prod_{j=2}^{K}I(d_j(y_{j-1},\cdots,y_1,Y_0,\bar{a}_{j-1})=a_j)\Bigg] \\
&+\mathbb{E}\Bigg[\sum_{y_K,\dots,y_1} y_K q_{11}(Y_0,Z_1,A_1)I(A_1=a_1)h_{K1}(y_K,\cdots,y_1,Y_0,W_1,\bar{a}_K) \\
&\quad\times S(W_1,Y_0) \\
&\quad\times I(d_1(Y_0)=a_1)\prod_{j=2}^{K}I(d_j(y_{j-1},\cdots,y_1,Y_0,\bar{a}_{j-1})=a_j)\Bigg] \\
&=-\mathbb{E}\Bigg[\sum_{y_K,\dots,y_1} y_K q_{11}(Y_0,Z_1,A_1)I(A_1=a_1)h_{K1}(y_K,\cdots,y_1,Y_0,W_1,\bar{a}_K) \\
&\quad\times S(A_1,Z_1,W_1,Y_0) \\
&\quad\times I(d_1(Y_0)=a_1)\prod_{j=2}^{K}I(d_j(y_{j-1},\cdots,y_1,Y_0,\bar{a}_{j-1})=a_j)\Bigg] \\
&+\mathbb{E}\Bigg[\sum_{y_K,\dots,y_1} y_K h_{K1}(y_K,\cdots,y_1,Y_0,W_1,\bar{a}_K)\prod_{j=2}^{K}I(d_j(y_{j-1},\cdots,y_1,Y_0,\bar{a}_{j-1})=a_j) \\
&\quad\times S(W_1,Y_0)I(d_1(Y_0)=a_1)\Bigg].
\end{aligned}
$$

The last equality follows from Equation (3). By these efforts, we have transformed all of the terms containing the derivative of the treatment confounding bridge functions into terms containing only the score functions. Also note the fact that for any arbitrary random variables $A, B$ and a square integrable function $g(\cdot)$,

$$\begin{aligned}\mathbb{E}[g(A)S(A)] &= \mathbb{E}[g(A)S(A)] + \mathbb{E}[g(A)S(B|A)] \\ &= \mathbb{E}[g(A)S(A,B)],\end{aligned}$$

and $\mathbb{E}[V(\bar{d}_K)S(\mathcal{O})] = 0$. For notational simplicity, for $k = 1, 2, \ldots, K$ we denote $\mathcal{Q}_{kk}(\bar{A}_k) = q_{kk}(\bar{Y}_{k-1}, \bar{Z}_k, \bar{A}_k)$ and

$$\begin{aligned}J_k(\bar{Y}_{k-1}, \bar{W}_k, \bar{a}_K)_{\bar{d}_K} = &\sum_{y_K,\cdots,y_k} y_K h_{Kk}(y_K, \cdots, y_k, \bar{Y}_{k-1}, \bar{W}_k, \bar{a}_K) \\ &\times \prod_{j=k+1}^{K} I(d_j(y_{j-1}, \cdots, y_k, \bar{Y}_{k-1}, \bar{a}_{j-1}) = a_j),\end{aligned}$$

we have:

$$\begin{aligned}&\frac{\partial}{\partial x} V_x(\bar{d}_K)|_{x=0} \\ &= \mathbb{E}\Bigg[\Bigg[\sum_{\bar{a}_K}\Bigg\{I(\bar{d}_K(\bar{Y}_{K-1}, \bar{A}_{K-1}) = \bar{A}_K = \bar{a}_K)\mathcal{Q}_{KK}(\bar{A}_K) \\ &\quad \times \Bigg[Y_K - \sum_{y_K} y_K h_{KK}(y_K, \bar{Y}_{K-1}, \bar{W}_K, \bar{a}_K)\Bigg] \\ &\quad + \sum_{k=1}^{K-1} I(\bar{d}_k(\bar{Y}_{k-1}, \bar{A}_{k-1}) = \bar{A}_k = \bar{a}_k)\mathcal{Q}_{kk}(\bar{A}_k) \\ &\quad \times \Bigg[J_{k+1}(\bar{Y}_k, \bar{W}_{k+1}, \bar{a}_K)_{\bar{d}_K} I(d_{k+1}(\bar{Y}_k, \bar{a}_k) = a_{k+1}) - J_k(\bar{Y}_{k-1}, \bar{W}_k, \bar{a}_K)_{\bar{d}_K}\Bigg] \\ &\quad + J_1(Y_0, W_1, \bar{a}_K)_{\bar{d}_K} I(d_1(Y_0) = a_1)\Bigg\} - V(\bar{d}_K)\Bigg] S(\mathcal{O})\Bigg]\end{aligned}$$

Therefore,

$$
\begin{aligned}
&\sum_{\bar{a}_K} \Big\{ I(\bar{d}_K(\bar{Y}_{K-1}, \bar{A}_{K-1}) = \bar{A}_K = \bar{a}_K) \mathcal{Q}_{KK}(\bar{A}_K) \\
&\quad \times \Big[ Y_K - \sum_{y_K} y_K h_{KK}(y_K, \bar{Y}_{K-1}, \bar{W}_K, \bar{a}_K) \Big] \\
&\quad + \sum_{k=1}^{K-1} I(\bar{d}_k(\bar{Y}_{k-1}, \bar{A}_{k-1}) = \bar{A}_k = \bar{a}_k) \mathcal{Q}_{kk}(\bar{A}_k) \\
&\quad \times \Big[ J_{k+1}(\bar{Y}_k, \bar{W}_{k+1}, \bar{a}_K)_{\bar{d}_K} I(d_{k+1}(\bar{Y}_k, \bar{a}_k) = a_{k+1}) - J_k(\bar{Y}_{k-1}, \bar{W}_k, \bar{a}_K)_{\bar{d}_K} \Big] \\
&\quad + J_1(Y_0, W_1, \bar{a}_K)_{\bar{d}_K} I(d_1(Y_0) = a_1) \Big\} - V(\bar{d}_K) \\
&=: IF_{V(\bar{d}_K)}
\end{aligned}
$$

is a valid influence function of $V(\bar{d}_K)$.

To show that $IF_{V(\bar{d}_K)} \in L_2(\mathcal{O})$ is the efficient influence function, it suffices to show that it belongs to the tangent space implied by the moment restrictions (36) and (37) on the scores. Specifically, the tangent space is comprised of the set of mean zero scores $S(\mathcal{O}) \in L_2(\mathcal{O})$ satisfying that:

For $t = 1$,

$$
\begin{aligned}
&\Big\{ \mathbb{E}[-q_{11}(Y_0, Z_1, A_1) f(A_1 | W_1, Y_0) S(\mathcal{O}) \\
&\qquad + q_{11}(Y_0, Z_1, A_1) f(A_1 | W_1, Y_0) S(\mathcal{O}) | A_1, W_1, Y_0] \Big\} \in cl(R(\Pi_1)),
\end{aligned}
$$

and for $t = 2, 3, \ldots, K$,

$$
\begin{aligned}
&\Big\{ \mathbb{E}[q_{tt}(\bar{Y}_{t-1}, \bar{Z}_t, \bar{A}_t) f(A_t | \bar{A}_{t-1}, \bar{W}_t, \bar{Y}_{t-1}) S(\mathcal{O}) | \bar{A}_t, \bar{W}_t, \bar{Y}_{t-1}] \\
&\quad - \mathbb{E}[q_{t-1,t-1}(\bar{Y}_{t-2}, \bar{Z}_{t-1}, \bar{A}_{t-1}) S(\mathcal{O}) | \bar{A}_{t-1}, \bar{W}_t, \bar{Y}_{t-1}] \Big\} \in cl(R(\Pi_t)),
\end{aligned}
$$

where $R()$ denotes the range space of an operator, and $cl()$ refers to the closure of a space. Evaluated at the submodel where the regularity Assumption $H.1$ holds, $R(\Pi_t)$ equals $L_2(\bar{A}_t, \bar{W}_t, \bar{Y}_{t-1})$ for $t = 1, 2, \cdots, K$, and thus the tangent space is essentially equal to $L_2(\mathcal{O})$ with zero mean, which completes the proof.

## D.5 Proof of Theorem 4.2

We first prove the multiple robustness of $\hat{V}_{PMR}(\bar{d}_K)$. The two key lemmas depicted below will be used to simplify the proof process. For notational simplicity, for $k = 1, 2, \ldots, K$ we denote $\mathcal{Q}_{kk}(\bar{A}_k) = q_{kk}(\bar{Y}_{k-1}, \bar{Z}_k, \bar{A}_k)$ and

$$J_k(\bar{Y}_{k-1}, \bar{W}_k, \bar{a}_K)_{\bar{d}_K} = \sum_{y_K,\cdots,y_k} y_K h_{Kk}(y_K, \cdots, y_k, \bar{Y}_{k-1}, \bar{W}_k, \bar{a}_K) \times \prod_{j=k+1}^{K} I(d_j(y_{j-1}, \cdots, y_k, \bar{Y}_{k-1}, \bar{a}_{j-1}) = a_j),$$

Under some regularity conditions, the nuisance estimators $\hat{\beta}_l, \hat{\gamma}_t$, $t, l = 1, 2, \ldots, K$ would converge in probability to some population limits, denoted by $\beta_l^*, \gamma_t^*$, $t, l = 1, 2, \ldots, K$, respectively. Cause we will mainly discuss the expectation about the bridge functions with converged parameters $\beta_l^*, \gamma_t^*$, $t, l = 1, 2, \ldots, K$ below, we will use $\mathcal{H}_{Kl}(\bar{a}_K), \mathcal{Q}_{tt}(\bar{a}_t)$ for $t, l = 1, 2, \ldots, K$ instead of $\mathcal{H}_{Kl}(\bar{a}_K; \beta_l^*), \mathcal{Q}_{tt}(\bar{a}_t; \gamma_t^*)$, $t, l = 1, 2, \ldots, K$ to simplify the notation, unless otherwise specified.

**Lemma D.4.** *If $\mathcal{H}_{Kl}(\bar{a}_K)$, $l = K, K-1, \ldots, k$ $(1 \le k \le K)$ that serve as solutions to Equations (1) and (2) exist, then for any treatment sequence $\bar{a}_K$, we have:*
*For $l = K$,*

$$\mathbb{E}[Y_K - \sum_{y_K} y_K h_{KK}(y_K, \bar{Y}_{K-1}, \bar{W}_K, \bar{a}_K) | \bar{Z}_K, \bar{Y}_{K-1}, \bar{A}_K] = 0,$$

*and for $l = K-1, \ldots, k$,*

$$\mathbb{E}[J_{l+1}(\bar{Y}_l, \bar{W}_{l+1}, \bar{a}_K)_{\bar{d}_K} I(d_{l+1}(\bar{Y}_l, \bar{a}_l) = a_{l+1}) - J_l(\bar{Y}_{l-1}, \bar{W}_l, \bar{a}_K)_{\bar{d}_K} | \bar{Z}_l, \bar{Y}_{l-1}, \bar{A}_l] = 0.$$

*Proof.* For $l = K$, we have:

$$\begin{aligned}
&\mathbb{E}[Y_K - \sum_{y_K} y_K h_{KK}(y_K, \bar{Y}_{K-1}, \bar{W}_K, \bar{a}_K) | \bar{Z}_K = \bar{z}_K, \bar{Y}_{K-1} = \bar{y}_{K-1}, \bar{A}_K = \bar{a}_K] \\
&= \sum_{\bar{y}_K} y_K (f(y_K | \bar{z}_K, \bar{y}_{K-1}, \bar{a}_K) - \sum_{\bar{w}_K} h_{KK}(\bar{y}_K, \bar{w}_K, \bar{a}_K) f(\bar{w}_K | \bar{z}_K, \bar{y}_{K-1}, \bar{a}_K)) \\
&= 0.
\end{aligned}$$

The second equality follows from Equation (1). For $l = K-1, \ldots, k$, we have:

$$
\begin{aligned}
&\mathbb{E}[J_{l+1}(\bar{Y}_l, \bar{W}_{l+1}, \bar{a}_K)_{\bar{d}_K} I(d_{l+1}(\bar{Y}_l, \bar{a}_l) = a_{l+1}) - J_l(\bar{Y}_{l-1}, \bar{W}_l, \bar{a}_K)_{\bar{d}_K} | \bar{Z}_l = \bar{z}_l, \bar{Y}_{l-1} = \bar{y}_{l-1}, \bar{A}_l = \bar{a}_l] \\
&= \sum_{\bar{y}_K} y_K \left[ \sum_{w_{l+1}} h_{K,l+1}(\bar{y}_K, \bar{w}_{l+1}, \bar{a}_K) f(y_l, \bar{w}_{l+1} | \bar{z}_l, \bar{y}_{l-1}, \bar{a}_l) - \sum_{\bar{w}_l} h_{Kl}(\bar{y}_K, \bar{w}_l, \bar{a}_K) f(\bar{w}_l | \bar{z}_l, \bar{y}_{l-1}, \bar{a}_l) \right] \\
&\quad \times \prod_{j=l+1}^{K} I(d_j(\bar{y}_{j-1}, \bar{a}_{j-1}) = a_j) \\
&= 0
\end{aligned}
$$

The second equality follows from Equation (2).

**Lemma D.5.** *If $\mathcal{Q}_{tt}(\bar{a}_t)$, $t = 1, 2, \ldots, k$ ($1 \leq k \leq K$) that serve as solutions to Equations* (3) *and* (4) *exist, then for any treatment sequence $\bar{a}_K$, we have:*

*For $t = 1$,*

$$
\mathbb{E}[I(d_1(Y_0) = a_1) - I(d_1(Y_0) = A_1 = a_1) \mathcal{Q}_{11}(a_1) | Y_0, W_1] = 0,
$$

*and for $t = 2, 3, \ldots, k$,*

$$
\begin{aligned}
&\mathbb{E}[I(\bar{d}_{t-1}(\bar{Y}_{t-2}, \bar{A}_{t-2}) = \bar{A}_{t-1} = \bar{a}_{t-1}) I(d_t(\bar{Y}_{t-1}, \bar{a}_{t-1}) = a_t) \mathcal{Q}_{t-1,t-1}(\bar{A}_{t-1}) \\
&\quad - I(\bar{d}_t(\bar{Y}_{t-1}, \bar{A}_{t-1}) = \bar{A}_t = \bar{a}_t) \mathcal{Q}_{tt}(\bar{A}_t) | \bar{Y}_{t-1}, \bar{W}_t, \bar{A}_{t-1}] = 0.
\end{aligned}
$$

*Proof.* For $t = 1$, we have:

$$
\begin{aligned}
&\mathbb{E}[I(d_1(Y_0) = a_1) - I(d_1(Y_0) = A_1 = a_1) \mathcal{Q}_{11}(a_1) | Y_0 = y_0, W_1 = w_1] \\
&= I(d_1(y_0) = a_1)(1 - \sum_{z_1} q_{11}(y_0, z_1, a_1) f(z_1, a_1 | y_0, w_1)) \\
&= 0.
\end{aligned}
$$

The second equality follows from Equation (3). For $t = 2, 3, \dots, K$, we have:

$$
\begin{aligned}
&\mathbb{E}[I(\bar{d}_{t-1}(\bar{Y}_{t-2}, \bar{A}_{t-2}) = \bar{A}_{t-1} = \bar{a}_{t-1}) I(d_t(\bar{Y}_{t-1}, \bar{a}_{t-1}) = a_t) \mathcal{Q}_{t-1,t-1}(\bar{A}_{t-1}) \\
&\quad - I(\bar{d}_t(\bar{Y}_{t-1}, \bar{A}_{t-1}) = \bar{A}_t = \bar{a}_t) \mathcal{Q}_{tt}(\bar{A}_t) | \bar{Y}_{t-1} = \bar{y}_{t-1}, \bar{W}_t = \bar{w}_t, \bar{A}_{t-1} = \bar{a}_{t-1}] \\
&= I(\bar{d}_t(\bar{y}_{t-1}, \bar{a}_{t-1}) = \bar{a}_t) \Bigg( \sum_{\bar{z}_{t-1}} q_{t-1,t-1}(\bar{y}_{t-2}, \bar{z}_{t-1}, \bar{a}_{t-1}) f(\bar{z}_{t-1} | \bar{w}_t, \bar{y}_{t-1}, \bar{a}_{t-1}) \\
&\quad - \sum_{\bar{z}_t} q_{tt}(\bar{y}_{t-1}, \bar{z}_t, \bar{a}_t) f(\bar{z}_t, a_t | \bar{w}_t, \bar{y}_{t-1}, \bar{a}_{t-1}) \Bigg) \\
&= 0
\end{aligned}
$$

The second equality follows from Equation (4).

We now prove the multiple robustness of $\hat{V}_{PMR}(\bar{d}_K)$. For its probability limitation with parameters $\beta_l^*, \gamma_t^*$, $t, l = 1, 2, \dots, K$, we follow the notation above to use $\mathcal{H}_{Kl}(\bar{a}_K), \mathcal{Q}_{tt}(\bar{a}_t)$ for $t, l = 1, 2, \dots, K$ instead of $\mathcal{H}_{Kl}(\bar{a}_K; \beta_l^*), \mathcal{Q}_{tt}(\bar{a}_t; \gamma_t^*)$, $t, l = 1, 2, \dots, K$ to simplify the notation, unless otherwise specified.

By Theorem 3.2 and Corollary 3.1, we have:

$$
\begin{aligned}
V(\bar{d}_K) &= \sum_{\bar{a}_K} \mathbb{E}[Y_K \mathcal{Q}_{KK}(\bar{A}_K) I(\bar{d}_K(\bar{Y}_{K-1}, \bar{A}_{K-1}) = \bar{A}_K = \bar{a}_K)] \\
&= \sum_{\bar{a}_K} \mathbb{E}[I(\bar{d}_{K-1}(\bar{Y}_{K-2}, \bar{A}_{K-2}) = \bar{A}_{K-1} = \bar{a}_{K-1}) \mathcal{Q}_{K-1,K-1}(\bar{A}_{K-1}) \\
&\quad \times J_K(\bar{Y}_{K-1}, \bar{W}_K, \bar{a}_K)_{\bar{d}_K} I(d_K(\bar{Y}_{K-1}, \bar{a}_{K-1}) = a_K)] \\
&= \cdots \\
&= \sum_{\bar{a}_K} \mathbb{E}[I(d_1(Y_0) = A_1 = a_1) \mathcal{Q}_{11}(\bar{A}_1) J_2(\bar{Y}_1, \bar{W}_2, \bar{a}_K)_{\bar{d}_K} I(d_2(\bar{Y}_1, a_1) = a_2)] \\
&= \sum_{\bar{a}_K} \mathbb{E}[J_1(Y_0, W_1, \bar{a}_K)_{\bar{d}_K} I(d_1(Y_0) = a_1)],
\end{aligned}
$$

where the $(k{+}1)^{th}$ equality holds at the submodel $\mathcal{M}_k$, for $k = 0, 1, 2, \cdots, K$.

Setting $k = K$, by Lemma D.5, we have:

$$
\begin{aligned}
&\mathbb{E}[I(\bar{d}_{t-1}(\bar{Y}_{t-2}, \bar{A}_{t-2}) = \bar{A}_{t-1} = \bar{a}_{t-1})I(d_t(\bar{Y}_{t-1}, \bar{a}_{t-1}) = a_t)\mathcal{Q}_{t-1,t-1}(\bar{A}_{t-1})J_t(\bar{Y}_{t-1}, \bar{W}_t, \bar{a}_K)_{\bar{d}_K} \\
&\quad - I(\bar{d}_t(\bar{Y}_{t-1}, \bar{A}_{t-1}) = \bar{A}_t = \bar{a}_t)\mathcal{Q}_{tt}(\bar{A}_t)J_t(\bar{Y}_{t-1}, \bar{W}_t, \bar{a}_K)_{\bar{d}_K}] \\
&= \mathbb{E}\Bigg[\mathbb{E}[I(\bar{d}_{t-1}(\bar{Y}_{t-2}, \bar{A}_{t-2}) = \bar{A}_{t-1} = \bar{a}_{t-1})I(d_t(\bar{Y}_{t-1}, \bar{a}_{t-1}) = a_t)\mathcal{Q}_{t-1,t-1}(\bar{A}_{t-1}) \\
&\quad - I(\bar{d}_t(\bar{Y}_{t-1}, \bar{A}_{t-1}) = \bar{A}_t = \bar{a}_t)\mathcal{Q}_{tt}(\bar{A}_t)|\bar{Y}_{t-1}, \bar{W}_t, \bar{A}_{t-1}]J_t(\bar{Y}_{t-1}, \bar{W}_t, \bar{a}_K)_{\bar{d}_K}\Bigg] \\
&= 0,
\end{aligned}
$$

for $t = 2, 3, \dots, K$. For $t = 1$, we have:

$$
\begin{aligned}
&\mathbb{E}[J_1(Y_0, W_1, \bar{a}_K)_{\bar{d}_K} I(d_1(Y_0) = a_1) - I(d_1(Y_0) = A_1 = a_1)\mathcal{Q}_{11}(\bar{A}_1)J_1(Y_0, W_1, \bar{a}_K)_{\bar{d}_K}] \\
&= \mathbb{E}\Bigg[\mathbb{E}[I(d_1(Y_0) = a_1) - I(d_1(Y_0) = a_1)I(A_1 = a_1)\mathcal{Q}_{11}(\bar{A}_1)|W_1, Y_0]J_1(Y_0, W_1, \bar{a}_K)_{\bar{d}_K}\Bigg] \\
&= 0,
\end{aligned}
$$

which proves the unbiased property of $\hat{V}_{PMR}(\bar{d}_K)$ under the submodel $\mathcal{M}_K$.

Setting $k = 0$, by Lemma D.4, we have:

$$
\begin{aligned}
&\mathbb{E}\Bigg[I(\bar{d}_l(\bar{Y}_{l-1}, \bar{A}_{l-1}) = \bar{A}_l = \bar{a}_l)\mathcal{Q}_{ll}(\bar{A}_l) \\
&\qquad \times \Bigg[J_{l+1}(\bar{Y}_l, \bar{W}_{l+1}, \bar{a}_K)_{\bar{d}_K} I(d_{l+1}(\bar{Y}_l, \bar{a}_l) = a_{l+1}) - J_l(\bar{Y}_{l-1}, \bar{W}_l, \bar{a}_K)_{\bar{d}_K}\Bigg]\Bigg] \\
&= \mathbb{E}\Bigg[I(\bar{d}_l(\bar{Y}_{l-1}, \bar{A}_{l-1}) = \bar{A}_l = \bar{a}_l)\mathcal{Q}_{ll}(\bar{A}_l) \\
&\qquad \times \mathbb{E}\Bigg[J_{l+1}(\bar{Y}_l, \bar{W}_{l+1}, \bar{a}_K)_{\bar{d}_K} I(d_{l+1}(\bar{Y}_l, \bar{a}_l) = a_{l+1}) - J_l(\bar{Y}_{l-1}, \bar{W}_l, \bar{a}_K)_{\bar{d}_K}|\bar{Z}_l, \bar{Y}_{l-1}, \bar{A}_l\Bigg]\Bigg] \\
&= 0,
\end{aligned}
$$

for $l = K-1, \ldots, 1$. For $l = K$, we have:

$$\begin{aligned}
&\mathbb{E}\Bigg[I(\bar{d}_K(\bar{Y}_{K-1}, \bar{A}_{K-1}) = \bar{A}_K = \bar{a}_K)\mathcal{Q}_{KK}(\bar{A}_K) \\
&\quad \times \Bigg[Y_K - \sum_{y_K} y_K h_{KK}(y_K, \bar{Y}_{K-1}, \bar{W}_K, \bar{a}_K)\Bigg]\Bigg] \\
&= \mathbb{E}\Bigg[I(\bar{d}_K(\bar{Y}_{K-1}, \bar{A}_{K-1}) = \bar{A}_K = \bar{a}_K)\mathcal{Q}_{KK}(\bar{A}_K) \\
&\quad \times \mathbb{E}\Bigg[Y_K - \sum_{y_K} y_K h_{KK}(y_K, \bar{Y}_{K-1}, \bar{W}_K, \bar{a}_K)|\bar{Z}_K, \bar{Y}_{K-1}, \bar{A}_K\Bigg]\Bigg] \\
&= 0,
\end{aligned}$$

which proves the unbiased property of $\hat{V}_{PMR}(\bar{d}_K)$ under the submodel $\mathcal{M}_0$.

For arbitrary $1 \leq k \leq K-1$, by Lemma D.5:

$$\begin{aligned}
&\mathbb{E}[I(\bar{d}_{t-1}(\bar{Y}_{t-2}, \bar{A}_{t-2}) = \bar{A}_{t-1} = \bar{a}_{t-1})I(d_t(\bar{Y}_{t-1}, \bar{a}_{t-1}) = a_t)\mathcal{Q}_{t-1,t-1}(\bar{A}_{t-1})J_t(\bar{Y}_{t-1}, \bar{W}_t, \bar{a}_K)_{\bar{d}_K} \\
&\quad - I(\bar{d}_t(\bar{Y}_{t-1}, \bar{A}_{t-1}) = \bar{A}_t = \bar{a}_t)\mathcal{Q}_{tt}(\bar{A}_t)J_t(\bar{Y}_{t-1}, \bar{W}_t, \bar{a}_K)_{\bar{d}_K}] \\
&= \mathbb{E}\Bigg[\mathbb{E}[I(\bar{d}_{t-1}(\bar{Y}_{t-2}, \bar{A}_{t-2}) = \bar{A}_{t-1} = \bar{a}_{t-1})I(d_t(\bar{Y}_{t-1}, \bar{a}_{t-1}) = a_t)\mathcal{Q}_{t-1,t-1}(\bar{A}_{t-1}) \\
&\quad - I(\bar{d}_t(\bar{Y}_{t-1}, \bar{A}_{t-1}) = \bar{A}_t = \bar{a}_t)\mathcal{Q}_{tt}(\bar{A}_t)|\bar{Y}_{t-1}, \bar{W}_t, \bar{A}_{t-1}]J_t(\bar{Y}_{t-1}, \bar{W}_t, \bar{a}_K)_{\bar{d}_K}\Bigg] \\
&= 0,
\end{aligned}$$

for $t = 2, 3, \ldots, k$. For $t = 1$,

$$\begin{aligned}
&\mathbb{E}[J_1(Y_0, W_1, \bar{a}_K)_{\bar{d}_K} I(d_1(Y_0) = a_1) - I(d_1(Y_0) = A_1 = a_1)\mathcal{Q}_{11}(\bar{A}_1)J_1(Y_0, W_1, \bar{a}_K)_{\bar{d}_K}] \\
&= \mathbb{E}\Bigg[\mathbb{E}[I(d_1(Y_0) = a_1) - I(d_1(Y_0) = a_1)I(A_1 = a_1)\mathcal{Q}_{11}(\bar{A}_1)|W_1, Y_0]J_1(Y_0, W_1, \bar{a}_K)_{\bar{d}_K}\Bigg] \\
&= 0.
\end{aligned}$$

Then, by Lemma D.4, we have:

$$
\begin{aligned}
&\mathbb{E}\Bigg[I(\bar{d}_l(\bar{Y}_{l-1}, \bar{A}_{l-1}) = \bar{A}_l = \bar{a}_l)\mathcal{Q}_{ll}(\bar{A}_l) \\
&\qquad \times \Big[J_{l+1}(\bar{Y}_l, \bar{W}_{l+1}, \bar{a}_K)_{\bar{d}_K} I(d_{l+1}(\bar{Y}_l, \bar{a}_l) = a_{l+1}) - J_l(\bar{Y}_{l-1}, \bar{W}_l, \bar{a}_K)_{\bar{d}_K}\Big]\Bigg] \\
&= \mathbb{E}\Bigg[I(\bar{d}_l(\bar{Y}_{l-1}, \bar{A}_{l-1}) = \bar{A}_l = \bar{a}_l)\mathcal{Q}_{ll}(\bar{A}_l) \\
&\qquad \times \mathbb{E}\Big[J_{l+1}(\bar{Y}_l, \bar{W}_{l+1}, \bar{a}_K)_{\bar{d}_K} I(d_{l+1}(\bar{Y}_l, \bar{a}_l) = a_{l+1}) - J_l(\bar{Y}_{l-1}, \bar{W}_l, \bar{a}_K)_{\bar{d}_K} | \bar{Z}_l, \bar{Y}_{l-1}, \bar{A}_l\Big]\Bigg] \\
&= 0,
\end{aligned}
$$

for $l = K-1, \dots, k+1$. For $l = K$, we have:

$$
\begin{aligned}
&\mathbb{E}\Bigg[I(\bar{d}_K(\bar{Y}_{K-1}, \bar{A}_{K-1}) = \bar{A}_K = \bar{a}_K)\mathcal{Q}_{KK}(\bar{A}_K) \\
&\qquad \times \Big[Y_K - \sum_{y_K} y_K h_{KK}(y_K, \bar{Y}_{K-1}, \bar{W}_K, \bar{a}_K)\Big]\Bigg] \\
&= \mathbb{E}\Bigg[I(\bar{d}_K(\bar{Y}_{K-1}, \bar{A}_{K-1}) = \bar{A}_K = \bar{a}_K)\mathcal{Q}_{KK}(\bar{A}_K) \\
&\qquad \times \mathbb{E}\Big[Y_K - \sum_{y_K} y_K h_{KK}(y_K, \bar{Y}_{K-1}, \bar{W}_K, \bar{a}_K) | \bar{Z}_K, \bar{Y}_{K-1}, \bar{A}_K\Big]\Bigg] \\
&= 0.
\end{aligned}
$$

By these facts, we have that under the submodel $\mathcal{M}_k$,

$$
\begin{aligned}
\mathbb{E}[\hat{V}_{PMR}(\bar{d}_K)] &= \sum_{\bar{a}_K} \mathbb{E}[I(\bar{d}_k(\bar{Y}_{k-1}, \bar{A}_{k-1}) = \bar{A}_k = \bar{a}_k)\mathcal{Q}_{kk}(\bar{A}_k) \\
&\qquad \times J_{k+1}(\bar{Y}_k, \bar{W}_k, \bar{a}_K)_{\bar{d}_K} I(d_{k+1}(\bar{Y}_k, \bar{a}_k) = a_{k+1})] \\
&= V(\bar{d}_K),
\end{aligned}
$$

which ends the proof of multiple robustness. To prove asymptotic normality and local efficiency, it suffices to show that $\hat{V}_{PMR}(\bar{d}_K)$ has an influence function equal to $IF_{V(\bar{d}_K)}$ under the intersection submodel model $\mathcal{M}_{int} = \mathcal{M}_0 \cap \mathcal{M}_K$. Following a Taylor expansion of $\hat{V}_{PMR}(\bar{d}_K)$ around $(\bar{\beta}_K^*, \bar{\gamma}_K^*)$ and standard M-estimation arguments, we have that:

$$
\begin{aligned}
&\sqrt{n}(\hat{V}_{PMR}(\bar{d}_K) - V_{PMR}(\bar{d}_K)) \\
&= \sqrt{n}\mathbb{P}_n\Bigg\{ I(\bar{d}_K(\bar{Y}_{K-1}, \bar{A}_{K-1}) = \bar{A}_K = \bar{a}_K)\mathcal{Q}_{KK}(\bar{A}_K; \gamma_K^*) \\
&\qquad \times \left[ Y_K - \sum_{y_K} y_K h_{KK}(y_K, \bar{Y}_{K-1}, \bar{W}_K, \bar{a}_K; \beta_K^*) \right] \\
&\qquad + \sum_{k=1}^{K-1} I(\bar{d}_k(\bar{Y}_{k-1}, \bar{A}_{k-1}) = \bar{A}_k = \bar{a}_k)\mathcal{Q}_{kk}(\bar{A}_k; \gamma_k^*) \\
&\qquad \times \left[ J_{k+1}(\bar{Y}_k, \bar{W}_{k+1}, \bar{a}_K; \beta_{k+1}^*)_{\bar{d}_K} I(d_{k+1}(\bar{Y}_k, \bar{a}_k) = a_{k+1}) - J_k(\bar{Y}_{k-1}, \bar{W}_k, \bar{a}_K; \beta_k^*)_{\bar{d}_K} \right] \\
&\qquad + J_1(Y_0, W_1, \bar{a}_K; \beta_1^*)_{\bar{d}_K} I(d_1(Y_0) = a_1) - V_{PMR}(\bar{d}_K) \Bigg\} \\
&+ \sqrt{n}(\hat{\gamma}_K - \gamma_K^*)\mathbb{E}\Bigg[ I(\bar{d}_K(\bar{Y}_{K-1}, \bar{A}_{K-1}) = \bar{A}_K = \bar{a}_K)\frac{\partial}{\partial \gamma_K}\mathcal{Q}_{KK}(\bar{A}_K; \gamma_K)|_{\gamma_K = \gamma_K^*} \\
&\qquad \times \left[ Y_K - \sum_{y_K} y_K h_{KK}(y_K, \bar{Y}_{K-1}, \bar{W}_K, \bar{a}_K; \beta_K^*) \right] \Bigg] \\
&+ \sqrt{n}\sum_{k=1}^{K-1}(\hat{\gamma}_k - \gamma_k^*)\mathbb{E}\Bigg[ I(\bar{d}_k(\bar{Y}_{k-1}, \bar{A}_{k-1}) = \bar{A}_k = \bar{a}_k)\frac{\partial}{\partial \gamma_k}\mathcal{Q}_{kk}(\bar{A}_k; \gamma_k)|_{\gamma_k = \gamma_k^*} \\
&\qquad \times \left[ J_{k+1}(\bar{Y}_k, \bar{W}_{k+1}, \bar{a}_K; \beta_{k+1}^*)_{\bar{d}_K} I(d_{k+1}(\bar{Y}_k, \bar{a}_k) = a_{k+1}) - J_k(\bar{Y}_{k-1}, \bar{W}_k, \bar{a}_K; \beta_k^*)_{\bar{d}_K} \right] \Bigg] \\
&+ \sqrt{n}\sum_{k=2}^{K}(\hat{\beta}_k - \beta_k^*)\mathbb{E}\Bigg[ I(\bar{d}_{k-1}(\bar{Y}_{k-2}, \bar{A}_{k-2}) = \bar{A}_{k-1} = \bar{a}_{k-1})\frac{\partial}{\partial \beta_k} J_k(\bar{Y}_{k-1}, \bar{W}_k, \bar{a}_K; \beta_k)_{\bar{d}_K}|_{\beta_k = \beta_k^*} \\
&\qquad \times \left[ I(d_k(\bar{Y}_{k-1}, \bar{a}_{k-1}) = a_k)\mathcal{Q}_{k-1,k-1}(\bar{A}_{k-1}) - I(d_k(\bar{Y}_{k-1}, \bar{A}_{k-1}) = A_k = a_k)\mathcal{Q}_{kk}(\bar{A}_k) \right] \Bigg] \\
&+ \sqrt{n}(\hat{\beta}_1 - \beta_1^*)\mathbb{E}\left[ \frac{\partial}{\partial \beta_1} J_1(Y_0, W_1, \bar{a}_K; \beta_1)_{\bar{d}_K}|_{\beta_1 = \beta_1^*} \left[ I(d_1(Y_0) = a_1) - I(d_1(Y_0) = A_1 = a_1)\mathcal{Q}_{11}(\bar{A}_1) \right] \right] \\
&+ o_p(1)
\end{aligned}
$$

According to the previous arguments in the proof of multiple robustness, by Lemmas D.5 and D.4, under the intersection model $\mathcal{M}_{int}$ where $\mathcal{H}_{Kl}(\bar{a}_K; \beta_l)$ and $\mathcal{Q}_{tt}(\bar{a}_t; \gamma)$, $t, l = 1, 2, \ldots, K$ are all correctly specified, for $t = 1, 2, 3, \cdots, K-1$ and $l = 2, 3, \cdots, K$, we have:

$$\mathbb{E}\Bigg[I(\bar{d}_K(\bar{Y}_{K-1}, \bar{A}_{K-1}) = \bar{A}_K = \bar{a}_K)\frac{\partial}{\partial \gamma_K}\mathcal{Q}_{KK}(\bar{A}_K; \gamma_K)|_{\gamma_K=\gamma_K^*}$$
$$\times \Bigg[Y_K - \sum_{y_K} y_K h_{KK}(y_K, \bar{Y}_{K-1}, \bar{W}_K, \bar{a}_K; \beta_K^*)\Bigg]\Bigg] = 0;$$

$$\mathbb{E}\Bigg[I(\bar{d}_t(\bar{Y}_{t-1}, \bar{A}_{t-1}) = \bar{A}_t = \bar{a}_t)\frac{\partial}{\partial \gamma_t}\mathcal{Q}_{tt}(\bar{A}_t; \gamma_t)|_{\gamma_t=\gamma_t^*}$$
$$\times \Bigg[J_{t+1}(\bar{Y}_t, \bar{W}_{t+1}, \bar{a}_K; \beta_{t+1}^*)_{\bar{d}_K} I(d_{t+1}(\bar{Y}_t, \bar{a}_t) = a_{t+1}) - J_t(\bar{Y}_{t-1}, \bar{W}_t, \bar{a}_K; \beta_t^*)_{\bar{d}_K}\Bigg]\Bigg] = 0;$$

$$\mathbb{E}\Bigg[I(\bar{d}_{l-1}(\bar{Y}_{l-2}, \bar{A}_{l-2}) = \bar{A}_{l-1} = \bar{a}_{l-1})\frac{\partial}{\partial \beta_l}J_l(\bar{Y}_{l-1}, \bar{W}_l, \bar{a}_K; \beta_l)_{\bar{d}_K}|_{\beta_l=\beta_l^*}$$
$$\times \Bigg[I(d_l(\bar{Y}_{l-1}, \bar{a}_{l-1}) = a_l)\mathcal{Q}_{l-1,l-1}(\bar{A}_{l-1}) - I(d_l(\bar{Y}_{l-1}, \bar{A}_{l-1}) = A_l = a_l)\mathcal{Q}_{ll}(\bar{A}_l)\Bigg]\Bigg] = 0;$$

$$\mathbb{E}\Bigg[\frac{\partial}{\partial \beta_1}J_1(Y_0, W_1, \bar{a}_K; \beta_1)_{\bar{d}_K}|_{\beta_1=\beta_1^*}\Bigg[I(d_1(Y_0) = a_1) - I(d_1(Y_0) = A_1 = a_1)\mathcal{Q}_{11}(\bar{A}_1)\Bigg]\Bigg] = 0.$$

These complete our proof.

## D.6 Proof of Proposition A.1

*Proof.* To prove the multiple robustness of this identification result, we again prove two key lemmas:

*Lemma* D.6. *If $\mathcal{H}_{Kl}(\bar{a}_K)$, $l = K, K-1, \ldots, k$ $(1 \le k \le K)$ that serve as solutions to Equations* (1) *and* (2) *exist, then for any treatment sequence $\bar{a}_K$, we have:*

*For $l = K$,*

$$f(\bar{y}_K, \bar{z}_K, \bar{a}_K|y_0) - \sum_{\bar{w}_K} \mathcal{H}_{KK}(\bar{a}_K) f(\bar{w}_K, \bar{z}_K, \bar{y}_{K-1}, \bar{a}_K|y_0) = 0,$$

*and for $l = K-1, \ldots, k$,*

$$\sum_{\bar{w}_{l+1}} \mathcal{H}_{K,l+1}(\bar{a}_K) f(\bar{w}_{l+1}, \bar{z}_l, \bar{y}_l, \bar{a}_l|y_0) - \sum_{\bar{w}_l} \mathcal{H}_{Kl}(\bar{a}_K) f(\bar{w}_l, \bar{z}_l, \bar{y}_{l-1}, \bar{a}_l|y_0) = 0.$$

*Proof.* By Equation (1),

$$\begin{aligned}
&f(\bar{y}_K, \bar{z}_K, \bar{a}_K|y_0) - \sum_{\bar{w}_K} \mathcal{H}_{KK}(\bar{a}_K) f(\bar{w}_K, \bar{z}_K, \bar{y}_{K-1}, \bar{a}_K|y_0) \\
&= \left[ f(y_K|\bar{y}_{K-1}, \bar{z}_K, \bar{a}_K) - \sum_{\bar{w}_K} \mathcal{H}_{KK}(\bar{a}_K) f(\bar{w}_K|\bar{z}_K, \bar{y}_{K-1}, \bar{a}_K|y_0) \right] f(\bar{z}_K, \bar{y}_{K-1}, \bar{a}_K|y_0) \\
&= 0.
\end{aligned}$$

By Equation (2),

$$\begin{aligned}
&\sum_{\bar{w}_{l+1}} \mathcal{H}_{K,l+1}(\bar{a}_K) f(\bar{w}_{l+1}, \bar{z}_l, \bar{y}_l, \bar{a}_l|y_0) - \sum_{\bar{w}_l} \mathcal{H}_{Kl}(\bar{a}_K) f(\bar{w}_l, \bar{z}_l, \bar{y}_{l-1}, \bar{a}_l|y_0) \\
&= f(\bar{z}_l, \bar{y}_{l-1}, \bar{a}_l|y_0) \left[ \sum_{\bar{w}_{l+1}} \mathcal{H}_{K,l+1}(\bar{a}_K) f(\bar{w}_{l+1}, y_l|\bar{z}_l, \bar{y}_{l-1}, \bar{a}_l) - \sum_{\bar{w}_l} \mathcal{H}_{Kl}(\bar{a}_K) f(\bar{w}_l|\bar{z}_l, \bar{y}_{l-1}, \bar{a}_l) \right] \\
&= 0
\end{aligned}$$

*Lemma D.7. If $\mathcal{Q}_{tt}(\bar{a}_t)$, $t = 1, 2, \ldots, k$ $(1 \leq k \leq K)$ that serve as solutions to Equations (3) and (4) exist, then for any treatment sequence $\bar{a}_K$, we have:*

*For $t = 1$,*

$$f(w_1|y_0) - \sum_{z_1} \mathcal{Q}_{11}(a_1) f(z_1, a_1, w_1|y_0) = 0$$

*and for $t = 2, 3, \ldots, k$,*

$$\sum_{\bar{z}_t} \mathcal{Q}_{tt}(\bar{a}_t) f(\bar{w}_t, \bar{z}_t, \bar{y}_{t-1}, \bar{a}_t|y_0) - \sum_{\bar{z}_{t-1}} \mathcal{Q}_{t-1,t-1}(\bar{a}_{t-1}) f(\bar{w}_t, \bar{z}_{t-1}, \bar{y}_{t-1}, \bar{a}_{t-1}|y_0) = 0.$$

*Proof.* By Equation (3),

$$\begin{aligned}
&f(w_1|y_0) - \sum_{z_1} \mathcal{Q}_{11}(a_1) f(z_1, a_1, w_1|y_0) \\
&= f(w_1|y_0) \left[ 1 - \sum_{z_1} \mathcal{Q}_{11}(a_1) f(z_1, a_1|w_1, y_0) \right] \\
&= 0.
\end{aligned}$$

By Equation (4),

$$
\begin{aligned}
&\sum_{\bar{z}_t} \mathcal{Q}_{tt}(\bar{a}_t) f(\bar{w}_t, \bar{z}_t, \bar{y}_{t-1}, \bar{a}_t | y_0) - \sum_{\bar{z}_{t-1}} \mathcal{Q}_{t-1,t-1}(\bar{a}_{t-1}) f(\bar{w}_t, \bar{z}_{t-1}, \bar{y}_{t-1}, \bar{a}_{t-1} | y_0) \\
&= \left[ \sum_{\bar{z}_t} \mathcal{Q}_{tt}(\bar{a}_t) f(\bar{z}_t, a_t | \bar{w}_t, \bar{y}_{t-1}, \bar{a}_{t-1}) - \sum_{\bar{z}_{t-1}} \mathcal{Q}_{t-1,t-1}(\bar{a}_{t-1}) f(\bar{z}_{t-1} | \bar{w}_t, \bar{y}_{t-1}, \bar{a}_{t-1}) \right] \\
&\times f(\bar{w}_t, \bar{y}_{t-1}, \bar{a}_{t-1} | y_0) \\
&= 0.
\end{aligned}
$$

Setting $k = K$, for submodel $\mathcal{M}_K$, it follows from Theorem 3.2 that

$$
f(\bar{Y}_K(\bar{a}_K) = \bar{y}_K | Y_0 = y_0) = \sum_{\bar{z}_K} \mathcal{Q}_{KK}(\bar{a}_K) f(\bar{y}_K, \bar{z}_K, \bar{a}_K | y_0).
$$

By Lemma D.7, we have that the remained terms

$$
\begin{aligned}
&\sum_{k=2}^{K} \sum_{\bar{w}_k} \mathcal{H}_{Kk}(\bar{a}_K) \\
&\quad \times \left( \sum_{\bar{z}_k} \mathcal{Q}_{kk}(\bar{a}_k) f(\bar{w}_k, \bar{z}_k, \bar{y}_{k-1}, \bar{a}_k | y_0) - \sum_{\bar{z}_{k-1}} \mathcal{Q}_{k-1,k-1}(\bar{a}_{k-1}) f(\bar{w}_k, \bar{z}_{k-1}, \bar{y}_{k-1}, \bar{a}_{k-1} | y_0) \right) \\
&+ \sum_{w_1} \mathcal{H}_{K1}(\bar{a}_K) \left( f(w_1 | y_0) - \sum_{z_1} \mathcal{Q}_{11}(a_1) f(z_1, a_1, w_1 | y_0) \right) \\
&= 0,
\end{aligned}
$$

which proves the identification result under the submodel $\mathcal{M}_K$.

Setting $k = 0$, for submodel $\mathcal{M}_0$, it also follows from Theorem 3.2 that

$$
f(\bar{Y}_K(\bar{a}_K) = \bar{y}_K | Y_0 = y_0) = \sum_{w_1} \mathcal{H}_{K1}(\bar{a}_K) f(w_1 | y_0).
$$

By Lemma D.6, we have that the remained terms

$$
\begin{aligned}
&\sum_{\bar{z}_K} \mathcal{Q}_{KK}(\bar{a}_K) \left( f(\bar{y}_K, \bar{z}_K, \bar{a}_K | y_0) - \sum_{\bar{w}_K} \mathcal{H}_{KK}(\bar{a}_K) f(\bar{w}_K, \bar{z}_K, \bar{y}_{K-1}, \bar{a}_K | y_0) \right) \\
&\quad + \sum_{k=1}^{K-1} \sum_{\bar{z}_k} \Bigg\{ \mathcal{Q}_{kk}(\bar{a}_k) \\
&\qquad \times \left( \sum_{\bar{w}_{k+1}} \mathcal{H}_{K,k+1}(\bar{a}_K) f(\bar{w}_{k+1}, \bar{z}_k, \bar{y}_k, \bar{a}_k | y_0) - \sum_{\bar{w}_k} \mathcal{H}_{Kk}(\bar{a}_K) f(\bar{w}_k, \bar{z}_k, \bar{y}_{k-1}, \bar{a}_k | y_0) \right) \Bigg\} \\
&= 0,
\end{aligned}
$$

which proves the identification result under the submodel $\mathcal{M}_0$.

For arbitrary $1 \leq k \leq K-1$, by Lemma D.6,

$$
\begin{aligned}
&\sum_{\bar{z}_K} \mathcal{Q}_{KK}(\bar{a}_K) \left( f(\bar{y}_K, \bar{z}_K, \bar{a}_K | y_0) - \sum_{\bar{w}_K} \mathcal{H}_{KK}(\bar{a}_K) f(\bar{w}_K, \bar{z}_K, \bar{y}_{K-1}, \bar{a}_K | y_0) \right) \\
&\quad + \sum_{l=k+1}^{K-1} \sum_{\bar{z}_l} \Bigg\{ \mathcal{Q}_{ll}(\bar{a}_l) \\
&\qquad \times \left( \sum_{\bar{w}_{l+1}} \mathcal{H}_{K,l+1}(\bar{a}_K) f(\bar{w}_{l+1}, \bar{z}_l, \bar{y}_l, \bar{a}_l | y_0) - \sum_{\bar{w}_l} \mathcal{H}_{Kl}(\bar{a}_K) f(\bar{w}_l, \bar{z}_l, \bar{y}_{l-1}, \bar{a}_l | y_0) \right) \Bigg\} \\
&= 0,
\end{aligned}
$$

and

$$
\begin{aligned}
&\sum_{t=2}^{k} \sum_{\bar{w}_t} \mathcal{H}_{Kt}(\bar{a}_K) \\
&\quad \times \left( \sum_{\bar{z}_t} \mathcal{Q}_{tt}(\bar{a}_t) f(\bar{w}_t, \bar{z}_t, \bar{y}_{t-1}, \bar{a}_t | y_0) - \sum_{\bar{z}_{t-1}} \mathcal{Q}_{t-1,t-1}(\bar{a}_{t-1}) f(\bar{w}_t, \bar{z}_{t-1}, \bar{y}_{t-1}, \bar{a}_{t-1} | y_0) \right) \\
&+ \sum_{w_1} \mathcal{H}_{K1}(\bar{a}_K) \left( f(w_1 | y_0) - \sum_{z_1} \mathcal{Q}_{11}(a_1) f(z_1, a_1, w_1 | y_0) \right) \\
&= 0.
\end{aligned}
$$

By Theorem 3.2, under the submodel $\mathcal{M}_k$, the only remaining term gives the identification:

$$
\begin{aligned}
&f(\bar{Y}_K(\bar{a}_K) = \bar{y}_K | Y_0 = y_0) \\
&= \sum_{\bar{w}_{k+1}} \sum_{\bar{z}_k} \mathcal{H}_{K,k+1}(\bar{a}_K) \mathcal{Q}_{kk}(\bar{a}_k) f(\bar{w}_{k+1}, \bar{z}_k, \bar{y}_k, \bar{a}_k | y_0).
\end{aligned}
$$

This ends the proof.

## D.7 Proof of Theorem 5.1

*Proof.* As the proof technique used in 2-stage setting has no essential difference from the general $K$-stage setting, we take $K = 2$ as an example. The proof can be readily generalized to any $K$-stage setting with $K \geq 3$.

We first present a useful Lemma from Zhao & Cui (2025), which illustrates the basic technique of cross-fitting (Schick 1986, Chernozhukov et al. 2018, Wager 2024).

*Lemma* D.8. *Consider two independent samples $\mathcal{O}_1 = (O_1, \cdots, O_n)$ and $\mathcal{O}_2 = (O_{n+1}, \cdots, O_{\tilde{n}})$, let $\hat{f}(o)$ be a function estimated from $\mathcal{O}_2$ and $\mathbb{P}_n$ be the empirical measure over $\mathcal{O}_1$, then we have*

$$(\mathbb{P}_n - \mathbb{P})(\hat{f} - f) = O_p(\frac{\|\hat{f} - f\|_{L_2}}{\sqrt{n}}).$$

*Proof.* The complete proof is akin to Zhao & Cui (2025), thus omitted.

The Lemma below ensures the finite VC (Vapnik-Chervonenkis)-dimension for the multiplication of VC-classes, which will be used to ensure the envelope function classes that will be defined later are all VC-classes:

*Lemma* D.9. *If function classes $\Pi_1$ and $\Pi_2$ both have finite VC-dimensions $\nu_1$ and $\nu_2$, then the function class $\Pi_3 = \{\pi_1(x) \times \pi_2(x) | \pi_1 \in \Pi_1, \pi_2 \in \Pi_2\}$ also has a finite VC-dimension $\nu_3$.*

*Proof.* Set $n$ as any positive integer. For a certain sample $x^n = \{x_1, x_2, \cdots, x_n\}$ with size $n$, the sets $\Pi_i(x^n)$, $i = 1, 2, 3$, respectively correspond to all those vectors in $\mathbb{R}^n$ that can be realized by applying a function $\pi_i \in \Pi_i$ to the collection $(x_1, \ldots, x_n)$, that is:

$$\Pi_i(x^n) := \{(\pi_i(x_1), \ldots, \pi_i(x_n)) | \pi_i \in \Pi_i\}.$$

We first note the fact that for a certain sample $x^n$, $card(\Pi_3(x^n)) \leq card(\Pi_1(x^n)) \times card(\Pi_2(x^n))$. From the definition of $\Pi_3$, we know that each element $\pi_3(x) \in \Pi_3$ corresponds to an element $\pi_1(x) \in \Pi_1$ and $\pi_2(x) \in \Pi_2$, respectively. Thus, the classification results of $x^n$ generated by $\pi_3(x)$ can be expressed as the classification results generated by $\pi_1(x)$ multiplied by those generated by $\pi_2(x)$. Due to this fact, we can get the point that each element of $\Pi_3(x^n)$ can be generated by the product of a certain element of $\Pi_1(x^n)$ and a certain element of $\Pi_2(x^n)$, and this fact leads to the conclusion that $card(\Pi_3(x^n)) \leq card(\Pi_1(x^n)) \times card(\Pi_2(x^n))$.

Next, according to Proposition 4.18 in Wainwright Wainwright (2019), we have

$$card(\Pi_1(x^n)) \leq (n+1)^{\nu_1}, card(\Pi_2(x^n)) \leq (n+1)^{\nu_2},$$

then

$$card(\Pi_3(x^n)) \leq (n+1)^{\nu_1} \times (n+1)^{\nu_2} = (n+1)^{\nu_1 + \nu_2}.$$

If the VC-dimension of $\Pi_3$ is infinite, then for any positive integer $n$ and arbitrary sample with size $n$, $card(\Pi_3(x^n)) = 2^n$. However, as $card(\Pi_3(x^n)) \leq (n+1)^{\nu_1+\nu_2}$, a contradiction when $n$ is large enough is apparent, thus ends the proof.

By cross-fitting algorithm, we randomly split the data set into $K$ non-intersecting folds, and get $\hat{V}_n(\theta)$ through

$$\hat{V}_n(\theta) = \frac{1}{K}\sum_{k=1}^{K} \hat{V}_k(\theta) =: \frac{1}{K}\sum_{k=1}^{K} \mathbb{P}_{n,k}\Big\{\phi(O;\hat{h}^{(-k)},\hat{q}^{(-k)},\theta)\Big\},$$

where $\mathbb{P}_{n,k}$ denotes the empirical average over the $k$-th fold, and $\hat{h}^{(-k)}, \hat{q}^{(-k)}$ denote estimators of the nuisance parameters constructed on the sample set excluding the $k$-th fold.

We divide the main proof into three steps below. At the first step, we will prove that for any $\theta \in \Theta$, $\sqrt{n}(\hat{V}_n(\theta) - \hat{V}_n^*(\theta)) = o_p(1)$. At the second step, we will prove that $n^{1/3}\|\hat{\theta} - \theta\| = O_p(1)$. At the final step, we will prove that $\sqrt{n}(V(\hat{\theta}) - V(\theta^*)) = o_p(1)$ and $\sqrt{n}(\hat{V}_n(\hat{\theta}) - V(\theta^*)) \xrightarrow{d} \mathcal{N}(0, \sigma^2)$, and $\sigma^2 = \mathbb{E}[EIF^2_{V(\theta^*)}]$.

**Step 1.** To prove that for any $\theta \in \Theta$, $\sqrt{n}(\hat{V}_n(\theta) - \hat{V}_n^*(\theta)) = o_p(1)$, it is sufficient to prove that $\sqrt{n}(\hat{V}_k(\theta) - \hat{V}_k^*(\theta)) = o_p(1)$, for $k = 1, 2, \cdots, K$, where $\hat{V}_k^*(\theta) = \mathbb{P}_{n,k}\{\phi(O; h, q, \theta)\}$. Note the following decomposition:

$$
\begin{aligned}
&\hat{V}_{n,k}(\theta) - \hat{V}^*_{n,k}(\theta) \\
&= \mathbb{P}_{n,k}\Big\{\phi(O; \hat{h}^{(-k)}, \hat{q}^{(-k)}, \theta) - \phi(O; h, q, \theta)\Big\} \\
&= \sum_{\bar{a}_2} \Bigg\{ \mathbb{P}_{n,k} \Bigg\{ I(A_2 = a_2) I(A_1 = a_1) I(d_1(Y_0) = a_1) I(d_2(\bar{Y}_1, A_1) = a_2) \\
&\quad \times \Big[\hat{q}_{22}^{(-k)}(\bar{A}_2, \bar{Y}_1, \bar{Z}_2) - q_{22}(\bar{A}_2, \bar{Y}_1, \bar{Z}_2)\Big] Y_2 \\
&\quad - I(A_2 = a_2) I(A_1 = a_1) I(d_1(Y_0) = a_1) I(d_2(\bar{Y}_1, A_1) = a_2) \\
&\quad \times \Bigg[\hat{q}_{22}^{(-k)}(\bar{A}_2, \bar{Y}_1, \bar{Z}_2) \sum_{y_2} y_2 \hat{h}_{22}^{(-k)}(y_2, a_2, A_1, \bar{Y}_1, \bar{W}_2) - q_{22}(\bar{A}_2, \bar{Y}_1, \bar{Z}_2) \sum_{y_2} y_2 h_{22}(y_2, a_2, A_1, \bar{Y}_1, \bar{W}_2)\Bigg] \\
&\quad + I(A_1 = a_1) I(d_1(Y_0) = a_1) I(d_2(\bar{Y}_1, A_1) = a_2) \\
&\quad \times \Bigg[\hat{q}_{11}^{(-k)}(Y_0, Z_1, A_1) \sum_{y_2} y_2 \hat{h}_{22}^{(-k)}(y_2, a_2, \bar{A}_1, \bar{Y}_1, \bar{W}_2) - q_{11}(Y_0, Z_1, A_1) \sum_{y_2} y_2 h_{22}(y_2, a_2, \bar{A}_1, \bar{Y}_1, \bar{W}_2)\Bigg] \\
&\quad - I(A_1 = a_1) I(d_1(Y_0) = a_1) \\
&\quad \times \sum_{y_1, y_2} y_2 \Big[\hat{q}_{11}^{(-k)}(Y_0, Z_1, A_1) \hat{h}_{21}^{(-k)}(\bar{y}_2, \bar{a}_2, Y_0, W_1) - q_{11}(Y_0, Z_1, A_1) h_{21}(\bar{y}_2, \bar{a}_2, Y_0, W_1)\Big] \\
&\quad \times I(d_2(y_1, Y_0, A_1) = a_2) \\
&\quad + \sum_{y_1, y_2} y_2 I(d_1(Y_0) = a_1) \\
&\quad \times \Big[\hat{h}_{21}^{(-k)}(\bar{y}_2, \bar{a}_2, Y_0, W_1) I(d_2(a_1, Y_0, y_1) = a_2) - h_{21}(\bar{y}_2, \bar{a}_2, Y_0, W_1) I(d_2(a_1, Y_0, y_1) = a_2)\Big] \Bigg\}\Bigg\}
\end{aligned}
$$

And through simple algebra, we can further get

$$
\begin{aligned}
&\hat{V}_{n,k}(\theta) - \hat{V}^*_{n,k}(\theta) \\
&= \sum_{\bar{a}_2} \Bigg\{ \mathbb{P}_{n,k} \Bigg\{ I(A_2 = a_2) I(A_1 = a_1) I(d_1(Y_0) = a_1) I(d_2(\bar{Y}_1, A_1) = a_2) \\
&\quad \times \left( \hat{q}_{22}^{(-k)}(\bar{A}_2, \bar{Y}_1, \bar{Z}_2) - q_{22}(\bar{A}_2, \bar{Y}_1, \bar{Z}_2) \right) \\
&\quad \times \Bigg[ \bigg( Y_2 - \sum_{y_2} y_2 h_{22}(y_2, a_2, A_1, \bar{Y}_1, \bar{W}_2) \bigg) \\
&\qquad - \bigg( \sum_{y_2} y_2 \left( \hat{h}_{22}^{(-k)}(y_2, a_2, A_1, \bar{Y}_1, \bar{W}_2) - h_{22}(y_2, a_2, A_1, \bar{Y}_1, \bar{W}_2) \right) \bigg) \Bigg] \\
&\quad + I(A_1 = a_1) I(d_1(Y_0) = a_1) I(d_2(\bar{Y}_1, A_1) = a_2) \\
&\quad \times \left( q_{11}(Y_0, Z_1, A_1) - I(A_2 = a_2) q_{22}(\bar{A}_2, \bar{Y}_1, \bar{Z}_2) \right) \\
&\quad \times \bigg( \sum_{y_2} y_2 \left( \hat{h}_{22}^{(-k)}(y_2, a_2, A_1, \bar{Y}_1, \bar{W}_2) - h_{22}(y_2, a_2, A_1, \bar{Y}_1, \bar{W}_2) \right) \bigg) \\
&\quad + I(A_1 = a_1) I(d_1(Y_0) = a_1) I(d_2(\bar{Y}_1, A_1) = a_2) \\
&\quad \times \Bigg[ \left( \hat{q}_{11}^{(-k)}(Y_0, Z_1, A_1) - q_{11}(Y_0, Z_1, A_1) \right) \\
&\quad \times \bigg( \sum_{y_2} y_2 \left( \hat{h}_{22}^{(-k)}(y_2, a_2, \bar{A}_1, \bar{Y}_1, \bar{W}_2) - h_{22}(y_2, a_2, \bar{A}_1, \bar{Y}_1, \bar{W}_2) \right) \bigg) \Bigg] \\
&\quad + I(A_1 = a_1) I(d_1(Y_0) = a_1) \left( \hat{q}_{11}^{(-k)}(Y_0, Z_1, A_1) - q_{11}(Y_0, Z_1, A_1) \right) \\
&\quad \times \bigg( \sum_{y_2} y_2 h_{22}(y_2, a_2, \bar{A}_1, \bar{Y}_1, \bar{W}_2) I(d_2(\bar{Y}_1, A_1) = a_2) \\
&\qquad - \sum_{y_1, y_2} y_2 h_{21}(\bar{y}_2, \bar{a}_2, Y_0, W_1) I(d_2(y_1, Y_0, A_1) = a_2) \bigg) \\
&\quad - I(A_1 = a_1) I(d_1(Y_0) = a_1) \\
&\quad \times \Bigg[ \left( \hat{q}_{11}^{(-k)}(Y_0, Z_1, A_1) - q_{11}(Y_0, Z_1, A_1) \right) \\
&\qquad \times \bigg( \sum_{y_1, y_2} y_2 \left( \hat{h}_{21}^{(-k)}(\bar{y}_2, \bar{a}_2, Y_0, W_1) - h_{21}(\bar{y}_2, \bar{a}_2, Y_0, W_1) \right) I(d_2(y_1, Y_0, A_1) = a_2) \bigg) \Bigg] \\
&\quad + I(d_1(Y_0) = a_1) \Big( 1 - I(A_1 = a_1) q_{11}(Y_0, Z_1, A_1) \Big) \\
&\quad \times \sum_{y_1, y_2} y_2 \left( \hat{h}_{21}^{(-k)}(\bar{y}_2, \bar{a}_2, Y_0, W_1) - h_{21}(\bar{y}_2, \bar{a}_2, Y_0, W_1) \right) I(d_2(a_1, Y_0, y_1) = a_2) \Bigg\} \Bigg\}
\end{aligned}
$$

In summary, this decomposition can be viewed as consisting of product terms and mean-zero terms (by multiple robustness property). According to the Cauchy-Schwartz Inequality and Condition 2.2, we can conclude that the product terms are all $o_p(\frac{1}{\sqrt{n}})$; through Lemma D.8, we can get that the mean zero terms are also all $o_p(\frac{1}{\sqrt{n}})$. Thus, we have proved that $\sqrt{n}\left(\hat{V}_n(\theta) - \hat{V}_n^*(\theta)\right) = o_p(1)$.

**Step 2.** In the Step 1, we have proved that $\hat{V}_n(\theta) = \hat{V}_n^*(\theta) + o_p(\frac{1}{\sqrt{n}})$. By the weak law of large number, we also have that $V_n^*(\theta) = V(\theta) + o_p(1)$, thus $\hat{V}_n(\theta) = V(\theta) + o_p(1)$. If $\hat{\theta}$ maximizes $\hat{V}_n(\theta)$, then we also have $\hat{V}_n(\hat{\theta}) \geq \sup_{\theta\in\Theta} \hat{V}_n(\theta) - o_p(1)$. By the Theorem 14.1 (Argmax Theorem) of Kosorok (2008), we have $\hat{\theta} \xrightarrow{p} \theta^*$ as $n \to \infty$.

Now we prove that $n^{1/3}\|\hat{\theta} - \theta\|_2 = O_p(1)$ through Theorem 14.4 (Rate of Convergence) of Kosorok (2008). Therefore, we need to verify the three propositions below:

*Proposition* D.1. $\|V(\theta) - V(\theta^*)\|_1 \leq c_0\|\theta - \theta^*\|_2^2$.

*Proof.* By Condition 1.3, $V(\theta)$ is twice continuously differentiable in a neighborhood of $\theta^*$, which allows the expansion for each $\theta$ in a neighborhood of $\theta^*$ such that $\|\theta - \theta^*\|_2 < \delta$:

$$V(\theta) - V(\theta^*) = (\theta - \theta^*)^T V'(\theta^*) + \frac{1}{2}(\theta - \theta^*)^T V''(\theta^*)(\theta - \theta^*) + o(\|\theta - \theta^*\|_2^2)$$
$$= \frac{1}{2}(\theta - \theta^*)^T V''(\theta^*)(\theta - \theta^*) + o(\|\theta - \theta^*\|_2^2),$$

where the second equality is due to the fact that $\theta^*$ is the unique maximum point of $V(\theta)$. Also according to this fact, $V''(\theta^*)$ is semi-negative definite. Thus, let $c_0$ denote the maximum eigenvalue of $-\frac{1}{2}V''(\theta^*)$, then $\|V(\theta) - V(\theta^*)\|_1 \leq c_0\|\theta - \theta^*\|_2^2$.

*Proposition* D.2. *For all large enough $n$ and sufficient small $\delta$, we consider the centered*

*process* $\hat{V}_n(\theta) - V(\theta)$*:*

$$
\begin{aligned}
&\mathbb{E}^*\left[\sqrt{n}\sup_{\|\theta-\theta^*\|_2<\delta}\left|\hat{V}_n(\theta) - V(\theta) - \left\{\hat{V}_n(\theta^*) - V(\theta^*)\right\}\right|\right] \\
&= \mathbb{E}^*\left[\sqrt{n}\sup_{\|\theta-\theta^*\|_2<\delta}\left|\hat{V}_n(\theta) - \hat{V}_n^*(\theta) + \hat{V}_n^*(\theta) - V(\theta) - \left\{\hat{V}_n(\theta^*) - \hat{V}_n^*(\theta^*) + \hat{V}_n^*(\theta^*) - V(\theta^*)\right\}\right|\right] \\
&= \mathbb{E}^*\left[\sqrt{n}\sup_{\|\theta-\theta^*\|_2<\delta}\left|\hat{V}_n(\theta) - \hat{V}_n^*(\theta) - \left\{\hat{V}_n(\theta^*) - \hat{V}_n^*(\theta^*)\right\} + \hat{V}_n^*(\theta) - V(\theta) - \left\{\hat{V}_n^*(\theta^*) - V(\theta^*)\right\}\right|\right] \\
&\leq \mathbb{E}^*\left[\sqrt{n}\sup_{\|\theta-\theta^*\|_2<\delta}\left|\hat{V}_n(\theta) - \hat{V}_n^*(\theta) - \left\{\hat{V}_n(\theta^*) - \hat{V}_n^*(\theta^*)\right\}\right|\right] \\
&+ \mathbb{E}^*\left[\sqrt{n}\sup_{\|\theta-\theta^*\|_2<\delta}\left|\hat{V}_n^*(\theta) - V(\theta) - \left\{\hat{V}_n^*(\theta^*) - V(\theta^*)\right\}\right|\right] \\
&=: (I) + (II).
\end{aligned}
$$

*Then as* $n \to \infty$*, when* $\|\theta - \theta^*\|_2 < \delta$ *there exists a constant* $k_2 < \infty$ *satisfies*

$$(I) + (II) < k_2\delta^{1/2}.$$

*Proof.* From **Step 1**, $(I)$ is $o(1)$. To bound $(II)$, we first note that

$$
\begin{aligned}
&\hat{V}_n^*(\theta) - \hat{V}_n^*(\theta^*) \\
&= \frac{1}{n}\sum_{i=1}^n \left(\phi(O_i; h, q, \theta) - \phi(O_i; h, q, \theta^*)\right) \\
&= \frac{1}{n}\sum_{i=1}^n \sum_{\bar{a}_2} \Bigg\{ \Big[ I(A_{2i} = a_2) I(A_{1i} = a_1) q_{22}(\bar{A}_{2i}, \bar{Y}_{1i}, \bar{Z}_{2i}) \Big[ Y_{2i} - \sum_{y_2} y_2 h_{22}(y_2, a_2, A_{1i}, \bar{Y}_{1i}, \bar{W}_{2i}) \Big] \Big] \\
&\qquad \times \Big[ I(d_1(Y_{0i}) = a_1) I(d_2(\bar{Y}_{1i}, d_1(Y_{0i})) = a_2) - I(d_1^*(Y_{0i}) = a_1) I(d_2^*(\bar{Y}_{1i}, d_1^*(Y_{0i})) = a_2) \Big] \\
&\quad + I(A_{1i} = a_1) q_{11}(Y_{0i}, Z_{1i}, A_{1i}) \sum_{y_2} y_2 h_{22}(y_2, a_2, \bar{A}_{1i}, \bar{Y}_{1i}, \bar{W}_{2i}) \\
&\qquad \times \Big[ I(d_1(Y_{0i}) = a_1) I(d_2(\bar{Y}_{1i}, d_1(Y_{0i})) = a_2) - I(d_1^*(Y_{0i}) = a_1) I(d_2^*(\bar{Y}_{1i}, d_1^*(Y_{0i})) = a_2) \Big] \\
&\quad + \Big[ 1 - I(A_{1i} = a_1) q_{11}(Y_{0i}, Z_{1i}, A_{1i}) \Big] \\
&\qquad \times \sum_{y_1, y_2} y_2 h_{21}(\bar{y}_2, \bar{a}_2, Y_{0i}, W_{1i}) \Big[ I(d_1(Y_{0i}) = a_1) I(d_2(d_1(Y_{0i}), Y_{0i}, y_1) = a_2) \\
&\qquad - I(d_1^*(Y_{0i}) = a_1) I(d_2^*(d_1^*(Y_{0i}), Y_{0i}, y_1) = a_2) \Big] \Bigg\} \\
&= \sum_{\bar{a}_2} \Bigg\{ \frac{1}{n}\sum_{i=1}^n \Delta_1(O_i) \\
&\qquad \times \Big[ I(d_1(Y_{0i}) = a_1) I(d_2(\bar{Y}_{1i}, d_1(Y_{0i})) = a_2) - I(d_1^*(Y_{0i}) = a_1) I(d_2^*(\bar{Y}_{1i}, d_1^*(Y_{0i})) = a_2) \Big] \\
&\quad + \Delta_2(O_i) \sum_{y_1, y_2} y_2 h_{21}(\bar{y}_2, \bar{a}_2, Y_{0i}, W_{1i}) \\
&\qquad \times \Big[ I(d_1(Y_{0i}) = a_1) I(d_2(d_1(Y_{0i}), Y_{0i}, y_1) = a_2) - I(d_1^*(Y_{0i}) = a_1) I(d_2^*(d_1(Y_{0i}), Y_{0i}, y_1) = a_2) \Big] \Bigg\} \\
&=: \sum_{\bar{a}_2} \Psi_{a_1, a_2},
\end{aligned}
$$

where

$$
\begin{aligned}
\Delta_1(O_i) &= I(A_{2i} = a_2) I(A_{1i} = a_1) q_{22}(\bar{A}_{2i}, \bar{Y}_{1i}, \bar{Z}_{2i}) \Big[ Y_{2i} - \sum_{y_2} y_2 h_{22}(y_2, a_2, A_{1i}, \bar{Y}_{1i}, \bar{W}_{2i}) \Big] \\
&\qquad + I(A_{1i} = a_1) q_{11}(Y_{0i}, Z_{1i}, A_{1i}) \sum_{y_2} y_2 h_{22}(y_2, a_2, A_{1i}, \bar{Y}_{1i}, \bar{W}_{2i}), \\
\Delta_2(O_i) &= 1 - I(A_{1i} = a_1) q_{11}(Y_{0i}, Z_{1i}, A_{1i}).
\end{aligned}
$$

Note the fact that

$$I(d_1(y_0) = 0)I(d_2(\bar{y}_1, d_1(y_0)) = 0)$$
$$= I(y_0^T\theta_1 \le 0)I((\bar{y}_1, I(y_0^T\theta_1 > 0))^T\theta_2 \le 0);$$

$$I(d_1(y_0) = 0)I(d_2(\bar{y}_1, d_1(y_0)) = 1)$$
$$= I(y_0^T\theta_1 \le 0)I((\bar{y}_1, I(y_0^T\theta_1 > 0))^T\theta_2 > 0);$$

$$I(d_1(y_0) = 1)I(d_2(\bar{y}_1, d_1(y_0)) = 0)$$
$$= I(y_0^T\theta_1 > 0)I((\bar{y}_1, I(y_0^T\theta_1 > 0))^T\theta_2 \le 0);$$

$$I(d_1(y_0) = 1)I(d_2(\bar{y}_1, d_1(y_0)) = 1)$$
$$= I(y_0^T\theta_1 > 0)I((\bar{y}_1, I(y_0^T\theta_1 > 0))^T\theta_2 > 0),$$

and

$$\begin{aligned}
&I(d_1(y_0) = a_1)I(d_2((\bar{y}_1, d_1(y_0))) = a_2) \\
&= \Big[a_1\Big(I(y_0^T\theta_1 > 0) - I(y_0^T\theta_1 \le 0)\Big) + I(y_0^T\theta_1 \le 0)\Big] \\
&\quad \times \Big[a_2\Big(I((\bar{y}_1, I(y_0^T\theta_1 > 0))^T\theta_2 > 0) - I((\bar{y}_1, I(y_0^T\theta_1 > 0))^T\theta_2 \le 0)\Big) \\
&\quad + I((\bar{y}_1, I(y_0^T\theta_1 > 0))^T\theta_2 \le 0)\Big] \\
&= a_1a_2I(y_0^T\theta_1 > 0)I((\bar{y}_1, I(y_0^T\theta_1 > 0))^T\theta_2 > 0) \\
&\quad + a_1(1-a_2)I(y_0^T\theta_1 > 0)I((\bar{y}_1, I(y_0^T\theta_1 > 0))^T\theta_2 \le 0) \\
&\quad + a_2(1-a_1)I(y_0^T\theta_1 \le 0)I((\bar{y}_1, I(y_0^T\theta_1 > 0))^T\theta_2 > 0) \\
&\quad + (1-a_1)(1-a_2)I(y_0^T\theta_1 \le 0)I((\bar{y}_1, I(y_0^T\theta_1 > 0))^T\theta_2 \le 0).
\end{aligned} \tag{38}$$

To prove that $\Psi_{a_1,a_2}$ is bounded, we consider

$$
\begin{aligned}
\Psi_{a_1,a_2} =& \frac{1}{n}\sum_{i=1}^{n}\Bigg\{\Delta_1(O_i)\Big[I(d_1(Y_{0i})=a_1)I(d_2(\bar{Y}_{1i},d_1(Y_{0i}))=a_2) \\
&\quad - I(d_1^*(Y_{0i})=0)I(d_2^*(\bar{Y}_{1i},d_1^*(Y_{0i}))=a_2)\Big] \\
&\quad + \Delta_2(O_i)\sum_{y_1,y_2} y_2 h_{21}(\bar{y}_2,\bar{a}_2,Y_{0i},W_{1i})\Big[I(d_1(Y_{0i})=a_1)I(d_2(d_1(Y_{0i}),Y_{0i},y_1)=a_2) \\
&\quad - I(d_1^*(Y_{0i})=a_1)I(d_2^*(d_1^*(Y_{0i}),Y_{0i},y_1)=a_2)\Big]\Bigg\} \\
=& \frac{1}{n}\sum_{i=1}^{n}\Bigg\{\Delta_1(O_i) \\
&\quad \times\Big[a_1a_2\Big(I(Y_{0i}^T\theta_1>0)I((\bar{Y}_{1i},I(Y_{0i}^T\theta_1>0))^T\theta_2>0) \\
&\qquad - I(Y_{0i}^T\theta_1^*>0)I((\bar{Y}_{1i},I(Y_{0i}^T\theta_1^*>0))^T\theta_2^*>0)\Big) \\
&\quad + a_1(1-a_2) \\
&\quad \times\Big(I(Y_{0i}^T\theta_1>0)I((\bar{Y}_{1i},I(Y_{0i}^T\theta_1>0))^T\theta_2\le 0) \\
&\quad - I(Y_{0i}^T\theta_1^*>0)I((\bar{Y}_{1i},I(Y_{0i}^T\theta_1^*>0))^T\theta_2^*\le 0)\Big) \\
&\quad + a_2(1-a_1)\Big(I(Y_{0i}^T\theta_1\le 0)I((\bar{Y}_{1i},I(Y_{0i}^T\theta_1>0))^T\theta_2>0) \\
&\quad - I(Y_{0i}^T\theta_1^*\le 0)I((\bar{Y}_{1i},I(Y_{0i}^T\theta_1^*>0))^T\theta_2^*>0)\Big) \\
&\quad + (1-a_1)(1-a_2)\Big(I(Y_{0i}^T\theta_1\le 0)I((\bar{Y}_{1i},I(Y_{0i}^T\theta_1>0))^T\theta_2\le 0) \\
&\qquad - I(Y_{0i}^T\theta_1^*\le 0)I((\bar{Y}_{1i},I(Y_{0i}^T\theta_1^*>0))^T\theta_2^*\le 0)\Big)\Big] \\
&\quad + \Delta_2(O_i)\sum_{y_1,y_2} y_2 h_{21}(\bar{y}_2,\bar{a}_2,Y_{0i},W_{1i})\Big[ \\
&\quad a_1a_2\Big(I(Y_{0i}^T\theta_1>0)I((y_1,Y_{0i},I(Y_{0i}^T\theta_1>0))^T\theta_2>0) \\
&\qquad - I(Y_{0i}^T\theta_1^*>0)I((y_1,Y_{0i},I(Y_{0i}^T\theta_1^*>0))^T\theta_2^*>0)\Big) \\
&\quad + a_1(1-a_2)\Big(I(Y_{0i}^T\theta_1>0)I((y_1,Y_{0i},I(Y_{0i}^T\theta_1>0))^T\theta_2\le 0) \\
&\qquad - I(Y_{0i}^T\theta_1^*>0)I((y_1,Y_{0i},I(Y_{0i}^T\theta_1^*>0))^T\theta_2^*\le 0)\Big) \\
&\quad + a_2(1-a_1)\Big(I(Y_{0i}^T\theta_1\le 0)I((y_1,Y_{0i},I(Y_{0i}^T\theta_1>0))^T\theta_2>0) \\
&\qquad - I(Y_{0i}^T\theta_1^*\le 0)I((y_1,Y_{0i},I(Y_{0i}^T\theta_1^*>0))^T\theta_2^*>0)\Big) \\
&\quad + (1-a_1)(1-a_2)\Big(I(Y_{0i}^T\theta_1\le 0)I((y_1,Y_{0i},I(Y_{0i}^T\theta_1>0))^T\theta_2\le 0) \\
&\qquad - I(Y_{0i}^T\theta_1^*\le 0)I((y_1,Y_{0i},I(Y_{0i}^T\theta_1^*>0))^T\theta_2^*\le 0)\Big)\Big]\Bigg\}
\end{aligned}
$$

Also note that $\max\{a_1a_2, a_1(1-a_2), a_2(1-a_1), (1-a_1)(1-a_2)\} \le 1$, we have

$$\Psi_{a_1,a_2} \le \frac{1}{n}\sum_{i=1}^{n}\Bigg\{\Bigg[\Bigg|\Delta_1(O_i)\Big(I(Y_{0i}^T\theta_1 > 0)I((\bar{Y}_{1i}, I(Y_{0i}^T\theta_1 > 0))^T\theta_2 > 0)$$

$$- I(Y_{0i}^T\theta_1^* > 0)I((\bar{Y}_{1i}, I(Y_{0i}^T\theta_1^* > 0))^T\theta_2^* > 0)\Big)\Bigg|$$

$$+ \Bigg|\Delta_1(O_i)\Big(I(Y_{0i}^T\theta_1 > 0)I((\bar{Y}_{1i}, I(Y_{0i}^T\theta_1 > 0))^T\theta_2 \le 0)$$

$$- I(Y_{0i}^T\theta_1^* > 0)I((\bar{Y}_{1i}, I(Y_{0i}^T\theta_1^* > 0))^T\theta_2^* \le 0)\Big)\Bigg|$$

$$+ \Bigg|\Delta_1(O_i)\Big(I(Y_{0i}^T\theta_1 \le 0)I((\bar{Y}_{1i}, I(Y_{0i}^T\theta_1 > 0))^T\theta_2 > 0)$$

$$- I(Y_{0i}^T\theta_1^* \le 0)I((\bar{Y}_{1i}, I(Y_{0i}^T\theta_1^* > 0))^T\theta_2^* > 0)\Big)\Bigg|$$

$$+ \Bigg|\Delta_1(O_i)\Big(I(Y_{0i}^T\theta_1 \le 0)I((\bar{Y}_{1i}, I(Y_{0i}^T\theta_1 > 0))^T\theta_2 \le 0)$$

$$- I(Y_{0i}^T\theta_1^* \le 0)I((\bar{Y}_{1i}, I(Y_{0i}^T\theta_1^* > 0))^T\theta_2^* \le 0)\Big)\Bigg|\Bigg]$$

$$+ \Bigg[\Bigg|\Delta_2(O_i)\sum_{y_1,y_2} y_2 h_{21}(\bar{y}_2, 1, 1, Y_{0i}, W_{1i})$$

$$\times \Big(I(Y_{0i}^T\theta_1 > 0)I((y_1, Y_{0i}, I(Y_{0i}^T\theta_1 > 0))^T\theta_2 > 0)$$

$$- I(Y_{0i}^T\theta_1^* > 0)I((y_1, Y_{0i}, I(Y_{0i}^T\theta_1^* > 0))^T\theta_2^* > 0)\Big)\Bigg|$$

$$+ \Bigg|\Delta_2(O_i)\sum_{y_1,y_2} y_2 h_{21}(\bar{y}_2, 1, 0, Y_{0i}, W_{1i})$$

$$\times \Big(I(Y_{0i}^T\theta_1 > 0)I((y_1, Y_{0i}, I(Y_{0i}^T\theta_1 > 0))^T\theta_2 \le 0)$$

$$- I(Y_{0i}^T\theta_1^* > 0)I((y_1, Y_{0i}, I(Y_{0i}^T\theta_1^* > 0))^T\theta_2^* \le 0)\Big)\Bigg|$$

$$+ \Bigg|\Delta_2(O_i)\sum_{y_1,y_2} y_2 h_{21}(\bar{y}_2, 0, 1, Y_{0i}, W_{1i})$$

$$\times \Big(I(Y_{0i}^T\theta_1 \le 0)I((y_1, Y_{0i}, I(Y_{0i}^T\theta_1 > 0))^T\theta_2 > 0)$$

$$- I(Y_{0i}^T\theta_1^* \le 0)I((y_1, Y_{0i}, I(Y_{0i}^T\theta_1^* > 0))^T\theta_2^* > 0)\Big)\Bigg|$$

$$+ \Bigg|\Delta_2(O_i)\sum_{y_1,y_2} y_2 h_{21}(\bar{y}_2, 0, 0, Y_{0i}, W_{1i})$$

$$\times \Big(I(Y_{0i}^T\theta_1 \le 0)I((y_1, Y_{0i}, I(Y_{0i}^T\theta_1 > 0))^T\theta_2 \le 0)$$

$$- I(Y_{0i}^T\theta_1^* \le 0)I((y_1, Y_{0i}, I(Y_{0i}^T\theta_1^* > 0))^T\theta_2^* \le 0)\Big)\Bigg|\Bigg]\Bigg\}$$

Let $B_1 = \sup(|\Delta_1(o)|)$, $B_2 = \sup(|\Delta_2(o)|)$ and $B_3 = \sup(|\sum_{y_1,y_2} y_2 h_{21}(\bar{y}_2, \bar{a}_2, w_1)|)$, by Conditions 1.1 and 1.2, we conclude that $\max\{B_1, B_2, B_3\} < \infty$. Note that under the context of this theorem, $d_1(y_0) = I(y_0^T\theta_1 > 0)$ and $d_1^*(y_0) = I(y_0^T\theta_1^* > 0)$, we define eight distinct classes of functions:

$$\begin{aligned}\mathcal{F}_\theta^1(o) = \Big\{\Delta_1(o)\Big[&I(y_0^T\theta_1 > 0)I((\bar{y}_1, d_1(y_0))^T\theta_2 > 0)\\ &- I(y_0^T\theta_1^* > 0)I((\bar{y}_1, d_1^*(y_0))^T\theta_2^* > 0)\Big] : \|\theta - \theta^*\|_2 < \delta\Big\}\end{aligned}$$

$$\begin{aligned}\mathcal{F}_\theta^2(o) = \Big\{\Delta_1(o)\Big[&I(y_0^T\theta_1 > 0)I((\bar{y}_1, d_1(y_0))^T\theta_2 \le 0)\\ &- I(y_0^T\theta_1^* > 0)I((\bar{y}_1, d_1^*(y_0))^T\theta_2^* \le 0)\Big] : \|\theta - \theta^*\|_2 < \delta\Big\}\end{aligned}$$

$$\begin{aligned}\mathcal{F}_\theta^3(o) = \Big\{\Delta_1(o)\Big[&I(y_0^T\theta_1 \le 0)I((\bar{y}_1, d_1(y_0))^T\theta_2 > 0)\\ &- I(y_0^T\theta_1^* \le 0)I((\bar{y}_1, d_1^*(y_0))^T\theta_2^* > 0)\Big] : \|\theta - \theta^*\|_2 < \delta\Big\}\end{aligned}$$

$$\begin{aligned}\mathcal{F}_\theta^4(o) = \Big\{\Delta_1(o)\Big[&I(y_0^T\theta_1 \le 0)I((\bar{y}_1, d_1(y_0))^T\theta_2 \le 0)\\ &- I(y_0^T\theta_1^* \le 0)I((\bar{y}_1, d_1^*(y_0))^T\theta_2^* \le 0)\Big] : \|\theta - \theta^*\|_2 < \delta\Big\}\end{aligned}$$

$$\begin{aligned}\mathcal{F}_\theta^5(o) = \Bigg\{\Delta_2(o)\sum_{y_1,y_2} y_2 h_{21}(\bar{y}_2, 1, 1, w_1)\Big[&I(y_0^T\theta_1 > 0)I((\bar{y}_1, d_1(y_0))^T\theta_2 > 0)\\ &- I(y_0^T\theta_1^* > 0)I((\bar{y}_1, d_1^*(y_0))^T\theta_2^* > 0)\Big] : \|\theta - \theta^*\|_2 < \delta\Bigg\}\end{aligned}$$

$$\begin{aligned}\mathcal{F}_\theta^6(o) = \Bigg\{\Delta_2(o)\sum_{y_1,y_2} y_2 h_{21}(\bar{y}_2, 1, 0, w_1)\Big[&I(y_0^T\theta_1 > 0)I((\bar{y}_1, d_1(y_0))^T\theta_2 \le 0)\\ &- I(y_0^T\theta_1^* > 0)I((\bar{y}_1, d_1^*(y_0))^T\theta_2^* \le 0)\Big] : \|\theta - \theta^*\|_2 < \delta\Bigg\}\end{aligned}$$

$$\mathcal{F}_\theta^7(o) = \Bigg\{\Delta_2(o)\sum_{y_1,y_2} y_2 h_{21}(\bar{y}_2, 0, 1, w_1)\Big[I(y_0^T\theta_1 \le 0)I((\bar{y}_1, d_1(y_0))^T\theta_2 > 0)$$
$$- I(y_0^T\theta_1^* \le 0)I((\bar{y}_1, d_1^*(y_0))^T\theta_2^* > 0)\Big] : \|\theta - \theta^*\|_2 < \delta\Bigg\}$$
$$\mathcal{F}_\theta^8(o) = \Bigg\{\Delta_2(o)\sum_{y_1,y_2} y_2 h_{21}(\bar{y}_2, 0, 0, w_1)\Big[I(y_0^T\theta_1 \le 0)I((\bar{y}_1, d_1(y_0))^T\theta_2 \le 0)$$
$$- I(y_0^T\theta_1^* \le 0)I((\bar{y}_1, d_1^*(y_0))^T\theta_2^* \le 0)\Big] : \|\theta - \theta^*\|_2 < \delta\Bigg\}$$

When $\|\theta - \theta^*\| < \delta$, by Condition 1.1, there exists a constant $0 < k_0 < \infty$ such that $|y_0^T(\theta_1 - \theta_1^*)| < k_0\delta$ and $|(\bar{y}_1, a_1)^T(\theta_2 - \theta_2^*)| < k_0\delta$. Furthermore, we will prove that

$$|I(y_0^T\theta_1 > 0)I((\bar{y}_1, d_1(y_0))^T\theta_2 > 0) - I(y_0^T\theta_1^* > 0)I((\bar{y}_1, d_1^*(y_0))^T\theta_2^* > 0)| \tag{39}$$
$$\le I(-k_0\delta \le y_0^T\theta_1^* \le k_0\delta) + I(-k_0\delta \le (\bar{y}_1, d_1^*(y_0))^T\theta_2^* \le k_0\delta)$$
$$|I(y_0^T\theta_1 > 0)I((\bar{y}_1, d_1(y_0))^T\theta_2 \le 0) - I(y_0^T\theta_1^* > 0)I((\bar{y}_1, d_1^*(y_0))^T\theta_2^* \le 0)| \tag{40}$$
$$\le I(-k_0\delta \le y_0^T\theta_1^* \le k_0\delta) + I(-k_0\delta \le (\bar{y}_1, d_1^*(y_0))^T\theta_2^* \le k_0\delta)$$
$$|I(y_0^T\theta_1 \le 0)I((\bar{y}_1, d_1(y_0))^T\theta_2 > 0) - I(y_0^T\theta_1^* \le 0)I((\bar{y}_1, d_1^*(y_0))^T\theta_2^* > 0)| \tag{41}$$
$$\le I(-k_0\delta \le y_0^T\theta_1^* \le k_0\delta) + I(-k_0\delta \le (\bar{y}_1, d_1^*(y_0))^T\theta_2^* \le k_0\delta)$$
$$|I(y_0^T\theta_1 \le 0)I((\bar{y}_1, d_1(y_0))^T\theta_2 \le 0) - I(y_0^T\theta_1^* \le 0)I((\bar{y}_1, d_1^*(y_0))^T\theta_2^* \le 0)| \tag{42}$$
$$\le I(-k_0\delta \le y_0^T\theta_1^* \le k_0\delta) + I(-k_0\delta \le (\bar{y}_1, d_1^*(y_0))^T\theta_2^* \le k_0\delta)$$

For the first inequality (39), we consider the following five cases:

(a) When $-k_0\delta \le y_0^T\theta_1^* \le k_0\delta$, then

$$|I(y_0^T\theta_1 > 0)I((\bar{y}_1, d_1(y_0))^T\theta_2 > 0) - I(y_0^T\theta_1^* > 0)I((\bar{y}_1, d_1^*(y_0))^T\theta_2^* > 0)|$$
$$\le 1$$
$$\le I(-k_0\delta \le y_0^T\theta_1^* \le k_0\delta) + I(-k_0\delta \le (\bar{y}_1, d_1^*(y_0))^T\theta_2^* \le k_0\delta).$$

(b) When $y_0^T\theta_1^* < -k_0\delta < 0$, then $y_0^T\theta_1 = y_0^T(\theta_1 - \theta^*) + y_0^T\theta^* < 0$, and

$$
\begin{aligned}
&|I(y_0^T\theta_1 > 0)I((\bar{y}_1, d_1(y_0))^T\theta_2 > 0) - I(y_0^T\theta_1^* > 0)I((\bar{y}_1, d_1^*(y_0))^T\theta_2^* > 0)| \\
&= 0 \\
&\leq I(-k_0\delta \leq y_0^T\theta_1^* \leq k_0\delta) + I(-k_0\delta \leq (\bar{y}_1, d_1^*(y_0))^T\theta_2^* \leq k_0\delta).
\end{aligned}
$$

(c) When $y_0^T\theta_1^* > k_0\delta > 0$ and $-k_0\delta \leq (\bar{y}_1, d_1^*(y_0))^T\theta_2^* = (\bar{y}_1, 1)^T\theta_2^* \leq k_0\delta$, then

$$
\begin{aligned}
&|I(y_0^T\theta_1 > 0)I((\bar{y}_1, d_1(y_0))^T\theta_2 > 0) - I(y_0^T\theta_1^* > 0)I((\bar{y}_1, d_1^*(y_0))^T\theta_2^* > 0)| \\
&\leq 1 \\
&\leq I(-k_0\delta \leq y_0^T\theta_1^* \leq k_0\delta) + I(-k_0\delta \leq (\bar{y}_1, d_1^*(y_0))^T\theta_2^* \leq k_0\delta).
\end{aligned}
$$

(d) When $y_0^T\theta_1^* > k_0\delta > 0$ and $(\bar{y}_1, 1)^T\theta_2^* < -k_0\delta < 0$, we have $y_0^T\theta_1 = y_0^T(\theta_1 - \theta^*) + y_0^T\theta^* > 0$ as well as $(\bar{y}_1, 1)^T\theta_2 = (\bar{y}_1, 1)^T(\theta_2 - \theta_2^*) + (\bar{y}_1, 1)^T\theta_2^* < 0$, then

$$
\begin{aligned}
&|I(y_0^T\theta_1 > 0)I((\bar{y}_1, d_1(y_0))^T\theta_2 > 0) - I(y_0^T\theta_1^* > 0)I((\bar{y}_1, d_1^*(y_0))^T\theta_2^* > 0)| \\
&= |I(y_0^T\theta_1 > 0)I((\bar{y}_1, 1)^T\theta_2 > 0) - I(y_0^T\theta_1^* > 0)I((\bar{y}_1, 1)^T\theta_2^* > 0)| \\
&= 0 \\
&\leq I(-k_0\delta \leq y_0^T\theta_1^* \leq k_0\delta) + I(-k_0\delta \leq (\bar{y}_1, d_1^*(y_0))^T\theta_2^* \leq k_0\delta).
\end{aligned}
$$

(e) When $y_0^T\theta_1^* > k_0\delta > 0$ and $(\bar{y}_1, 1)^T\theta_2^* > k_0\delta > 0$, we have $y_0^T\theta_1 = y_0^T(\theta_1 - \theta^*) + y_0^T\theta^* > 0$ as well as $(\bar{y}_1, 1)^T\theta_2 = (\bar{y}_1, 1)^T(\theta_2 - \theta_2^*) + (\bar{y}_1, 1)^T\theta_2^* > 0$, then

$$
\begin{aligned}
&|I(y_0^T\theta_1 > 0)I((\bar{y}_1, 1)^T\theta_2 > 0) - I(y_0^T\theta_1^* > 0)I((\bar{y}_1, 1)^T\theta_2^* > 0)| \\
&= 0 \\
&\leq I(-k_0\delta \leq y_0^T\theta_1^* \leq k_0\delta) + I(-k_0\delta \leq (\bar{y}_1, d_1^*(y_0))^T\theta_2^* \leq k_0\delta).
\end{aligned}
$$

For the last inequality (42), we consider the following five cases similarly:

(a) When $-k_0\delta \leq y_0^T\theta_1^* \leq k_0\delta$, then

$$
\begin{aligned}
&|I(y_0^T\theta_1 \leq 0)I((\bar{y}_1, d_1(y_0))^T\theta_2 \leq 0) - I(y_0^T\theta_1^* \leq 0)I((\bar{y}_1, d_1(y_0))^T\theta_2^* \leq 0)| \\
&\leq 1 \\
&\leq I(-k_0\delta \leq y_0^T\theta_1^* \leq k_0\delta) + I(-k_0\delta \leq (\bar{y}_1, d_1^*(y_0))^T\theta_2^* \leq k_0\delta).
\end{aligned}
$$

(b) When $y_0^T\theta_1^* > k_0\delta > 0$, then $y_0^T\theta_1 = y_0^T(\theta_1 - \theta^*) + y_0^T\theta^* > 0$, and

$$
\begin{aligned}
&|I(y_0^T\theta_1 \le 0)I((\bar{y}_1, d_1(y_0))^T\theta_2 \le 0) - I(y_0^T\theta_1^* \le 0)I((\bar{y}_1, d_1^*(y_0))^T\theta_2^* \le 0)| \\
&= 0 \\
&\le I(-k_0\delta \le y_0^T\theta_1^* \le k_0\delta) + I(-k_0\delta \le (\bar{y}_1, d_1^*(y_0))^T\theta_2^* \le k_0\delta).
\end{aligned}
$$

(c) When $y_0^T\theta_1^* < -k_0\delta < 0$ and $-k_0\delta \le (\bar{y}_1, d_1^*(y_0))^T\theta_2^* = (\bar{y}_1, 0)^T\theta_2^* \le k_0\delta$, then

$$
\begin{aligned}
&|I(y_0^T\theta_1 \le 0)I((\bar{y}_1, d_1(y_0))^T\theta_2 \le 0) - I(y_0^T\theta_1^* \le 0)I((\bar{y}_1, d_1^*(y_0))^T\theta_2^* \le 0)| \\
&\le 1 \\
&\le I(-k_0\delta \le y_0^T\theta_1^* \le k_0\delta) + I(-k_0\delta \le (\bar{y}_1, d_1^*(y_0))^T\theta_2^* \le k_0\delta).
\end{aligned}
$$

(d) When $y_0^T\theta_1^* < -k_0\delta < 0$ and $(\bar{y}_1, 0)^T\theta_2^* < -k_0\delta < 0$, we have $y_0^T\theta_1 = y_0^T(\theta_1 - \theta^*) + y_0^T\theta^* < 0$ as well as $(\bar{y}_1, 0)^T\theta_2 = (\bar{y}_1, 0)^T(\theta_2 - \theta_2^*) + (\bar{y}_1, 0)^T\theta_2^* < 0$, then

$$
\begin{aligned}
&|I(y_0^T\theta_1 \le 0)I((\bar{y}_1, d_1(y_0))^T\theta_2 \le 0) - I(y_0^T\theta_1^* \le 0)I((\bar{y}_1, d_1^*(y_0))^T\theta_2^* \le 0)| \\
&= |I(y_0^T\theta_1 \le 0)I((\bar{y}_1, 0)^T\theta_2 \le 0) - I(y_0^T\theta_1^* \le 0)I((\bar{y}_1, 1)^T\theta_2^* \le 0)| \\
&= 0 \\
&\le I(-k_0\delta \le y_0^T\theta_1^* \le k_0\delta) + I(-k_0\delta \le (\bar{y}_1, d_1^*(y_0))^T\theta_2^* \le k_0\delta).
\end{aligned}
$$

(e) When $y_0^T\theta_1^* < -k_0\delta < 0$ and $(\bar{y}_1, 0)^T\theta_2^* > k_0\delta > 0$, we have $y_0^T\theta_1 = y_0^T(\theta_1 - \theta^*) + y_0^T\theta^* < 0$ as well as $(\bar{y}_1, 0)^T\theta_2 = (\bar{y}_1, 0)^T(\theta_2 - \theta_2^*) + (\bar{y}_1, 0)^T\theta_2^* < 0$, then

$$
\begin{aligned}
&|I(y_0^T\theta_1 \le 0)I((\bar{y}_1, 0)^T\theta_2 \le 0) - I(y_0^T\theta_1^* \le 0)I((\bar{y}_1, 0)^T\theta_2^* \le 0)| \\
&= 0 \\
&\le I(-k_0\delta \le y_0^T\theta_1^* \le k_0\delta) + I(-k_0\delta \le (\bar{y}_1, d_1^*(y_0))^T\theta_2^* \le k_0\delta).
\end{aligned}
$$

The proof of the other 2 inequalities (40),(41) is similar, thus omitted.

Now that we have proved the inequality (39), (40), (41) and (42), we can develop the envelop functions of $\mathcal{F}_\theta^1 \sim \mathcal{F}_\theta^8$ respectively:

$$
\begin{aligned}
F_1 &= B_1 I(-k_0\delta \le y_0^T\theta_1^* \le k_0\delta) + B_1 I(-k_0\delta \le (\bar{y}_1, d_1^*(y_0))^T\theta_2^* \le k_0\delta) \\
F_2 &= B_1 I(-k_0\delta \le y_0^T\theta_1^* \le k_0\delta) + B_1 I(-k_0\delta \le (\bar{y}_1, d_1^*(y_0))^T\theta_2^* \le k_0\delta) \\
F_3 &= B_1 I(-k_0\delta \le y_0^T\theta_1^* \le k_0\delta) + B_1 I(-k_0\delta \le (\bar{y}_1, d_1^*(y_0))^T\theta_2^* \le k_0\delta) \\
F_4 &= B_1 I(-k_0\delta \le y_0^T\theta_1^* \le k_0\delta) + B_1 I(-k_0\delta \le (\bar{y}_1, d_1^*(y_0))^T\theta_2^* \le k_0\delta) \\
F_5 &= B_2 \Big|\sum_{y_1,y_2} y_2 h_{21}(\bar{y}_2, 1, 1, w_1)\Big| I(-k_0\delta \le y_0^T\theta_1^* \le k_0\delta) \\
&\quad + B_2 \sum_{y_1,y_2} \Big|y_2 h_{21}(\bar{y}_2, 1, 1, w_1)\Big| I(-k_0\delta \le (\bar{y}_1, d_1^*(y_0))^T\theta_2^* \le k_0\delta) \\
F_6 &= B_2 \Big|\sum_{y_1,y_2} y_2 h_{21}(\bar{y}_2, 1, 0, w_1)\Big| I(-k_0\delta \le y_0^T\theta_1^* \le k_0\delta) \\
&\quad + B_2 \sum_{y_1,y_2} \Big|y_2 h_{21}(\bar{y}_2, 1, 0, w_1)\Big| I(-k_0\delta \le (\bar{y}_1, d_1^*(y_0))^T\theta_2^* \le k_0\delta) \\
F_7 &= B_2 \Big|\sum_{y_1,y_2} y_2 h_{21}(\bar{y}_2, 0, 1, w_1)\Big| I(-k_0\delta \le y_0^T\theta_1^* \le k_0\delta) \\
&\quad + B_2 \sum_{y_1,y_2} \Big|y_2 h_{21}(\bar{y}_2, 0, 1, w_1)\Big| I(-k_0\delta \le (\bar{y}_1, d_1^*(y_0))^T\theta_2^* \le k_0\delta) \\
F_8 &= B_2 \Big|\sum_{y_1,y_2} y_2 h_{21}(\bar{y}_2, 0, 0, w_1)\Big| I(-k_0\delta \le y_0^T\theta_1^* \le k_0\delta) \\
&\quad + B_2 \sum_{y_1,y_2} \Big|y_2 h_{21}(\bar{y}_2, 0, 0, w_1)\Big| I(-k_0\delta \le (\bar{y}_1, d_1^*(y_0))^T\theta_2^* \le k_0\delta)
\end{aligned}
$$

By Cauchy-Schwartz Inequality, Triangle Inequality, and Condition 2.1, we have

$$
\begin{aligned}
&\|F_1\|_{P,2} = \|F_2\|_{P,2} = \|F_3\|_{P,2} = \|F_4\|_{P,2} \\
&\le B_1\sqrt{P(-k_0\delta \le Y_0^T\theta_1^* \le k_0\delta)} + B_1 \max_{\bar{a}_1} \sqrt{P(-k_0\delta \le (\bar{Y}_1(\bar{a}_1), d_1^*(Y_0))^T\theta_2^* \le k_0\delta)} \\
&\le B_1\sqrt{c_1 k_0 + c_2 k_0}\,\delta^{1/2} \\
&= \tilde{B}_1\delta^{1/2}.
\end{aligned}
$$

By Condition 1.1 and 1.2, there exist constants $M_1, M_2, M_3 < \infty$ that for any $a_1, a_2 = 0, 1$

satisfy

$$\max\{|Y_1(a_1)|, |Y_2(a_1,a_2)|\} \leq M_1$$
$$|h_{21}(\bar{Y}_2, \bar{a}_2, W_1)| \leq M_2$$
$$|\sum_{y_1,y_2} y_2 h_{21}(\bar{y}_2, Y_0, \bar{a}_2, W_1)| \leq M_3$$

Then, we have

$$\begin{aligned}
&\mathbb{E}\left[\left(\sum_{y_1,y_2}\left|y_2 h_{21}(\bar{y}_2, Y_0, \bar{a}_2, W_1)\right| I(-k_0\delta \leq (\bar{y}_1, d_1^*(Y_0))^T\theta_2^* \leq k_0\delta)\right)^2\right] \\
&= \sum_{w_1,y_0}\left(\sum_{y_1,y_2}\left|y_2 h_{21}(\bar{y}_2, \bar{a}_2, w_1)\right| I(-k_0\delta \leq (\bar{y}_1, d_1^*(y_0))^T\theta_2^* \leq k_0\delta)\right)^2 f(w_1, y_0) \\
&\leq \sum_{w_1,y_0}\left(\sum_{y_1,y_2} y_2^2 h_{21}(\bar{y}_2, \bar{a}_2, w_1)^2 I(-k_0\delta \leq (\bar{y}_1, d_1^*(y_0))^T\theta_2^* \leq k_0\delta)\right)\left(\sum_{y_1,y_2} 1^2\right) f(w_1, y_0) \\
&\leq 4M_1^2 \sum_{w_1,y_0}\left(\sum_{y_1,y_2} y_2^2 h_{21}(\bar{y}_2, \bar{a}_2, w_1)^2 I(-k_0\delta \leq (\bar{y}_1, d_1^*(y_0))^T\theta_2^* \leq k_0\delta)\right) f(w_1, y_0) \\
&\leq 4M_1^2\left(\sum_{\bar{y}_2} y_2^2 I(-k_0\delta \leq (\bar{y}_1, d_1^*(y_0))^T\theta_2^* \leq k_0\delta) \sum_{w_1} h_{21}(\bar{y}_2, \bar{a}_2, w_1)^2 f(w_1, y_0)\right).
\end{aligned}$$

If $h_{21}(\bar{y}_2, \bar{a}_2, w_1) \geq 0$, then

$$\begin{aligned}
&\sum_{w_1} h_{21}(\bar{y}_2, \bar{a}_2, w_1)^2 f(w_1, y_0) \\
&\leq M_2 \sum_{w_1} h_{21}(\bar{y}_2, \bar{a}_2, w_1) f(w_1, y_0) \\
&= M_2 f(Y_2(a_1,a_2) = y_2, Y_1(a_1) = y_1, Y_0 = y_0).
\end{aligned}$$

Also note that $\sum_{w_1} h_{21}(\bar{y}_2, \bar{a}_2, w_1)^2 f(w_1|y_0) \leq M_2^2$ for every $(y_0, y_1, y_2, a_1, a_2)$, if there exist a positive constant $c$ serve as the lower bound for $f(Y_2(a_1,a_2) = y_2, Y_1(a_1) = y_1|Y_0 = y_0)$, i.e. $f(Y_2(a_1,a_2) = y_2, Y_1(a_1) = y_1|Y_0 = y_0) \geq c > 0$ for every $(y_0, y_1, y_2)$, then

$$\begin{aligned}
&\sum_{w_1} h_{21}(\bar{y}_2, \bar{a}_2, w_1)^2 f(w_1, y_0) \\
&\leq M_2^2 f(y_0) \\
&\leq \frac{M_2^2}{c} f(Y_2(a_1,a_2) = y_2, Y_1(a_1) = y_1, Y_0 = y_0) \\
&=: M_4 f(Y_2(a_1,a_2) = y_2, Y_1(a_1) = y_1, Y_0 = y_0).
\end{aligned}$$

Let $\tilde{M}_4 = \max\{M_2, M_4\}$, then:

$$4M_1^2\left(\sum_{\bar{y}_2} y_2^2 I(-k_0\delta \le (\bar{y}_1, d_1^*(y_0))^T\theta_2^* \le k_0\delta)\sum_{w_1} h_{21}(\bar{y}_2, \bar{a}_2, w_1)^2 f(w_1, y_0)\right)$$
$$\le 4M_1^2\tilde{M}_4\sum_{\bar{y}_2} y_2^2 I(-k_0\delta \le (\bar{y}_1, d_1^*(y_0))^T\theta_2^* \le k_0\delta) f(Y_2(a_1,a_2)=y_2, Y_1(a_1)=y_1, Y_0=y_0)$$
$$\le 4M_1^4\tilde{M}_4\sum_{\bar{y}_2} I(-k_0\delta \le (\bar{y}_1, d_1^*(y_0))^T\theta_2^* \le k_0\delta) f(Y_2(a_1,a_2)=y_2, Y_1(a_1)=y_1, Y_0=y_0)$$
$$= 4M_1^4\tilde{M}_4 P(|(\bar{Y}_1(a_1), d_1^*(Y_0))^T\theta_2^*| \le k_0\delta),$$

and

$$\mathbb{E}\left[\left(\sum_{y_1,y_2} y_2 h_{21}(\bar{y}_2, Y_0, \bar{a}_2, W_1) I(-k_0\delta \le Y_0^T\theta_1^* \le k_0\delta)\right)^2\right]$$
$$= \sum_{w_1,y_0}\left(\sum_{y_1,y_2} y_2 h_{21}(\bar{y}_2, \bar{a}_2, w_1)\right)^2 I(-k_0\delta \le y_0^T\theta_1^* \le k_0\delta) f(w_1, y_0)$$
$$\le M_3^2 \sum_{w_1,y_0} I(-k_0\delta \le y_0^T\theta_1^* \le k_0\delta) f(w_1, y_0)$$
$$\le M_3^2 P(|Y_0^T\theta_1^*| \le k_0\delta).$$

Thus,

$$\max\{\|F_5\|_{P,2}, \|F_6\|_{P,2}, \|F_7\|_{P,2}, \|F_8\|_{P,2}\}$$
$$\le B_2M_3\sqrt{P(|Y_0^T\theta_1^*| \le k_0\delta)} + 2B_2M_1^2\tilde{M}_4^{1/2}\sqrt{P(|(\bar{Y}_1(a_1), d_1^*(Y_0))^T\theta_2^*| \le k_0\delta)}$$
$$\le \tilde{B}_2\delta^{1/2}.$$

By (38) and some simple algebra, we have

$$
\begin{aligned}
&I(d_1(y_0) = a_1)I(d_2(\bar{y}_1, d_1(y_0)) = a_2)\\
&= a_1a_2I(y_0^T\theta_1 > 0)I((\bar{y}_1, I(y_0^T\theta_1 > 0))^T\theta_2 > 0)\\
&\quad + a_1(1-a_2)I(y_0^T\theta_1 > 0)I((\bar{y}_1, I(y_0^T\theta_1 > 0))^T\theta_2 \leq 0)\\
&\quad + a_2(1-a_1)I(y_0^T\theta_1 \leq 0)I((\bar{y}_1, I(y_0^T\theta_1 > 0))^T\theta_2 > 0)\\
&\quad + (1-a_1)(1-a_2)I(y_0^T\theta_1 \leq 0)I((\bar{y}_1, I(y_0^T\theta_1 > 0))^T\theta_2 \leq 0)\\
&= a_1a_2I(y_0^T\theta_1 > 0)I((\bar{y}_1, 1)^T\theta_2 > 0)\\
&\quad + a_1(1-a_2)I(y_0^T\theta_1 > 0)I((\bar{y}_1, 1)^T\theta_2 \leq 0)\\
&\quad + a_2(1-a_1)I(y_0^T\theta_1 \leq 0)I((\bar{y}_1, 0)^T\theta_2 > 0)\\
&\quad + (1-a_1)(1-a_2)I(y_0^T\theta_1 \leq 0)I((\bar{y}_1, 0)^T\theta_2 \leq 0)\\
&= a_1a_2d_1(y_0)d_2(\bar{y}_1, 1) + a_1(1-a_2)d_1(y_0)(1-d_2(\bar{y}_1, 1))\\
&\quad + a_2(1-a_1)(1-d_1(y_0))d_2(\bar{y}_1, 0) + (1-a_1)(1-a_2)(1-d_1(y_0))(1-d_2(\bar{y}_1, 0)).
\end{aligned}
$$

According to Lemma D.9 and the fact that the Vapnik–Chervonenkis (VC) dimension of the linear function class is finite, we can conclude that function classes $\mathcal{F}_\theta^1 \sim \mathcal{F}_\theta^8$ are all VC-classes. By Theorem 2.6.7 of van der Vaart & Wellner (1996), the uniform-entropy condition

$$
\mathcal{J}(1, \mathcal{F}_\theta{}^i) = \sup_Q \int_0^1 \sqrt{1 + \log N(\epsilon \|F_i\|_{Q,2}, \mathcal{F}, L_2(Q))} d\epsilon < \infty
$$

is satisfied for $i = 1, 2, \cdots, 8$, respectively. Also Note that

$$
\mathbb{G}_n \mathcal{F}_\theta^i = n^{-1/2} \sum_{i=1}^n \{\mathcal{F}_\theta^i - E[\mathcal{F}_\theta^i]\},
$$

By Theorem 2.14.1 of van der Vaart & Wellner (1996), we have that there exists a constant

$k_1 < \infty$,

$$
\begin{aligned}
(II) &= E^*\left[\sqrt{n} \sup_{\|\theta-\theta^*\|_2<\delta} \left|\hat{V}_n^*(\theta) - V(\theta) - \left\{\hat{V}_n^*(\theta^*) - V(\theta^*)\right\}\right|\right] \\
&\leq \sum_{i=1}^{8} E^*\left[\sup_{\|\theta-\theta^*\|_2<\delta} \left|\mathbb{G}_n \mathcal{F}_\theta^i\right|\right] \\
&\leq \sum_{i=1}^{8} k_1 \mathcal{J}(1, {\mathcal{F}_\theta}^i)\|F_i\|_{P,2} \\
&\leq \sum_{i=1}^{4} k_1 \tilde{B}_1 \mathcal{J}(1, {\mathcal{F}_\theta}^i)\delta^{1/2} + \sum_{i=5}^{8} k_1 \tilde{B}_2 \mathcal{J}(1, {\mathcal{F}_\theta}^i)\delta^{1/2} \\
&= k_2 \delta^{1/2}.
\end{aligned}
$$

Finally, we obtain that the centered process satisfies

$$
\begin{aligned}
&E^*\left[\sqrt{n} \sup_{\|\theta-\theta^*\|_2<\delta} \left|\hat{V}_n(\theta) - V(\theta) - \left\{\hat{V}_n(\theta^*) - V(\theta^*)\right\}\right|\right] \\
&\leq (I) + (II) \qquad (43) \\
&\leq k_2 \delta^{1/2},
\end{aligned}
$$

as $n \to \infty$.

Let $\phi_n(\delta) = \delta^{1/2}$ and $b = \frac{3}{2} < 2$, thus we have $\frac{\phi_n(\delta)}{\delta^b} = \delta^{-1}$ is decreasing, and $b$ does not depend on $n$.

*Proposition* D.3. *$r_n = n^{1/3}$ satisfies that $r_n^2 \phi_n(r_n^{-1}) = n^{1/2}$.*

*Proof.* We can easily verify that when $r_n = n^{1/3}$, $r_n^2 \phi_n(r_n^{-1}) = n^{2/3}\phi_n(n^{-1/3}) = n^{1/2}$.

In summary, by Propositions D.1,D.2, and D.3, $r_n = n^{1/3}$ meets the conditions required in Theorem 14.4 (Rate of Convergence) of Kosorok (2008). By the facts that $\theta - \theta^* \xrightarrow{p} 0$ as $n \to \infty$, and that $\hat{V}_n(\hat{\theta}) \geq \sup_{\theta\in\Theta} \hat{V}_n(\theta) \geq \sup_{\theta\in\Theta} \hat{V}_n(\theta) - O_p(n^{-2/3})$, we conclude that $n^{1/3}\|\theta - \theta^*\|_2 = O_p(1)$, which completes the proof of (1) of Theorem 5.1.

**Step 3.** Now we prove (b) and (c) of Theorem 5.1. Note that by Proposition D.1, for $\hat{\theta}$ obtained at sufficiently large $n$, we have that

$$\sqrt{n}(V(\hat{\theta}) - V(\theta^*)) \leq c_0 \sqrt{n}\|\hat{\theta} - \theta^*\|_2^2,$$

where $c_0$ is the maximum eigenvalue of $-\frac{1}{2}V''(\theta^*)$. As $n^{1/3}\|\theta-\theta\|_2 = O_p(1)$, we have

$$\sqrt{n}(V(\hat{\theta}) - V(\theta^*)) \leq c_0\sqrt{n}O_p(n^{-2/3}) = o_p(1).$$

Thus, $\sqrt{n}(V(\hat{\theta}) - V(\theta^*)) = o_p(1)$, which proves (b) of Theorem 5.1.

To prove (c), we first prove that $\sqrt{n}(\hat{V}_n(\hat{\theta}) - \hat{V}_n^*(\theta^*)) = o_p(1)$. Also by the fact that $n^{1/3}\|\theta-\theta\|_2 = O_p(1)$, for sufficiently large $n$, there exists a constant $k_3 < \infty$, satisfies $\|\theta-\theta\|_2 \leq k_3 n^{-1/3} = \tilde{\delta}$. By Markov Inequality, $\forall\epsilon > 0$, we have

$$\begin{aligned}
&P\left(\sqrt{n}\left|\hat{V}_n(\hat{\theta}) - \hat{V}_n(\theta^*) - \left\{V(\hat{\theta}) - V(\theta^*)\right\}\right| \geq \epsilon\right) \\
&\leq \frac{\mathbb{E}\left[\sqrt{n}\left|\hat{V}_n(\hat{\theta}) - V(\hat{\theta}) - \left\{\hat{V}_n(\theta^*) - V(\theta^*)\right\}\right|\right]}{\epsilon} \\
&\leq \frac{\mathbb{E}^*\left[\sqrt{n}\sup_{\|\theta-\theta^*\|_2<\tilde{\delta}}\left|\hat{V}_n(\theta) - V(\theta) - \left\{\hat{V}_n(\theta^*) - V(\theta^*)\right\}\right|\right]}{\epsilon} \\
&\leq \frac{k_2}{\epsilon}\tilde{\delta}^{1/2} + o(1) \\
&= O(n^{-1/6}) + o(1) \\
&= o(1) \to 0, \text{ as } n \to \infty
\end{aligned}$$

where the third inequality comes from Equation (43). Thus,

$$\sqrt{n}(\hat{V}_n(\theta) - \hat{V}_n(\theta^*) - \{V(\hat{\theta}) - V(\theta^*)\}) = o_p(1),$$

which implies that $\sqrt{n}(\hat{V}_n(\hat{\theta}) - \hat{V}_n(\theta^*)) = o_p(1)$. By **Step 1**, we have $\sqrt{n}(\hat{V}_n(\theta^*) - \hat{V}_n^*(\theta^*)) = o_p(1)$. By Conditions 1.1, 1.2, 1.4, Slutsky Lemma Van der Vaart (2000) and Lindeberg-Feller Central Limit Theory Van der Vaart (2000),

$$\begin{aligned}
&\sqrt{n}(\hat{V}_n(\hat{\theta}) - V(\theta^*)) \\
&= \sqrt{n}(\hat{V}_n(\hat{\theta}) - \hat{V}_n(\theta^*) + \hat{V}_n(\theta^*) - \hat{V}_n^*(\theta^*) + \hat{V}_n^*(\theta^*) - V(\theta^*)) \\
&= \sqrt{n}(\hat{V}_n(\hat{\theta}) - \hat{V}_n(\theta^*)) + \sqrt{n}(\hat{V}_n(\theta^*) - \hat{V}_n^*(\theta^*)) + \sqrt{n}(\hat{V}_n^*(\theta^*) - V(\theta^*)) \\
&= \sqrt{n}(\hat{V}_n^*(\theta^*) - V(\theta^*)) + o_p(1) \\
&\xrightarrow{d} \mathcal{N}(0, \sigma^2),
\end{aligned}$$

where $\sigma^2 = \mathbb{E}[IF^2_{V(\theta^*)}]$. This proves Theorem 5.1 (c).

## D.8 proof of Theorem A.1

*Proof.* As the proof technique used in 2-stage setting has no essential difference from the general $K$-stage setting, we take $K = 2$ as an example. We first prove that the plug-in estimators of Q functions are consistent pointwise.

By the definition of Q functions:

$$
\begin{aligned}
Q_2(\bar{y}_1, a_1; a_2) &= \sum_{y_2} y_2 f(Y_2(a_1, a_2) = y_2 | \bar{Y}_1(a_1) = \bar{y}_1) \\
&= \frac{\sum_{y_2} y_2 f(Y_2(a_1, a_2) = y_2, Y_1(a_1) = y_1 | Y_0 = y_0)}{f(Y_1(a_1) = y_1 | Y_0 = y_0)} \\
&=: \frac{g_2(\bar{y}_1, \bar{a}_2)}{g_1(\bar{y}_1, a_1)} \\
Q_1(y_0; a_1) &= \sum_{y_1} \max_{a_2} \left[ \frac{\sum_{y_2} y_2 f(Y_2(a_1, a_2) = y_2, Y_1(a_1) = y_1 | Y_0 = y_0)}{f(Y_1(a_1) = y_1 | Y_0 = y_0)} \right] f(Y_1(a_1) = y_1 | Y_0 = y_0) \\
&= \sum_{y_1} \max_{a_2} g_2(\bar{y}_1, \bar{a}_2)
\end{aligned}
$$

and the plug-in estimator $\hat{g}_2(\bar{y}_1, \bar{a}_2)$ of the molecular:

$$
\begin{aligned}
\hat{g}_2(\bar{y}_1, \bar{a}_2) = \hat{\mathbb{E}} &\left[ I(\bar{A}_2 = \bar{a}_2) \hat{q}_{22}(\bar{A}_2, \bar{y}_1, \bar{Z}_2) Y_2 | \bar{Y}_1 = \bar{y}_1 \right] \hat{f}(Y_1 = y_1 | Y_0 = y_0) \\
&- \hat{\mathbb{E}}[\sum_{y_2} y_2 I(\bar{A}_2 = \bar{a}_2) \hat{q}_{22}(\bar{A}_2, \bar{y}_1, \bar{Z}_2) \hat{h}_{22}(\bar{y}_2, \bar{W}_2, \bar{A}_2) | \bar{Y}_1 = \bar{y}_1] \hat{f}(Y_1 = y_1 | Y_0 = y_0) \\
&+ \hat{\mathbb{E}}[\sum_{y_2} y_2 I(A_1 = a_1) \hat{q}_{11}(A_1, y_0, Z_1) \hat{h}_{22}(\bar{y}_2, \bar{W}_2, \bar{A}_2) | \bar{Y}_1 = \bar{y}_1] \hat{f}(Y_1 = y_1 | Y_0 = y_0) \\
&- \hat{\mathbb{E}}[\sum_{y_2} y_2 I(A_1 = a_1) \hat{q}_{11}(A_1, y_0, Z_1) \hat{h}_{21}(\bar{y}_2, W_1, \bar{A}_2) | Y_0 = y_0] \\
&+ \hat{\mathbb{E}}[\sum_{y_2} y_2 \hat{h}_{21}(\bar{y}_2, W_1, \bar{a}_2) | Y_0 = y_0],
\end{aligned}
$$

as well as the plug-in estimator $\hat{g}_1(\bar{y}_1, a_1)$ of the denominator:

$$
\begin{aligned}
\hat{g}_1(\bar{y}_1, a_1) = \hat{\mathbb{E}}&[I(A_1 = a_1) \hat{q}_{11}(A_1, y_0, Z_1) | \bar{Y}_1 = \bar{y}_1] \hat{f}(Y_1 = y_1 | Y_0 = y_0) \\
&- \hat{\mathbb{E}}[I(A_1 = a_1) \hat{q}_{11}(A_1, y_0, Z_1) \hat{h}_{11}(\bar{y}_1, W_1, A_1) | Y_0 = y_0] \\
&+ \hat{\mathbb{E}}[\hat{h}_{11}(\bar{y}_1, W_1, a_1) | Y_0 = y_0].
\end{aligned}
$$

The conditional expectation can be estimated through the kernel technique. For example, the Nadaraya-Watson kernel regression estimator Nadaraya (1964) can be used.

Furthermore, the conditional density can also be estimated through kernel estimation, which ensures point-wise consistency for sufficiently large sample sizes. Then under some mild conditions (such as $\frac{1}{\hat{g}_1(\bar{y}_1,a_1)g_1(\bar{y}_1,a_1)}$ is bounded), by the Cross-fitting technique Wager (2024) and the Continuous Mapping Theorem Van der Vaart (2000), we have

$$\hat{Q}_2(\bar{y}_1, a_1; a_2) = \frac{\hat{g}_2(\bar{y}_1, \bar{a}_2)}{\hat{g}_1(\bar{y}_1, a_1)} \xrightarrow{p} \frac{g_2(\bar{y}_1, \bar{a}_2)}{g_1(\bar{y}_1, a_1)} = Q_2(\bar{y}_1, a_1; a_2)$$

point-wisely for every $(\bar{y}_1, a_1, a_2)$ if $f(Y_1(a_1) = y_1 | Y_0 = y_0) > 0$. Note that $\max(\cdot)$ is a continuous function, and $\max_{a_2} \hat{g}_2(\bar{y}_1, \bar{a}_2)$ is bounded, then by the Dominant Control Theorem and the Continuous Mapping Theorem, we can also get

$$\hat{Q}_1(y_0; a_1) = \sum_{y_1} \max_{a_2} \hat{g}_2(\bar{y}_1, \bar{a}_2) \xrightarrow{p} \sum_{y_1} \max_{a_2} g_2(\bar{y}_1, \bar{a}_2) = Q_1(y_0; a_1).$$

Note that

$$\begin{aligned} \hat{V}(\hat{d}_1, \hat{d}_2) &= \mathbb{P}_n \max_{a_1} \hat{Q}_1(Y_0; a_1) \\ &= \frac{1}{n} \sum_{i=1}^{n} \max_{a_1} \hat{Q}_1(Y_{0i}; a_1). \end{aligned}$$

Thus, by the Dominant Control Theorem and the Continuous Mapping Theorem Van der Vaart (2000) again, we get

$$\hat{V}(\hat{d}_1, \hat{d}_2) \xrightarrow{p} V(d_1^*, d_2^*).$$

By Condition1.1, there exist a constant $M < \infty$ such that $|Y_2(a_1, a_2)| < M, a_1, a_2 = 0, 1$.

Note that

$$
\begin{aligned}
&V(d_1^*, d_2^*) - V(\hat{d}_1, \hat{d}_2) \\
&= \sum_{\bar{a}_2} \sum_{\bar{y}_2} y_2 \left( I(d_1^*(y_0) = a_1) I(d_2^*(\bar{y}_1, a_1) = a_2) - I(\hat{d}_1(y_0) = a_1) I(\hat{d}_2(\bar{y}_1, a_1) = a_2) \right) \\
&\qquad \times f(Y_2(a_1, a_2) = y_2, Y_1(a_1) = y_1, Y_0 = y_0) \\
&= \sum_{\bar{y}_2} y_2 \left( d_1^*(y_0) d_2^*(\bar{y}_1, 1) - \hat{d}_1(y_0) \hat{d}_2(\bar{y}_1, 1) \right) f(Y_2(1,1) = y_2, Y_1(1) = y_1, Y_0 = y_0) \\
&\qquad + \sum_{\bar{y}_2} y_2 \left( (1 - d_1^*(y_0)) d_2^*(\bar{y}_1, 0) - (1 - \hat{d}_1(y_0)) \hat{d}_2(\bar{y}_1, 0) \right) f(Y_2(0,1) = y_2, Y_1(0) = y_1, Y_0 = y_0) \\
&\qquad + \sum_{\bar{y}_2} y_2 \left( d_1^*(y_0)(1 - d_2^*(\bar{y}_1, 1)) - \hat{d}_1(y_0)(1 - \hat{d}_2(\bar{y}_1, 1)) \right) f(Y_2(1,0) = y_2, Y_1(1) = y_1, Y_0 = y_0) \\
&\qquad + \sum_{\bar{y}_2} y_2 \left( (1 - d_1^*(y_0))(1 - d_2^*(\bar{y}_1, 0)) - (1 - \hat{d}_1(y_0))(1 - \hat{d}_2(\bar{y}_1, 0)) \right) \\
&\qquad \times f(Y_2(0,0) = y_2, Y_1(0) = y_1, Y_0 = y_0) \\
&= \sum_{y_0} d_1^*(y_0) \sum_{\bar{y}_2} y_2 \Big( d_2^*(\bar{y}_1, 1) f(Y_2(1,1) = y_2 | Y_1(1) = y_1, Y_0 = y_0) \\
&\qquad + (1 - d_2^*(\bar{y}_1, 1) f(Y_2(1,0) = y_2 | \bar{Y}_1(1) = \bar{y}_1)) \Big) f(\bar{Y}_1(1) = \bar{y}_1) \\
&\qquad - \sum_{y_0} \hat{d}_1(y_0) \sum_{\bar{y}_2} y_2 \Big( \hat{d}_2(\bar{y}_1, 1) f(Y_2(1,1) = y_2 | Y_1(1) = y_1, Y_0 = y_0) \\
&\qquad + (1 - \hat{d}_2(\bar{y}_1, 1) f(Y_2(1,0) = y_2 | \bar{Y}_1(1) = \bar{y}_1)) \Big) f(\bar{Y}_1(1) = \bar{y}_1) \\
&\qquad + \sum_{y_0} (1 - d_1^*(y_0)) \sum_{\bar{y}_2} y_2 \Big( d_2^*(\bar{y}_1, 0) f(Y_2(0,1) = y_2 | \bar{Y}_1(0) = \bar{y}_1) \\
&\qquad + (1 - d_2^*(\bar{y}_1, 0) f(Y_2(0,0) = y_2 | \bar{Y}_1(0) = \bar{y}_1)) \Big) f(\bar{Y}_1(0) = \bar{y}_1) \\
&\qquad - \sum_{y_0} (1 - \hat{d}_1(y_0)) \sum_{\bar{y}_2} y_2 \Big( \hat{d}_2(\bar{y}_1, 0) f(Y_2(0,1) = y_2 | \bar{Y}_1(0) = \bar{y}_1) \\
&\qquad + (1 - \hat{d}_2(\bar{y}_1, 0) f(Y_2(0,0) = y_2 | \bar{Y}_1(0) = \bar{y}_1)) \Big) f(\bar{Y}_1(0) = \bar{y}_1)
\end{aligned}
$$

$$
\begin{aligned}
&= \sum_{y_0} d_1^*(y_0) \sum_{y_1,y_2} y_2 \Big( d_2^*(\bar{y}_1, 1) f(Y_2(1,1) = y_2 | Y_1(1) = y_1, Y_0 = y_0) \\
&\quad + (1 - d_2^*(\bar{y}_1, 1) f(Y_2(1,0) = y_2 | \bar{Y}_1(1) = \bar{y}_1)) \Big) f(\bar{Y}_1(1) = \bar{y}_1) \\
&\quad - \sum_{y_0} \hat{d}_1(y_0) \sum_{y_1,y_2} y_2 \Big( \hat{d}_2(\bar{y}_1, 1) f(Y_2(1,1) = y_2 | Y_1(1) = y_1, Y_0 = y_0) \\
&\quad + (1 - \hat{d}_2(\bar{y}_1, 1) f(Y_2(1,0) = y_2 | \bar{Y}_1(1) = \bar{y}_1)) \Big) f(\bar{Y}_1(1) = \bar{y}_1) \\
&\quad + \sum_{y_0} (1 - d_1^*(y_0)) \sum_{y_1,y_2} y_2 \Big( d_2^*(\bar{y}_1, 0) f(Y_2(0,1) = y_2 | \bar{Y}_1(0) = \bar{y}_1) \\
&\quad + (1 - d_2^*(\bar{y}_1, 0) f(Y_2(0,0) = y_2 | \bar{Y}_1(0) = \bar{y}_1)) \Big) f(\bar{Y}_1(0) = \bar{y}_1) \\
&\quad - \sum_{y_0} (1 - \hat{d}_1(y_0)) \sum_{y_1,y_2} y_2 \Big( \hat{d}_2(\bar{y}_1, 0) f(Y_2(0,1) = y_2 | \bar{Y}_1(0) = \bar{y}_1) \\
&\quad + (1 - \hat{d}_2(\bar{y}_1, 0) f(Y_2(0,0) = y_2 | \bar{Y}_1(0) = \bar{y}_1)) \Big) f(\bar{Y}_1(0) = \bar{y}_1).
\end{aligned}
$$

By simple algebra, we have

$$
\begin{aligned}
&\sum_{y_0} d_1^*(y_0) \sum_{y_1,y_2} y_2 \Big( d_2^*(\bar{y}_1, 1) f(Y_2(1,1) = y_2 | Y_1(1) = y_1, Y_0 = y_0) \\
&\qquad + (1 - d_2^*(\bar{y}_1, 1) f(Y_2(1,0) = y_2 | \bar{Y}_1(1) = \bar{y}_1)) \Big) f(\bar{Y}_1(1) = \bar{y}_1) \\
&\quad - \sum_{y_0} \hat{d}_1(y_0) \sum_{y_1,y_2} y_2 \Big( \hat{d}_2(\bar{y}_1, 1) f(Y_2(1,1) = y_2 | Y_1(1) = y_1, Y_0 = y_0) \\
&\qquad + (1 - \hat{d}_2(\bar{y}_1, 1) f(Y_2(1,0) = y_2 | \bar{Y}_1(1) = \bar{y}_1)) \Big) f(\bar{Y}_1(1) = \bar{y}_1) \\
&= \sum_{y_0} d_1^*(y_0) \sum_{y_1,y_2} y_2 \left( f(Y_2(1,1) = y_2 | \bar{Y}_1(1) = y_1) - f(Y_2(1,0) = y_2 | \bar{Y}_1(1) = y_1) \right) \\
&\qquad \times \left( d_2^*(\bar{y}_1, 1) - \hat{d}_2(\bar{y}_1, 1) \right) f(\bar{Y}_1(1) = \bar{y}_1) \\
&\quad + \sum_{y_0} \left( d_1^*(y_0) - \hat{d}_1(y_0) \right) \sum_{y_1,y_2} y_2 \left( f(Y_2(1,1) = y_2 | \bar{Y}_1(1) = y_1) - f(Y_2(1,0) = y_2 | \bar{Y}_1(1) = y_1) \right) \\
&\qquad \times \left( d_2^*(\bar{y}_1, 1) - \hat{d}_2(\bar{y}_1, 1) \right) f(\bar{Y}_1(1) = \bar{y}_1) \\
&\quad + \sum y_0 \left( d_1^*(y_0) - \hat{d}_1(y_0) \right) \sum_{y_1,y_2} y_2 \Big( d_2^*(\bar{y}_1, 1) f(Y_2(1,1) = y_2 | Y_1(1) = y_1, Y_0 = y_0) \\
&\qquad + (1 - d_2^*(\bar{y}_1, 1) f(Y_2(1,0) = y_2 | \bar{Y}_1(1) = \bar{y}_1)) \Big) f(\bar{Y}_1(1) = \bar{y}_1) \\
&\leq \Big| \sum_{y_0} d_1^*(y_0) \sum_{y_1,y_2} y_2 \Big( f(Y_2(1,1) = y_2 | \bar{Y}_1(1) = y_1) - f(Y_2(1,0) = y_2 | \bar{Y}_1(1) = y_1) \Big) \\
&\qquad \times \left( d_2^*(\bar{y}_1, 1) - \hat{d}_2(\bar{y}_1, 1) \right) f(\bar{Y}_1(1) = \bar{y}_1) \Big| \\
&\quad + \Big| \sum_{y_0} \left( d_1^*(y_0) - \hat{d}_1(y_0) \right) \sum_{y_1,y_2} y_2 \left( f(Y_2(1,1) = y_2 | \bar{Y}_1(1) = y_1) - f(Y_2(1,0) = y_2 | \bar{Y}_1(1) = y_1) \right) \\
&\qquad \times \left( d_2^*(\bar{y}_1, 1) - \hat{d}_2(\bar{y}_1, 1) \right) f(\bar{Y}_1(1) = \bar{y}_1) \Big| \\
&\quad + \sum y_0 \left( d_1^*(y_0) - \hat{d}_1(y_0) \right) \sum_{y_1,y_2} y_2 \Big( d_2^*(\bar{y}_1, 1) f(Y_2(1,1) = y_2 | Y_1(1) = y_1, Y_0 = y_0) \\
&\qquad + (1 - d_2^*(\bar{y}_1, 1) f(Y_2(1,0) = y_2 | \bar{Y}_1(1) = \bar{y}_1)) \Big) f(\bar{Y}_1(1) = \bar{y}_1) \\
&=: (I) + (II) + (III)
\end{aligned}
$$

For the first item, note that $|d_1^*(y_0)| \leq 1$ and

$$
\sum_{y_2} y_2 \left( f(Y_2(1,1) = y_2 | \bar{Y}_1(1) = \bar{y}_1) - f(Y_2(1,0) = y_2 | \bar{Y}_1(1) = \bar{y}_1) \right) = Q_2^*(\bar{y}_1, 1; 1) - Q_2^*(\bar{y}_1, 1; 0),
$$

then

$$
\begin{aligned}
(I) &\leq \sum_{\bar{y}_1} \Big| Q_2^*(\bar{y}_1,1;1) - Q_2^*(\bar{y}_1,1;0) \Big| \Big| \Big( d_2^*(\bar{y}_1,1) - \hat{d}_2(\bar{y}_1,1) \Big) \Big| f(\bar{Y}_1(1) = \bar{y}_1) \\
&= \sum_{\bar{y}_1} \Big| Q_2^*(\bar{y}_1,1;1) - Q_2^*(\bar{y}_1,1;0) \Big| \cdot \Big| I(\hat{Q}_2(\bar{y}_1,1;1) > \hat{Q}_2(\bar{y}_1,1;0)) I(Q_2^*(\bar{y}_1,1;1) \leq Q_2^*(\bar{y}_1,1;0)) \Big| \\
&\qquad \times f(\bar{Y}_1(1) = \bar{y}_1) \\
&\quad + \sum_{\bar{y}_1} \Big| Q_2^*(\bar{y}_1,1;1) - Q_2^*(\bar{y}_1,1;0) \Big| \cdot \Big| I(\hat{Q}_2(\bar{y}_1,1;1) \leq \hat{Q}_2(\bar{y}_1,1;0)) I(Q_2^*(\bar{y}_1,1;1) > Q_2^*(\bar{y}_1,1;0)) \Big| \\
&\qquad \times f(\bar{Y}_1(1) = \bar{y}_1).
\end{aligned}
$$

For any $\delta > 0$, we have

$$
\begin{aligned}
& I(\hat{Q}_2(\bar{y}_1,1;1) > \hat{Q}_2(\bar{y}_1,1;0)) I(Q_2^*(\bar{y}_1,1;1) \leq Q_2^*(\bar{y}_1,1;0)) \\
&= I\Big( \hat{Q}_2(\bar{y}_1,1;1) > \hat{Q}_2(\bar{y}_1,1;0) \Big) \\
&\quad \times \Big( I(\delta \leq Q_2^*(\bar{y}_1,1;0) - Q_2^*(\bar{y}_1,1;1)) + I(0 \leq Q_2^*(\bar{y}_1,1;0) - Q_2^*(\bar{y}_1,1;1) \leq \delta) \Big).
\end{aligned}
$$

We also have that

$$
\begin{aligned}
& I(\hat{Q}_2(\bar{y}_1,1;1) > \hat{Q}_2(\bar{y}_1,1;0)) I(\delta \leq Q_2^*(\bar{y}_1,1;0) - Q_2^*(\bar{y}_1,1;1)) \\
& \leq I(Q_2^*(\bar{y}_1,1;0) - \hat{Q}_2(\bar{y}_1,1;0) + \hat{Q}_2(\bar{y}_1,1;1) - Q_2^*(\bar{y}_1,1;1) > \delta) \\
& \leq I(\Big| Q_2^*(\bar{y}_1,1;0) - \hat{Q}_2(\bar{y}_1,1;0) \Big| + \Big| \hat{Q}_2(\bar{y}_1,1;1) - Q_2^*(\bar{y}_1,1;1) \Big| > \delta) \\
& \leq I(\Big| Q_2^*(\bar{y}_1,1;0) - \hat{Q}_2(\bar{y}_1,1;0) \Big| > \frac{\delta}{2}) + I(\Big| \hat{Q}_2(\bar{y}_1,1;1) - Q_2^*(\bar{y}_1,1;1) \Big| > \frac{\delta}{2}),
\end{aligned}
$$

and

$$
\begin{aligned}
& I(\hat{Q}_2(\bar{y}_1,1;1) > \hat{Q}_2(\bar{y}_1,1;0)) I(0 \leq Q_2^*(\bar{y}_1,1;0) - Q_2^*(\bar{y}_1,1;1) \leq \delta) \\
& \leq I(0 \leq Q_2^*(\bar{y}_1,1;0) - Q_2^*(\bar{y}_1,1;1) \leq \delta) \\
& \leq I(\Big| Q_2^*(\bar{y}_1,1;0) - Q_2^*(\bar{y}_1,1;1) \Big| \leq \delta)
\end{aligned}
$$

Thus,

$$\sum_{\bar{y}_1} \left| Q_2^*(\bar{y}_1, 1; 1) - Q_2^*(\bar{y}_1, 1; 0) \right| \cdot \left| I(\hat{Q}_2(\bar{y}_1, 1; 1) > \hat{Q}_2(\bar{y}_1, 1; 0)) I(Q_2^*(\bar{y}_1, 1; 1) \leq Q_2^*(\bar{y}_1, 1; 0)) \right| f(\bar{Y}_1(1) = \bar{y}_1)$$

$$\leq 2M \left[ P(\left| Q_2^*(\bar{Y}_1(1), 1; 0) - \hat{Q}_2(\bar{Y}_1(1), 1; 0) \right| > \frac{\delta}{2}) + P(\left| \hat{Q}_2(\bar{Y}_1(1), 1; 1) - Q_2^*(\bar{Y}_1(1), 1; 1) \right| > \frac{\delta}{2}) \right]$$

$$+ \delta P(\left| Q_2^*(\bar{Y}_1(1), 1; 0) - Q_2^*(\bar{Y}_1(1), 1; 1) \right| \leq \delta)$$

$$\leq 2M \left[ P(\left| Q_2^*(\bar{Y}_1(1), 1; 0) - \hat{Q}_2(\bar{Y}_1(1), 1; 0) \right| > \frac{\delta}{2}) + P(\left| \hat{Q}_2(\bar{Y}_1(1), 1; 1) - Q_2^*(\bar{Y}_1(1), 1; 1) \right| > \frac{\delta}{2}) \right]$$

$$+ \delta.$$

We first fix $\delta = \frac{1}{k}$, $k \in \mathbb{N}$ and there exists a finite number $N_k$, satisfying:

When $n \geq N_k$,

$$2M \left[ P(\left| Q_2^*(\bar{Y}_1(1), 1; 0) - \hat{Q}_2(\bar{Y}_1(1), 1; 0) \right| > \frac{1}{2k}) + P(\left| \hat{Q}_2(\bar{Y}_1(1), 1; 1) - Q_2^*(\bar{Y}_1(1), 1; 1) \right| > \frac{1}{2k}) \right] \leq \frac{1}{k}.$$

Then, we let $n \geq N_k \to \infty$ with $k \to \infty$ and get $(I) = o_p(1)$. This technique will be referred to as the "diagonalization" method (Kallenberg Kallenberg (1997) Lemma 5.13, P108) hereafter. Also, by the fact that $\left| d_1^*(y_0) - \hat{d}_1(y_0) \right| \leq 1$, we can similarly use the diagonalization method to prove that $(II) = o_p(1)$. Thus, we have

$$\begin{aligned}
& V(d_1^*, d_2^*) - V(\hat{d}_1, \hat{d}_2) \\
& \leq \sum_{y_0} \Big( d_1^*(y_0) - \hat{d}_1(y_0) \Big) \sum_{y_1, y_2} y_2 \Big( d_2^*(\bar{y}_1, 1) f(Y_2(1,1) = y_2 | \bar{Y}_1(1) = \bar{y}_1) \\
& \quad + (1 - d_2^*(\bar{y}_1, 1) f(Y_2(1,0) = y_2 | \bar{Y}_1(1) = \bar{y}_1)) \Big) f(\bar{Y}_1(1) = \bar{y}_1) \\
& \quad + \sum_{y_0} \Big( \hat{d}_1(y_0) - d_1^*(y_0) \Big) \sum_{y_1, y_2} y_2 \Big( d_2^*(\bar{y}_1, 0) f(Y_2(0,1) = y_2 | \bar{Y}_1(0) = \bar{y}_1) \\
& \quad + (1 - d_2^*(\bar{y}_1, 0) f(Y_2(0,0) = y_2 | \bar{Y}_1(0) = \bar{y}_1)) \Big) f(\bar{Y}_1(0) = \bar{y}_1) \\
& \quad + o_p(1)
\end{aligned}$$

Then the problem is reduced to proving that

$$
\begin{aligned}
&\sum y_0 \left(d_1^*(y_0) - \hat{d}_1(y_0)\right) \\
&\quad \times \sum_{y_1,y_2} y_2 \Big( d_2^*(\bar{y}_1, 1) f(Y_2(1,1) = y_2 | \bar{Y}_1(1) = \bar{y}_1) \\
&\qquad + (1 - d_2^*(\bar{y}_1, 1) f(Y_2(1,0) = y_2 | \bar{Y}_1(1) = \bar{y}_1)) \Big) f(\bar{Y}_1(1) = \bar{y}_1) \\
&\quad + \sum y_0 \left(\hat{d}_1(y_0) - d_1^*(y_0)\right) \sum_{y_1,y_2} y_2 \Big( d_2^*(\bar{y}_1, 0) f(Y_2(0,1) = y_2 | \bar{Y}_1(0) = \bar{y}_1) \\
&\qquad + (1 - d_2^*(\bar{y}_1, 0) f(Y_2(0,0) = y_2 | \bar{Y}_1(0) = \bar{y}_1)) \Big) f(\bar{Y}_1(0) = \bar{y}_1) \\
&= o_p(1).
\end{aligned}
$$

Note that

$$
\begin{aligned}
&\sum_{y_0} \left(d_1^*(y_0) - \hat{d}_1(y_0)\right) \\
&\quad \times \sum_{y_1,y_2} y_2 \left(d_2^*(\bar{y}_1,1) f(Y_2(1,1) = y_2|\bar{Y}_1(1) = \bar{y}_1) + (1 - d_2^*(\bar{y}_1,1) f(Y_2(1,0) = y_2|\bar{Y}_1(1) = \bar{y}_1))\right) \\
&\quad \times f(\bar{Y}_1(1) = \bar{y}_1) \\
&+ \sum_{y_0} \left(\hat{d}_1(y_0) - d_1^*(y_0)\right) \\
&\quad \times \sum_{y_1,y_2} y_2 \left(d_2^*(\bar{y}_1,0) f(Y_2(0,1) = y_2|\bar{Y}_1(0) = \bar{y}_1) + (1 - d_2^*(\bar{y}_1,0) f(Y_2(0,0) = y_2|\bar{Y}_1(0) = \bar{y}_1))\right) \\
&\quad \times f(\bar{Y}_1(0) = \bar{y}_1) \\
&= \sum_{y_0} \left(d_1^*(y_0) - \hat{d}_1(y_0)\right) \\
&\quad \times \Big[ \sum_{y_1,y_2} y_2 \Big( d_2^*(\bar{y}_1,1) f(Y_2(1,1) = y_2|\bar{Y}_1(1) = \bar{y}_1) \\
&\qquad + (1 - d_2^*(\bar{y}_1,1) f(Y_2(1,0) = y_2|\bar{Y}_1(1) = \bar{y}_1)) \Big) f(\bar{Y}_1(1) = \bar{y}_1) \\
&\qquad - \sum_{y_1,y_2} y_2 \Big( d_2^*(\bar{y}_1,0) f(Y_2(0,1) = y_2|\bar{Y}_1(0) = \bar{y}_1) \\
&\qquad + (1 - d_2^*(\bar{y}_1,0) f(Y_2(0,0) = y_2|\bar{Y}_1(0) = \bar{y}_1)) \Big) f(\bar{Y}_1(0) = \bar{y}_1) \Big] \\
&= \sum_{y_0} \left(d_1^*(y_0) - \hat{d}_1(y_0)\right) \left[Q_1^*(y_0;1) - Q_1^*(y_0;0)\right] f(Y_0 = y_0) \\
&\leq \sum_{y_0} \left|d_1^*(y_0) - \hat{d}_1(y_0)\right| \left|Q_1^*(y_0;1) - Q_1^*(y_0;0)\right| f(Y_0 = y_0) \\
&\leq \sum_{y_0} I(\left|d_1^*(y_0) - \hat{d}_1(y_0)\right| = 1) \left|Q_1^*(y_0;1) - Q_1^*(y_0;0)\right| f(Y_0 = y_0)
\end{aligned}
$$

For $I(\left|\hat{d}_1(Y_0) - d_1^*(Y_0)\right| = 1)$, we have

$$
\begin{aligned}
&I(\left|\hat{d}_1(Y_0) - d_1^*(Y_0)\right| = 1) \\
&= I(\hat{Q}_1(Y_0;1) > \hat{Q}_1(Y_0;0)) I(Q_1^*(Y_0;1) \leq Q_1^*(Y_0;0)) \\
&\quad + I(\hat{Q}_1(Y_0;0) > \hat{Q}_1(Y_0;1)) I(Q_1^*(Y_0;0) \leq Q_1^*(Y_0;1))
\end{aligned}
$$

Similarly, for any $\delta > 0$, we have

$$I(\hat{Q}_1(Y_0;1) > \hat{Q}_1(Y_0;0))I(Q_1^*(Y_0;1) \le Q_1^*(Y_0;0))$$
$$= I(\hat{Q}_1(Y_0;1) - \hat{Q}_1(Y_0;0) > 0)(I(\delta \le Q_1^*(Y_0;0) - Q_1^*(Y_0;1)) + I(0 \le Q_1^*(Y_0;0) - Q_1^*(Y_0;1) < \delta)).$$

For $I(\hat{Q}_1(Y_0;1) - \hat{Q}_1(Y_0;0) > 0)I(\delta \le Q_1^*(Y_0;0) - Q_1^*(Y_0;1))$, we have

$$\begin{aligned}
&I(\hat{Q}_1(Y_0;1) - \hat{Q}_1(Y_0;0) > 0)I(\delta \le Q_1^*(Y_0;0) - Q_1^*(Y_0;1)) \\
&\le I(Q_1^*(Y_0;0) - Q_1^*(Y_0;1) + \hat{Q}_1(Y_0;1) - \hat{Q}_1(Y_0;0) \ge \delta) \\
&\le I(\left|Q_1^*(Y_0;0) - \hat{Q}_1(Y_0;0)\right| + \left|\hat{Q}_1(Y_0;1) - Q_1^*(Y_0;1)\right| \ge \delta) \\
&\le I(\left|Q_1^*(Y_0;0) - \hat{Q}_1(Y_0;0)\right| \ge \frac{\delta}{2}) + I(\left|\hat{Q}_1(Y_0;1) - Q_1^*(Y_0;1)\right| \ge \frac{\delta}{2}),
\end{aligned}$$

and for $I(\hat{Q}_1(Y_0;1) - \hat{Q}_1(Y_0;0) > 0)I(0 \le Q_1^*(Y_0;0) - Q_1^*(Y_0;1) < \delta)$, we have

$$\begin{aligned}
&I(\hat{Q}_1(Y_0;1) - \hat{Q}_1(Y_0;0) > 0)I(0 \le Q_1^*(Y_0;0) - Q_1^*(Y_0;1) < \delta) \\
&\le I(0 \le Q_1^*(Y_0;0) - Q_1^*(Y_0;1) < \delta) \\
&\le I(\left|Q_1^*(Y_0;0) - Q_1^*(Y_0;1)\right| < \delta).
\end{aligned}$$

Through the same trick, we can also get

$$\begin{aligned}
&\sum_{y_0} I(\left|d_1^*(y_0) - \hat{d}_1(y_0)\right| = 1) \left|Q_1^*(Y_0;1) - Q_1^*(Y_0;0)\right| f(Y_0 = y_0) \\
&\le 2M\left[P(\left|Q_1^*(Y_0;0) - \hat{Q}_1(Y_0;0)\right| > \frac{\delta}{2}) + P(\left|Q_1^*(Y_0;1) - \hat{Q}_1(Y_0;1)\right| > \frac{\delta}{2})\right] \\
&\quad + \delta P(\left|Q_1^*(Y_0;0) - Q_1^*(Y_0;1)\right| < \delta) \\
&\le 2M\left[P(\left|Q_1^*(Y_0;0) - \hat{Q}_1(Y_0;0)\right| > \frac{\delta}{2}) + P(\left|Q_1^*(Y_0;1) - \hat{Q}_1(Y_0;1)\right| > \frac{\delta}{2})\right] \\
&\quad + \delta.
\end{aligned}$$

Again, by the diagonalization method, we get

$$
\begin{aligned}
&\sum y_0 \left(d_1^*(y_0) - \hat{d}_1(y_0)\right) \\
&\quad \times \sum_{y_1,y_2} y_2 \left(d_2^*(\bar{y}_1, 1) f(Y_2(1,1) = y_2 | \bar{Y}_1(1) = \bar{y}_1) + (1 - d_2^*(\bar{y}_1, 1) f(Y_2(1,0) = y_2 | \bar{Y}_1(1) = \bar{y}_1))\right) \\
&\quad \times f(\bar{Y}_1(1) = \bar{y}_1) \\
&+ \sum y_0 \left(\hat{d}_1(y_0) - d_1^*(y_0)\right) \\
&\quad \times \sum_{y_1,y_2} y_2 \left(d_2^*(\bar{y}_1, 0) f(Y_2(0,1) = y_2 | \bar{Y}_1(0) = \bar{y}_1) + (1 - d_2^*(\bar{y}_1, 0) f(Y_2(0,0) = y_2 | \bar{Y}_1(0) = \bar{y}_1))\right) \\
&\quad \times f(\bar{Y}_1(0) = \bar{y}_1) \\
&= o_p(1).
\end{aligned}
$$

Thus, we finally prove that

$$V(d_1^*, d_2^*) - V(\hat{d}_1, \hat{d}_2) = o_p(1).$$

# E Existence and Uniqueness of Bridge Functions

## E.1 Existence

We first list the existence conditions for outcome and treatment confounding bridge functions below, according to Zhang & Tchetgen Tchetgen (2024) and Ying et al. (2023), for the sake of completeness.

Let $L_2\{F(t)\}$ denote the space of all square integrable functions of $t$ concerning a cumulative distribution function $F(t)$, which is a Hilbert space with inner product $\langle g, h\rangle = \int g(t)h(t)dF(t)$. Define $T_{\bar{a}_k,\bar{y}_{k-1}}$ as the conditional expectation operator, mapping $L_2\{F(\bar{w}_k|\bar{a}_k,\bar{y}_{k-1})\} \to L_2\{F(\bar{z}_k|\bar{a}_k,\bar{y}_{k-1})\}$ and $T_{\bar{a}_k,\bar{y}_{k-1}}h = \mathbb{E}[h(\bar{W}_k)|\bar{z}_k,\bar{a}_k,\bar{y}_{k-1}]$. Let $(\lambda_{\bar{a}_k,\bar{y}_{k-1},l}, \psi_{\bar{a}_k,\bar{y}_{k-1},l}, \phi_{\bar{a}_k,\bar{y}_{k-1},l})_{l=1}^{\infty}$ denote a singular value decomposition for $T_{\bar{a}_k,\bar{y}_{k-1}}$. That is, $T_{\bar{a}_k,\bar{y}_{k-1}}\psi_{\bar{a}_k,\bar{y}_{k-1},l} = \lambda_{\bar{a}_k,\bar{y}_{k-1},l}\phi_{\bar{a}_k,\bar{y}_{k-1},l}$. Also let $T'_{\bar{a}_k,\bar{y}_{k-1}}$ denote the conditional expectation operator: $L_2\{F(\bar{z}_k|\bar{a}_k,\bar{y}_{k-1})\} \to L_2\{F(\bar{w}_k|\bar{a}_k,\bar{y}_{k-1})\}$, $T'_{\bar{a}_k,\bar{y}_{k-1}}q = \mathbb{E}[q(\bar{Z}_k)|\bar{w}_k,\bar{a}_k,\bar{y}_{k-1}]$ and let

$(\lambda'_{\bar{a}_k,\bar{y}_{k-1},l}, \psi'_{\bar{a}_k,\bar{y}_{k-1},l}, \phi'_{\bar{a}_k,\bar{y}_{k-1},l})_{l=1}^{\infty}$ denote a singular value decomposition of $T'_{\bar{a}_k,\bar{y}_{k-1}}$. We assume the following:

*Assumption* 1 (The Existence of outcome confounding bridge functions).

1. $\sum_{k=1}^{K}\iint f(\overline{w}_k \mid \overline{z}_k, \bar{a}_k, \overline{y}_{k-1}) f(\overline{z}_k \mid \overline{w}_k, \bar{a}_k, \overline{y}_{k-1}) d\overline{w}_k d\overline{z}_k < \infty$.
2. For fixed $\bar{y}_K, \bar{a}_K$,

$$\int f^2(y_K \mid \overline{z}_K, \overline{a}_K, \overline{y}_{K-1}) f(\overline{z}_K \mid \overline{a}_K, \overline{y}_{K-1}) d\overline{z}_K < \infty,$$
$$\sum_{k=1}^{K-1}\int\left(\sum_{\overline{w}_{k+1}} h_{K,k+1}\left(\overline{y}_K, \overline{w}_{k+1}, \overline{a}_K\right) f\left(\overline{w}_{k+1}, y_k|\overline{y}_{k-1}, \bar{a}_k, \bar{z}_k\right)\right)^2 f(\bar{z}_k \mid \bar{a}_k, \bar{y}_{k-1}) d\bar{z}_k < \infty.$$

3. For fixed $\bar{y}_K, \bar{a}_K$,

$$\sum_{l=1}^{\infty}\lambda^{-2}_{\overline{a}_K,\overline{y}_{K-1},l}\langle f(y_K \mid \overline{z}_K, \overline{a}_K, \overline{y}_{K-1}), \phi_{\overline{a}_K,\overline{y}_{K-1},l}\rangle^2 < \infty,$$
$$\sum_{k=1}^{K-1}\sum_{l=1}^{\infty}\lambda^{-2}_{\bar{a}_k,\bar{y}_{k-1},l}\left\langle\sum_{\overline{w}_{k+1}} h_{K,k+1}\left(\overline{y}_K, \overline{w}_{k+1}, \overline{a}_K\right) f\left(\overline{w}_{k+1}, y_k|\overline{y}_{k-1}, a_k, z_k\right), \phi_{\bar{a}_k,\bar{y}_{k-1},l}\right\rangle^2 < \infty.$$

4. For any square integrable function $g$, we have that

$$\mathbb{E}[g(\bar{Z}_k)|\bar{y}_{k-1}, \bar{W}_k, \bar{a}_k] = 0 \text{ implies } g(\bar{Z}_k) = 0 \text{ almost surely} 1 \leq k \leq K.$$

*Assumption* 2 (The Existence of Treatment Confounding Bridge Functions).

1. $\sum_{k=1}^{K}\iint f(\overline{w}_k \mid \overline{z}_k, \bar{a}_k, \overline{y}_{k-1}) f(\overline{z}_k \mid \overline{w}_k, \bar{a}_k, \overline{y}_{k-1}) d\overline{w}_k d\overline{z}_k < \infty$.
2. For fixed $\bar{y}_K, \bar{a}_K$,

$$\int f^{-2}(a_1 \mid w_1, y_0) f(w_1 \mid a_1, y_0) dw_1 < \infty,$$
$$\sum_{k=1}^{K-1}\int\left(\frac{\sum_{\bar{z}_k} q_{kk}(\bar{y}_{k-1}, \bar{z}_k, \bar{a}_k) f(\bar{z}_k|\bar{a}_k, \bar{w}_{k+1}, \bar{y}_k)}{f(a_{k+1}|\bar{a}_k, \bar{w}_{k+1}, \bar{y}_k)}\right)^2 f(\bar{w}_k \mid \bar{a}_k, \bar{y}_{k-1}) d\bar{z}_k < \infty.$$

3. For fixed $\bar{y}_K, \bar{a}_K$,

$$\sum_{l=1}^{\infty}{\lambda'}^{-2}_{a_1,y_0,l}\langle f^{-1}(a_1 \mid w_1, y_0), \phi_{a_1,y_0,l}\rangle^2 < \infty,$$
$$\sum_{k=1}^{K-1}\sum_{l=1}^{\infty}\lambda^{-2}_{\bar{a}_k,\bar{y}_{k-1},l}\left\langle\frac{\sum_{\bar{z}_k} q_{kk}(\bar{y}_{k-1}, \bar{z}_k, \bar{a}_k) f(\bar{z}_k|\bar{a}_k, \bar{w}_{k+1}, \bar{y}_k)}{f(a_{k+1}|\bar{a}_k, \bar{w}_{k+1}, \bar{y}_k)}, \phi_{\bar{a}_k,\bar{y}_{k-1},l}\right\rangle^2 < \infty.$$

4. For any square integrable function $g$, we have that

$$\mathbb{E}[g(\bar{W}_k)|\bar{y}_{k-1}, \bar{Z}_k, \bar{a}_k] = 0 \text{ implies } g(\bar{Z}_k) = 0 \text{ almost surely} 1 \leq k \leq K.$$

**Lemma E.1.** *Under Assumption 1, there exist functions $\mathcal{H}_{Kl}(\bar{a}_K)$, $l = 1, 2, \dots, K$ such that*

$$f(Y_K = y_K|\bar{y}_{K-1}, \bar{z}_K, \bar{a}_K) = \sum_{\bar{w}_K} \mathcal{H}_{KK}(\bar{a}_K) f(\bar{w}_K|\bar{y}_{K-1}, \bar{z}_K, \bar{a}_K),$$

*and*

$$\sum_{\bar{w}_{l+1}} \mathcal{H}_{K,l+1}(\bar{a}_K) f(\bar{w}_{l+1}, y_l|\bar{y}_{l-1}, \bar{z}_l, \bar{a}_l) = \sum_{\bar{w}_l} \mathcal{H}_{Kl}(\bar{a}_K) f(\bar{w}_l|\bar{y}_{l-1}, \bar{z}_l, \bar{a}_l),$$

*for any $1 \leq l \leq K-1$.*

*Proof.* The proof follows straightforwardly from Picard's theorem (Kress Kress (1989)) and Lemma 2 of Miao et al. (2018).

**Lemma E.2.** *Under Assumption 2, there exist functions $\mathcal{Q}_{tt}(\bar{a}_t)$, $t = 1, 2, \dots, K$ such that*

$$\frac{1}{f(a_1|w_1, y_0)} = \sum_{z_1} \mathcal{Q}_{11}(\bar{a}_t) f(z_1|a_1, w_1, y_0),$$

*and for $t = 2, \dots, k$,*

$$\frac{\sum_{\bar{z}_{t-1}} \mathcal{Q}_{t-1,t-1}(\bar{a}_{t-1}) f(\bar{z}_{t-1}|\bar{a}_{t-1}, \bar{w}_t, \bar{y}_{t-1})}{f(a_t|\bar{a}_{t-1}, \bar{w}_t, \bar{y}_{t-1})} = \sum_{\bar{z}_t} \mathcal{Q}_{tt}(\bar{a}_t) f(\bar{z}_t|\bar{a}_t, \bar{w}_t, \bar{y}_{t-1}).$$

*Proof.* The proof also follows straightforwardly from Picard's theorem (Kress Kress (1989)) and Lemma 2 of Miao et al. (2018).

## E.2 Uniqueness

The uniqueness of outcome and treatment confounding bridges hinges on Assumptions 2.4 and 1.4, respectively. For instance, by Assumption 2.4, suppose that there exist $\mathcal{H}'_{Kl}(\bar{a}_K)$ and $\mathcal{H}_{Kl}(\bar{a}_K)$, $l = 1, 2, \dots, K$ that both solve the integral equation, we have:

$$\begin{aligned}
&\sum_{\bar{w}_K} [\mathcal{H}_{KK}(\bar{a}_K) - \mathcal{H}'_{KK}(\bar{a}_K)]\, f(\bar{w}_K|\bar{y}_{K-1}, \bar{z}_K, \bar{a}_K) \\
&= \mathbb{E}[h_{KK}(\bar{y}_K, \bar{W}_K, \bar{a}_K) - h'_{KK}(\bar{y}_K, \bar{W}_K, \bar{a}_K)|\bar{y}_{K-1}, \bar{z}_K, \bar{a}_K] \\
&= 0,
\end{aligned}$$

which by Assumption 2.4 implies that

$$h_{KK}(\bar{y}_K, \bar{W}_K, \bar{a}_K) = h'_{KK}(\bar{y}_K, \bar{W}_K, \bar{a}_K).$$

We can similarly prove that $\mathcal{H}_{Kl}(\bar{a}_K) = \mathcal{H}'_{Kl}(\bar{a}_K)$ for $l = K-1, \ldots, 1$ by repeatedly use Assumption 2.4. The proof of the uniqueness of treatment confounding bridge functions under Assumption 1.4 is also similar, thus omitted.

# F More about the Completeness Assumption

Notably, Newey & Powell (2003) employ completeness conditions to establish identification in nonparametric regression with instrumental variables, and Andrews (2017) formalizes the generalized concept of $L_2$-completeness. Specifically, they define that a bivariate distribution $F_{XV}$ of random elements $X$ and $V$ is $L_2$-complete concerning $X$ if $\forall h \in L_2(F_X)$,

$$\mathbb{E}(h(X)|V) = 0 \text{ a.s.}[F_V] \iff h(X) = 0 \text{ a.s.}[F_X],$$

where the expectation is taken under $F_{XV}$, and "a.s.$[F_V]$" is the abbreviation of "almost surely with respect to $F_V$", similar to "a.s.$[F_X]$". Assumptions $3'.1$ at the $l^{th}$ stage and $3'.2$ at the $t^{th}$ stage both fall within the class of $L_2$-completeness conditions, which essentially rule out the conditional independence between the unmeasured confounders and the corresponding proxy variables, given $(\bar{Y}_{l-1}, \bar{A}_l)$ in the case of Assumption $3'.1$, or $(\bar{Y}_{t-1}, \bar{A}_t)$ in the case of Assumption $3'.2$. Take Assumption $3'.1$ at the $l^{th}$ stage for an example, as the discussion for another is essentially the same. By linking the range of $\bar{U}_{l-1}$ to the range of $\bar{Z}_l$, Assumption $3'.1$ can also be interpreted as the requirement that the variability of $\bar{Z}_l$ is sufficiently rich relative to that of $\bar{U}_{l-1}$. Therefore, any variation in the unmeasured confounders $\bar{U}_{l-1}$ is expected to be captured by the proxy variables $\bar{Z}_l$, given $\bar{Y}_{l-1}$ and $\bar{A}_l$. To develop further intuition for this explanation, it is helpful to consider the case of categorical variables. Following the discussion in Newey & Powell (2003), Zhang & Tchetgen Tchetgen (2024), Darolles et al. (2011), Miao et al. (2018), Cui et al. (2024), Ying et al. (2023), if we suppose that $U$ and $Z$ are both categorical, then

Assumption 3′.1 requires that $\dim(\bar{U}_{l-1}) \leq \dim(\bar{Z}_l)$. This implies that the proxy variables $Z$ must have at least the same categories as the unmeasured confounders $U$. While completeness is often viewed as a technical condition that can be challenging to verify in practice, several studies have provided constructive approaches to assess its plausibility (Lehmann & Romano 2005, Newey & Powell 2003, Andrews 2017, Hu et al. 2017, Freyberger 2017). These works demonstrate specific conditions for distributions where the completeness assumption can be formally verified. For a more comprehensive discussion of such conditions and their applications in proximal causal inference, we refer interested readers to Ying et al. (2023), Appendix C.

# G More about the Assumption 5

**Proposition G.1.** *Under Assumptions 1 and 2, if we admit that in a causal model, all independencies in the data are determined by the underlying causal structure, rather than by accidental cancellations resulting from specific parameter values, which is also known as the faithfulness assumption in causal DAG related literature Pearl (1995, 2009), then for $k = 1, 2, \ldots, K$, Assumption 5 with several additional conclusions:*

$$\{Y_k, \bar{W}_{k+1}\} \perp\!\!\!\perp \bar{Z}_k | \bar{U}_{k-1}, \bar{Y}_{k-1}, \bar{A}_k. \tag{44}$$

*holds.*

*Proof.* For a given $k \in \{1, 2, \ldots, K-1\}$, we will prove Equation (44) part by part. For the first two parts, we will prove that $Y_k \perp\!\!\!\perp \bar{Z}_k | \bar{U}_{k-1}, \bar{Y}_{k-1}, \bar{A}_k$ and $\bar{W}_k \perp\!\!\!\perp \bar{Z}_k | \bar{U}_{k-1}, \bar{Y}_{k-1}, \bar{A}_k$, respectively. Notably, the proof of these parts does not need the fathifulness assumption.

To prove that $Y_k \perp\!\!\!\perp \bar{Z}_k | \bar{U}_{k-1}, \bar{Y}_{k-1}, \bar{A}_k$, we note that for arbitrarily given $(\bar{u}_{k-1}, \bar{y}_{k-1}, \bar{a}_k)$,

$$
\begin{aligned}
& f(Y_k, \bar{Z}_k | \bar{u}_{k-1}, \bar{y}_{k-1}, \bar{a}_k) \\
&= f(Y_k(\bar{a}_k), \bar{Z}_k | \bar{u}_{k-1}, \bar{y}_{k-1}, \bar{a}_k) \\
&= f(Y_k(\bar{a}_k) | \bar{Z}_k, \bar{u}_{k-1}, \bar{y}_{k-1}, \bar{a}_k) f(\bar{Z}_k | \bar{u}_{k-1}, \bar{y}_{k-1}, \bar{a}_k) \\
&= f(Y_k(\bar{a}_k) | \bar{u}_{k-1}, \bar{y}_{k-1}, \bar{a}_k) f(\bar{Z}_k | \bar{u}_{k-1}, \bar{y}_{k-1}, \bar{a}_k) \\
&= f(Y_k | \bar{u}_{k-1}, \bar{y}_{k-1}, \bar{a}_k) f(\bar{Z}_k | \bar{u}_{k-1}, \bar{y}_{k-1}, \bar{a}_k),
\end{aligned}
$$

where the first as well as the fourth equalities follow from the consistency, and the third equality follows from Assumption 1.3.

To prove that $\bar{W}_k \perp\!\!\!\perp \bar{Z}_k | \bar{U}_{k-1}, \bar{Y}_{k-1}, \bar{A}_k$, we note that for arbitrarily given $(\bar{u}_{k-1}, \bar{y}_{k-1}, \bar{a}_k)$,

$$
\begin{aligned}
& f(\bar{W}_k, \bar{Z}_k | \bar{u}_{k-1}, \bar{y}_{k-1}, \bar{a}_k) \\
&= f(\bar{W}_k | \bar{Z}_k, \bar{u}_{k-1}, \bar{y}_{k-1}, \bar{a}_k) f(\bar{Z}_k | \bar{u}_{k-1}, \bar{y}_{k-1}, \bar{a}_k) \\
&= f(\bar{W}_k | \bar{u}_{k-1}, \bar{y}_{k-1}, \bar{a}_k) f(\bar{Z}_k | \bar{u}_{k-1}, \bar{y}_{k-1}, \bar{a}_k),
\end{aligned}
$$

where the second equality follows from Assumption 1.3.

Now that we have proved that $\{Y_k, \bar{W}_k\} \perp\!\!\!\perp \bar{Z}_k | \bar{U}_{k-1}, \bar{Y}_{k-1}, \bar{A}_k$, the remained task is to show that $W_{k+1} \perp\!\!\!\perp \bar{Z}_k | \bar{U}_{k-1}, \bar{Y}_{k-1}, \bar{A}_k$. Trying to prove it by contradiction, we first assume that $W_{k+1} \not\perp\!\!\!\perp \bar{Z}_k | \bar{U}_{k-1}, \bar{Y}_{k-1}, \bar{A}_k$. Under this condition, we will prove that $\bar{Z}_k \not\perp\!\!\!\perp U_k | \bar{U}_{k-1}, \bar{Y}_{k-1}, \bar{A}_k$, which will lead to the violation of Assumption 1.3.

Now, we have:

$$W_{k+1} \not\perp\!\!\!\perp \bar{Z}_k | \bar{U}_{k-1}, \bar{Y}_{k-1}, \bar{A}_k, \tag{45}$$

$$W_{k+1} \perp\!\!\!\perp \bar{Z}_k | \bar{U}_k, \bar{Y}_k, \bar{A}_k, \tag{46}$$

$$\bar{Z}_k \perp\!\!\!\perp Y_k | \bar{U}_{k-1}, \bar{Y}_{k-1}, \bar{A}_k. \tag{47}$$

The validity of Equation (46) follows from Assumption 1.3. Under these conditions, we will prove that $\bar{Z}_k \not\perp\!\!\!\perp U_k | \bar{U}_{k-1}, \bar{Y}_{k-1}, \bar{A}_k$ by contradiction again. We suppose that $\bar{Z}_k \perp\!\!\!\perp U_k | \bar{U}_{k-1}, \bar{Y}_{k-1}, \bar{A}_k$.

Then, by Equation (47) we have

$$\bar{Z}_k \perp\!\!\!\perp \{U_k, Y_k\} | \bar{U}_{k-1}, \bar{Y}_{k-1}, \bar{A}_k. \tag{48}$$

Thus,

$$\begin{aligned}
&f(\bar{Z}_k, U_k, Y_k, W_{k+1} | \bar{U}_{k-1}, \bar{Y}_{k-1}, \bar{A}_k) \\
&= f(\bar{Z}_k | \bar{U}_k, \bar{Y}_k, \bar{A}_k, W_{k+1}) f(W_{k+1}, U_k, Y_k | \bar{U}_{k-1}, \bar{Y}_{k-1}, \bar{A}_k) \\
&= f(\bar{Z}_k | \bar{U}_k, \bar{Y}_k, \bar{A}_k) f(W_{k+1}, U_k, Y_k | \bar{U}_{k-1}, \bar{Y}_{k-1}, \bar{A}_k) \\
&= f(\bar{Z}_k | \bar{U}_{k-1}, \bar{Y}_{k-1}, \bar{A}_k) f(W_{k+1}, U_k, Y_k | \bar{U}_{k-1}, \bar{Y}_{k-1}, \bar{A}_k),
\end{aligned} \tag{49}$$

where the second equality follows from Equation (46) and the third equality follows from Equation (48). If we sum over $U_k, Y_k$ for the both sides of Equation (49), we conclude that

$$f(\bar{Z}_k, W_{k+1} | \bar{U}_{k-1}, \bar{Y}_{k-1}, \bar{A}_k) = f(\bar{Z}_k | \bar{U}_{k-1}, \bar{Y}_{k-1}, \bar{A}_k) f(W_{k+1} | \bar{U}_{k-1}, \bar{Y}_{k-1}, \bar{A}_k),$$

which implies $\bar{Z}_k \perp\!\!\!\perp W_{k+1} | \bar{U}_{k-1}, \bar{Y}_{k-1}, \bar{A}_k$ by the definition of probability independence. This is a contradition to Equation (45), thus $\bar{Z}_k \not\perp\!\!\!\perp U_k | \bar{U}_{k-1}, \bar{Y}_{k-1}, \bar{A}_k$.

However, as $U_k$ is an unmeasured confounder between $A_{k+1}$ and $Y_{k+1}$, it will directly affect $Y_{k+1}$, which implies that $\bar{Z}_k$ and $Y_{k+1}$ can be related through $U_k$ even if we have set the value of $\bar{A}_k$. In other words, there is an unblocked path between $\bar{Z}_k$ and $Y_{k+1}$ through $U_k$ in each possible causal DAG that satisfies Assumption 1 and $\bar{W}_{k+1} \not\perp\!\!\!\perp \bar{Z}_k | \bar{U}_{k-1}, \bar{Y}_{k-1}, \bar{A}_k$. By Assumption 1.3, $Y_{k+1}(\bar{a}_{k+1}) \perp\!\!\!\perp \bar{Z}_k | \bar{U}_{k-1}, \bar{Y}_{k-1}, \bar{A}_k$, which will lead to an immediate contradiction under the faithful assumption. These prove Equation (44).

We admit that there indeed exist some extreme cases where the faithfulness assumption may fail to hold. For example, when $K = 2$, we note that even if we have obtained that $Z_1 \not\perp\!\!\!\perp U_1 | U_0, Y_0, A_1$, $U_1 \not\perp\!\!\!\perp Y_2(\bar{a}_2) | U_0, Y_0, A_1$, and $U_1 \not\perp\!\!\!\perp W_2 | U_0, Y_0, A_1$, the possibility that $Z_1 \not\perp\!\!\!\perp Y_2(\bar{a}_2) | U_0, Y_0$ and $W_2(\bar{a}_2, \bar{z}_2) = W_2(a_1)$ still can not be denied. As an example, suppose that the conditional joint distribution of $(Z_1, U_1, Y_2(\bar{a}_2))$ (or the conditional joint distribution of $(Z_1, U_1, W_2))$ under the condition of $(U_0, Y_0, A_1)$ follows the normal distribution $\mathcal{N}(0, D)$, where

$$D = \begin{bmatrix} 1 & \rho & 0 \\ \rho & 1 & \rho \\ 0 & \rho & 1 \end{bmatrix},$$

and $\rho = \frac{1}{2}$. Then, as $D$ is positive-definite, this is a valid normal distribution. However, as $Z_1$ and $Y_2(\bar{a}_2)$ (or $W_2$) is not correlated, we conclude that $Z_1$ and $Y_2(\bar{a}_2)$ (or $W_2$) is still conditional independent.

Nevertheless, we argue that the faithfulness assumption is generally reasonable because the pathological parameter constellations that would violate it are measure-zero cases in the underlying parameter space and thus rarely occur in practice. Moreover, our primary interest typically lies in uncovering the core causal mechanisms rather than modeling highly contrived dependencies, making faithfulness, and thus Assumption 5, practical and defensible working assumptions.

# H More Details about the Semiparametric Theory and the Multiply Robust Estimator

## H.1 Regularity Assumptions

We denote by $\Pi_t : L_2(\bar{Y}_{t-1}, \bar{Z}_t, \bar{A}_t) \to L_2(\bar{Y}_{t-1}, \bar{W}_t, \bar{A}_t)$, $t = 1, 2, \ldots, K$ the conditional expectation operator given by $\Pi_t(g) = \mathbb{E}[g(\bar{Y}_{t-1}, \bar{Z}_t, \bar{A}_t) | \bar{Y}_{t-1}, \bar{W}_t, \bar{A}_t]$.

*Assumption H.1.* At the true data-generating law,

1. The conditional expectation operators $\Pi_t$, $t = 1, 2, \ldots, K$ are all surjective.
2. There exists a positive constant $\eta > 0$, such that $\eta < f(A_t | \bar{A}_{t-1}, \bar{W}_t, \bar{Y}_{t-1}) < 1 - \eta$, $t = 1, 2, \ldots, K$.

Assumption H.1.1 essentially requires that for $t = 1, 2, \ldots, K$, $L_2(\bar{Y}_{t-1}, \bar{Z}_t, \bar{A}_t)$ is sufficiently rich, such that its mapping into $L_2(\bar{Y}_{t-1}, \bar{W}_t, \bar{A}_t)$ via conditional expectation operator $\Pi_t$ can generate all elements of the latter space. As noted by Ying et al. (2023), it is a sufficient condition for a closed-form expression of the semiparametric efficiency bound. Without imposing this assumption, a closed-form expression for the semiparametric efficiency bound is generally unavailable. Assumption *H*.1.2 is often referred to as the strong positivity or strict overlap assumption (Wager 2024, Ding 2024), which we adopt here to bound the reciprocal of propensity score $\frac{1}{f(A_t|\bar{A}_{t-1},\bar{W}_t,\bar{Y}_{t-1})}$ away from 0 by a fixed amount $\min\{\frac{1}{\eta},\frac{1}{1-\eta}\}$.

## H.2 The Underlying Connections between the Proximal Multiply Robust Estimator and Other Works

We will present the connection between the PMR estimator of the value function of a sequence of given regimes $\bar{d}_K$ and the doubly robust estimator proposed in Zhang et al. (2013), which will induce another closed-form expression of the PMR estimator. Moreover, as the identification methods respectively implied by $\mathcal{M}_k,\ k = 0,1,2,\ldots,K$ can also be viewed as a monotone coarsening process, we note that the closed-form of the efficient influence function presented in Equation 11 is not a coincidence, according to Tsiatis (2006), Chapter 10 and 11.

In Theorem 4.2, we directly derive the closed-form of the multiply robust estimator from the influence function depicted in Equation (11). For simplicity, we denote $\mathcal{Q}_{kk}(\bar{A}_k) = q_{kk}(\bar{Y}_{k-1},\bar{Z}_k,\bar{A}_k)$ and

$$
\begin{aligned}
J_k(\bar{Y}_{k-1},\bar{W}_k,\bar{a}_K)_{\bar{d}_K} = &\sum_{y_K,\cdots,y_k} y_K h_{Kk}(y_K,\cdots,y_k,\bar{Y}_{k-1},\bar{W}_k,\bar{a}_K)\\
&\times\prod_{j=k+1}^{K} I(d_j(y_{j-1},\cdots,y_k,\bar{Y}_{k-1},\bar{a}_{j-1}) = a_j),
\end{aligned}
$$

for $k = 1, 2, \dots, K$. Then the PMR estimator for $V(\bar{d}_K)$ can be also rewritten as:

$$\begin{aligned} &\hat{V}_{PMR}(\bar{d}_K) \\ &= \sum_{\bar{a}_K} \mathbb{P}_n \Bigg\{ I(\bar{d}_K(\bar{Y}_{K-1}, \bar{A}_{K-1}) = \bar{A}_K = \bar{a}_K) \mathcal{Q}_{KK}(\bar{A}_K; \hat{\gamma}_K) Y_K \\ &\quad + \sum_{k=1}^{K} \Big[ I(\bar{d}_{k-1}(\bar{Y}_{k-2}, \bar{A}_{k-2}) = \bar{A}_{k-1} = \bar{a}_{k-1}) \mathcal{Q}_{k-1,k-1}(\bar{A}_{k-1}; \hat{\gamma}_{k-1}) \\ &\qquad - I(\bar{d}_k(\bar{Y}_{k-1}, \bar{A}_{k-1}) = \bar{A}_k = \bar{a}_k) \mathcal{Q}_{kk}(\bar{A}_k; \hat{\gamma}_k) \Big] J_k(\bar{Y}_{k-1}, \bar{W}_k, \bar{a}_K)_{\bar{d}_K} \Bigg\}, \end{aligned} \tag{50}$$

where the term $I(\bar{d}_k(\bar{Y}_{k-1}, \bar{A}_{k-1}) = \bar{A}_k = \bar{a}_k) \mathcal{Q}_{kk}(\bar{A}_k; \hat{\gamma}_k)$ is defined as $I(d_1(Y_0) = a_1)$ when $k = 0$. Comparing between Equation (50) and the form of Equation (6) in Zhang et al. (2013) motivates the following lemma:

**Lemma H.1.**

$$\begin{aligned} &\sum_{\bar{a}_K} \mathbb{E}\Big[ I(\bar{d}_k(\bar{Y}_{k-1}, \bar{A}_{k-1}) = \bar{A}_k = \bar{a}_k) \mathcal{Q}_{kk}(\bar{A}_k) J_k(\bar{Y}_{k-1}, \bar{W}_k, \bar{a}_K)_{\bar{d}_K} \Big] \\ &= \mathbb{E}\Big[ I(\bar{d}_k(\bar{Y}_{k-1}, \bar{A}_{k-1}) = \bar{A}_k) \mathcal{Q}_{kk}(\bar{A}_k) \mathbb{E}[Y_K(\bar{d}_K) | \bar{Y}_{k-1}(\bar{d}_{k-1})] \Big], \end{aligned} \tag{51}$$

*and*

$$\begin{aligned} &\sum_{\bar{a}_K} \mathbb{E}\Big[ I(\bar{d}_{k-1}(\bar{Y}_{k-2}, \bar{A}_{k-2}) = \bar{A}_{k-1} = \bar{a}_{k-1}) \mathcal{Q}_{k-1,k-1}(\bar{A}_{k-1}) J_k(\bar{Y}_{k-1}, \bar{W}_k, \bar{a}_K)_{\bar{d}_K} \Big] \\ &= \mathbb{E}\Big[ I(\bar{d}_{k-1}(\bar{Y}_{k-2}, \bar{A}_{k-2}) = \bar{A}_{k-1}) \mathcal{Q}_{k-1,k-1}(\bar{A}_{k-1}) \mathbb{E}[Y_K(\bar{d}_K) | \bar{Y}_{k-1}(\bar{d}_{k-1})] \Big]. \end{aligned} \tag{52}$$

*Proof.* For Equation (51), we note that

$$\begin{aligned} &\sum_{\bar{a}_K} \mathbb{E}\Big[ I(\bar{d}_k(\bar{Y}_{k-1}, \bar{A}_{k-1}) = \bar{A}_k = \bar{a}_k) \mathcal{Q}_{kk}(\bar{A}_k) J_k(\bar{Y}_{k-1}, \bar{W}_k, \bar{a}_K)_{\bar{d}_K} \Big] \\ &= \sum_{\bar{a}_K} \sum_{\bar{y}_K} \sum_{\bar{z}_k} \sum_{\bar{w}_k} y_K \mathcal{Q}_{kk}(\bar{a}_k) \mathcal{H}_{Kk}(\bar{a}_K) f(\bar{z}_k, \bar{w}_k, \bar{y}_{k-1}, \bar{a}_k) \prod_{j=1}^{K} I(d_j(\bar{y}_{j-1}, \bar{a}_{j-1}) = a_j) \\ &= \sum_{\bar{y}_{k-1}} \frac{\sum_{y_K, \dots, y_k} \sum_{\bar{a}_K} \sum_{\bar{z}_k} \sum_{\bar{w}_k} y_K \mathcal{Q}_{kk}(\bar{a}_k) \mathcal{H}_{Kk}(\bar{a}_K) f(\bar{z}_k, \bar{w}_k, \bar{y}_{k-1}, \bar{a}_k) \prod_{j=1}^{K} I(d_j(\bar{y}_{j-1}, \bar{a}_{j-1}) = a_j)}{\sum_{\bar{a}_k} \sum_{\bar{z}_k} \mathcal{Q}_{kk}(\bar{a}_k) f(\bar{z}_k, \bar{y}_{k-1}, \bar{a}_k) \prod_{j=1}^{k-1} I(d_j(\bar{y}_{j-1}, \bar{a}_{j-1}) = a_j)} \\ &\quad \times \sum_{\bar{a}_k} \sum_{\bar{z}_k} \mathcal{Q}_{kk}(\bar{a}_k) f(\bar{z}_k, \bar{y}_{k-1}, \bar{a}_k) \prod_{j=1}^{k-1} I(d_j(\bar{y}_{j-1}, \bar{a}_{j-1}) = a_j). \end{aligned}$$

Also note that

$$
\begin{aligned}
&\sum_{\bar{z}_k} \mathcal{Q}_{kk}(\bar{a}_k) f(\bar{z}_k, \bar{y}_{k-1}, \bar{a}_k) \prod_{j=1}^{k-1} I(d_j(\bar{y}_{j-1}, \bar{a}_{j-1}) = a_j) \\
&= \sum_{\bar{u}_{k-1}} \sum_{\bar{z}_k} \mathcal{Q}_{kk}(\bar{a}_k) f(\bar{z}_k | \bar{y}_{k-1}, \bar{a}_k, \bar{u}_{k-1}) f(\bar{y}_{k-1}, \bar{a}_k, \bar{u}_{k-1}) \prod_{j=1}^{k-1} I(d_j(\bar{y}_{j-1}, \bar{a}_{j-1}) = a_j) \\
&= \sum_{\bar{u}_{k-1}} \frac{\sum_{\bar{z}_k} \mathcal{Q}_{k-1,k-1}(\bar{a}_{k-1}) f(\bar{z}_{k-1} | \bar{y}_{k-1}, \bar{a}_{k-1}, \bar{u}_{k-1})}{f(a_k | \bar{a}_{k-1}, \bar{u}_{k-1}, \bar{y}_{k-1})} f(\bar{y}_{k-1}, \bar{a}_k, \bar{u}_{k-1}) \prod_{j=1}^{k-1} I(d_j(\bar{y}_{j-1}, \bar{a}_{j-1}) = a_j) \\
&= \sum_{\bar{u}_{k-1}} \sum_{\bar{z}_k} \mathcal{Q}_{k-1,k-1}(\bar{a}_{k-1}) f(\bar{z}_{k-1} | \bar{y}_{k-1}, \bar{a}_{k-1}, \bar{u}_{k-1}) f(\bar{y}_{k-1}, \bar{a}_{k-1}, \bar{u}_{k-1}) \prod_{j=1}^{k-1} I(d_j(\bar{y}_{j-1}, \bar{a}_{j-1}) = a_j) \\
&= \sum_{\bar{z}_{k-1}} \mathcal{Q}_{k-1,k-1}(\bar{a}_k) f(\bar{z}_{k-1}, \bar{y}_{k-1}, \bar{a}_{k-1}) \prod_{j=1}^{k-1} I(d_j(\bar{y}_{j-1}, \bar{a}_{j-1}) = a_j)
\end{aligned}
$$

By Theorem 3.2, we have

$$
\begin{aligned}
&\sum_{\bar{z}_{k-1}} \mathcal{Q}_{k-1,k-1}(\bar{a}_k) f(\bar{z}_{k-1}, \bar{y}_{k-1}, \bar{a}_{k-1}) \prod_{j=1}^{k-1} I(d_j(\bar{y}_{j-1}, \bar{a}_{j-1}) = a_j) \\
&= f(\bar{Y}_{k-1}(\bar{a}_{k-1}) = \bar{y}_{k-1}) \prod_{j=1}^{k-1} I(d_j(\bar{y}_{j-1}, \bar{a}_{j-1}) = a_j),
\end{aligned}
$$

and

$$\sum_{y_K,\dots,y_k}\sum_{\bar z_k}\sum_{\bar w_k} y_K \mathcal{Q}_{kk}(\bar a_k)\mathcal{H}_{Kk}(\bar a_K) f(\bar z_k,\bar w_k,\bar y_{k-1},\bar a_k)\prod_{j=1}^{K} I(d_j(\bar y_{j-1},\bar a_{j-1})=a_j)$$

$$= \sum_{y_K,\dots,y_k}\sum_{\bar z_k}\sum_{\bar w_k} y_K \mathcal{Q}_{kk}(\bar a_k)\mathcal{H}_{Kk}(\bar a_K) f(\bar z_k|\bar w_k,\bar y_{k-1},\bar a_k) f(\bar w_k,\bar y_{k-1},\bar a_k)\prod_{j=1}^{K} I(d_j(\bar y_{j-1},\bar a_{j-1})=a_j)$$

$$= \sum_{y_K,\dots,y_k} \frac{\sum_{\bar z_{k-1}} \mathcal{Q}_{k-1,k-1}(\bar a_{k-1}) f(\bar z_{k-1}|\bar w_k,\bar y_{k-1},\bar a_{k-1})}{f(a_k|\bar w_k,\bar y_{k-1},\bar a_{k-1})}$$

$$\times \sum_{\bar w_k} y_K \mathcal{H}_{Kk}(\bar a_K) f(\bar w_k,\bar y_{k-1},\bar a_k)\prod_{j=1}^{K} I(d_j(\bar y_{j-1},\bar a_{j-1})=a_j)$$

$$= \sum_{y_K,\dots,y_k}\sum_{\bar z_{k-1}} \mathcal{Q}_{k-1,k-1}(\bar a_{k-1}) f(\bar z_{k-1}|\bar w_k,\bar y_{k-1},\bar a_{k-1})$$

$$\times \sum_{\bar w_k} y_K \mathcal{H}_{Kk}(\bar a_K) f(\bar w_k,\bar y_{k-1},\bar a_{k-1})\prod_{j=1}^{K} I(d_j(\bar y_{j-1},\bar a_{j-1})=a_j)$$

$$= \sum_{y_K,\dots,y_k}\sum_{\bar z_{k-1}}\sum_{\bar w_k} y_K \mathcal{Q}_{k-1,k-1}(\bar a_{k-1})\mathcal{H}_{Kk}(\bar a_K) f(\bar z_{k-1},\bar w_k,\bar y_{k-1},\bar a_{k-1})\prod_{j=1}^{K} I(d_j(\bar y_{j-1},\bar a_{j-1})=a_j)$$

$$= \sum_{y_K,\dots,y_k} y_K f(\bar Y_K(\bar a_K)=\bar y_K)\prod_{j=1}^{K} I(d_j(\bar y_{j-1},\bar a_{j-1})=a_j),$$

we have:

$$\sum_{\bar y_{k-1}} \frac{\sum_{y_K,\dots,y_k}\sum_{\bar a_K}\sum_{\bar z_k}\sum_{\bar w_k} y_K \mathcal{Q}_{kk}(\bar a_k)\mathcal{H}_{Kk}(\bar a_K) f(\bar z_k,\bar w_k,\bar y_{k-1},\bar a_k)\prod_{j=1}^{K} I(d_j(\bar y_{j-1},\bar a_{j-1})=a_j)}{\sum_{\bar a_k}\sum_{\bar z_k} \mathcal{Q}_{kk}(\bar a_k) f(\bar z_k,\bar y_{k-1},\bar a_k)\prod_{j=1}^{k-1} I(d_j(\bar y_{j-1},\bar a_{j-1})=a_j)}$$

$$\times \sum_{\bar a_k}\sum_{\bar z_k} \mathcal{Q}_{kk}(\bar a_k) f(\bar z_k,\bar y_{k-1},\bar a_k)\prod_{j=1}^{k-1} I(d_j(\bar y_{j-1},\bar a_{j-1})=a_j)$$

$$= \sum_{\bar y_{k-1}} \frac{\sum_{y_K,\dots,y_k} y_K f(\bar Y_K(\bar d_K)=\bar y_K)}{f(\bar Y_{k-1}(\bar d_{k-1})=\bar y_{k-1})}\sum_{\bar a_k}\sum_{\bar z_k} \mathcal{Q}_{kk}(\bar a_k) f(\bar z_k,\bar y_{k-1},\bar a_k)\prod_{j=1}^{k-1} I(d_j(\bar y_{j-1},\bar a_{j-1})=a_j)$$

$$= \sum_{\bar y_{k-1}}\sum_{\bar a_k}\sum_{\bar z_k} \mathcal{Q}_{kk}(\bar a_k) f(\bar z_k,\bar y_{k-1},\bar a_k)\mathbb{E}[Y_K(\bar d_K)|\bar Y_{k-1}(\bar d_{k-1})=\bar y_{k-1}]\prod_{j=1}^{k-1} I(d_j(\bar y_{j-1},\bar a_{j-1})=a_j)$$

$$= \mathbb{E}\Big[I(\bar d_k(\bar Y_{k-1},\bar A_{k-1})=\bar A_k)\mathcal{Q}_{kk}(\bar A_k)\mathbb{E}[Y_K(\bar d_K)|\bar Y_{k-1}(\bar d_{k-1})]\Big].$$

The proof of Equation (52) follows the same logic, thus omitted.

By Lemma H.1, we can rewrite Equation (50) as:

$$
\begin{aligned}
&\hat{V}_{PMR}(\bar{d}_K) \\
&= \mathbb{P}_n \Bigg\{ I(\bar{d}_K(\bar{Y}_{K-1}, \bar{A}_{K-1}) = \bar{A}_K)\mathcal{Q}_{KK}(\bar{A}_K; \hat{\gamma}_K)Y_K \\
&\quad + \sum_{k=1}^{K} \Big[ I(\bar{d}_{k-1}(\bar{Y}_{k-2}, \bar{A}_{k-2}) = \bar{A}_{k-1})\mathcal{Q}_{k-1,k-1}(\bar{A}_{k-1}; \hat{\gamma}_{k-1}) \\
&\qquad - I(\bar{d}_k(\bar{Y}_{k-1}, \bar{A}_{k-1}) = \bar{A}_k)\mathcal{Q}_{kk}(\bar{A}_k; \hat{\gamma}_k) \Big] \mathbb{E}[Y_K(\bar{d}_K)|\bar{Y}_{k-1}(\bar{d}_{k-1})] \Bigg\},
\end{aligned} \tag{53}
$$

where the term $I(\bar{d}_k(\bar{Y}_{k-1}, \bar{A}_{k-1}) = \bar{A}_k)\mathcal{Q}_{kk}(\bar{A}_k; \hat{\gamma}_k)$ is defined as 1 when $k = 0$. Equation (53) corresponds to the doubly robust estimator proposed in Zhang et al. (2013) term by term if we regard the treatment confounding bridge functions $\mathcal{Q}_{kk}(\bar{A}_k)$ as the $\frac{1}{\prod_{j=1}^{k} P(A_k|\bar{X}_k)}$. It also follows the semiparametric efficiency structure suggested by Tsiatis (2006), Chapters 10 and 11, for the longitudinal monotone coarsening. The term $I(\bar{d}_k(\bar{Y}_{k-1}, \bar{A}_{k-1}) = \bar{A}_k)$ actually serves as the indicator for coarsening, according to which only information of individuals who follow the treatment regimes $\bar{d}_K$ until the $k^{th}$ stage can be exploited at the corresponding term.

# I Identify Optimal DTRs on $(\bar{Y}_{K-1}, \bar{A}_{K-1}, \bar{W}_K)$

In Qi et al. (2024), three types of individualized decision regimes with different decision spaces are considered: one depends on $X$, one depends on $(X, Z)$, and the other depends on $(X, W)$. However, under the longitudinal proximal causal inference framework, taking the 2-stage setting as an example, we note that there is only one possible extension of the decision space without adding additional assumptions. To be specific, by PIPW, we can identify the following joint density:

$$
\begin{aligned}
&f(Y_2(a_1, a_2) = y_2, Y_1(a_1) = y_1, W_2(a_1) = w_2 | w_1, y_0) \\
&= \sum_{\bar{z}_2} q_{22}(\bar{a}_2, \bar{y}_1, \bar{z}_2) f(\bar{y}_2, \bar{z}_2, \bar{a}_2, w_2 | w_1, y_0).
\end{aligned}
$$

*Proof.* To prove this identification result, we need the assumption that

$$W_2(a_1) \perp\!\!\!\perp A_1 | U_0, Y_0, W_1.$$

However, by Assumptions 1.1-5, as $W_2(a_1) = W_2(a_1, z_1)$, $W_2 \perp\!\!\!\perp Z_1 | U_0, Y_0, A_1$, and the other possible confounders $(Y_0, U_0)$ is adjusted, this assumption holds based on original assumptions.

Thus, we have:

$$
\begin{aligned}
&f(Y_2(a_1,a_2)=y_2, Y_1(a_1)=y_1, W_2(a_1)=w_2|w_1,y_0)\\
&=\sum_{u_0} f(Y_2(a_1,a_2)=y_2, Y_1(a_1)=y_1, W_2(a_1)=w_2|u_0,y_0,w_1,a_1)f(u_0|y_0,w_1)\\
&=\sum_{u_0} f(Y_2(a_1,a_2)=y_2|u_0,\bar{y}_1,\bar{w}_2,a_1)f(y_1,w_2|u_0,y_0,w_1,a_1)f(u_0|y_0,w_1)\\
&=\sum_{u_0} f(Y_2(a_1,a_2)=y_2|u_0,\bar{y}_1,\bar{w}_2,a_1)f(y_1,w_2|u_0,y_0,w_1,a_1)f(u_0|y_0,w_1)f(a_1|u_0,y_0,w_1)\\
&\qquad\times\sum_{z_1} q_{11}(a_1,y_0,z_1)f(z_1|a_1,u_0,y_0,w_1)\\
&=\sum_{u_0}\sum_{z_1} f(Y_2(a_1,a_2)=y_2, Y_1(a_1)=y_1, W_2(a_1)=w_2|u_0,y_0,w_1,a_1)f(u_0|y_0,w_1)f(a_1|u_0,y_0,w_1)\\
&\qquad\times q_{11}(a_1,y_0,z_1)f(z_1|a_1,u_0,y_0,w_1)\\
&=\sum_{u_0}\sum_{z_1} f(Y_2(a_1,a_2)=y_2, Y_1(a_1)=y_1, z_1|u_0,y_0,\bar{w}_2,a_1)f(w_2|u_0,y_0,w_1,a_1)f(u_0|y_0,w_1)\\
&\qquad\times f(a_1|u_0,y_0,w_1)q_{11}(a_1,y_0,z_1)\\
&=\sum_{u_0}\sum_{u_1} f(Y_2(a_1,a_2)=y_2|\bar{u}_1,\bar{y}_1,\bar{w}_2,a_1)f(u_1|u_0,\bar{y}_1,\bar{w}_2,a_1)f(y_1|u_0,y_0,\bar{w}_2,a_1)\\
&\qquad\times f(w_2|u_0,y_0,w_1,a_1)f(u_0|y_0,w_1)f(a_1|u_0,y_0,w_1)f(a_2|a_1,\bar{u}_1,\bar{y}_1)\frac{\sum_{z_1} q_{11}(a_1,y_0,z_1)f(z_1|\bar{u}_1,\bar{y}_1,\bar{w}_2,a_1)}{f(a_2|a_1,\bar{u}_1,\bar{y}_1)}\\
&=\sum_{u_0}\sum_{u_1} f(Y_2(a_1,a_2)=y_2|\bar{u}_1,\bar{y}_1,\bar{w}_2,a_1)f(u_1|u_0,\bar{y}_1,\bar{w}_2,a_1)f(y_1|u_0,y_0,\bar{w}_2,a_1)\\
&\qquad\times f(w_2|u_0,y_0,w_1,a_1)f(u_0|y_0,w_1)f(a_1|u_0,y_0,w_1)f(a_2|a_1,\bar{u}_1,\bar{y}_1)\frac{\sum_{z_1} q_{11}(a_1,y_0,z_1)f(z_1|\bar{u}_1,\bar{y}_1,a_1)}{f(a_2|a_1,\bar{u}_1,\bar{y}_1)}\\
&=\sum_{u_0}\sum_{u_1} f(Y_2(a_1,a_2)=y_2|\bar{u}_1,\bar{y}_1,\bar{w}_2,a_1)f(u_1|u_0,\bar{y}_1,\bar{w}_2,a_1)f(y_1|u_0,y_0,\bar{w}_2,a_1)\\
&\qquad\times f(w_2|u_0,y_0,w_1,a_1)f(u_0|y_0,w_1)f(a_1|u_0,y_0,w_1)f(a_2|a_1,\bar{u}_1,\bar{y}_1)\sum_{\bar{z}_2} q_{22}(\bar{a}_2,\bar{y}_1,\bar{z}_2)f(\bar{z}_2|\bar{a}_2,\bar{u}_1,\bar{y}_1)\\
&=\sum_{u_0}\sum_{u_1} f(Y_2=y_2|\bar{u}_1,\bar{y}_1,\bar{w}_2,\bar{a}_2)f(u_1|u_0,\bar{y}_1,\bar{w}_2,a_1)f(y_1|u_0,y_0,\bar{w}_2,a_1)f(w_2|u_0,y_0,w_1,a_1)\\
&\qquad\times f(u_0|y_0,w_1)f(a_1|u_0,y_0,w_1)f(a_2|a_1,\bar{u}_1,\bar{y}_1)\sum_{\bar{z}_2} q_{22}(\bar{a}_2,\bar{y}_1,\bar{z}_2)f(\bar{z}_2|\bar{a}_2,\bar{u}_1,\bar{y}_1)\\
&=\sum_{u_0}\sum_{u_1}\sum_{\bar{z}_2} f(Y_2=y_2|\bar{u}_1,\bar{y}_1,\bar{w}_2,\bar{a}_2,\bar{z}_2)f(\bar{z}_2|\bar{a}_2,\bar{u}_1,\bar{y}_1,\bar{w}_2)f(a_2|a_1,\bar{u}_1,\bar{y}_1,\bar{w}_2)\\
&\qquad\times f(y_1|u_0,y_0,\bar{w}_2,a_1)f(w_2|u_0,y_0,w_1,a_1)f(u_1|u_0,\bar{y}_1,\bar{w}_2,a_1)q_{22}(\bar{a}_2,\bar{y}_1,\bar{z}_2)f(u_0|y_0,w_1)f(a_1|u_0,y_0,w_1)\\
&=\sum_{\bar{z}_2} q_{22}(\bar{a}_2,\bar{y}_1,\bar{z}_2)f(\bar{y}_2,\bar{z}_2,\bar{a}_2,w_2|w_1,y_0),
\end{aligned}
$$

which ends the proof.

By this result, we can identify the optimal DTRs $\bar{d}_2^*$ based on $(\bar{Y}_1, \bar{W}_2, A_1)$ by:

$$d_2^*(\bar{y}_1, \bar{w}_2, a_1) = \arg\max_{a_2} \mathbb{E}[Y_2(a_1, a_2)|\bar{Y}_1(a_1) = \bar{y}_1, \bar{W}_2(a_1) = \bar{w}_2];$$
$$d_1^*(y_0, w_1) = \arg\max_{a_2} \mathbb{E}[Y_2(a_1, d_2^*(\bar{Y}_1(a_1), \bar{W}_2(a_1), a_1))|Y_0 = y_0, W_1 = w_1],$$

or

$$\bar{d}_2^* = \arg\max_{\bar{d}_2} E[Y_2\Big(d_1(W_1, Y_0), d_2(\bar{W}_2(d_1), \bar{Y}_2(d_1), d_1)\Big)]$$

For regimes with decision space of $(\bar{Y}_{K-1}, \bar{A}_{K-1}, \bar{Z}_K)$, the symmetric additional assumption that

$$Z_2(a_1) \perp\!\!\!\perp A_1|U_0, Y_0, Z_1$$

is needed, and the outcome confounding bridge functions need to be modified. According to the DAG depicted in Figure 1, this additional assumption can hold in some special situations. However, the modification of outcome bridge functions needs to absorb the densities of $Z$ variables into the iteration Equation (2), which will lead to structural contradiction. Thus, the identification of optimal DTRs with decision space of $(\bar{Y}_{K-1}, \bar{A}_{K-1}, \bar{Z}_K)$ remains an open question.

# J More about the Outcome Confounding Bridge Functions

## J.1 The Relationships across the Outcome Confounding Bridge Functions in Settings with Different Stages

As we have pointed out in Section 3.1, a significant advantage of the identification strategy via the treatment confounding bridge functions is that if there are auxiliary data for the future $(K+1)^{th}$ stage, the treatment confounding bridge functions for the $k^{th}$ stage, $k = 1, 2, \ldots, K$ can be reused for the $(K+1)$-stage identification, which implies the potential for incremental learning Van de Ven et al. (2022) and continual lifelong learning Parisi et al. (2019). In fact, as suggested by Zhang & Tchetgen Tchetgen (2024), there are also relationships between $\mathcal{H}_{Kl}(\bar{a}_K)$ and $\mathcal{H}_{K-1,l}(\bar{a}_{K-1})$, for $l = 1, 2, \ldots, K$. Specifically, suppose that $\mathcal{H}_{Kl}(\bar{a}_K)$, $l = 1, 2, \ldots, K$ that solve Equations (1) and (2) exist, then under Assumption 3′.1 for $l = 1, 2, \ldots, K$, $\tilde{\mathcal{H}}_{K-1,l}(\bar{a}_K) = \sum_{y_K} \mathcal{H}_{Kl}(\bar{a}_K)$, $l = 1, 2, \ldots, K$ can also solve the corresponding latent equation for $\mathcal{H}_{K-1,l}(\bar{a}_{K-1})$, which implies that $\tilde{\mathcal{H}}_{K-1,l}(\bar{a}_K)$ can also be viewed as $\mathcal{H}_{K-1,l}(\bar{a}_{K-1})$. This relationship is formalized in the following proposition:

**Proposition J.1.** *Suppose that $\mathcal{H}_{Kl}(\bar{a}_K)$, $l = 1, 2, \ldots, K$ that solve Equations (1) and (2) exist. Then, under Assumption 3′.1 for $l = 1, 2, \ldots, K$, $\tilde{\mathcal{H}}_{K-1,l}(\bar{a}_K) = \sum_{y_K} \mathcal{H}_{Kl}(\bar{a}_K)$, $l = 1, 2, \ldots, K-1$ can also solve the following (K-1) equations:*

*For $l = K-1$,*

$$f(Y_{K-1} = y_{K-1}|\bar{y}_{K-2}, \bar{u}_{K-2}, \bar{a}_{K-1}) = \sum_{\bar{w}_{K-1}} \mathcal{H}_{K-1,K-1}(\bar{a}_{K-1}) f(\bar{w}_{K-1}|\bar{y}_{K-2}, \bar{u}_{K-2}, \bar{a}_{K-1}), \tag{54}$$

*and for $l = K-2, \ldots, 1$,*

$$\sum_{\bar{w}_{l+1}} \mathcal{H}_{K-1,l+1}(\bar{a}_{K-1}) f(\bar{w}_{l+1}, y_l|\bar{y}_{l-1}, \bar{u}_{l-1}, \bar{a}_l) = \sum_{\bar{w}_l} \mathcal{H}_{K-1,l}(\bar{a}_{K-1}) f(\bar{w}_l|\bar{y}_{l-1}, \bar{u}_{l-1}, \bar{a}_l). \tag{55}$$

*Proof.* Note that under Assumption 3′.1 for $l = K, K-1$, $\mathcal{H}_{KK}(\bar{a}_K)$ and $\mathcal{H}_{K,K-1}(\bar{a}_K)$ satisfies the following equations:

$$f(Y_K = y_K|\bar{y}_{K-1}, \bar{u}_{K-1}, \bar{a}_K) = \sum_{\bar{w}_K} \mathcal{H}_{KK}(\bar{a}_K) f(\bar{w}_K|\bar{y}_{K-1}, \bar{u}_{K-1}, \bar{a}_K) \tag{56}$$

$$\sum_{\bar{w}_K} \mathcal{H}_{KK}(\bar{a}_K) f(\bar{w}_K, y_{K-1}|\bar{y}_{K-2}, \bar{u}_{K-2}, \bar{a}_{K-1}) = \sum_{\bar{w}_{K-1}} \mathcal{H}_{K,K-1}(\bar{a}_K) f(\bar{w}_{K-1}|\bar{y}_{K-2}, \bar{u}_{K-2}, \bar{a}_{K-1}). \tag{57}$$

Then, we note that

$$\begin{aligned}
&\sum_{\bar{w}_K} \mathcal{H}_{KK}(\bar{a}_K) f(\bar{w}_K, y_{K-1}|\bar{y}_{K-2}, \bar{u}_{K-2}, \bar{a}_{K-1}) \\
&= \sum_{u_{K-1}} \sum_{\bar{w}_K} \mathcal{H}_{KK}(\bar{a}_K) f(\bar{w}_K|\bar{y}_{K-1}, \bar{u}_{K-1}, \bar{a}_{K-1}) f(u_{K-1}|\bar{y}_{K-1}, \bar{u}_{K-2}, \bar{a}_{K-1}) f(y_{K-1}|\bar{y}_{K-2}, \bar{u}_{K-2}, \bar{a}_{K-1}) \\
&= \sum_{u_{K-1}} \sum_{\bar{w}_K} \mathcal{H}_{KK}(\bar{a}_K) f(\bar{w}_K|\bar{y}_{K-1}, \bar{u}_{K-1}, \bar{a}_K) f(u_{K-1}|\bar{y}_{K-1}, \bar{u}_{K-2}, \bar{a}_{K-1}) f(y_{K-1}|\bar{y}_{K-2}, \bar{u}_{K-2}, \bar{a}_{K-1}) \\
&= \sum_{u_{K-1}} f(y_K|\bar{y}_{K-1}, \bar{u}_{K-1}, \bar{a}_K) f(u_{K-1}|\bar{y}_{K-1}, \bar{u}_{K-2}, \bar{a}_{K-1}) f(y_{K-1}|\bar{y}_{K-2}, \bar{u}_{K-2}, \bar{a}_{K-1}).
\end{aligned}$$

Thus,

$$\sum_{y_K}\sum_{\bar{w}_K}\mathcal{H}_{KK}(\bar{a}_K)f(\bar{w}_K, y_{K-1}|\bar{y}_{K-2}, \bar{u}_{K-2}, \bar{a}_{K-1})$$
$$\sum_{y_K}\sum_{u_{K-1}} f(y_K|\bar{y}_{K-1}, \bar{u}_{K-1}, \bar{a}_K)f(u_{K-1}|\bar{y}_{K-1}, \bar{u}_{K-2}, \bar{a}_{K-1})f(y_{K-1}|\bar{y}_{K-2}, \bar{u}_{K-2}, \bar{a}_{K-1})$$
$$= f(y_{K-1}|\bar{y}_{K-2}, \bar{u}_{K-2}, \bar{a}_{K-1})$$
$$= \sum_{\bar{w}_{K-1}}\left\{\sum_{y_K}\mathcal{H}_{K,K-1}(\bar{a}_K)\right\}f(\bar{w}_{K-1}|\bar{y}_{K-2}, \bar{u}_{K-2}, \bar{a}_{K-1})$$
$$= \sum_{\bar{w}_{K-1}}\tilde{\mathcal{H}}_{K-1,K-1}(\bar{a}_K)f(\bar{w}_{K-1}|\bar{y}_{K-2}, \bar{u}_{K-2}, \bar{a}_{K-1}).$$

This proves that $\tilde{\mathcal{H}}_{K-1,K-1}(\bar{a}_K)$ can solve Equation (54). By Equation (55), we note that

$$\sum_{y_K}\sum_{\bar{w}_{K-1}}\mathcal{H}_{K,K-1}(\bar{a}_K)f(\bar{w}_{K-1}, y_{K-2}|\bar{y}_{K-3}, \bar{u}_{K-3}, \bar{a}_{K-2})$$
$$= \sum_{\bar{w}_{K-1}}\tilde{\mathcal{H}}_{K-1,K-1}(\bar{a}_K)f(\bar{w}_{K-1}, y_{K-2}|\bar{y}_{K-3}, \bar{u}_{K-3}, \bar{a}_{K-2})$$
$$= \sum_{\bar{w}_{K-2}}\sum_{y_K}\mathcal{H}_{K,K-2}(\bar{a}_K)f(\bar{w}_{K-2}|\bar{y}_{K-3}, \bar{u}_{K-3}, \bar{a}_{K-2})$$
$$= \sum_{\bar{w}_{K-2}}\tilde{\mathcal{H}}_{K-1,K-2}(\bar{a}_K)f(\bar{w}_{K-2}|\bar{y}_{K-3}, \bar{u}_{K-3}, \bar{a}_{K-2}),$$

which implies that $\tilde{\mathcal{H}}_{K-1,K-2}(\bar{a}_K)$ can also solve Equation (55) for $l = K-1$. Repeatedly use Equation (55) for $l = K-2, \ldots, 1$, we get the conclusion that $\tilde{\mathcal{H}}_{K-1,l}(\bar{a}_K)$, $l = K-1, K-2, \ldots, 1$ solve Equations (54) and (55).

By repeatedly use Proposition J.1, we further note the conclusion that $\sum_{y_K,\ldots,y_{K-k}}\mathcal{H}_{Kl}(\bar{a}_K)$ for $l = K-k, \ldots, 1$ can also be used as outcome confounding bridge functions $\mathcal{H}_{K-k,l}(\bar{a}_{K-k})$, $l = K-k, \ldots, 1$ for the $(K-k)$-stage setting under Assumption $3'.1$ for $l = 1, 2, \ldots, K$, which demystifies the relationships between the outcome confounding bridge functions across settings with different stages.

## J.2 A Slight Modification of the Outcome Confounding Bridge Functions

For the estimation of optimal DTRs aimed at maximizing the mean of the outcome, there is a trick to modify the outcome confounding bridge functions slightly while the identification still holds. Although this modification may lose generality in transferring to other criteria, it can relieve the pressure of estimating bridge functions to some extent. To be specific, we note that the following outcome bridge functions $\mathcal{H}^m_{Kl}(\bar{a}_K) = h^m_{Kl}(\bar{y}_{K-1}, \bar{w}_l, \bar{a}_K)$, $l = 1, 2, \ldots, K$ can also be used to identify and estimate the optimal DTRs:
For $l = K$,

$$\mathbb{E}[Y_K|\bar{Y}_{K-1}, \bar{Z}_K, \bar{A}_K = \bar{a}_K] = \mathbb{E}[\mathcal{H}^m_{KK}(\bar{a}_K)|\bar{Y}_{K-1}, \bar{Z}_K, \bar{A}_K = \bar{a}_K] \tag{58}$$

and for $l = K-1, K-2, \ldots, 1$:

$$\sum_{y_K}\sum_{\bar{w}_{l+1}} y_K \mathcal{H}^m_{K,l+1}(\bar{a}_K) f(\bar{w}_{l+1}, y_l|\bar{y}_{l-1}, \bar{z}_l, \bar{a}_l) = \sum_{y_K}\sum_{\bar{w}_l} y_K \mathcal{H}^m_{Kl}(\bar{a}_K) f(\bar{w}_l|\bar{y}_{l-1}, \bar{z}_l, \bar{a}_l),$$

which, by Assumption 3′.1 for $l = 1, 2, \ldots, K$, can also solve:

$$\mathbb{E}[Y_K|\bar{Y}_{K-1}, \bar{U}_{K-1}, \bar{A}_K = \bar{a}_K] = \mathbb{E}[\mathcal{H}^m_{KK}(\bar{a}_K)|\bar{Y}_{K-1}, \bar{U}_{K-1}, \bar{A}_K = \bar{a}_K],$$

and for $l = K-1, \ldots, 1$,

$$\sum_{y_K}\sum_{\bar{w}_{l+1}} y_K \mathcal{H}^m_{K,l+1}(\bar{a}_K) f(\bar{w}_{l+1}, y_l|\bar{y}_{l-1}, \bar{u}_{l-1}, \bar{a}_l) = \sum_{y_K}\sum_{\bar{w}_l} y_K \mathcal{H}^m_{Kl}(\bar{a}_K) f(\bar{w}_l|\bar{y}_{l-1}, \bar{u}_{l-1}, \bar{a}_l).$$

We only prove the identification result implied by $\mathcal{H}_{Kl}^m(\bar{a}_K)$, $l = 1, 2, \ldots, K$ under the special setting with $K = 2$, and the general setting can be done similarly. We note that

$$
\begin{aligned}
&\mathbb{E}[Y_2(d_1, d_2)] \\
&= \sum_{\bar{a}_2} \sum_{\bar{y}_2} y_2 f(Y_2(\bar{a}_2) = y_2, Y_1(a_1) = y_1, Y_0 = y_0) I(d_1(y_0) = a_1) I(d_2(\bar{y}_1, a_1) = a_2) \\
&= \sum_{\bar{a}_2} \sum_{\bar{y}_2} \sum_{\bar{u}_1} y_2 f(y_2|\bar{u}_1, \bar{y}_1, \bar{a}_2) f(u_1|\bar{y}_1, u_0, a_1) f(y_1|u_0, y_0, a_1) f(u_0, y_0) I(d_1(y_0) = a_1) I(d_2(\bar{y}_1, a_1) = a_2) \\
&= \sum_{\bar{a}_2} \sum_{\bar{y}_1} \sum_{\bar{u}_1} \sum_{\bar{w}_2} h_{22}(\bar{y}_1, \bar{w}_2, \bar{a}_2) f(\bar{w}_2|\bar{y}_1, \bar{u}_1, \bar{a}_2) f(u_1|\bar{y}_1, u_0, a_1) f(y_1|u_0, y_0, a_1) f(u_0, y_0) \\
&\qquad \times I(d_1(y_0) = a_1) I(d_2(\bar{y}_1, a_1) = a_2) \\
&= \sum_{\bar{a}_2} \sum_{\bar{y}_1} \sum_{u_0} \sum_{\bar{w}_2} h_{22}(\bar{y}_1, \bar{w}_2, \bar{a}_2) f(\bar{w}_2|\bar{y}_1, u_0, a_1) f(y_1|u_0, y_0, a_1) f(u_0, y_0) I(d_1(y_0) = a_1) I(d_2(\bar{y}_1, a_1) = a_2) \\
&= \sum_{\bar{a}_2} \sum_{\bar{y}_1} \sum_{u_0} \sum_{w_1} h_{21}(\bar{y}_1, w_1, \bar{a}_2) f(w_1|u_0, y_0, a_1) f(u_0, y_0) I(d_1(y_0) = a_1) I(d_2(\bar{y}_1, a_1) = a_2) \\
&= \sum_{\bar{a}_2} \sum_{\bar{y}_1} \sum_{w_1} h_{21}(\bar{y}_1, w_1, \bar{a}_2) f(w_1, y_0) I(d_1(y_0) = a_1) I(d_2(\bar{y}_1, a_1) = a_2).
\end{aligned}
$$

Thus, the optimal DTRs can be obtained through:

$$
\bar{d}_2^* = \arg\max_{\bar{d}_2} \sum_{\bar{a}_2} \sum_{\bar{y}_1} \sum_{w_1} h_{21}(\bar{y}_1, w_1, \bar{a}_2) f(w_1, y_0) I(d_1(y_0) = a_1) I(d_2(\bar{y}_1, a_1) = a_2).
$$

By this modification, we simplify the estimation procedure of $\mathcal{H}_{KK}(\bar{a}_K)$ and still allow the identification of optimal DTRs by value maximization. Also note that when $k = K$, the value function can be identified via PHA as:

$$
\begin{aligned}
V(\bar{d}_K) &= \sum_{\bar{a}_K} \sum_{\bar{w}_K} \sum_{z_{K-1}} \sum_{\bar{y}_K} y_K \mathcal{H}_{KK}(\bar{a}_K) \mathcal{Q}_{K-1,K-1}(\bar{a}_{K-1}) \\
&\qquad \times \prod_{j=1}^{K} I(d_j(\bar{y}_{j-1}, \bar{a}_{j-1}) = a_j) f(\bar{w}_K, z_{K-1}, \bar{y}_{K-1}, \bar{a}_{K-1}) \\
&= \sum_{\bar{a}_K} \sum_{\bar{w}_K} \sum_{z_{K-1}} \sum_{\bar{y}_{K-1}} \mathcal{H}_{KK}^m(\bar{a}_K) \mathcal{Q}_{K-1,K-1}(\bar{a}_{K-1}) \\
&\qquad \times \prod_{j=1}^{K} I(d_j(\bar{y}_{j-1}, \bar{a}_{j-1}) = a_j) f(\bar{w}_K, z_{K-1}, \bar{y}_{K-1}, \bar{a}_{K-1})
\end{aligned} \tag{59}
$$

Specifically, for a two-stage setting, we can identify optimal DTRs without solving the nested integral equations or problems with continuous moment restrictions by this method, as $\mathcal{H}_{22}(\bar{a}_2)$ can be estimated from Equation (58) by many existing methods (Mastouri et al. 2021, Kallus et al. 2022, Ghassami et al. 2022, Kompa et al. 2022).

However, the summation over $y_K$ destroys the generality of the identification results. Neither can these modified outcome bridge functions be applied to identifying the optimal DTRs via Q-learning, nor can they be directly used for estimating the optimal DTRs aimed at maximizing other distributional functionals of $Y_K(\bar{d}_K)$.

## J.3 The Relationships between Outcome Confounding Bridge Functions in Ying et al. (2023) and Zhang & Tchetgen Tchetgen (2024)

We have mentioned that the treatment confounding bridge functions used to identify the optimal DTRs in this article are essentially the same as those defined in Ying et al. (2023). It is easy to understand because the treatment confounding bridge functions only relate to the reciprocal of propensity scores, but not the joint distribution of potential outcomes. For outcome confounding bridge functions, although the recursive equations (1) and (2) are different from those defined in Ying et al. (2023), we still note some relationships between them, which may help better understand the working mechanism behind the outcome confounding bridge functions. We will use the 2-stage setting to support the intuition further. First of all, the recursive equations in Ying et al. (2023) for the outcome confounding bridges $(h^o_{22}, h^o_{21})$ are:

$$\sum_{y_2} y_2 f(y_2|\bar{a}_2, \bar{z}_2, \bar{y}_1) = \sum_{\bar{w}_2} h^o_{22}(\bar{y}_1, \bar{w}_2, \bar{a}_2) f(\bar{w}_2|\bar{a}_2, \bar{z}_2, \bar{y}_1)$$

$$\sum_{y_1}\sum_{\bar{w}_2} h^o_{22}(\bar{y}_1, \bar{w}_2, \bar{a}_2) f(\bar{w}_2, y_1|a_1, z_1, y_0) = \sum_{w_1} h^o_{21}(y_0, w_1, \bar{a}_2) f(w_1|a_1, z_1, y_0).$$

By the outcome bridge functions defined by Equations (1), (2), we can restate these equations as:

$$\sum_{y_2} y_2 f(y_2|\bar{a}_2, \bar{z}_2, \bar{y}_1) = \sum_{\bar{w}_2}\sum_{y_2} y_2 h_{22}(\bar{y}_2, \bar{w}_2, \bar{a}_2) f(\bar{w}_2|\bar{a}_2, \bar{z}_2, \bar{y}_1)$$

$$\sum_{y_1}\sum_{\bar{w}_2}\sum_{y_2} y_2 h_{22}(\bar{y}_2, \bar{w}_2, \bar{a}_2) f(\bar{w}_2, y_1|a_1, z_1, y_0) = \sum_{w_1}\sum_{y_1,y_2} y_2 h_{21}(\bar{y}_2, w_1, \bar{a}_2) f(w_1|a_1, z_1, y_0).$$

Thus, we conclude that $h^o_{22}(\bar{y}_1, \bar{w}_2, \bar{a}_2)$ can be viewed as $\sum_{y_2} y_2 h_{22}(\bar{y}_2, \bar{w}_2, \bar{a}_2)$, and $h^o_{21}(y_0, w_1, \bar{a}_2)$ can be viewed as $\sum_{y_1,y_2} y_2 h_{21}(\bar{y}_2, w_1, \bar{a}_2)$, which implies that we can derive the identification result of Ying et al. (2023) from the newly designed outcome confounding bridge functions in Zhang & Tchetgen Tchetgen (2024). This observation verifies the generality of $\mathcal{H}_{Kl}(\bar{a}_K)$, for $l = 1, 2, \ldots, K$.

# K Additional Simulation Details

In our numerical simulation, we consider the 2-stage categorical setting for simplicity. We will present the closed-form expression of the treatment confounding bridge functions and use them to construct the plug-in estimators of the value function and Q function by POR, PHA, PIPW, PMR, along with the closed-form of outcome confounding bridge functions $h_{22}, h_{21}$ presented in Appendix D of Zhang & Tchetgen Tchetgen (2024) in Section K.1. We will also describe the Oracle and the SRA estimators. To mitigate the extra error caused by overfitting, we introduce the cross-fitting algorithm into the estimation process, which will be elaborated in Section K.2. The estimation procedure for the optimal linear decision rule by value maximization will be stated in Section K.3.

## K.1 Closed Forms of Confounding Bridge Functions and Corresponding Estimators under Categorical Setting

In this section, we consider in detail the canonical case where variables $U_{t-1}, W_t$, and $Z_t$ are categorical with $|U_{t-1}| = |W_t| = |Z_t| = k$, so that all proxies and unmeasured confounders have the same cardinality. The more general setting is $min(|W_t|, |Z_t|) \geq |U_{t-1}|$, which will be discussed in the supplementary material. We proceed with the following notation: we write $P(W|u) = \{pr(w_1|u), \cdots, pr(w_k|u)\}^T$, $P(w|U) = \{pr(w|u_1), \cdots, pr(w|u_k)\}$ and $P(W|U) = \{P(W|u_1), \cdots, P(W|u_k)\}$ for a column vector, a row vector, and a matrix that consist of conditional probabilities $pr(w|u)$, respectively. For all other variables, vectors and matrices consisting of conditional probabilities are analogously defined. We also define the element-wise product $\odot$ as follows: for $\mathbf{T} = \mathbf{v} \odot \mathbf{M}$, the resulting $n \times m$ matrix $\mathbf{T}$ is obtained by broadcasting the row vector $\mathbf{v}$ across each row of $\mathbf{M}$, such that the elements of $\mathbf{T}$ are given by $T_{ij}$: $T_{ij} = M_{ij} \cdot v_j$, for $i = 1, 2, \cdots, n$ and $j = 1, 2, \cdots, m$. This operator is commutative, meaning that $\mathbf{M} \odot \mathbf{v} = \mathbf{T}$ as well.

As an example, we consider the closed form of treatment bridge functions under the 2-stage setting, which we use in the simulation. In the categorical case, the completeness Assumptions 3′.1, 3′.2, 1.4, and 2.4 reduce to the following rank conditions:

*Assumption* 3. For all $(\bar{y}_1, \bar{a}_2)$, the matrices $P(\bar{Z}_2|\bar{U}_1, \bar{y}_1, \bar{a}_2)$ and $P(\bar{U}_1|\bar{W}_2, y_1, a_2)$ are invertible. In addition, for all $(y_0, a_1)$, $P(Z_1|U_0, y_0, a_1)$ and $P(U_1|W_1, y_0, a_1)$ are invertible.

*Assumption* 4. For all $(y_1, a_2)$, the matrices $P(\bar{Z}_2|\bar{U}_1, \bar{y}_1, \bar{a}_2)$ and $P(\bar{U}_1|\bar{W}_2, y_1, a_2)$ are of rank $|\bar{U}_1|$. In addition, for all $(y_0, a_1)$, $P(Z_1|U_0, y_0, a_1)$ and $P(U_1|W_1, y_0, a_1)$ are of rank $|U_0|$.

The invertibility Assumption 3 requires that the corresponding matrices are of full rank, which ensures both the existence and uniqueness of the treatment confounding bridge functions. We can now state a result about identification of $f(Y_2(a_1, a_2) = y_2, Y_1(a_1) = y_1|y_0)$ and $f(Y_1(a_1) = y_1|y_0)$ in the categorical setting by treatment confounding bridge functions.

**Proposition K.1.** *Let the cardinality of each of the $U$ , $Z$, and $W$ variables equal $k$ for $2 \leq k < \infty$. Suppose that Assumptions 1.3 and 3 hold, then:*

$$
\begin{aligned}
&f(Y_2(a_1, a_2) = y_2, Y_1(a_1) = y_1|y_0) \\
&= (P(a_1|W_1, y_0)^{-1} P(Z_1|W_1, a_1, y_0)^{-1} P(Z_1|a_1, \bar{W}_2, \bar{y}_1) \odot P(a_2|\bar{W}_2, a_1, \bar{y}_1)^{-1}) \\
&\quad P(\bar{Z}_2|\bar{W}_2, \bar{a}_2, \bar{y}_1)^{-1} P(\bar{y}_2, \bar{a}_2, \bar{Z}_2|y_0) \\
&f(Y_1(a_1) = y_1|y_0) \\
&= P(a_1|W_1, y_0)^{-1} P(Z_1|W_1, a_1, y_0)^{-1} P(\bar{y}_1, a_1, Z_1|y_0)
\end{aligned}
$$

*Proof.* Under Assumption 1.3, we know that $\{Z_1, A_1\} \perp\!\!\!\perp W_1|U_0, Y_0$, so

$$
\begin{aligned}
P(Z_1|W_1, a_1, y_0) &= P(Z_1|U_0, a_1, y_0) P(U_0|W_1, a_1, y_0) \\
P(a_1|W_1, y_0)^{-1} &= P(a_1|U_0, y_0)^{-1} P(U_0|W_1, a_1, y_0).
\end{aligned}
$$

We first prove a more general result under Assumption 4. By this weaker assumption, we have $min(|Z_t|, |W_t|) \geq |U_{t-1}|$, so $P(Z_1|U_0, a_1, y_0)$ and $P(U_0|W_1, a_1, y_0)$ have left and right inverses,

respectively. Moreover, the left inverse equals the Moore-Penrose pseudoinverse. We denote the Moore-Penrose pseudoinverse of a matrix $A$ by $A^{-}$. We obtain

$$P(U_0|W_1, a_1, y_0) = P(Z_1|U_0, a_1, y_0)^{-}P(Z_1|W_1, a_1, y_0)$$
$$P(a_1|W_1, y_0)^{-1} = P(a_1|U_0, y_0)^{-1}P(Z_1|U_0, a_1, y_0)^{-}P(Z_1|W_1, a_1, y_0)$$

Thus, there exists an $q_{11}$ (which may not be unique) that solves

$$P(a_1|W_1, y_0)^{-1} = q_{11}P(Z_1|W_1, a_1, y_0), \tag{60}$$

where we regard the vector $q_{11}$ (of dimension $1 \times |Z_1|$), as holding the values of the function $q_{11}(y_0, a_1, z_1)$ for the different values $z_1$ cam take. Since $P(U_0|W_1, a_1, y_0)$ has a right inverse, we obtain

$$P(Z_1|W_1, a_1, y_0)P(U_0|W_1, a_1, y_0)^{-} = P(Z_1|U_0, a_1, y_0)$$
$$P(a_1|W_1, y_0)^{-1}P(U_0|W_1, a_1, y_0)^{-} = P(a_1|U_0, y_0)^{-1}.$$

Thus,

$$\begin{aligned} P(a_1|U_0, y_0)^{-1} &= P(a_1|W_1, y_0)^{-1}P(U_0|W_1, a_1, y_0)^{-} \\ &= q_{11}P(Z_1|W_1, a_1, y_0)P(U_0|W_1, a_1, y_0)^{-} \\ &= q_{11}P(Z_1|U_0, a_1, y_0). \end{aligned}$$

Similarly, for any fixed $(\bar{y}_1, \bar{a}_2)$, we may also view $P(\bar{Z}_2|\bar{a}_2, \bar{U}_1, \bar{y}_1)$ as a $|Z_1| \times |Z_2|$by $|U_0| \times |U_1|$ matrix. Using Assumption 1.3 and through simple algebra, we know that

$$P(\bar{Z}_2|\bar{a}_2, \bar{W}_2, \bar{y}_1) = P(\bar{Z}_2|\bar{a}_2, \bar{U}_1, \bar{y}_1)P(\bar{U}_1|\bar{a}_2, \bar{W}_2, \bar{y}_1)$$
$$P(Z_1|a_1, \bar{W}_2, \bar{y}_1) \odot P(a_2|\bar{W}_2, a_1, \bar{y}_1)^{-1} = (P(Z_1|a_1, \bar{U}_1, \bar{y}_1) \odot P(a_2|\bar{U}_1, a_1, \bar{y}_1)^{-1})P(\bar{U}_1|\bar{a}_2, \bar{W}_2, \bar{y}_1)$$

Since $P(\bar{Z}_2|\bar{a}_2, \bar{U}_1, \bar{y}_1)$ has rank $|U_0| \times |U_1|$ and thus has a left inverse, we obtain

$$\begin{aligned} &P(Z_1|a_1, \bar{W}_2, \bar{y}_1) \odot P(a_2|\bar{W}_2, a_1, \bar{y}_1)^{-1} \\ &= (P(Z_1|a_1, \bar{U}_1, \bar{y}_1) \odot P(a_2|\bar{U}_1, a_1, \bar{y}_1)^{-1})P(\bar{Z}_2|\bar{a}_2, \bar{U}_1, \bar{y}_1)^{-}P(\bar{Z}_2|\bar{a}_2, \bar{W}_2, \bar{y}_1) \end{aligned}$$

which also implies

$$
\begin{aligned}
&q_{11}P(Z_1|a_1,\bar{W}_2,\bar{y}_1)\odot P(a_2|\bar{W}_2,a_1,\bar{y}_1)^{-1}\\
&=q_{11}(P(Z_1|a_1,\bar{U}_1,\bar{y}_1)\odot P(a_2|\bar{U}_1,a_1,\bar{y}_1)^{-1})P(\bar{Z}_2|\bar{a}_2,\bar{U}_1,\bar{y}_1)^{-}P(\bar{Z}_2|\bar{a}_2,\bar{W}_2,\bar{y}_1)
\end{aligned}
$$

for any $q_{11}$ vector, so there exists a $q_{22}$ (which may not be unique) vector that solves

$$
q_{11}P(Z_1|a_1,\bar{W}_2,\bar{y}_1)\odot P(a_2|\bar{W}_2,a_1,\bar{y}_1)^{-1}=q_{22}P(\bar{Z}_2|\bar{a}_2,\bar{W}_2,\bar{y}_1) \tag{61}
$$

where we regard the vector $q_{22}$ (of dimension $1\times(|Z_1|\times|Z_2|)$) as holding the values of the function $q_{22}(\bar{y}_2,\bar{z}_2,\bar{a}_2)$ for all of the different values $\bar{z}_2$ can take. We also know that since $P(\bar{U}_1|\bar{W}_2,\bar{a}_2,\bar{y}_1)$ has a right inverse, we can get

$$
\begin{gathered}
P(\bar{Z}_2|\bar{a}_2,\bar{W}_2,\bar{y}_1)P(\bar{U}_1|\bar{a}_2,\bar{W}_2,\bar{y}_1)^{-}=P(\bar{Z}_2|\bar{a}_2,\bar{U}_1,\bar{y}_1)\\
P(Z_1|a_1,\bar{W}_2,\bar{y}_1)\odot P(a_2|\bar{W}_2,a_1,\bar{y}_1)^{-1}P(\bar{U}_1|\bar{a}_2,\bar{W}_2,\bar{y}_1)^{-}=P(Z_1|a_1,\bar{U}_1,\bar{y}_1)\odot P(a_2|\bar{U}_1,a_1,\bar{y}_1)^{-1}.
\end{gathered}
$$

Thus, we get that

$$
\begin{aligned}
&q_{11}P(Z_1|a_1,\bar{U}_1,\bar{y}_1)\odot P(a_2|\bar{U}_1,a_1,\bar{y}_1)^{-1}\\
&=q_{11}P(Z_1|a_1,\bar{W}_2,\bar{y}_1)\odot P(a_2|\bar{W}_2,a_1,\bar{y}_1)^{-1}P(\bar{U}_1|\bar{a}_2,\bar{W}_2,\bar{y}_1)^{-}\\
&=q_{22}P(\bar{Z}_2|\bar{a}_2,\bar{W}_2,\bar{y}_1)P(\bar{U}_1|\bar{a}_2,\bar{W}_2,\bar{y}_1)^{-}\\
&=q_{22}P(\bar{Z}_2|\bar{a}_2,\bar{U}_1,\bar{y}_1)
\end{aligned}
$$

Lastly, due to the fact that

$$
P(a_1|U_0,y_0)^{-1}=q_{11}P(Z_1|U_0,a_1,y_0),
$$

by Assumption 1.3 $\{Y_1 \perp\!\!\!\perp Z_1|U_0,Y_0,A_1\}$, we obtain

$$
\begin{aligned}
P(y_1|U_0,y_0,a_1)&=q_{11}P(Z_1|U_0,a_1,y_0)\odot P(a_1|U_0,y_0)\odot P(y_1|U_0,y_0,a_1)\\
&=q_{11}P(y_1,a_1,Z_1|U_0,y_0)
\end{aligned}
$$

These results above show that in the categorical confounding scenario, there exist $q_{11}$ and $q_{22}$ that solve Equations (30) and (31), and thus can be used to identify the joint density of the potential outcomes. In the special case implied by Assumption 3 where the cardinalities of the $W, U$, and $Z$ variables are the same, the Moore-Penrose pseudo-inverses are ordinary inverses. Then

$$q_{11} = P(a_1|W_1, y_0)^{-1} P(Z_1|W_1, a_1, y_0)^{-1}$$

$$q_{22} = P(a_1|W_1, y_0)^{-1} P(Z_1|W_1, a_1, y_0)^{-1} P(Z_1|a_1, \bar{W}_2, \bar{y}_1) \odot P(a_2|\bar{W}_2, a_1, \bar{y}_1)^{-1} P(\bar{Z}_2|\bar{a}_2, \bar{W}_2, \bar{y}_1)^{-1}$$

Putting everything together and applying Theorem 3.1.2, we get that

$$\begin{aligned}
&P(Y_2(a_1, a_2) = y_2, Y_1(a_1) = y_1|y_0) \\
&= P(a_1|W_1, y_0)^{-1} P(Z_1|W_1, a_1, y_0)^{-1} P(Z_1|a_1, \bar{W}_2, \bar{y}_1) \odot P(a_2|\bar{W}_2, a_1, \bar{y}_1)^{-1} \\
&\quad P(\bar{Z}_2|\bar{a}_2, \bar{W}_2, \bar{y}_1)^{-1} P(\bar{y}_2, \bar{a}_2, \bar{Z}_2|y_0),
\end{aligned}$$

and

$$\begin{aligned}
&P(Y_1(a_1) = y_1|y_0) \\
&= P(a_1|W_1, y_0)^{-1} P(Z_1|W_1, a_1, y_0)^{-1} P(y_1, a_1, Z_1|y_0)
\end{aligned}$$

Given the preceding developments, simple plug-in estimators are immediate. We can replace the unknown probabilities with empirical estimates. Then, following the above derivation process, we obtain plug-in estimators $\hat{f}(Y_2(a_1, a_2) = y_2, Y_1(a_1) = y_1|y_0)$ and $\hat{f}(Y_1(a_1) = y_1|y_0)$. Using these plug-in estimators, we can find the optimal DTRs by value maximization:

$$\begin{aligned}
&\hat{\bar{d}}_2^{PIPW} \\
&= \underset{\bar{d}_2}{\operatorname{argmax}} \sum_{\bar{a}_2} \sum_{\bar{y}_2} y_2 (\hat{P}(a_1|W_1, y_0)^{-1} \hat{P}(Z_1|W_1, a_1, y_0)^{-1} \hat{P}(Z_1|a_1, \bar{W}_2, \bar{y}_1) \odot \hat{P}(a_2|\bar{W}_2, a_1, \bar{y}_1)^{-1}) \\
&\quad \hat{P}(\bar{Z}_2|\bar{W}_2, \bar{a}_2, \bar{y}_1)^{-1} \hat{P}(\bar{y}_2, \bar{a}_2, \bar{Z}_2) I(d_2(\bar{y}_1, a_1) = a_2) I(d_1(y_0) = a_1)
\end{aligned}$$

We can also get estimators of the optimal DTRs by Q-learning:

$$
\begin{aligned}
&\hat{d}_2^{PIPW}(\bar{y}_1, a_1) \\
&= \operatorname*{argmax}_{a_2} \sum_{y_2} \frac{(\hat{P}(a_1|W_1,y_0)^{-1}\hat{P}(Z_1|W_1,a_1,y_0)^{-1}\hat{P}(Z_1|a_1,\bar{W}_2,\bar{y}_1)\odot\hat{P}(a_2|\bar{W}_2,a_1,\bar{y}_1)^{-1})\hat{P}(\bar{Z}_2|\bar{W}_2,\bar{a}_2,\bar{y}_1)^{-1}\hat{P}(\bar{y}_2,\bar{a}_2,\bar{Z}_2|y_0)}{\hat{P}(a_1|W_1,y_0)^{-1}\hat{P}(Z_1|W_1,a_1,y_0)^{-1}\hat{P}(\bar{y}_1,a_1,Z_1|y_0)} y_2 \\
&\hat{Q}_2^{PIPW}(\bar{y}_1, a_1; a_2) \\
&= \sum_{y_2} \frac{(\hat{P}(a_1|W_1,y_0)^{-1}\hat{P}(Z_1|W_1,a_1,y_0)^{-1}\hat{P}(Z_1|a_1,\bar{W}_2,\bar{y}_1)\odot\hat{P}(a_2|\bar{W}_2,a_1,\bar{y}_1)^{-1})\hat{P}(\bar{Z}_2|\bar{W}_2,\bar{a}_2,\bar{y}_1)^{-1}\hat{P}(\bar{y}_2,\bar{a}_2,\bar{Z}_2|y_0)}{\hat{P}(a_1|W_1,y_0)^{-1}\hat{P}(Z_1|W_1,a_1,y_0)^{-1}\hat{P}(\bar{y}_1,a_1,Z_1|y_0)} y_2 \\
&\hat{d}_1^{PIPW}(y_0) \\
&= \operatorname*{argmax}_{a_1} \sum_{y_1} \hat{P}(a_1|W_1,y_0)^{-1}\hat{P}(Z_1|W_1,a_1,y_0)^{-1}\hat{P}(\bar{y}_1,a_1,Z_1|y_0) \max_{a_2} \hat{Q}_2^{PIPW}(\bar{y}_1, a_1; a_2).
\end{aligned}
$$

With categorical variables, all (conditional) probabilities can be estimated from the observed cell counts.

By the PHA Method, we can also get a result about the identification of $f(Y_2(a_1, a_2) = y_2, Y_1(a_1) = y_1|y_0)$ and $f(Y_1(a_1) = y_1|y_0)$ in the categorical setting given the already obtained results of $h_{22}$ and $q_{11}$. Here, we list the closed-form expression of $h_{22}$, $h_{21}$ and $h_{11}$ under Assumption 4 (Inverbility) of Zhang & Tchetgen Tchetgen (2024), which states that for all $(\bar{y}_1, \bar{a}_2)$, the matrices $P(\bar{W}_2|\bar{U}_1, \bar{y}_1, \bar{a}_2)$ and $P(\bar{U}_1|\bar{Z}_2, \bar{y}_1, \bar{a}_2)$ are invertible. In addition, for all $(y_0, a_1)$, $P(W_1|U_0, y_0, a_1)$ and $P(U_1|Z_1, y_0, a_1)$ are invertible.

$$
\begin{aligned}
h_{22} &= P(y_2|\bar{Z}_2, \bar{y}_1, \bar{a}_2)P(\bar{W}_2|\bar{Z}_2, \bar{y}_1, \bar{a}_2)^{-1} \\
h_{21} &= P(y_2|\bar{Z}_2, \bar{y}_1, \bar{a}_2)P(\bar{W}_2|\bar{Z}_2, \bar{y}_1, \bar{a}_2)^{-1}P(\bar{W}_2, y_1|Z_1, y_0, a_1)P(W_1|Z_1.y_0, a_1)^{-1} \\
h_{11} &= P(y_1|Z_1, y_0, a_1)P(W_1|Z_1, y_0, a_1)^{-1}.
\end{aligned}
$$

By the closed-form expressions of $q_{11}$ and $h_{22}$, we have:

**Proposition K.2.** *Let the cardinality of each of the $U$, $Z$, and $W$ variables equal $k$ for $2 \leq k < \infty$.*

*Also, suppose that Assumptions 1.3 and 3 hold. Then*

$$
\begin{aligned}
& f(Y_2(a_1,a_2)=y_2, Y_1(a_1)=y_1|y_0) \\
&= P(y_2|\bar{Z}_2,\bar{y}_1,\bar{a}_2)P(\bar{W}_2|\bar{Z}_2,\bar{y}_1,\bar{a}_2)^{-1}P(\bar{W}_2,y_1,a_1|Z_1,y_0)\odot P(Z_1|y_0)^T \\
&\quad (P(a_1|W_1,y_0)^{-1}P(Z_1|W_1,a_1,y_0)^{-1})^T \\
& f(Y_1(a_1)=y_1|y_0) \\
&= P(a_1|W_1,y_0)^{-1}P(Z_1|W_1,a_1,y_0)^{-1}P(\bar{y}_1,a_1,Z_1|y_0)
\end{aligned}
$$

Similarly, using the plug-in estimators, we can find the optimal DTRs by value maximization:

$$
\begin{aligned}
& \hat{\bar{d}}_2^{PHA} \\
&= \operatorname*{argmax}_{\bar{d}_2} \sum_{\bar{a}_2}\sum_{\bar{y}_2} y_2 P(y_2|\bar{Z}_2,\bar{y}_1,\bar{a}_2)P(\bar{W}_2|\bar{Z}_2,\bar{y}_1,\bar{a}_2)^{-1}P(\bar{W}_2,y_1,a_1|Z_1,y_0)\odot P(Z_1,y_0)^T \\
&\quad (P(a_1|W_1,y_0)^{-1}P(Z_1|W_1,a_1,y_0)^{-1})^T I(d_2(\bar{y}_1,a_1)=a_2)I(d_1(y_0)=a_1)
\end{aligned}
$$

or Q-learning:

$$
\begin{aligned}
& \hat{d}_2^{PHA}(\bar{y}_1,a_1) \\
&= \operatorname*{argmax}_{a_2}\sum_{y_2} \frac{\hat{P}(y_2|\bar{Z}_2,\bar{y}_1,\bar{a}_2)\hat{P}(\bar{W}_2|\bar{Z}_2,\bar{y}_1,\bar{a}_2)^{-1}\hat{P}(\bar{W}_2,y_1,a_1|Z_1,y_0)\odot\hat{P}(Z_1|y_0)^T(\hat{P}(a_1|W_1,y_0)^{-1}\hat{P}(Z_1|W_1,a_1,y_0)^{-1})^T}{\hat{P}(a_1|W_1,y_0)^{-1}\hat{P}(Z_1|W_1,a_1,y_0)^{-1}\hat{P}(\bar{y}_1,a_1,Z_1|y_0)} y_2 \\
& \hat{Q}_2^{PHA}(\bar{y}_1,a_1;a_2) \\
&= \sum_{y_2} \frac{\hat{P}(y_2|\bar{Z}_2,\bar{y}_1,\bar{a}_2)\hat{P}(\bar{W}_2|\bar{Z}_2,\bar{y}_1,\bar{a}_2)^{-1}\hat{P}(\bar{W}_2,y_1,a_1|Z_1,y_0)\odot\hat{P}(Z_1|y_0)^T(\hat{P}(a_1|W_1,y_0)^{-1}\hat{P}(Z_1|W_1,a_1,y_0)^{-1})^T}{\hat{P}(a_1|W_1,y_0)^{-1}\hat{P}(Z_1|W_1,a_1,y_0)^{-1}\hat{P}(\bar{y}_1,a_1,Z_1|y_0)} y_2 \\
& \hat{d}_1^{PHA}(y_0) \\
&= \operatorname*{argmax}_{a_1}\sum_{y_1} \hat{P}(a_1|W_1,y_0)^{-1}\hat{P}(Z_1|W_1,a_1,y_0)^{-1}\hat{P}(\bar{y}_1,a_1,Z_1|y_0)\max_{a_2}\hat{Q}_2^{PHA}(\bar{y}_1,a_1;a_2)
\end{aligned}
$$

For the multiply robust estimator, by Theorem 4.2 and Corollary A.3, we can similarly obtain the plug-in estimators for the value function and the Q functions respectively follow the same steps above, as we have already derived the clsed-form expressions for the outcome and treatment confounding bridge functions $h_{22}, h_{21}, q_{22}, q_{11}$. For the construction of Oracle estimators and SRA estimators, we follow the steps presented in Appendix E of Zhang & Tchetgen Tchetgen (2024): For the oracle estimator, we assume that we have access to the unmeasured confounders. By the value maximization method, we obtain:

$$\begin{aligned}\hat{\bar{d}}_2^{Oracle} &= \arg\max_{\bar{d}_2} \sum_{\bar{a}_2}\sum_{\bar{y}_2} \hat{P}\left(Y_2\left(a_1,a_2\right)=y_2, Y_1\left(a_1\right)=y_1, Y_0=y_0\right) I(\bar{d}_2(\bar{y}_1,a_1)=\bar{a}_2)\\ &= \arg\max_{\bar{d}_2} \sum_{\bar{a}_2}\sum_{\bar{y}_2}\sum_{\bar{u}_1} \hat{P}\left(Y_2=y_2|\bar{y}_1,\bar{u}_1,\bar{a}_2\right)\hat{P}\left(u_1|\bar{y}_1,u_0,a_1\right)\hat{P}(y_1|y_0,u_0,a_1)P(u_0,y_0)\\ &\qquad I(\bar{d}_2(\bar{y}_1,a_1)=\bar{a}_2)\end{aligned}$$

By the Q-learning method, these follow directly from the g-formulas:

$$\begin{aligned}\hat{d}_2^{Oracle}\left(\bar{y}_1,a_1\right) &= \arg\max_{a_2}\hat{\mathbb{E}}\left\{Y_2\left(a_1,a_2\right)|\overline{Y}_1\left(a_1\right)=\bar{y}_1\right\}\\ &= \arg\max_{a_2}\sum_{y_2}\frac{\hat{P}\left(Y_2\left(a_1,a_2\right)=y_2, Y_1\left(a_1\right)=y_1|y_0\right)}{\hat{P}\left(Y_1\left(a_1\right)=y_1|y_0\right)}y_2\\ &= \arg\max_{a_2}\sum_{y_2}\frac{\sum_{\bar{u}_1}\hat{P}\left(Y_2=y_2|\bar{y}_1,\bar{u}_1,\bar{a}_2\right)\hat{P}\left(U_1=u_1|\bar{y}_1,u_0,a_1\right)\hat{P}(y_1|y_0,u_0,a_1)\hat{P}(u_0|y_0)}{\sum_{u_0}\hat{P}\left(Y_1=y_1|a_1,y_0,u_0\right)\hat{P}\left(u_0|y_0\right)}y_2\end{aligned}$$

$$\hat{Q}_2^{Oracle}\left(a_1,\bar{y}_1;a_2\right)=\sum_{y_2}\frac{\sum_{\bar{u}_1}\hat{P}\left(Y_2=y_2|\bar{y}_1,\bar{u}_1,\bar{a}_2\right)\hat{P}\left(U_1=u_1|\bar{y}_1,u_0,a_1\right)\hat{P}(y_1|y_0,u_0,a_1)\hat{P}(u_0|y_0)}{\sum_{u_0}\hat{P}\left(Y_1=y_1|a_1,y_0,u_0\right)\hat{P}\left(u_0|y_0\right)}y_2$$

$$\begin{aligned}\hat{d}_1^{Oracle}(y_0) &= \arg\max_{a_1}\mathbb{E}\left\{\max_{a_2}\hat{Q}_2^{Oracle}(a_1,Y_1(a_1),y_0;a_2)|Y_0=y_0\right\}\\ &= \arg\max_{a_1}\sum_{y_1}\hat{P}\left(Y_1\left(a_1\right)=y_1|y_0\right)\max_{a_2}\hat{Q}_2^{Oracle}\left(a_1,\bar{y}_1;a_2\right)\\ &= \arg\max\sum_{y_1}\sum_{u_0}\hat{P}\left(Y_1\left(a_1\right)=y_1|y_0,u_0\right)\hat{P}\left(u_0|y_0\right)\max_{a_2}\hat{Q}_2^{Oracle}\left(a_1,\bar{y}_1;a_2\right)\\ &= \arg\max_{a_1}\sum_{u_1}\hat{P}\left(Y_1=y_1|a_1,y_0,u_0\right)\hat{P}\left(u_0|y_0\right)\max_{a_2}\hat{Q}_2^{Oracle}\left(a_1,\bar{y}_1;a_2\right)\end{aligned}$$

For the SRA estimator, we assume that there are no unmeasured confounders. By the value maximization method, we obtain:

$$\begin{aligned}\hat{\bar{d}}_2^{SRA} &= \arg\max_{\bar{d}_2}\sum_{\bar{a}_2}\sum_{\bar{y}_2}\hat{P}\left(Y_2\left(a_1,a_2\right)=y_2, Y_1\left(a_1\right)=y_1, Y_0=y_0\right) I(\bar{d}_2(\bar{y}_1,a_1)=\bar{a}_2)\\ &= \arg\max_{\bar{d}_2}\sum_{\bar{a}_2}\sum_{\bar{y}_2}\hat{P}\left(Y_2=y_2|\bar{y}_1,\bar{a}_2\right)\hat{P}(y_1|y_0,a_1)I(\bar{d}_2(\bar{y}_1,a_1)=\bar{a}_2)\end{aligned}$$

By the Q-learning method, these follow directly from the g-formulas:

$$\begin{aligned}\hat{d}_2^{SRA}\left(\overline{y}_1, a_1\right) &= \arg\max_{a_2} \hat{\mathbb{E}}\left\{Y_2|\overline{y}_1, \overline{a}_2\right\} \\ &= \arg\max_{a_2} \sum_{y_2} \hat{P}\left(Y_2 = y_2|\overline{y}_1, \overline{a}_2\right) y_2\end{aligned}$$

$$Q_2^{SRA}\left(a_1, \overline{y}_1; a_2\right) = \sum_{y_2} \hat{P}\left(Y_2 = y_2|\overline{y}_1, \overline{a}_2\right) y_2$$

$$\hat{d}_1^{SRA}(y_0) = \arg\max_{a_1} \sum_{y_1} \hat{P}\left(Y_1 = y_1|a_1, y_0\right) \max_{a_2} \hat{Q}_2^{SRA}\left(a_1, \bar{y}_1; a_2\right)$$

## K.2 Corss-Fitting Algorithm

Sample splitting is crucial to mitigate the bias caused by overfitting, thereby ensuring the $\sqrt{n}$-consistency and asymptotic normality of $\hat{V}_{PMR}(\hat{\theta})$, which is the foundation for consistent estimation of $\theta^*$. However, direct sample splitting will reduce the efficiency of the final estimator, as each subset only partially utilizes the information contained in the original sample. To preserve the efficiency of the final estimator, we have to achieve this by cross-fitting (Schick 1986, Chernozhukov et al. 2018, Wager 2024, Bai et al. 2025). To clarify, suppose that we have randomly split the samples into $L$ disjoint subsets of equal size: $\{I_1, I_2, \ldots, I_L\}$. For each fold $l \in \{1, 2, \ldots, L\}$, the nuisance functions are estimated using data excluding $I_l$, and the corresponding estimator contributed by $I_l$ is constructed using these nuisance estimates and the observations in $I_l$. The final estimator for $V(\theta)$ is then calculated by averaging the $L$ fold-specific estimators, that is:

$$\hat{V}_{PMR}^{cf}(\theta) = \frac{1}{L}\sum_{l=1}^{L} \hat{V}_{PMR,l}(\theta; \hat{\hat{\mathcal{H}}}_{KK}^{(-l)}, \hat{\hat{\mathcal{Q}}}_{KK}^{(-l)}),$$

where $\hat{V}_{PMR,l}(\theta; \hat{\hat{\mathcal{H}}}_{KK}^{(-l)}, \hat{\hat{\mathcal{Q}}}_{KK}^{(-l)})$ is the estimator contributed by the $l$-fold sample, and $(\hat{\hat{\mathcal{H}}}_{KK}^{(-l)}, \hat{\hat{\mathcal{Q}}}_{KK}^{(-l)})$ is confounding bridge functions estimated on the data excluding $I_l$. With a slight notation abuse, we still use $\hat{V}_{PMR}(\theta)$ to denote the multiply robust estimator constructed by cross-fitting in the main text for simplicity.

## K.3 Estimation of the Optimal Linear DTRs

We estimate the optimal linear DTRs using a mixed-integer program formulation, according to Zhang and Imai Zhang & Imai (2023) and Kitagawa and Tetenov Kitagawa & Tetenov (2018). Consider the linear policy rules of the following forms:

$$\mathcal{D}_1 = \{d_1 : \mathcal{Y}_0 \to \{0,1\} | \pi_1(Y_0) = I(Y_0^T\theta_1 \geq 0), \theta_1 \in \Theta_1\},$$
$$\mathcal{D}_2 = \{d_2 : \mathcal{Y}_0 \times \mathcal{Y}_1 \times \mathcal{A}_1 \to \{0,1\} | \pi_2(\bar{Y}_1, A_1) = I((Y_0, Y_1, A_1)^T\theta_2 \geq 0), \theta_2 \in \Theta_2\},$$

where $\Pi_1$ and $\Pi_2$ are the candidate linear decision function classes for $d_1$ and $d_2$ with $\|\theta_1\|_{L_2} = \|\theta_2\|_{L_2} = 1$, respectively. We introduce binary variables $p_{1i}$ and $p_{2i}$ and write

$$\frac{Y_{0i}^T\theta_1}{M_i} < p_{1i} \leq 1 + \frac{Y_{0i}^T\theta_1}{M_i}$$
$$\frac{(\bar{Y}_{1i}, \bar{A}_{1i})^T\theta_2}{M_i} < p_{2i} \leq 1 + \frac{(\bar{Y}_{1i}, \bar{A}_{1i})^T\theta_2}{M_i},$$

where

$$M_i > \max\{\sup_{\theta_1 \in \Theta_1} |Y_{0i}^T\theta_1|, \sup_{\theta_2 \in \Theta_2} |(\bar{Y}_{1i}, A_{1i})^T(\theta_2)|\},\ p_{1i}, p_{2i} \in \{0,1\}.$$

Then, $p_{1i}$ and $p_{2i}$ will equal to 1 if $Y_{0i}^T\theta_1$ and $(\bar{Y}_{1i}, \bar{A}_{1i})^T\theta_2$ are non-negative, respectively, and 0 otherwise. That is, $p_{1i} = d_1(Y_{0i})$ and $p_{2i} = d_2(\bar{Y}_{1i}, A_{1i})$. We can now write the objective function (up to some constants) for the POR, PHA, and PIPW methods as,

$$\hat{V}(\theta)_{POR} = \sum_{\bar{a}_2} \mathbb{P}_n\left\{\sum_{y_2, y_1}\sum_{w_1} h_{21}(y_2, y_1, Y_0, W_1, \bar{a}_2) I(p_1 = a_1) I(p_2(y_1) = a_2)\right\},$$
$$\hat{V}(\theta)_{PHA} = \sum_{a_2} \mathbb{P}_n\left\{\sum_{y_2} y_2 h_{22}(y_2, a_2, A_1, \bar{Y}_1, \bar{W}_2) q_{11}(Y_0, Z_1, A_1) I(p_1 = A_1) I(p_2 = a_2)\right\},$$
$$\hat{V}(\theta)_{PIPW} = \mathbb{P}_n\left\{Y_2 q_{22}(\bar{A}_2, \bar{Y}_1, \bar{Z}_2) I(p_1 = A_1) I(p_2 = A_2)\right\}$$

and for the PMR method,

$$
\begin{aligned}
&\hat{V}(\theta)_{PMR} \\
&= \sum_{\bar{a}_2} \mathbb{P}_n \Bigg[ I(p_1 = A_1 = a_1) I(p_2 = A_2 = a_2) q_{22}(\bar{A}_2, \bar{Y}_1, \bar{Z}_2) \\
&\quad\quad \times (Y_2 - \sum_{y_2} y_2 h_{22}(y_2, a_2, A_1, \bar{Y}_1, \bar{W}_2)) \\
&\quad + I(p_1 = A_1 = a_1) q_{11}(Y_0, Z_1, A_1) \Big( \sum_{y_2} y_2 h_{22}(y_2, a_2, A_1, \bar{Y}_1, \bar{W}_2) I(p_2 = a_2) \\
&\quad\quad - \sum_{y_1, y_2} y_2 h_{21}(\bar{y}_2, \bar{a}_2, Y_0, W_1) I(p_2(y_1) = a_2) \Big) \\
&\quad + \sum_{y_1, y_2} y_2 h_{21}(\bar{y}_2, \bar{a}_2, Y_0, W_1) I(p_1 = a_1) I(p_2(y_1) = a_2) \Bigg]
\end{aligned}
$$

Here, we use $p_2(y_1)$ to emphasize that $p_2$ will be involved in the summation for $y_1$ in the estimation of the value function by the POR and PMR methods. For Oracle and SRA estimators, similar objective functions can be immediately written following the same methods above, and thus are omitted.

These imply that estimation of optimal DTRs via value maximization can be equivalently represented as the following mixed-integer linear program (MILP), which can be solved using an off-the-shelf algorithm by Gurobi:

$$
\max_{\theta \in \Theta, \{p_i\} \mathbb{R}} \hat{V}(\theta)_m.
$$

$$
s.t. \begin{cases} \frac{Y_{0i}^T \theta_1}{M_i} < p_{1i} \leq 1 + \frac{Y_{0i}^T \theta_1}{M_i}; \\ \frac{(\bar{Y}_{1i}, \bar{A}_{1i})^T \theta_2}{M_i} < p_{2i} \leq 1 + \frac{(\bar{Y}_{1i}, \bar{A}_{1i})^T \theta_2}{M_i}; \quad , \text{for } i = 1, 2, \ldots, n, \\ p_{1i}, p_{2i} \in \{0, 1\}. \end{cases}
$$

where constants $M_i$ should satisfy:

$$
M_i > \max\{ \sup_{\theta_1 \in \Theta_1} |Y_{0i}^T \theta_1|, \sup_{\theta_2 \in \Theta_2} |(\bar{Y}_{1i}, A_{1i})^T (\theta_2)| \}.
$$

While this MILP formulation enables exact optimization for many policy classes, solving large-scale MILP problems can be computationally demanding. However, in our binary categorical setting, the number of constraints above is essentially reduced to 10 for finite combinations of sample items. The complete optimization program can be found on GitHub.